\documentclass[a4paper,11pt]{article}
\pdfoutput=1

\usepackage{jheppub_mod} 
\usepackage[T1]{fontenc} 

\usepackage{hyperref}
\usepackage{booktabs,tabulary}
\usepackage{caption}
\usepackage{subcaption}

\usepackage{natbib}

\usepackage{amssymb,amsbsy,amsmath,amsfonts,dsfont}
\usepackage{graphicx}
\usepackage{epsf,epsfig,float,latexsym,amsthm,fancyhdr,rotating}
\usepackage{graphics,psfrag,longtable}
\usepackage{slashed}
\usepackage{scrextend}

\usepackage{enumerate}
\usepackage{array}

\DeclareMathAlphabet{\mathpzc}{OT1}{pzc}{m}{it}

\usepackage{diagbox}

\usepackage{mathrsfs} 

\usepackage{nicefrac}

\usepackage{upgreek}
\usepackage{multirow}

\makeatletter
\newcommand{\Cr}[2]{{\mathcal{C}_{#1}\big[#2\big]}}
\makeatother

\def\beq{\begin{equation}}
\def\eeq{\end{equation}}
\def\beqa{\begin{equation}\begin{array}{l}}
\def\eeqa{\end{array}\end{equation}}
\def\label#1{\label{eq:#1}}
\def\figlab#1{\label{fig:#1}}

\def\seclab#1{\label{sec:#1}}
\def\Eqref#1{(\ref{eq:#1})}

\def\Fref.~#1{figure~\ref{fig:#1}}

\def\secref#1{section~\ref{sec:#1}}

\def\barr{\left(\begin{array}{c}}
\def\earr{\end{array}\right)}
\def\bmat{\left(\begin{array}{cc}}
\def\emat{\end{array}\right)}
\def\al{\alpha}
\def\be{\beta}
\def\ga{\gamma} 
 
  \def\eps{\epsilon}

\def\si{\sigma}

\def\pa{\partial}

\def\pa{\partial}

\def\nn{\nonumber}

\def\3d{3-D}

\newcommand{\Tr}{\text{Tr}}
\newcommand{\unity}{\mathds{1}}
\newcommand{\GeV}{\,\text{GeV}}
\newcommand{\MeV}{\,\text{MeV}}
\newcommand{\keV}{\,\text{keV}}
\newcommand{\eV}{\,\text{eV}}
\newcommand{\BR}{\text{BR}}
\newcommand{\Cl}{\text{Cl}_2}

\newcommand{\Order}{\mathcal{O}}

\newcommand{\Q}{\mathcal{Q}}
\newcommand{\M}{\mathcal{M}}

\usepackage{mathtools}

\newcommand*{\bfrac}[2]{\genfrac{}{}{0pt}{}{#1}{#2}}

\DeclarePairedDelimiter\floor{\lfloor}{\rfloor}

\allowdisplaybreaks[1]

\preprint{INT-PUB-19-051}

\title{Longitudinal short-distance constraints for the hadronic light-by-light contribution to $\boldsymbol{(g-2)_\mu}$ with large-$\boldsymbol{N_c}$ Regge models}

\author[a]{Gilberto Colangelo,}
\author[a]{Franziska Hagelstein,}
\author[b,a]{Martin Hoferichter,}
\author[a]{Laetitia Laub,}
\author[c]{and Peter Stoffer}

\affiliation[a]{
Albert Einstein Center for Fundamental Physics, Institute for Theoretical Physics, University of Bern, Sidlerstrasse 5, CH--3012 Bern, Switzerland}

\affiliation[b]{Institute for Nuclear Theory, University of Washington, Seattle, WA 98195-1550, USA}

\affiliation[c]{Department of Physics, University of California at San Diego, La Jolla, CA 92093, USA}

\emailAdd{gilberto@itp.unibe.ch}
\emailAdd{hagelstein@itp.unibe.ch}
\emailAdd{mhofer@uw.edu}
\emailAdd{laub@itp.unibe.ch}
\emailAdd{pstoffer@ucsd.edu}

\abstract{While the low-energy part of the hadronic light-by-light (HLbL) tensor can be constrained from data using dispersion relations, 
  for a full evaluation of its contribution to the anomalous magnetic moment of the muon $(g-2)_\mu$ also mixed- and high-energy regions 
  need to be estimated. Both can be addressed within the operator product expansion (OPE), either for configurations where all photon virtualities become large or one of them remains finite. Imposing such short-distance constraints (SDCs) on the HLbL tensor is thus a major aspect
  of a model-independent approach towards HLbL scattering. Here, we focus on longitudinal SDCs, which concern the amplitudes containing the pseudoscalar-pole contributions from $\pi^0$, $\eta$, $\eta'$. Since these conditions cannot be fulfilled by a finite number of pseudoscalar poles, we
  consider a tower of excited pseudoscalars, constraining their masses and transition form factors from Regge theory, the OPE, 
  and phenomenology. Implementing a matching of the resulting expressions for the HLbL tensor onto the perturbative QCD quark loop, we are able to further constrain our calculation and significantly reduce its model dependence.
  We find that especially for the $\pi^0$ the corresponding increase of the HLbL contribution is much smaller than 
  previous prescriptions in the literature would imply. Overall, we estimate
  that longitudinal SDCs increase the HLbL contribution by $\Delta a_\mu^\text{LSDC}=13(6)\times 10^{-11}$. This number does not include the contribution from the charm quark, for which we find
  $a_\mu^{c\text{-quark}}=3(1)\times 10^{-11}$.
}

\begin{document}

\maketitle

\begin{figure}[h]
\center
\includegraphics[width=0.4\textwidth]{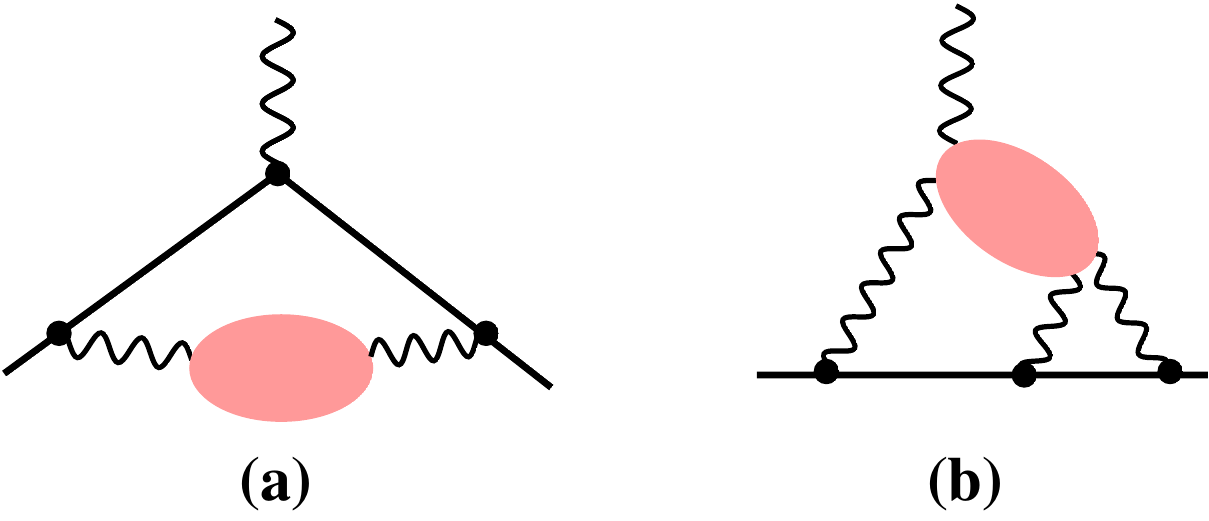}
\caption{Hadronic contributions to $(g-2)_\mu$: (a) HVP, (b) HLbL. The pink blobs symbolize hadronic intermediate states.}
\label{fig:Hamm}
\end{figure}

\section{Introduction}
\label{sec:intro}

Current Standard Model (SM) evaluations of the anomalous magnetic moment of the muon, $a_\mu=(g-2)_\mu/2$, differ from the value measured at the Brookhaven National Laboratory~\cite{Bennett:2006fi}
\beq
a_\mu^\text{exp}=116\ 592\ 089(63)\times 10^{-11},
\eeq
by around $3.5\,\sigma$.
In the near future, the new Fermilab E989 experiment~\cite{Grange:2015fou} will be able to reduce the experimental uncertainty by a factor $4$, and the E34 experiment at J-PARC~\cite{Abe:2019thb} will provide an important cross check, see~ref.~\cite{Gorringe:2015cma} for a comparison of the experimental methods. 
Therefore, the theoretical calculation of $a_\mu$ needs to be improved accordingly.
 
The uncertainty of the SM prediction mainly stems from
hadronic contributions, such as hadronic vacuum polarization (HVP), see figure~\ref{fig:Hamm} (a), and HLbL scattering, see figure~\ref{fig:Hamm} (b).
Since the HVP contribution can be systematically calculated with a data-driven dispersive approach~\cite{Keshavarzi:2018mgv,Jegerlehner:2018gjd,Colangelo:2018mtw,Hoferichter:2019gzf,Davier:2019can}, 
lattice QCD~\cite{Borsanyi:2017zdw,Blum:2018mom,Giusti:2018mdh,Shintani:2019wai,Davies:2019efs,Gerardin:2019rua,Aubin:2019usy},
and potentially be accessed independently by the proposed MUonE experiment~\cite{Calame:2015fva,Abbiendi:2016xup}, which aims to measure the space-like fine-structure constant $\alpha(t)$ in elastic electron--muon scattering,
the HLbL contribution may end up dominating the 
theoretical error.\footnote{Note that higher-order insertions of HVP~\cite{Calmet:1976kd,Keshavarzi:2018mgv,Kurz:2014wya} and HLbL~\cite{Colangelo:2014qya} are already under sufficient control, as are hadronic corrections in the anomalous magnetic moment of the electron, where recently a $2.5\,\sigma$ tension between the direct measurement~\cite{Hanneke:2008tm}
and the SM prediction~\cite{Aoyama:2017uqe} using the fine-structure constant from Cs interferometry~\cite{Parker:2018vye} emerged~\cite{Davoudiasl:2018fbb,Crivellin:2018qmi}.}

\begin{figure}[t]
\centering
\includegraphics[width=0.2\textwidth]{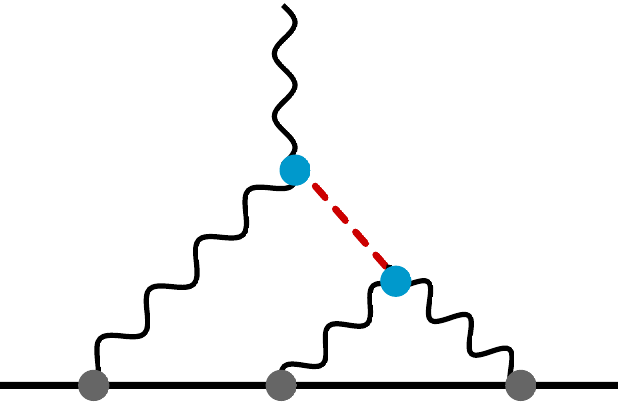}
\caption{Pseudoscalar-pole contribution to $(g-2)_\mu$. The cyan dots indicate the TFF of the pseudoscalar meson.}
\label{fig:HLbLpi0}
\end{figure}

Apart from lattice QCD~\cite{Blum:2016lnc,Blum:2017cer,Asmussen:2018oip}, recent data-driven approaches towards HLbL scattering are again 
rooted in dispersion theory, either for the HLbL tensor~\cite{Hoferichter:2013ama,Colangelo:2014dfa,Colangelo:2014pva,Colangelo:2015ama,Colangelo:2017qdm,Colangelo:2017fiz},
the Pauli form factor~\cite{Pauk:2014rfa}, or in terms of sum rules~\cite{Pascalutsa:2012pr,Green:2015sra,Danilkin:2016hnh,Hagelstein:2017obr,Hagelstein:2019tvp}.
In all approaches, the most important HLbL contributions are the $\pi^0$-pole and other pseudoscalar-meson-pole contributions, see figure~\ref{fig:HLbLpi0}. 
The strength of these pseudoscalar poles is determined by the transition form factors (TFFs), which in turn can be reconstructed from dispersion theory~\cite{Schneider:2012ez,Hoferichter:2012pm,Hoferichter:2014vra,Hoferichter:2017ftn,Hoferichter:2018dmo,Hoferichter:2018kwz}, leading to~\cite{Hoferichter:2018dmo,Hoferichter:2018kwz}
\beq
\label{pi0Hoferichter}
a_\mu^{\pi^0\text{-pole}}=62.6^{+3.0}_{-2.5}\times 10^{-11},
\eeq
in agreement with determinations from lattice QCD~\cite{Gerardin:2019vio}, Canterbury approximants (CA) \cite{Masjuan:2017tvw}, and Dyson--Schwinger equations (DSE)~\cite{Eichmann:2019tjk}.
Since the central value~\eqref{pi0Hoferichter} is close to earlier model-based calculations, e.g., within lowest-meson-dominance+vector (LMD+V) models~\cite{Knecht:2001qf}, 
the second-most important aspect of the dispersive approach apart from rigorous uncertainty estimates is the clear definition 
of the pseudoscalar intermediate states in terms of physical, on-shell form factors, in contrast to earlier 
notions of a pion-exchange contribution, see ref.~\cite{Jegerlehner:2009ry}, which involve the model-dependent concept of an off-shell pion.
This becomes particularly important when combined with other intermediate states, ensuring that the pseudoscalar poles 
are consistent with, for instance, the dispersive definition of two-pion intermediate states~\cite{Colangelo:2017qdm,Colangelo:2017fiz},
which in turn are determined by the corresponding on-shell quantities, in this case the helicity amplitudes
for $\gamma^*\gamma^*\to\pi\pi$~\cite{GarciaMartin:2010cw,Hoferichter:2011wk,Moussallam:2013una,Danilkin:2018qfn,Hoferichter:2019nlq,Danilkin:2019opj}. 

However, in contrast to HVP there is no closed formula that resums all possible intermediate states (in terms of the cross section for $e^+e^-\to\text{hadrons}$~\cite{Bouchiat:1961,Brodsky:1967sr}), in such a way that the consideration of exclusive channels will break down eventually, 
irrespective of the complications when extending the dispersive formalism to higher-multiplicity intermediate states. 
Therefore, to control the regions in the $(g-2)_\mu$ integral where either two or all three independent photon virtualities become large, additional constraints are required. In close analogy to HVP, 
where perturbative QCD (pQCD) becomes applicable in the high-energy tail of the dispersive integral, such constraints arise from the OPE and pQCD. 
In the regime where all three virtualities are large it was shown recently~\cite{Bijnens:2019ghy} that the pQCD quark loop indeed arises as the first term in a controlled OPE, with the next order 
suppressed by small quark masses and condensates. For the case in which one virtuality remains small, the leading OPE constraint was derived in~ref.~\cite{Melnikov:2003xd},
by reducing the HLbL tensor in this limit to the triangle anomaly and its known non-renormalization theorems~\cite{Vainshtein:2002nv,Knecht:2003xy,Knecht:2002hr,Czarnecki:2002nt}.\footnote{These non-renormalization theorems strictly apply only in the chiral limit and to the non-singlet components. Instances where additional corrections for the singlet component arise are pointed out throughout the discussion of the various SDCs in sections~\ref{sec:HLbL_tensor} and~\ref{sec:OPE}.}

The latter constraint decomposes into longitudinal and transversal contributions.
As noted in~ref.~\cite{Melnikov:2003xd}, the longitudinal part is intimately related to the pseudoscalar poles, but cannot be saturated by $\pi^0$, $\eta$, $\eta'$ alone,
nor by any finite number of poles. 
As a remedy it was suggested to drop the momentum dependence of the TFF at the vertex to which the external photon is attached, see figure~\ref{fig:HLbLpi0}, which leads to a substantial increase
of the pseudoscalar-pole HLbL contribution. Based on an LMD+V model for the $\pi^0$ and vector-meson-dominance (VMD) models for $\eta$, $\eta'$ from ref.~\cite{Knecht:2001qf}, 
this increase was found to be $13.5\times 10^{-11}$ for the $\pi^0$ and $5\times 10^{-11}$ each 
for $\eta$ and $\eta'$. This shift has been included, in one way or another, in subsequent estimates of the total HLbL value~\cite{Prades:2009tw,Jegerlehner:2009ry}.  
In fact, as we will show below, with modern input for the TFFs the corresponding increase would become even larger.
While there is no doubt that the SDC is important---it is, in fact, one of the few constraints on the mixed-energy regions in which one photon virtuality remains small---modifying the expression 
for the pseudoscalar poles in this way is not compatible with the dispersive description of the four-point HLbL tensor~\cite{Colangelo:2014dfa,Colangelo:2014pva,Colangelo:2015ama,Colangelo:2017qdm,Colangelo:2017fiz} and spoils consistency with other intermediate states in the same framework. 

In this work we address the question from a different standpoint: already in ref.~\cite{Melnikov:2003xd} it was observed that while a finite number of poles cannot saturate the SDC, 
an infinite tower of them potentially can---dropping the TFF at the external vertex has in fact been described in ref.~\cite{Melnikov:2003xd} as a model for the resummation of the tower of pseudoscalar states. Here we present explicit constructions that implement the Melnikov--Vainshtein (MV) constraint in terms of an infinite tower of
excited pseudoscalars, constraining their properties from Regge theory, all available SDCs, and phenomenology wherever possible~\cite{Colangelo:2019}. 
Given all these constraints the resulting models for the HLbL tensor prove remarkably rigid, without altering the low-energy properties. Since phenomenological information on excited pseudoscalars, especially their TFFs, is scarce, we do not attempt to construct TFF representations that apply for arbitrary kinematics, but concentrate on minimal models that cover the space-like region needed for $(g-2)_\mu$ and at the same time are able to fulfill all SDCs. Systematic uncertainties are estimated by comparing two such representations, either based on a truncated or untruncated Regge sum for the TFFs themselves, as well as the available phenomenological constraints, see appendices~\ref{app:photon_couplings} and~\ref{sec:systematics}. 
Moreover, our model is only needed for the low-energy part of the $(g-2)_\mu$ integral: above the energy where the matching occurs, we calculate the integral with the quark loop. 
This strategy also ensures that the estimate of the asymptotic region still applies in the chiral limit, in which the excited pseudoscalar states decouple, see section~\ref{sec:chiral_limit}, while at low energies phenomenological input is needed either way to account for the effect of quark-mass corrections. 
All this leads to a more reliable estimate for the impact of the OPE constraints on the total HLbL contribution. 
To this end, we first review the expression for the pQCD quark loop and the known OPE constraints on the HLbL tensor in sections~\ref{sec:HLbL_tensor} and~\ref{sec:OPE}, 
adapting the conventions to the language suitable for the decomposition of the HLbL tensor from refs.~\cite{Colangelo:2015ama,Colangelo:2017fiz}, in which the expressions for both the pseudoscalar poles 
and the pQCD quark loop become remarkably simple. Next, we present in section~\ref{sec:Regge} the explicit construction of large-$N_c$-inspired Regge models\footnote{For brevity we call our large-$N_c$-inspired Regge models simply large-$N_c$ Regge models.} implementing the OPE constraints
and derive the consequences for HLbL scattering and $(g-2)_\mu$. In section~\ref{sec:matching}, we match the resulting expressions for the HLbL tensor to the pQCD loop 
to obtain a first estimate of the scale where the description of the HLbL tensor in terms of hadronic intermediate states 
and its asymptotic properties should meet. 
A more detailed comparison to the results obtained with the MV model is provided in section~\ref{sec:MV},
before we summarize our main results and discuss future developments in section~\ref{sec:summary}. 
Technical details and alternative evaluations that are used to estimate the systematic uncertainty are collected in the appendices.

\section{The hadronic light-by-light tensor}
\label{sec:HLbL_tensor}

\subsection[Lorentz decomposition and $(g-2)_\mu$ integral]{Lorentz decomposition and $\boldsymbol{(g-2)_\mu}$ integral}
\label{sec:BTT}

Throughout, we follow the conventions for the decomposition of the HLbL tensor and its contribution to $(g-2)_\mu$ from ref.~\cite{Colangelo:2017fiz}. Starting point is 
the HLbL tensor defined as the four-point function 
\beq
	\label{eq:HLbLTensorDefinition}
	\Pi^{\mu\nu\lambda\sigma}(q_1,q_2,q_3) = -i \int d^4x \, d^4y \, d^4z \, e^{-i(q_1 \cdot x + q_2 \cdot y + q_3 \cdot z)} \langle 0 | T \{ j^\mu(x) j^\nu(y) j^\lambda(z) j^\sigma(0) \} | 0 \rangle
\eeq
of the electromagnetic current
\beq
	j^\mu = \bar \psi \mathcal{Q} \gamma^\mu \psi , \quad \psi = ( u , d, s )^T , \quad \mathcal{Q} = \mathrm{diag}\left(\frac{2}{3}, -\frac{1}{3}, -\frac{1}{3}\right),
\eeq
and momenta assigned as $q_1+q_2+q_3=q_4\to 0$. Its Lorentz decomposition in terms of scalar functions $\Pi_i$ is written following the Bardeen--Tung--Tarrach (BTT) prescription~\cite{Bardeen:1969aw,Tarrach:1975tu}  
\beq
\label{eqn:HLbLTensorKinematicFreeStructures}
\Pi^{\mu\nu\lambda\sigma} = \sum_{i=1}^{54} T_i^{\mu\nu\lambda\sigma} \Pi_i ,
\eeq
where the $\Pi_i$ are free of kinematic singularities and thus amenable to a dispersive treatment. However, this decomposition does not allow for a projection onto independent Lorentz structures,
given that there are only $41$ independent helicity amplitudes for fully off-shell photon--photon scattering. Moreover, two of these redundancies only occur in four space-time dimensions~\cite{Eichmann:2014ooa}. A given expression for the HLbL tensor is thus most conveniently projected onto 
a subset of $43$ Lorentz structures
\beq
\label{eqn:HLbLTensorBasisDecomposition}
	\Pi^{\mu\nu\lambda\sigma} = \sum_{i=1}^{43} \mathcal{B}_i^{\mu\nu\lambda\sigma} \tilde\Pi_i.
\eeq
The functions $\tilde\Pi_i$ are no longer free of kinematic singularities, but the form of their singularities follows from the projection of the BTT decomposition.
The necessary projectors are provided in ref.~\cite{Colangelo:2015ama}. Next, only a subset of the structures $T_i^{\mu\nu\lambda\sigma}$ actually contributes to $(g-2)_\mu$. To make this explicit it is convenient to perform another basis change
\beq
	\label{eq:HLbLTensorPiHatDecomposition}
	\Pi^{\mu\nu\lambda\sigma} = \sum_{i=1}^{54} T_i^{\mu\nu\lambda\sigma} \Pi_i =  \sum_{i=1}^{54} \hat T_i^{\mu\nu\lambda\sigma} \hat \Pi_i ,
\eeq
in such a way that in the limit $q_4\to0$ the derivative of 35 structures
$\hat T_i^{\mu\nu\lambda\sigma}$ vanishes. The 19 structures $\hat T_i^{\mu\nu\lambda\sigma}$ that do contribute to
$(g-2)_\mu$ can be chosen as~\cite{Colangelo:2017fiz}
\begin{align}
		\label{eq:ThatStructures}
		\hat T_i^{\mu\nu\lambda\sigma} &=
                T_i^{\mu\nu\lambda\sigma}, \quad i=1,\ldots, 11, 13, 14,
                16, 17, 50, 51, 54, \notag\\ 
		\hat T_{39}^{\mu\nu\lambda\sigma} &= \frac{1}{3} \left(
                  T_{39}^{\mu\nu\lambda\sigma} +
                  T_{40}^{\mu\nu\lambda\sigma} +
                  T_{46}^{\mu\nu\lambda\sigma} \right) . 
\end{align}
In this way, the $19$ relevant linear combinations of scalar functions are
\begin{align}
	\label{eq:PiHatFunctions}
	\hat\Pi_1 &= \Pi_1 + q_1 \cdot q_2 \Pi_{47} ,\notag \\
	\hat\Pi_4 &= \Pi_4 - q_1 \cdot q_3 \left( \Pi_{19} - \Pi_{42}
        \right) - q_2 \cdot q_3 \left( \Pi_{20} - \Pi_{43} \right) + q_1
        \cdot q_3  q_2 \cdot q_3 \Pi_{31} , \notag\\ 
	\hat \Pi_7 &= \Pi_7 - \Pi_{19} + q_2 \cdot q_3 \Pi_{31} ,\notag \\
	\hat \Pi_{17} &= \Pi_{17} + \Pi_{42} + \Pi_{43} - \Pi_{47} , \notag\\
	\hat \Pi_{39} &= \Pi_{39} + \Pi_{40} + \Pi_{46} , \notag\\
	\hat \Pi_{54} &= \Pi_{42} - \Pi_{43} + \Pi_{54} ,
\end{align}
together with the crossed versions 
\begin{align}
	\label{eq:CrossingRelationsPiHat}
	\hat \Pi_2 &= \Cr{23}{\hat \Pi_1} , \quad \hat \Pi_3 = \Cr{13}{\hat
          \Pi_1} , \quad 
	\hat \Pi_5 = \Cr{23}{\hat \Pi_4} , \quad \hat \Pi_6 = \Cr{13}{\hat
          \Pi_4} ,\quad 
	\hat \Pi_8 = \Cr{12}{\hat \Pi_7} , \notag\\ 
	\hat \Pi_9 &= \Cr{12}{\Cr{13}{\hat \Pi_7}} , \quad 
	\hat \Pi_{10} = \Cr{23}{\hat
          \Pi_7} , \quad \hat \Pi_{13} = \Cr{13}{\hat \Pi_7} , 
          \quad \hat\Pi_{14} = \Cr{12}{\Cr{23}{\hat \Pi_7}} ,\notag \\ 
	\hat \Pi_{11} &= \Cr{13}{\hat \Pi_{17}} , \quad \hat \Pi_{16} =
        \Cr{23}{\hat \Pi_{17}} , \quad 
	\hat \Pi_{50} = -\Cr{23}{\hat \Pi_{54}} , \quad \hat \Pi_{51} = 
        \Cr{13}{\hat \Pi_{54}} , 
\end{align}
where the crossing operators $\mathcal{C}_{ij}$ exchange momenta and
Lorentz indices of the photons $i$ and $j$
\begin{align}
	\label{eq:DefCrossingOperator}
	\mathcal{C}_{12}[f] := f( \mu \leftrightarrow \nu, q_1
        \leftrightarrow q_2 ) , \quad \mathcal{C}_{14}[f] := f( \mu
        \leftrightarrow \sigma, q_1 \leftrightarrow -q_4 ), 
\end{align}
and multiple operations are understood to act as in the example 
$\mathcal{C}_{12}[\mathcal{C}_{23}[f(q_1,q_2,q_3,q_4)]] = \mathcal{C}_{12}[f(q_1,q_3,q_2,q_4)] =
  f(q_2,q_3,q_1,q_4)$. 
 In addition, the $\hat\Pi_i$ preserve the crossing symmetries
\begin{align}
	\label{eq:InternalCrossingSymmetriesPiHat}
	\hat \Pi_1 &= \Cr{12}{\hat \Pi_1} , \quad \hat \Pi_4 = \Cr{12}{\hat
          \Pi_4} , \quad \hat \Pi_{17} = \Cr{12}{\hat \Pi_{17}} , \notag\\ 
	\hat \Pi_{39} &= \Cr{12}{\hat \Pi_{39}} = \Cr{13}{\hat \Pi_{39}} =
        \ldots , \quad \hat \Pi_{54} = - \Cr{12}{\hat \Pi_{54}}. 
\end{align}
The $\hat \Pi_i$ defined in this way display all crossing symmetries that survive in the limit $q_4\to 0$ and are thus particularly well suited for the HLbL application. 
In consequence, only the six functions~\eqref{eq:PiHatFunctions} need to be specified, with all the rest following from crossing symmetry.

In terms of these functions the HLbL contribution to $(g-2)_\mu$ becomes
\beq
\label{eq:MasterFormula3Dim}
\hspace{-0.1cm}
a_\mu^\mathrm{HLbL} = \frac{2 \alpha^3}{3 \pi^2} \int_0^\infty dQ_1 \int_0^\infty dQ_2 \int_{-1}^1 d\tau \sqrt{1-\tau^2} Q_1^3 Q_2^3 \sum_{i=1}^{12} T_i(Q_1,Q_2,\tau) \bar \Pi_i(Q_1,Q_2,\tau),
\eeq
where $Q_1 = |Q_1|$ and $Q_2 = |Q_2|$ denote the norm of the Euclidean four-vectors and we have used the symmetry of the kernel functions
under $q_1 \leftrightarrow -q_2$ to reduce the sum to only $12$ terms. The remaining kernel functions $T_i(Q_1,Q_2,\tau)$ are listed in ref.~\cite{Colangelo:2017fiz}
and the $12$ scalar function $\bar \Pi_i$ simply correspond to a subset of the $\hat \Pi_i$
\begin{align}
		\label{eq:PibarFunctions}
		\bar \Pi_1 &= \hat\Pi_1 , \quad &
		\bar \Pi_2 &= \hat\Pi_2 , \quad &
		\bar \Pi_3 &= \hat\Pi_4 , \quad &
		\bar \Pi_4 &= \hat\Pi_5 , \quad \notag \\
		\bar \Pi_5 &= \hat\Pi_7 , \quad &
		\bar \Pi_6 &= \hat\Pi_9 ,&
		\bar \Pi_7 &= \hat\Pi_{10} , \quad &
		\bar \Pi_8 &= \hat\Pi_{11} , \quad \notag\\
		\bar \Pi_9 &= \hat\Pi_{17} , \quad &
		\bar \Pi_{10} &= \hat\Pi_{39} , \quad &
		\bar \Pi_{11} &= \hat\Pi_{50} , \quad &
		\bar \Pi_{12} &= \hat\Pi_{54} .
\end{align}
They are evaluated for the kinematics
\beq
	\label{eq:Gm2Kinematics}
	s = q_3^2 = - Q_3^2 = - Q_1^2 - 2 Q_1 Q_2 \tau - Q_2^2 , \quad
	t = q_2^2 = -Q_2^2 , \quad
	u = q_1^2 = -Q_1^2 , \quad
	q_4^2 = 0,
\eeq
where $s$, $t$, $u$ are the Mandelstam variables of the original HLbL scattering process. Finally, we quote an alternative formulation of~\eqref{eq:MasterFormula3Dim}
based on the parameterization~\cite{Eichmann:2015nra}
\begin{align}\label{eq:g2coordinates}
		Q_1^2 &= \frac{\Sigma}{3} \left( 1 - \frac{r}{2} \cos\phi - \frac{r}{2}\sqrt{3} \sin\phi \right) , \notag\\
		Q_2^2 &= \frac{\Sigma}{3} \left( 1 - \frac{r}{2} \cos\phi + \frac{r}{2}\sqrt{3} \sin\phi \right) , \notag\\
		Q_3^2 &= Q_1^2 + 2 Q_1 Q_2 \tau + Q_2^2 = \frac{\Sigma}{3} \left( 1 + r \cos\phi \right) .
\end{align}
This variable transformation leads to
\beq
	\label{eq:MasterFormulaPolarCoord}
a_\mu^\mathrm{HLbL} = \frac{\alpha^3}{432\pi^2} \int_0^\infty d\Sigma\, \Sigma^3 \int_0^1 dr\, r\sqrt{1-r^2} \int_0^{2\pi} d\phi \,\sum_{i=1}^{12} T_i(Q_1,Q_2,\tau) \bar\Pi_i(Q_1,Q_2,\tau) ,
\eeq
which often facilitates the numerical evaluation.

\subsection{Pseudoscalar poles}
\label{sec:pseudoscalar}

The pseudoscalar poles only appear in $\hat\Pi_1$ (and by crossing symmetry in $\hat\Pi_{2,3}$)
\beq
\label{Pi_PS}
\hat\Pi_1^{P\text{-pole}}=\frac{F_{P\gamma^*\gamma^*}(q_1^2,q_2^2)F_{P\gamma\gamma^*}(q_3^2)}{q_3^2-M_P^2},
\eeq
with $F_{P\gamma^*\gamma ^*}(q_1^2,q_2^2)$ the doubly-virtual TFF, $F_{P\gamma\gamma ^*}(q^2)=F_{P\gamma ^* \gamma ^*}(q^2,0)$ the singly-virtual TFF, 
and $P=\pi^0,\eta,\eta'$ (see ref.~\cite{Colangelo:2015ama} for the detailed derivation). The TFFs are normalized to the two-photon decays according to
\beq
\Gamma(P\to\gamma\gamma)=\frac{\pi\alpha^2M_P^3}{4}F_{P\gamma\gamma}^2,\qquad F_{P\gamma\gamma}=F_{P\gamma^*\gamma^*}(0,0).\label{anomDef}
\eeq
They are defined by the matrix element
\beq
\label{eq:defpiTFF}
i\int d^4x \, e^{iq_1\cdot x}  \langle 0|T \, \{j_\mu(x)j_\nu(0)\}|P(q_1+q_2)\rangle 
 = \epsilon_{\mu\nu\alpha\beta} \, q_1^\alpha \, q_2^\beta \, F_{P\gamma^*\gamma^*}(q_1^2,q_2^2).
\eeq
In the chiral limit, the non-singlet normalizations are determined by the Adler--Bell--Jackiw anomaly \cite{Adler:1969gk,Bell:1969ts,Bardeen:1969md}
\beq
\sum_P F_P^a F_{P\gamma\gamma}=\frac{3}{2\pi^2}C_a,\qquad C_a=\frac{1}{2}\Tr(\mathcal{Q}^2\lambda_a),
\label{eq:ABJ}
\eeq
with Gell-Mann matrices $\lambda_a$, $\lambda_0=\sqrt{2/3}\,\unity$, 
\beq
\label{weights}
C_3=\frac{1}{6},\qquad C_8=\frac{1}{6\sqrt{3}},\qquad C_0=\frac{2}{3\sqrt{6}},
\eeq
and decay constants defined through $F_P^a$:
\beq
\langle 0 | A^a_\mu(0) | P(p) \rangle =: i p_\mu F^a_P, \label{decayconstantsDef}
\eeq
which is in general a $3\times 3$ matrix. Ignoring for simplicity any possible mixing between the $\pi^0$ and the other two states, this takes the form:
\beq
F^a_P=\left( \begin{array}{lll} F^3_\pi & 0 & 0 \\ 0 & F^8_\eta & F^8_{\eta^\prime} \\ 0 & F^0_{\eta} & F^0_{\eta^\prime} \end{array} \right)=\left( \begin{array}{lll} F^3_\pi & 0 & 0 \\ 0 & F^8 \cos \theta_8 & F^8 \sin \theta_8 \\ 0 &-F^0 \sin \theta_0 & F^0 \cos \theta_0 \end{array} \right),
\label{eq:Pmixing}
\eeq
where, after the second equality sign, we have already introduced the standard two-angle mixing scheme between $\eta$ and $\eta^\prime$.
For the pion $a=3$, the corresponding low-energy theorem 
\beq
F_{\pi^0\gamma\gamma}=\frac{3}{2\pi^2F_\pi}C_3=\frac{1}{4\pi^2F_\pi}\label{AnomalySec2}
\eeq
is very close to phenomenology, while for $\eta$, $\eta'$ chiral corrections and mixing effects need to be taken into account.
In particular, we stress that due to the renormalization of the singlet current $F_P^0$ is not actually an observable quantity, and the corresponding $\alpha_s$ corrections~\cite{Kodaira:1979pa,Espriu:1982bw}
need to be considered when relating the normalization, asymptotic constraints, and $\eta$--$\eta'$ mixing parameters~\cite{Leutwyler:1997yr,Kaiser:1998ds,Kaiser:2000gs,Escribano:2015yup,Escribano:2015nra}.
In the present work, we will take the $\eta$, $\eta'$ normalizations from experiment, so that the singlet corrections become most relevant when 
comparing the asymptotic constraints and $\eta$--$\eta'$ mixing parameters in different schemes. As described in section~\ref{sec:ModelEta}, we studied the impact of different such determinations 
on the numerics, with the result that the corresponding variations were numerically irrelevant 
in view of the accuracy anticipated for the pseudoscalar TFF models discussed in the following sections.

In addition to the normalizations, the pseudoscalar TFFs are subject to the (leading) asymptotic constraint \cite{Lepage:1979zb,Lepage:1980fj,Brodsky:1981rp}
\beq
 F_{P\gamma^*\gamma^*}(-Q_1^2,-Q_2^2) = 4\sum_a C_a F_P^a \int_0^1 \text{d}x \frac{\phi_P^a(x)}{x Q_1^2 +(1-x)Q_2^2},\label{FeynmanParameterNotation}
\eeq
which for the asymptotic wave function $\phi_P^a(x)=6x(1-x)$, and again ignoring $\alpha_s$ corrections for the singlet component, produces the limits\footnote{As argued in ref.~\cite{Manohar:1990hu}, the first limit goes beyond a strict OPE, but is consistent with the phenomenology of the ground-state TFFs, see, e.g., refs.~\cite{Hoferichter:2018kwz,Masjuan:2017tvw}.}
\begin{align}
 \lim_{Q^2\to\infty}Q^2F_{P\gamma\gamma^*}(-Q^2) &= 12\sum_a C_a F_P^a,\nn\\
  \lim_{Q^2\to\infty}Q^2F_{P\gamma^*\gamma^*}(-Q^2,-Q^2) &= 4\sum_a C_a F_P^a.\label{BLOPESec2}
\end{align}
In view of~\eqref{eq:ABJ}, multiplying these limits by $F_{P \gamma \gamma}$ and summing over $P$ one obtains an expression which depends neither on decay constants nor on mixing angles. Moreover, the block form of the matrix~\eqref{eq:Pmixing} leads to two separate combinations with such a property:
\begin{align}
F_{\pi^0\gamma\gamma}\lim_{Q^2\to\infty}Q^2F_{\pi^0\gamma^*\gamma^*}(-Q^2,-Q^2)&=\frac{6}{\pi^2}C_3^2=\frac{1}{6\pi^2},\nn\\
\sum_{P=\eta,\eta'}F_{P\gamma\gamma}\lim_{Q^2\to\infty}Q^2F_{P\gamma^*\gamma^*}(-Q^2,-Q^2)&=4\sum_{P,a} F_P^a F_{P\gamma\gamma} C_a
=\frac{6}{\pi^2}\sum_{a=0,8} C_a^2=\frac{1}{2\pi^2},\label{F_MV}
\end{align}
and similarly for the asymptotic limit of the singly-virtual TFF. Beyond the singlet $\alpha_s$ corrections that describe the scale dependence of $F_P^0$~\cite{Leutwyler:1997yr,Kaiser:1998ds,Kaiser:2000gs},
there are genuine pQCD corrections to the TFFs suppressed by $\alpha_s$ at scales related to the photon virtualities~\cite{Braaten:1982yp}. The impact of such 
next-to-leading-order pQCD corrections was studied in ref.~\cite{Hoferichter:2018kwz} in the context of the pion TFF, with the result 
that even for the ground-state pion the effect is small and safely covered by the uncertainty estimated from the onset of the asymptotic region.

\subsection{The perturbative QCD quark loop}
\label{sec:pQCD}

\begin{figure}[t]
\centering
\includegraphics[width=0.8\textwidth]{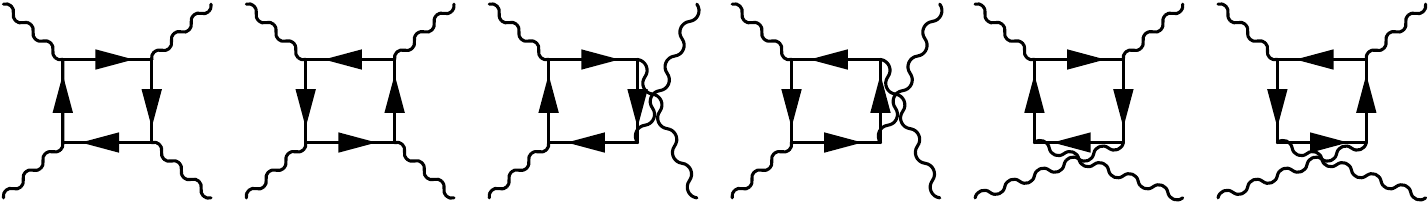}
\caption{Quark-loop contribution to HLbL scattering.}
\label{fig:quark_loop}
\end{figure} 

The quark-loop contribution to HLbL scattering is shown in figure~\ref{fig:quark_loop}, indicating the different permutations that need to be considered.
Compact expressions for the BTT scalar functions can be obtained as follows: 
we use a Feynman parameterization for the loop integrals and project the result onto the scalar basis functions $\tilde\Pi_i$~\cite{Colangelo:2015ama,Colangelo:2017fiz}. 
We find all necessary BTT functions $\Pi_i$ in the limit $q_4\to0$ by taking this limit in the appropriate order, so that the Tarrach poles drop out. 
Then we combine the functions $\Pi_i$ according to~\eqref{eq:PiHatFunctions} to obtain the scalar functions $\hat\Pi_i$. 
Due to the limit $q_4\to0$, one integral can be carried out and we are left with a two-dimensional Feynman-parameter integral. 
The result for the integrands contains spurious kinematic singularities, but the residues of these poles vanish when the Feynman integrals are carried out. 
Therefore, we can subtract these poles and obtain a representation that is manifestly free of kinematic singularities
\beq
\hat \Pi_i^{\text{quark loop}}= \sum_{q} N_c \, Q_q^4 \frac{1}{16\pi^2} \int_0^1 dx \int_0^{1-x} dy \, I_i(x,y) ,
\eeq
where
\begin{align}
\label{Pihat_quark_loop}
	I_1(x,y) &= - \frac{16 x(1-x-y)}{\Delta_{132}^2} - \frac{16 xy (1-2x) (1-2y)}{\Delta_{132} \Delta_{32}}, \notag\\
	I_4(x,y) &= \frac{32xy(1-2x)(x+y)(1-x-y)^2(q_1^2-q_2^2+q_3^2)}{\Delta_{312}^3} - \frac{32(1-x)x(x+y)(1-x-y)}{\Delta_{312}^2}\notag \\
		& - \frac{32xy(1-2x)(1-2y)}{\Delta_{312}\Delta_{12}}, \notag\\
	I_7(x,y) &= -\frac{64 x y^2 (1-x-y) (1-2x)(1-y)}{\Delta_{132}^3}, \notag\\
	I_{17}(x,y) &= - \frac{32x^2y^2(1-2x)(1-2y)}{\Delta_{312}^2 \Delta_{12}}, \notag\\
	I_{39}(x,y) &= \frac{64xy(1-x-y)\bigl((2x-1)y^2 +xy(2x-3)+x(1-x)+y\bigr)}{\Delta_{132}^3},\notag\\
	I_{54}(x,y) &= - \frac{16xy(1-x-y)(1-2x)(1-2y)(x-y)}{\Delta_{312} \Delta_{12}} \left( \frac{1}{\Delta_{312}} + \frac{1}{\Delta_{12}} \right),
\end{align}
and
\begin{align}
		\Delta_{ijk} &= m_q^2 - x y q_i^2 - x (1-x-y) q_j^2 - y(1-x-y) q_k^2 , \notag\\
		\Delta_{ij} &= m_q^2 - x (1-x) q_i^2 - y (1-y) q_j^2 .
\end{align}
In principle, it is also possible to extract the results by projecting onto the singly-on-shell basis $\check\Pi_i$~\cite{Colangelo:2017fiz}. 
However, it turns out that this method is less straightforward, because different spurious kinematic singularities appear, 
which have to be subtracted again and make the calculation more complicated.

As a cross check of~\eqref{Pihat_quark_loop} we have evaluated light-quark loops for $q=u,d,s$ with (constituent) quark mass $m_q$, 
including a factor $N_c \sum_{q=u,d,s} Q_q^4 = 2/3$ as well as the lepton loops. The latter agree well with the known analytic expressions~\cite{Laporta:1992pa},
while apart from the electron loop the results are well reproduced from the heavy-mass expansion~\cite{Kuhn:2003pu}. Throughout, for the matching to our Regge models in section~\ref{sec:matching}, we use the pQCD quark loop with $m_q=0$, given that even in configurations 
where chiral corrections for the light quarks $q=u,d,s$ can be controlled
within pQCD, they only enter at subleading orders.   

As a first application, we consider the contribution from the charm quark. 
Assuming that this contribution is fully perturbative, with mass $m_c=1.27(2)\GeV$~\cite{Tanabashi:2018oca}, the quark loop evaluates to
$a_\mu^{c\text{-loop}}=3.1(1)\times 10^{-11}$. In analogy to the light quarks, one would expect 
the most important non-perturbative effect to be related to the pole contribution from
the lowest-lying $c \bar c$ resonance, the $\eta_c(1S)$ with mass $m_{\eta_c(1S)}=2.9839(5)\GeV$
and two-photon width $\Gamma(\eta_c(1S)\to\gamma\gamma)=5.0(4)\keV$~\cite{Tanabashi:2018oca}.
Using a VMD-type form factor with scale set by the $J/\Psi$ (as suggested by a significant branching fraction $\BR(J/\Psi\to\eta_c(1S)\gamma)=1.7(4)\%$~\cite{Tanabashi:2018oca}), 
this leads to the estimate
$a_\mu^{\eta_c(1S)}=0.8\times 10^{-11}$ (this estimate agrees with the LMD result $a_\mu^{\eta_c(1S)}=0.9(1)\times 10^{-11}$ from~\cite{Raya:2019dnh}). Given the relatively low scale set by $m_c$ one may also expect
$\alpha_s$ corrections in a similar ballpark. Altogether, we estimate 
\beq
a_\mu^{c\text{-quark}}=3(1)\times 10^{-11},    
\eeq
while the $b$-quark contribution is already suppressed to the level of $10^{-13}$ and the $t$-quark loop to $10^{-15}$.

\section{OPE constraints for the hadronic light-by-light tensor}
\label{sec:OPE}

\subsection{OPE for the asymptotic region}
\label{sec:asym}

The first term in the OPE for the kinematic configuration in which all three momenta are large coincides with the pQCD quark loop. 
This has long been suspected in the literature, including ref.~\cite{Melnikov:2003xd}, but was only demonstrated recently in ref.~\cite{Bijnens:2019ghy}, by working out the 
next order in the expansion. While at leading order all quark masses can simply be put to zero, this is no longer true at subleading orders. 
In fact, it is the presence of quark masses and condensates that numerically suppresses the next-to-leading order corrections.

In the limit $q_1^2=q_2^2=q_3^2\equiv q^2$ the expressions for the pQCD quark loop simplify to
\begin{align}
\label{pQCDloop}
 \hat \Pi_1^\text{pQCD}&=-\frac{4}{9\pi^2 q^4},\qquad \hat \Pi_{54}^\text{pQCD}=0,\notag\\
 \hat \Pi_4^\text{pQCD}&=-\frac{8}{243\pi^2q^4}\bigg[33-16\sqrt{3}\,\Cl\Big(\frac{\pi}{3}\bigg)\bigg],\notag\\
 \hat \Pi_7^\text{pQCD}&=\frac{4}{243\pi^2q^6}\bigg[33-16\sqrt{3}\,\Cl\Big(\frac{\pi}{3}\bigg)\bigg],\notag\\
 \hat \Pi_{17}^\text{pQCD}&=\frac{16}{81\pi^2q^6}\bigg[3-2\sqrt{3}\,\Cl\Big(\frac{\pi}{3}\bigg)\bigg],\notag\\
 \hat \Pi_{39}^\text{pQCD}&=-\frac{8}{243\pi^2q^6}\bigg[15-4\sqrt{3}\,\Cl\Big(\frac{\pi}{3}\bigg)\bigg],
\end{align}
where the Clausen function is defined as 
\beq
\Cl(x)=-\int_0^x dt\log\left|2\sin\frac{t}{2}\right|.
\eeq
This result again includes the factor $2/3$ due to $N_c$ and quark charges, after summing over $q=u,d,s$.

\subsection{OPE for the mixed regions}
\label{sec:mixed}

The OPE constraint derived in ref.~\cite{Melnikov:2003xd} applies to the case where one virtuality remains smaller than the others, $Q_3^2\ll Q_1^2\sim Q_2^2$, also referred to as the mixed regions.
This constraint traces back to non-renormalization theorems for the $VVA$ correlator~\cite{Vainshtein:2002nv,Knecht:2003xy}, which had been used before in the context of the electroweak contributions to $(g-2)_\mu$~\cite{Knecht:2002hr,Czarnecki:2002nt}. 
Explicit pQCD calculations at two- and three-loop order exist~\cite{Jegerlehner:2005fs,Mondejar:2012sz}, but the main argument in ref.~\cite{Melnikov:2003xd} was that the non-renormalization theorems 
allow one to address the regions in which both perturbative and non-perturbative aspects might be important. We first review this derivation, while casting the results in a form suitable for the BTT decomposition of the HLbL tensor.

The central object is the OPE for two electromagnetic currents:
\begin{align}
\label{Pi_OPE}
	\Pi^{\mu\nu}(q_1,q_2)&=i \int d^4x \, d^4y \, e^{-i(q_1 \cdot x + q_2 \cdot y)}T \{ j^\mu(x) j^\nu(y)\} .
\end{align}
We consider large momenta $\hat q=(q_1-q_2)/2$ flowing through the currents and expand the operator product into a series of local operators. For $|\hat q| \gg \Lambda_\mathrm{QCD}$, the coefficients can be calculated in perturbation theory. At leading order in $\alpha_s$, only two-quark operators are generated, hence the matching can be easily obtained by inserting the operator~\eqref{Pi_OPE} into external quark states and expanding the diagrams for large momenta $\hat q$:
\begin{align}
	i &\langle \psi_q(p_2) | \Pi^{\mu\nu}(q_1,q_2) | \psi_q(p_1) \rangle \nn\\
	&= (2\pi)^4 \delta^{(4)}(q_1+q_2+p_1-p_2) \begin{aligned}[t]
		& \bigg[ \bar u_q(p_2) i^2 \Q_q^2 \gamma^\nu \frac{i(\slashed p_1 + \slashed q_1+m_q)}{(p_1+q_1)^2 - m_q^2} \gamma^\mu u_q(p_1) \nn\\
		& + \bar u_q(p_2) i^2 \Q_q^2 \gamma^\mu \frac{i(\slashed p_2 - \slashed q_1+m_q)}{(p_2-q_1)^2-m_q^2} \gamma^\nu u_q(p_1) \bigg] \end{aligned} \nn\\
	&= (2\pi)^4 \delta^{(4)}(q_1+q_2+p_1-p_2)  \bar u_q(p_2) i \Q_q^2 \Bigg[ - 2 i \epsilon^{\mu\nu\lambda\sigma} \frac{\hat q_\lambda}{\hat q^2} \gamma_\sigma \gamma_5  - 2 g^{\mu\nu} \frac{m_q}{\hat q^2}  \nn\\
		& - ( \gamma^\mu g^{\nu\sigma} + \gamma^\nu g^{\mu\sigma} - \gamma^\sigma g^{\mu\nu} ) \bigg(  \frac{1}{\hat q^2}(p_1 + p_2)_\sigma - \frac{2(p_1+p_2)\cdot \hat q}{(\hat q^2)^2} \hat q_\sigma \bigg) 
		+ \Order\big(\hat q^{-3}\big) \Bigg] u_q(p_1) ,
\end{align}
where we used
\beq
\gamma^\mu\gamma^\alpha\gamma^\nu=g^{\mu\alpha}\gamma^\nu+g^{\nu\alpha}\gamma^\mu-g^{\mu\nu}\gamma^\alpha-i\eps^{\mu\alpha\nu\beta}\gamma_5\gamma_\beta
\eeq
with $\eps^{0123}=+1$.
Introducing the scalar density $S(x)$, the axial vector $j_5^\mu(x)$, and the energy-momentum tensor $\theta^{\mu\nu}(x)$ with flavors weighted by the squared electric charges,
\beq
\hspace{-0.2cm}	S(x) = \bar\psi(x) \mathcal{Q}^2 \M \psi(x) , \quad
	j_5^\mu(x)=\bar \psi(x)\mathcal{Q}^2\gamma^\mu\gamma_5\psi(x) , \quad
	\theta^{\mu\nu}(x)=\frac{i}{2}\bar\psi(x)\mathcal{Q}^2\gamma^\mu\partial_-^\nu\psi(x),
\eeq
with the quark-mass matrix $\M = \mathrm{diag}(m_u,m_d,m_s)$ as well as the derivative $\partial_-=\overrightarrow{\partial}-\overleftarrow{\partial}$, we read off the matching for the OPE:
\begin{align}
    \label{eq:OPEtwoCurrents}
\Pi_{\mu\nu}(q_1,q_2)&=\int d^4z \, e^{-i(q_1 + q_2)\cdot z}\bigg[-\frac{2i}{\hat q^2}\eps_{\mu\nu\alpha\beta}\hat q^\alpha j_5^\beta(z)
-\frac{2}{\hat q^2}\Big(\theta_{\mu\nu}(z)+\theta_{\nu\mu}(z)-g_{\mu\nu}\theta^\alpha_\alpha(z)\Big)\notag\qquad\\
&+ \frac{4}{(\hat q^2)^2}\Big(\hat q_\mu\hat q^\alpha\theta_{\nu\alpha}+\hat q_\nu\hat q^\alpha\theta_{\mu\alpha}-g_{\mu\nu}\hat q^\alpha \hat q^\beta \theta_{\alpha\beta}\Big)
- \frac{2}{\hat q^2} g_{\mu\nu} S(z)
+\Order\big(\hat q^{-3}\big)\bigg].
\end{align}
The first term reproduces the expansion given in ref.~\cite{Melnikov:2003xd}, but differs in sign just because of different conventions (they use $\eps^{0123}=-1$). 

Applying the OPE to the HLbL tensor in the limit $Q_1^2\sim Q_2^2\gg Q_3^2,Q_4^2$ we then find at leading order
\begin{align}
\label{HLbL_LO}
\Pi_{\mu\nu\lambda\sigma}(q_1,q_2,q_3) &=\frac{2i}{\hat q^2}\epsilon_{\mu\nu\alpha\beta}\hat q^\alpha\int d^4 x\, d^4 y\, 
e^{-i(q_1 + q_2)\cdot x} e^{-iq_3\cdot y} \langle 0 | T \{ j^\beta_5(x) j_\lambda(y)j_\sigma(0) \} | 0 \rangle\notag\qquad\\
&=\frac{2i}{\hat q^2}\epsilon_{\mu\nu\alpha\beta}\hat q^\alpha\int d^4 x\, d^4 y\, 
e^{-iq_3\cdot x} e^{iq_4\cdot y} \langle 0 | T \{j_\lambda(x)j_\sigma(y)j_5^\beta(0) \} | 0 \rangle\notag\\
&=\frac{2}{\hat q^2}\epsilon_{\mu\nu\alpha\beta}\hat q^\alpha {W_{\lambda\sigma}}^\beta(-q_3,q_4),
\end{align}
where the correlator $W_{\mu\nu\rho}$ is defined as
\begin{align}
	\label{eq:jjj5corr}
	W_{\mu\nu\rho}(q_1,q_2)&=i\int d^4 x\, d^4 y\, e^{i(q_1\cdot x + q_2\cdot y)} \langle 0 | T \{j_\mu(x)j_\nu(y)j^5_\rho(0) \} | 0 \rangle .
\end{align}
Introducing the vector and axial-vector currents
\begin{align}
\label{norm_flavor_dec}
	V_\mu^a(x)=\bar \psi(x) \gamma_\mu \frac{\lambda_a}{2} \psi(x), \quad A_\mu^a(x)=\bar \psi(x) \gamma_\mu\gamma_5 \frac{\lambda_a}{2} \psi(x),
\end{align}
where $\{\lambda_a,\lambda_b\}=4/3\delta_{ab}+2d_{abc}\lambda_c$, we also define the correlator
\begin{align}
	W^{abc}_{\mu\nu\rho}(q_1,q_2)&= i\int d^4 x\, d^4 y\, e^{i(q_1\cdot x + q_2\cdot y)}\langle 0 | T \{V^a_\mu(x)V^b_\nu(y)A^c_\rho(0) \} | 0 \rangle .
\end{align}
Performing the flavor decompositions
\begin{align}
	\label{eq:CurrentDecomp}
	j^5_\mu(x) = \sum_{a=0,3,8} 2 C_a A^a_\mu(x) , \quad j_\mu(x) = \sum_{a=3,8} 2 D_a V_\mu^a(x)
\end{align}
with $C_a$ defined in~\eqref{eq:ABJ} and $D_a = \frac{1}{2}\Tr(\Q\lambda_a)$, we write the correlator~\eqref{eq:jjj5corr} as
\begin{align}
	\label{eq:CorrSum}
	W_{\mu\nu\rho}(q_1,q_2) = 4 \sum_{a=0,3,8} C_a^2 \, W^{(a)}_{\mu\nu\rho}(q_1,q_2) ,
\end{align}
where
\begin{align}
	W^{(a)}_{\mu\nu\rho}(q_1,q_2) := \frac{2}{C_a} \sum_{b,c=3,8} D_b D_c W^{bca}_{\mu\nu\rho}(q_1,q_2) .
\end{align}
The Lorentz decomposition of the $VVA$ correlator is chosen as~\cite{Knecht:2003xy}
\begin{align}
 W^{(a)}_{\mu\nu\rho}(q_1,q_2)&=-\frac{1}{8\pi^2} \bigg[ -w_L^{(a)}\big(q_1^2,q_2^2,(q_1+q_2)^2\big)\eps_{\mu\nu\alpha\beta}q_1^\alpha q_2^\beta(q_1+q_2)_\rho \nn\\
 &+ w_T^{+(a)}\big(q_1^2,q_2^2,(q_1+q_2)^2\big)t_{\mu\nu\rho}^+ + w_T^{-(a)}\big(q_1^2,q_2^2,(q_1+q_2)^2\big)t_{\mu\nu\rho}^- \nn\\
 &+ \tilde w_T^{-(a)} \big(q_1^2,q_2^2,(q_1+q_2)^2\big)\tilde t_{\mu\nu\rho}^-\bigg],
\end{align}
with the following Lorentz structures:
\begin{align}
 t_{\mu\nu\rho}^+&=\eps_{\mu\rho\alpha\beta}q_{1\nu}q_1^\alpha q_2^\beta-\eps_{\nu\rho\alpha\beta}q_{2\mu}q_1^\alpha q_2^\beta-q_1\cdot q_2 \eps_{\mu\nu\rho\alpha}(q_1-q_2)^\alpha\notag\\
 &  +\frac{q_1^2+q_2^2-(q_1+q_2)^2}{(q_1+q_2)^2}\eps_{\mu\nu\alpha\beta}q_1^\alpha q_2^\beta(q_1+q_2)_\rho,\notag\\
 t_{\mu\nu\rho}^-&=\bigg((q_1-q_2)_\rho-\frac{q_1^2-q_2^2}{(q_1+q_2)^2}(q_1+q_2)_\rho\bigg)\eps_{\mu\nu\alpha\beta}q_1^\alpha q_2^\beta,\notag\\
 \tilde t_{\mu\nu\rho}^-&=\eps_{\mu\rho\alpha\beta}q_{1\nu}q_1^\alpha q_2^\beta+\eps_{\nu\rho\alpha\beta}q_{2\mu}q_1^\alpha q_2^\beta-q_1\cdot q_2 \eps_{\mu\nu\rho\alpha}(q_1+q_2)^\alpha.
\end{align}
In the massless limit, one finds at one loop the contribution of the axial anomaly~\cite{Adler:1969gk,Bell:1969ts}
\begin{align}
	\label{eq:AxialAnomaly}
	w_L^{(a)}(q_1^2,q_2^2,(q_1+q_2)^2) = \frac{2N_c}{(q_1+q_2)^2}.
\end{align}
For the non-singlet contributions $a=3,8$, this result is modified neither by higher-order perturbative~\cite{Adler:1969er} nor non-perturbative contributions~\cite{tHooft:1979rat}, while the singlet contribution is affected by the gluonic $U(1)$ anomaly.

In the chiral limit, the factor $C_a^2$ in~\eqref{eq:CorrSum} arises naturally due to the flavor decomposition. One factor of $C_a$ stems from~\eqref{eq:CurrentDecomp}, the second factor emerges as follows. We consider singlet and octet parts of the axial current and define
\begin{align}
	W^{bc0}_{\mu\nu\rho}(q_1,q_2) &=: \sqrt{\frac{2}{3}} \delta^{bc} \overline W_{\mu\nu\rho}(q_1,q_2) , \quad W^{bca}_{\mu\nu\rho}(q_1,q_2) =: d^{abc} \overline{\overline W}_{\mu\nu\rho}(q_1,q_2) , 
\end{align}
which implies
\begin{align}
	W^{(0)}_{\mu\nu\rho} = \frac{2}{C_0} \bigg( \sqrt{\frac{2}{3}}  \sum_{b=3,8} D_b^2 \bigg) \overline W_{\mu\nu\rho}, \quad W^{(a\neq0)}_{\mu\nu\rho} = \frac{2}{C_a}\Big( \sum_{b,c=3,8} D_b D_c d^{abc} \Big) \overline{\overline W}_{\mu\nu\rho}.
\end{align}
The coefficients can be simplified to
\begin{align}
	\sqrt{\frac{2}{3}} \sum_{b=3,8} D_b^2 &= \frac{1}{\sqrt{6}} \sum_{b=3,8} D_b \Tr(\Q \lambda_b) = \frac{1}{\sqrt{6}} \Tr(\Q^2) = C_0 , \nn\\
	\sum_{b,c=3,8} D_b D_c d^{abc} &= \frac{1}{2} \sum_{b,c=3,8} D_b\Tr(\Q \lambda_c) d^{abc} = \frac{1}{4} \sum_{b=3,8} D_b \Tr\bigg(\Q\Big( \{\lambda_a,\lambda_b\} - \frac{4}{3}\delta_{ab} \Big) \bigg) \nn\\
		&= \frac{1}{4} \Tr\Big(\Q \{\lambda_a,\Q\} \Big) = C_a,
\end{align}
hence both singlet and octet components lead to another factor $C_a$.

In pQCD and in the chiral limit, the following non-renormalization theorems were derived in~\cite{Knecht:2003xy} for the non-singlet part of the axial current:
\begin{align}
\label{non_renormalization}
 0&=(w_T^+ + w_T^-)\big(q_1^2,q_2^2,(q_1+q_2)^2\big)-(w_T^+ + w_T^-)\big((q_1+q_2)^2,q_2^2,q_1^2\big),\notag\\
 0&=(\tilde w_T^- + w_T^-)\big(q_1^2,q_2^2,(q_1+q_2)^2\big)+(\tilde w_T^- + w_T^-)\big((q_1+q_2)^2,q_2^2,q_1^2\big),\notag\\
 w_L\big((q_1+q_2)^2,q_2^2,q_1^2\big)&=(w_T^+ + \tilde w_T^-)\big(q_1^2,q_2^2,(q_1+q_2)^2\big)+(w_T^+ + \tilde w_T^-)\big((q_1+q_2)^2,q_2^2,q_1^2\big)\notag\qquad\\
 & +\frac{2q_2\cdot(q_1+ q_2)}{q_1^2}w_T^+\big((q_1+q_2)^2,q_2^2,q_1^2\big) \notag\\
 & - \frac{2q_1\cdot q_2}{q_1^2}w_T^-\big((q_1+q_2)^2,q_2^2,q_1^2\big).
\end{align}
The transversal functions in these relations are subject to non-perturbative corrections.

In the following, we will use the OPE constraints as they arise at leading order and in the chiral limit. Both the anomaly constraint~\eqref{eq:AxialAnomaly} and the non-renormalization theorems~\eqref{non_renormalization} receive quark-mass corrections~\cite{Melnikov:2006qb}.

\subsection{Projection onto BTT}
\label{sec:mixed_BTT}

In this section, we derive the asymptotic constraints that the leading-order expression~\eqref{HLbL_LO} of the OPE imposes on the scalar BTT functions $\hat\Pi_i$~\eqref{eq:PiHatFunctions} entering the master formula for $a_\mu$. One might be tempted to simply project the OPE expression~\eqref{HLbL_LO} onto the BTT scalar functions. However, there are several problems with such an approach. First of all, the leading-order expression of the OPE is not manifestly gauge invariant: the contraction with $(q_1-q_2)^\mu$ vanishes, but the one with $(q_1+q_2)^\mu$ does not. Due to $q_1=-q_2+\Order(1)$ this does ensure gauge invariance at $\Order(1/\hat q)$, while for the subleading orders relations with the matrix elements of the energy-momentum tensor are needed to restore gauge invariance.  At leading order, gauge invariance could be restored by applying a gauge projector
\beq
\label{gauge_invariance}
\eps_{\mu\nu\alpha\beta}(q_1^\alpha-q_2^\alpha)
\to
\eps_{\mu\nu'\alpha\beta}q_1^\alpha\bigg(g^{\nu'}_\nu-\frac{q_{2\nu} q_2^{\nu'}}{q_2^2}\bigg)
-\eps_{\mu'\nu\alpha\beta}q_2^\alpha\bigg(g^{\mu'}_\mu-\frac{q_{1\mu} q_1^{\mu'}}{q_1^2}\bigg) ,
\eeq
which does not alter the $\Order(\hat q^{-1})$ terms of the OPE expression. The subsequent projection onto BTT and extraction of the scalar functions $\hat\Pi_i$ could then be performed immediately. However, this procedure is not uniquely defined: the BTT structures themselves become degenerate depending on the order of the expansion for large $\hat q$. This implies that the leading-order OPE only constrains certain linear combinations of scalar functions $\hat\Pi_i$. In an assignment of these constraints to individual scalar functions ambiguities are introduced. For the longitudinal amplitudes the linear combination of scalar functions that is uniquely constrained happens to coincide with $\hat\Pi_{1\text{--}3}$, but for the transversal amplitudes the situation becomes more complicated. 
We proceed as follows in order to determine this ambiguity explicitly and to work out the exact form of the OPE constraint at the level of BTT functions.

First, we remember that the HLbL tensor is linear in the external momentum $q_4$. Due to the relation
\begin{align}
	\Pi_{\mu\nu\lambda\rho} = - q_4^\sigma \frac{\pa}{\pa q_4^\rho} \Pi_{\mu\nu\lambda\sigma}
\end{align}
following from gauge invariance, it is enough to consider the derivative with respect to $q_4^\rho$ and then take the limit $q_4\to0$, as required for $(g-2)_\mu$ kinematics. The BTT functions $\hat\Pi_i$ in this limit are unambiguously defined, hence they have their own proper expansion in $1/\hat q$, which we would like to constrain using the OPE. The derivatives of the Lorentz structures $\hat T_i$ multiplying the functions $\hat\Pi_i$ in the tensor decomposition~\eqref{eq:HLbLTensorPiHatDecomposition} however contain several terms with different scaling for large $\hat q$. For instance, for the tensor structure $\hat T_5^{\mu\nu\lambda\sigma}$, one finds
\begin{align}
	\frac{\pa}{\pa q_4^\rho} \hat T_5^{\mu\nu\lambda\sigma} \bigg|_{q_4 = 0} &= \frac{1}{4} \Big( q_3^\mu q_3^\lambda - g^{\mu\lambda} q_3^2 \Big) \Big( q_3^\sigma g^{\nu\rho} - q_3^\rho g^{\nu\sigma} \Big)
		 + \frac{1}{2} \Big( q_3^\mu q_3^\lambda - g^{\mu\lambda} q_3^2 \Big) \Big( \hat q^\sigma g^{\nu\rho} - \hat q^\rho g^{\nu\sigma} \Big) \nn\\
		& - \frac{1}{2}  \Big( q_3^\mu \hat q^\lambda - g^{\mu\lambda} q_3 \cdot \hat q \Big) \Big( q_3^\sigma g^{\nu\rho} - q_3^\rho g^{\nu\sigma} \Big)\notag\\
		 &- \Big( q_3^\mu \hat q^\lambda - g^{\mu\lambda} q_3 \cdot \hat q \Big) \Big( \hat q^\sigma g^{\nu\rho} - \hat q^\rho g^{\nu\sigma} \Big),
\end{align}
where the first term scales as $\Order(\hat q^0)$, the second and third terms are of $\Order(\hat q)$, and the last term is of $\Order(\hat q^2)$. This illustrates that a certain coefficient of the expansion in $1/\hat q$ of a scalar function $\hat\Pi_i$ can contribute to different orders in the expansion in $1/\hat q$ of the full HLbL tensor, i.e., to different orders of the OPE. Vice versa, in order to determine the leading-order OPE constraint on the BTT functions, we e.g.\ have to consider terms up to and including $\Order(\hat q^{-3})$ in $\hat\Pi_5$.
We now write the scalar functions $\hat\Pi_i$ as a generic expansion in $1/\hat q$ and sum up the scalar functions times (derivatives of) tensor structures. Collecting in the resulting tensor terms according to the scaling with $\hat q$ and requiring equality with the leading-order OPE limit~\eqref{HLbL_LO} determines the expansion coefficients of the 19 BTT functions relevant in the $(g-2)_\mu$ limit as
\begin{align}
	\label{OPE}
	\hat\Pi_1 &= 2 w_L(q_3^2,0,q_3^2)f({\hat q}^2) + \Order(\hat q^{-3}) , \nn\\
	\hat\Pi_5 &= \frac{4}{3} \big(w_T^+ +\tilde w_T^-\big)(q_3^2,0,q_3^2) f({\hat q}^2) + c_5^{(2)} + c_5^{(3)} + \Order(\hat q^{-4}) , \nn\\
	\hat\Pi_6 &= \frac{4}{3} \big(w_T^+ +\tilde w_T^-\big)(q_3^2,0,q_3^2) f({\hat q}^2) + c_6^{(2)} + c_6^{(3)} + \Order(\hat q^{-4}) , \nn\\
	\hat\Pi_7 &= c_7^{(5)} + \Order(\hat q^{-6}) , \nn\\
	\hat\Pi_8 &= -c_7^{(5)} + \Order(\hat q^{-6}) , \nn\\
	\hat\Pi_9 &= c_9^{(3)} + \Order(\hat q^{-4}) , \nn\\
	\hat\Pi_{10} &= -\frac{4}{3 \hat q^2} \big(w_T^+ +\tilde w_T^-\big)(q_3^2,0,q_3^2) f({\hat q}^2) - \frac{1}{\hat q^2} \Big( c_5^{(2)} + c_5^{(3)} \Big) - c_{16}^{(5)} - c_{54}^{(5)} + \Order(\hat q^{-6}) , \nn\\
	\hat\Pi_{11} &= \frac{1}{2 \hat q^2} \Big( c_5^{(2)} - c_6^{(2)} - 2 c_6^{(3)} \Big) - c_{14}^{(5)} + c_{54}^{(5)} + \Order(\hat q^{-6}) , \nn\\
	\hat\Pi_{13} &= -c_9^{(3)} + \Order(\hat q^{-4}) , \nn\\
	\hat\Pi_{14} &= -\frac{4}{3 \hat q^2} \big(w_T^+ +\tilde w_T^-\big)(q_3^2,0,q_3^2) f({\hat q}^2) - \frac{1}{\hat q^2} c_6^{(2)} + c_{14}^{(5)} + \Order(\hat q^{-6}) , \nn\\
	\hat\Pi_{16} &= -\frac{1}{2 \hat q^2} \Big( c_5^{(2)} - c_6^{(2)} \Big) + c_{16}^{(5)} + \Order(\hat q^{-6}) , \nn\\
	\hat\Pi_{39} &= \frac{4}{3 \hat q^2} \big(w_T^+ +\tilde w_T^-\big)(q_3^2,0,q_3^2) f({\hat q}^2) + \frac{1}{2\hat q^2} \Big( c_5^{(2)} + c_6^{(2)} \Big) + \Order(\hat q^{-5}) , \nn\\
	\hat\Pi_{50} &= \hat\Pi_{51} = \frac{2}{3 \hat q^2} \big(w_T^+ +\tilde w_T^-\big)(q_3^2,0,q_3^2) f({\hat q}^2) - \frac{1}{2\hat q^2} \Big( c_5^{(2)} + c_6^{(2)} - (q_1^2 - q_2^2) c_9^{(3)} \Big) + \Order(\hat q^{-5}) , \nn\\
	\hat\Pi_{54} &= \frac{1}{2\hat q^2} \Big( c_5^{(2)} - c_6^{(2)} \Big) + c_{54}^{(5)} + \Order(\hat q^{-4}) , \nn\\
	\hat\Pi_i &= \Order(\hat q^{-4}) ,  \qquad  i\in\{2,3,4,17\} ,
\end{align}
where
\beq
f({\hat q}^2)=-\frac{1}{2\pi^2{\hat q}^2}\sum_a C_a^2=-\frac{1}{18\pi^2{\hat q}^2}
\eeq
and the remaining ambiguities are parameterized by functions $c_i^{(n)}$ behaving as $c_i^{(n)} \sim 1/\hat q^n$, which are subject to certain crossing-symmetry relations following from~\eqref{eq:CrossingRelationsPiHat} and~\eqref{eq:InternalCrossingSymmetriesPiHat}. Note that the small dimensional quantity that makes the expansion parameter dimensionless can be any of the small scales, e.g., the small momentum or matrix elements of the operators in~\eqref{eq:OPEtwoCurrents}. Due to the scaling of the tensor structures, the neglected terms in~\eqref{OPE} affect the HLbL tensor first at $\Order(1/\hat q^2)$ and therefore cannot interfere with the leading-order OPE.
This result specifies the configuration $Q_1^2\sim Q_2^2\equiv -q^2=-\hat q^2\gg Q_3^2$. The related limits for small $q_1^2$ or $q_2^2$ follow directly from crossing symmetry.

Since the longitudinal amplitude $w_L$ only contributes to $\hat \Pi_{1\text{--}3}$, we will refer to these scalar functions as the longitudinal ones,
and accordingly to the remaining $\hat \Pi_i$ as the transversal contribution. The non-trivial constraint on the non-singlet part of the latter emerges
from the corresponding limit of~\eqref{non_renormalization}
\beq
\label{wLwT}
w_L(q_3^2,0,q_3^2)=2\big(w_T^+ +\tilde w_T^-\big)(q_3^2,0,q_3^2),
\eeq
but in contrast to the anomaly condition~\eqref{eq:AxialAnomaly}, which is exact in the chiral limit, this relation does receive non-perturbative corrections. As noted above,
the projection~\eqref{OPE} shows that only the OPE constraints on the longitudinal amplitudes are free from ambiguities, whereas all those on the transversal ones are affected by them. The presence of these ambiguities is not a problem per se: it simply means that at leading order the OPE only constrains certain linear combinations of BTT functions. We also note that the ambiguities would be moved to higher orders if the next terms in the OPE were included.

For asymptotic values of $q_3^2$, the OPE constraints can be compared with the pQCD quark loop evaluated in the chiral limit and for $q_1^2=q_2^2\equiv q^2$, $q_3^2/q^2\to 0$, 
\begin{align}
\label{quark_loop_MV_limit}
 \hat \Pi_1^\text{pQCD}&=-\frac{2}{3\pi^2q^2 q_3^2},\notag\\
 \hat \Pi_5^\text{pQCD}&=\hat \Pi_6^\text{pQCD}=-\frac{2}{9\pi^2q^2 q_3^2},\notag\\
 \hat \Pi_{10}^\text{pQCD}&=\hat \Pi_{14}^\text{pQCD}=-\hat \Pi_{17}^\text{pQCD}=-\hat \Pi_{39}^\text{pQCD}
 =-2\hat \Pi_{50}^\text{pQCD}=-2\hat \Pi_{51}^\text{pQCD}=\frac{2}{9\pi^2q^4 q_3^2}.
\end{align}
These expressions perfectly agree with~\eqref{OPE} if we use the non-renormalization theorem in the form~\eqref{wLwT} and set the ambiguities $c_i^{(n)}$ to zero, which demonstrates that the OPE constraint and the pQCD quark loop coincide in the appropriate kinematic limit~\cite{Crewther:1972kn} (neither chiral effects nor $\alpha_s$ corrections related to the gluon anomaly in the singlet channel matter in this limit). We stress that one could impose the OPE constraints on the transversal functions without having to deal with these ambiguities by first building linear combinations of the BTT functions that are free from them. 
In principle, one could even use the freedom in the projection at a given order to simplify expressions, e.g., at leading order one could choose the $c_i^{(n)}$ in such a way that the only non-vanishing contribution arises in $\hat \Pi_{50}=\hat \Pi_{51}=2/\hat q^2 \big(w_T^+ +\tilde w_T^-\big)(q_3^2,0,q_3^2) f({\hat q}^2)$. However, such a simplification would no longer hold at subleading orders, therefore, we keep here the general form~\eqref{OPE} that shows directly how the OPE limit corresponds to the pQCD quark loop~\eqref{quark_loop_MV_limit} evaluated in the same kinematics.  
In the following we will focus on the OPE constraint on the longitudinal contribution, which can be unambiguously assigned to the BTT functions $\hat\Pi_{1-3}$ already at leading order.

\subsection{Relation to pseudoscalar poles}
\label{sec:mixed_BTT_PS}

Separating the longitudinal OPE constraint into flavor components one finds
\begin{align}
\hat \Pi_1^3&=-\frac{6}{\pi^2 q^2 q_3^2}C_3^2=-\frac{1}{6\pi^2 q^2 q_3^2},\label{MVconstraint}\\
\hat \Pi_1^{0,8}&=-\frac{6}{\pi^2 q^2 q_3^2}\big(C_8^2+C_0^2\big)=-\frac{1}{2\pi^2 q^2 q_3^2},\label{OPE_MV}
\end{align}
which due to~\eqref{F_MV} matches precisely onto~\eqref{Pi_PS} when the meson masses and, crucially, the momentum dependence of the singly-virtual form factor are neglected. 
This is the basic premise of the model suggested in ref.~\cite{Melnikov:2003xd}. We stress that for the non-singlet component these relations are exact in the chiral limit, see section~\ref{sec:chiral_limit}
for an extensive discussion of this point. For non-vanishing quark masses and, in the case of the singlet, due to the gluon anomaly they do receive corrections. However, at low energies, 
where such corrections matter most, we always use the full dispersive result that automatically corresponds to physical quantities, while~\eqref{MVconstraint} and \eqref{OPE_MV}
are only implemented for asymptotic values of the virtualities $q_i^2$.

The OPE constraint becomes potentially valuable in the context of the mixed-energy regions, where both a description in terms of hadronic intermediate states and pQCD have limited applicability. 
In practice, the constraint is rigorous once all momenta are large compared to $\Lambda_\text{QCD}$ to ensure that quark-mass corrections can be neglected. The Regge approach in the next section is our proposal for an explicit 
implementation of the OPE constraint, following a remark made in ref.~\cite{Melnikov:2003xd}: while the $1/q_3^2$ behavior 
in~\eqref{MVconstraint} and \eqref{OPE_MV} cannot be obtained with any finite number of pseudoscalar poles, an infinite sum over excited states can produce the required asymptotics.

\section{Regge models for the pseudoscalar-pole contribution}
\label{sec:Regge}

Assuming confinement, in the large-$N_c$ limit of QCD \cite{tHooft:1973alw} the spectrum of the theory in any sector (set of quantum numbers) reduces to an infinite number of narrow resonances. One should not expect the spectral functions in this limit to be close to those of QCD with $N_c=3$ locally: a series of $\delta$-functions does not look like the continuum observed in nature for any spectral function. On the other hand, one expects the large-$N_c$ limit to provide a good approximation to QCD {\em on average}, and in particular to reproduce to a reasonable accuracy its global properties 
such as asymptotic limits. There is a vast literature on the subject that shows that these theoretical considerations can be used with good success to build large-$N_c$ models that simultaneously satisfy low- and high-energy constraints~\cite{Peris:1998nj,Knecht:1998sp,Bijnens:2003rc,Golterman:2001pj,Golterman:2001nk,DAmbrosio:2019xph}.

The aim of the present section is to construct a large-$N_c$ Regge model in the pseudoscalar and vector-meson sectors of QCD that allows us to satisfy the SDCs discussed above via an infinite tower of pseudoscalar-pole contributions. 
The logic we follow in the construction of the model is very simple: we seek minimal models, in terms of algebraic form and number of free parameters, that are able 
to satisfy all known constraints, both of experimental as well as of theoretical nature, i.e., phenomenological constraints wherever available and all known
high- and low-energy limits. Accordingly, we construct these large-$N_c$ Regge models with the application to HLbL scattering in mind and thus work with physical quark masses.  
We will comment on the chiral limit and the potential role of axial-vector resonances in section~\ref{sec:chiral_limit}.

\subsection[Large-$N_c$ Regge model for the pion transition form factor]{Large-$\boldsymbol{N_c}$ Regge model for the pion transition form factor} 
\seclab{Model}

\begin{figure}[t]
\centering
\includegraphics[height=3cm]{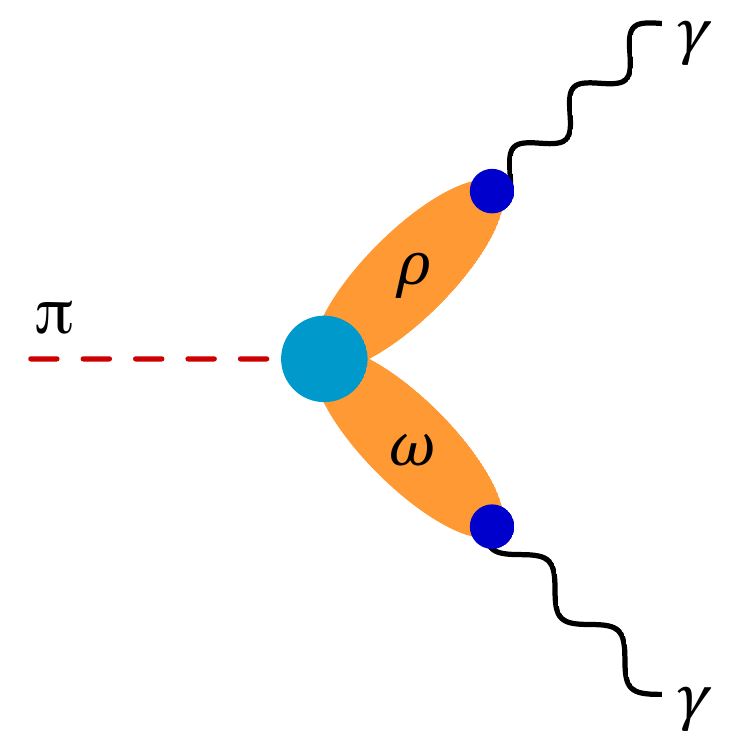}
\caption{Pion TFF in the large-$N_c$ limit. }
\label{fig:TFFlargeNc}
\end{figure}

The pion TFF describes the transition of a pion into two photons. VMD, LMD, and LMD+V models for the pseudoscalar TFFs are widely used~\cite{Knecht:2001qf,Nyffeler:2009tw}, cf.\ figures~\ref{fig:TFFlargeNc} and \ref{fig:TFFlargeNceta}. In this work, we use an untruncated large-$N_c$ model for the TFF, in which the pion couples to the photons through a tower of isovector, $I^G=1^+$, and a tower of isoscalar, $I^G=0^-$, vector mesons, $J^{PC}=1^{--}$, e.g., the $\rho$ and $\omega$, respectively. 
Here, a tower of $\rho$ ($\omega$) mesons means an infinite sum over radially-excited $\rho$ ($\omega$) mesons. The contributions from a $\phi$ instead of an $\omega$ are subdominant, 
see appendix~\ref{CouplingsApp}, and thus will be neglected for the pion.

\begin{figure}[t]
\centering 
 \includegraphics[height=5cm]{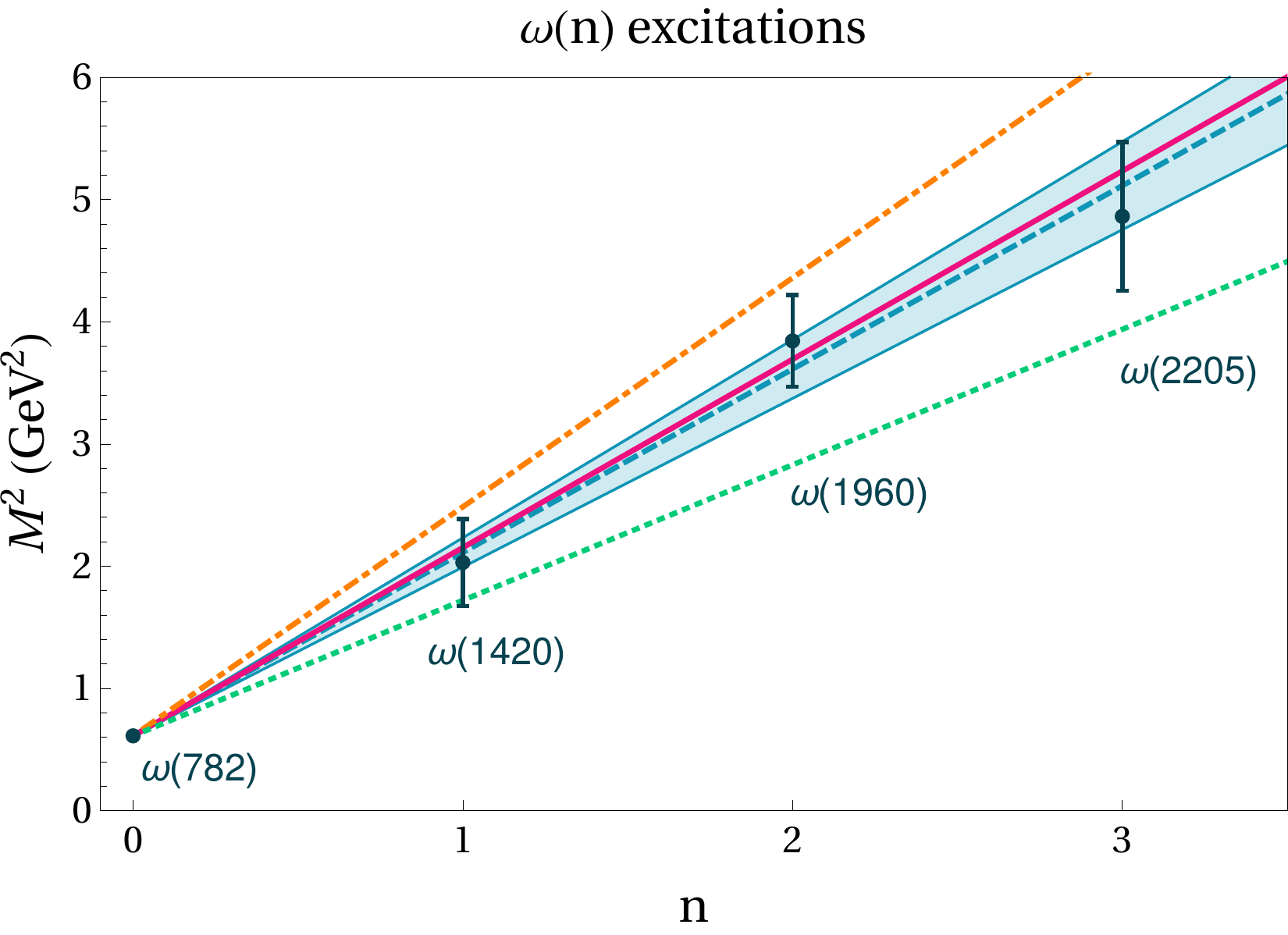}
 \hspace{0.3 cm}
 \includegraphics[height=5cm]{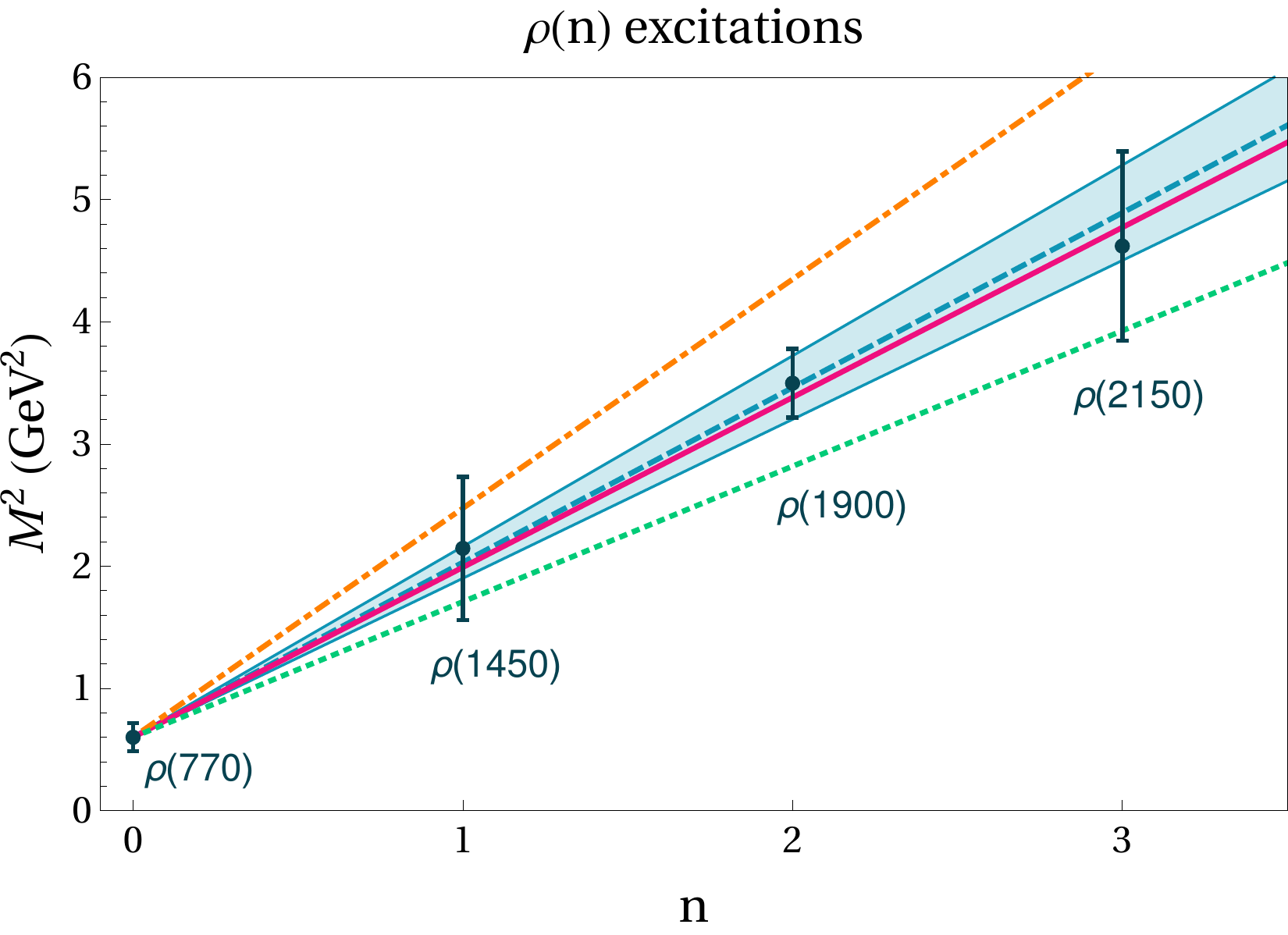}
 \includegraphics[height=5cm]{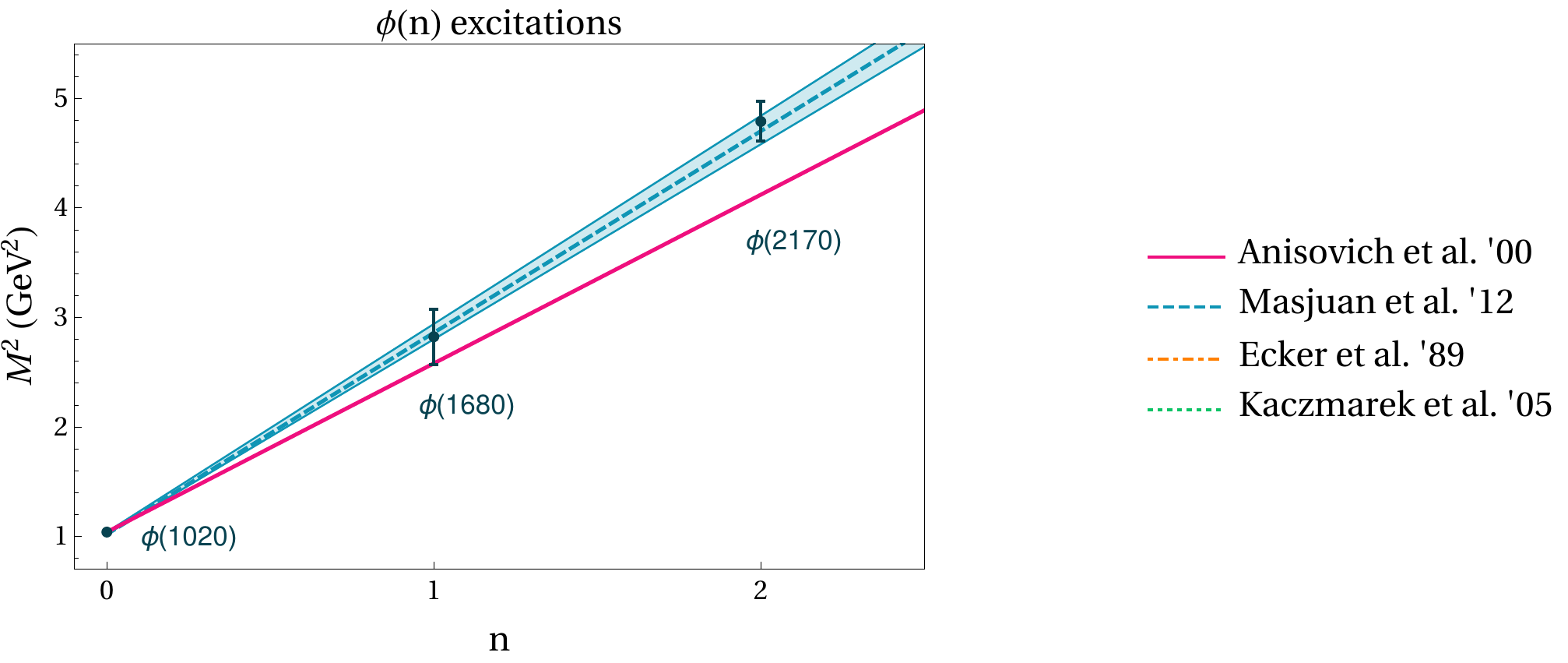}
 \hspace{2.5cm}
\caption{Radial Regge trajectories of the isovector $\rho$ and isoscalar $\omega$ and $\phi$ vector mesons. The states $\omega(782)$, $\omega(1420)$, $\rho(770)$, $\rho(1450)$, $\phi (1020)$, $\phi(1680)$, and $\phi(2170)$ are from PDG~\cite{Tanabashi:2018oca}. The states $\omega (1960)$, $\omega (2205)$, $\rho (1900)$, and $\rho(2150)$ are extracted from ref.~\cite{Masjuan:2012gc}. 
The errors are defined as $\Delta M^2=\Gamma M$~\cite{Masjuan:2012gc}. The solid magenta lines are from ref.~\cite{Anisovich:2000kxa} with $\sigma^2_\omega=1.54$, $\sigma^2_\rho=1.39$ GeV$^2$, and $\sigma^2_\phi=1.54$ GeV$^2$. The turquoise bands are from ref.~\cite{Masjuan:2012gc} with $\sigma^2_\omega=1.50(12)$ GeV$^2$, $\sigma^2_\rho=1.43(13)$ GeV$^2$, and $\sigma^2_\phi=1.84(6)$ GeV$^2$. 
The green dotted lines with slope $\sigma^2=1.11$ GeV$^2$ are based on the lattice calculation of ref.~\cite{Kaczmarek:2005ui}. The orange dot-dashed lines with slope $\sigma^2=1.87$ GeV$^2$ are based on the $\rho \rightarrow 2 \pi$ decay~\cite{RuizArriola:2006jge,Ecker:1988te}.}
\label{fig:TrajectoriesPlot}
\end{figure}

The standard large-$N_c$ ansatz for the pion TFF (see refs.~\cite{RuizArriola:2006jge,Arriola:2010aq}) reads:
\beq
F_{\pi^0 \ga^* \ga^*}(-Q_1^2,-Q_2^2)=\sum_{V_\rho, V_\omega}G_{\pi V_\rho V_\omega} \, F_{V_\rho}\,F_{V_\omega} \left[ \frac{1}{D^1_{V_\rho}D^2_{V_\omega}} + \frac{1}{D^1_{V_\omega}D^2_{V_\rho}}\right],\label{GeneralVMD}
\eeq
where 
\beq
D^i_X:=Q_i^2+M_X^2,
\eeq
$F_{V_\rho}$ and $F_{V_\omega}$, represented by blue dots in figure~\ref{fig:TFFlargeNc}, are the current--vector-meson couplings and $G_{\pi V_\rho V_\omega}$, the cyan dot in figure~\ref{fig:TFFlargeNc}, is the coupling of two vector mesons to the neutral pion. 
We stress that the couplings in~\eqref{GeneralVMD} are $Q^2$ independent as required by the large-$N_c$ approach (combined with analyticity): 
the contribution of a given intermediate state is fixed by its imaginary part, which for narrow resonances is a $\delta$-function, which freezes any $Q^2$ dependence. Indeed the latter could be interpreted as coming from the continuum between resonances, which is suppressed in the large-$N_c$ limit.  
Another potential source of $Q^2$-dependent corrections to~\eqref{GeneralVMD} is related to subtraction terms: while~\eqref{GeneralVMD} follows from an unsubtracted double-spectral representation
for the TFF, introducing subtractions would produce single-propagator terms and, eventually, a polynomial. However, for a $\delta$-function subtractions are not necessary, and even 
before taking the large-$N_c$ limit the pQCD behavior of the TFF implies that an unsubtracted representation holds. 
As argued in ref.~\cite{Hoferichter:2018kwz}, the advantage of using an unsubtracted dispersion relation, in favor of a subtracted variant that could suppress some of the high-energy input, 
is precisely that it allows one to manifestly incorporate the correct pQCD asymptotics. The large-$N_c$ ansatz~\eqref{GeneralVMD} corresponds to this scenario. 
In the following, the vector-meson spectra are assumed to obey a radial Regge model, see figure~\ref{fig:TrajectoriesPlot}:
\begin{align}
M_{V_\rho}^2&=M_{\rho(n_\rho)}^2=M_{\rho}^2+n_\rho\,\sigma_\rho^2,\nn\\
M_{V_\omega}^2&=M_{\omega(n_\omega)}^2=M_{\omega}^2+n_\omega\,\sigma_\omega^2,\label{nonIdenticalSpectra}
\end{align}
where  $\sigma_\rho$ and $\sigma_\omega$ are the slope parameters of the Regge trajectories, $n_\rho$ and $n_\omega$ are radial excitation numbers, and the ground-state masses are $M_{\rho}=M_{\rho(770)}=775.26(25)\MeV$ and $M_{\omega}=M_{\omega(782)}=782.65(12)\MeV$~\cite{Tanabashi:2018oca}.

\begin{figure}[t]
  \includegraphics[height=5cm]{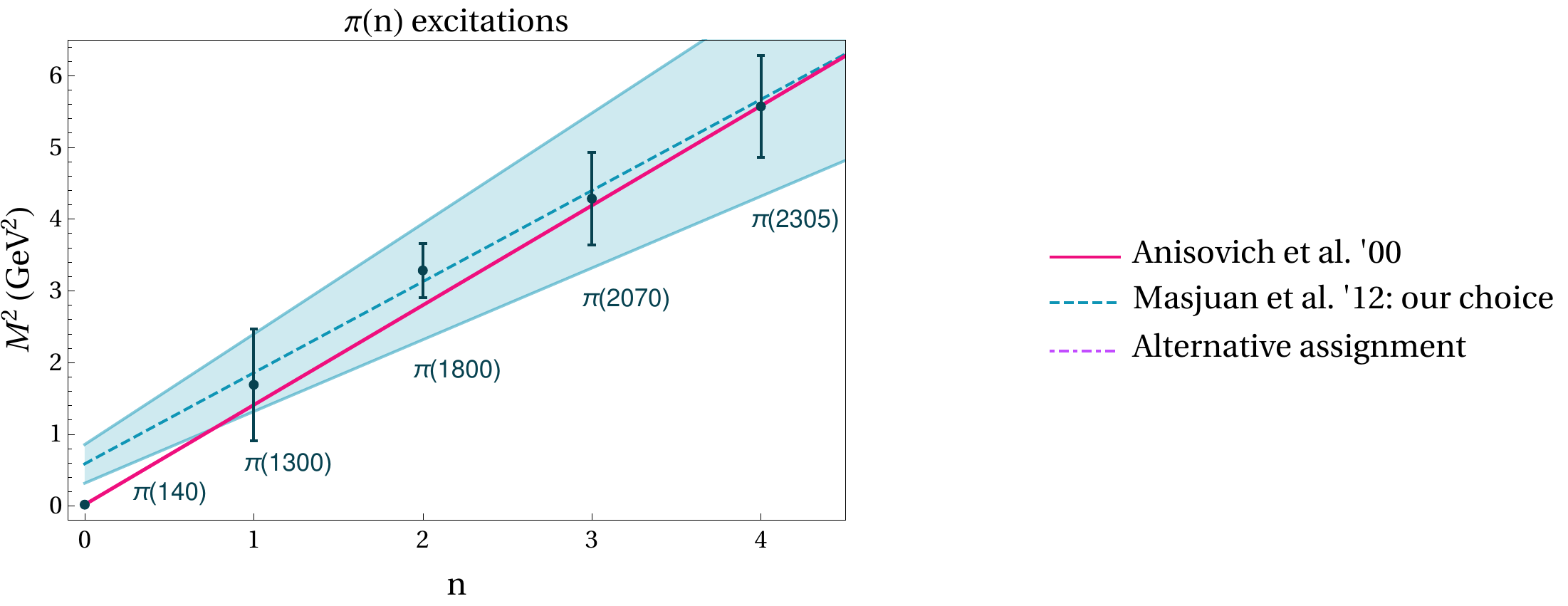}\hspace{2.5cm}
  
   \includegraphics[height=5cm]{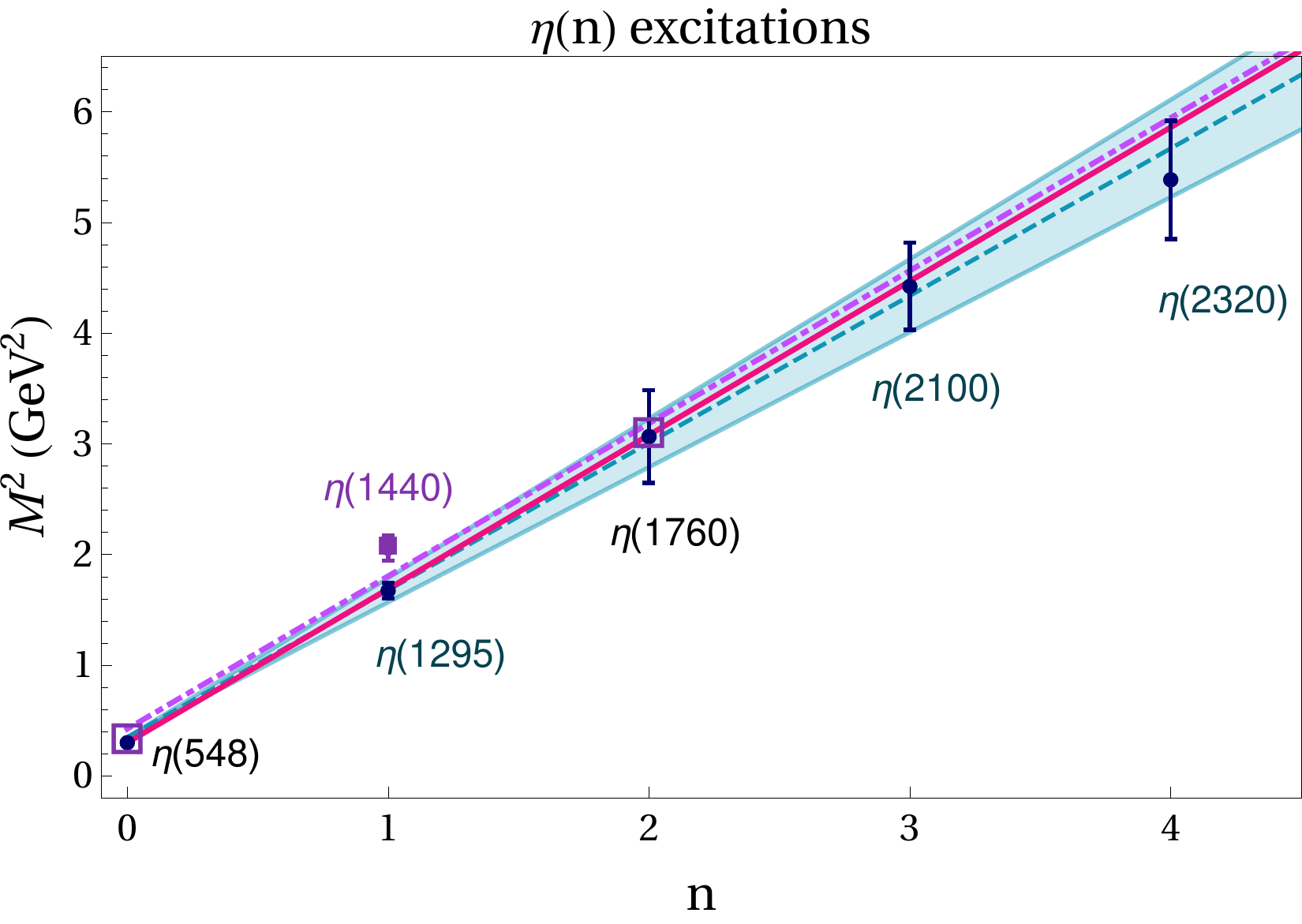}
   \hspace{0.4cm}
    \includegraphics[height=5cm]{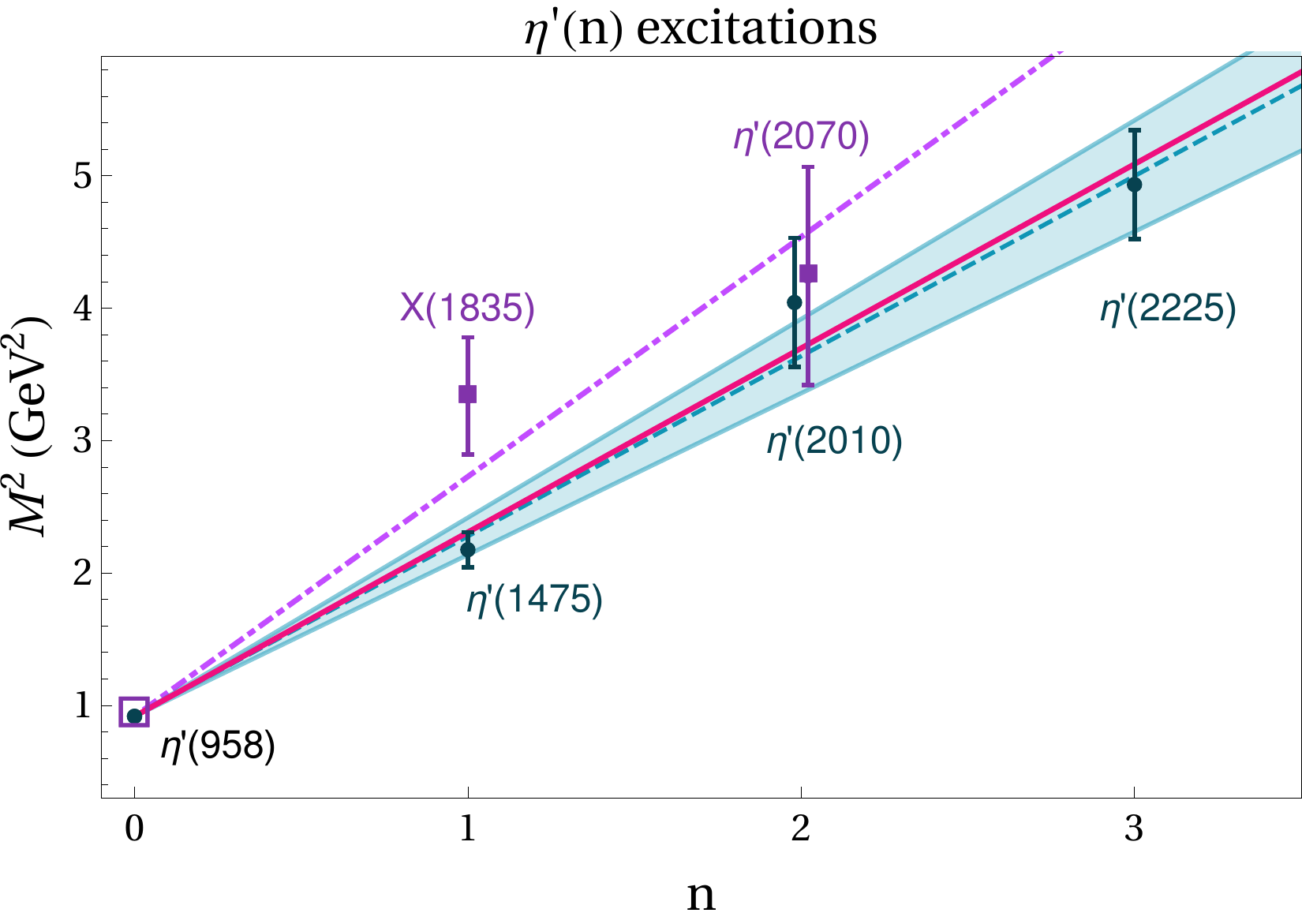}
\caption{Radial Regge trajectories of the $\pi$, $\eta$, and $\eta '$ pseudoscalar mesons. The states $\pi(140)$, $\pi(1300)$, $\pi (1800)$, $\eta (548)$, $\eta (1295)$, $\eta (1760)$, $\eta ^\prime (958)$, $\eta ^\prime (1475)$, $X(1835)$, and $\eta ^\prime (2225)$ are from PDG~\cite{Tanabashi:2018oca}. Note that the states $\eta (1760)$, $X(1835)$, and $\eta ^\prime (2225)$ are omitted from the PDG summary tables. The state $\eta(1440)$ is from PDG '00 \cite{Groom:2000in}. The states $\pi(2070)$, $\pi(2305)$, $\eta (2100)$, $\eta (2320)$, and $\eta ^\prime (2010)$ are extracted 
from ref.~\cite{Masjuan:2012gc}. The state $\eta'(2070)$ is taken from \cite[Table 26]{Klempt:2007cp}. The errors are defined as $\Delta M^2=\Gamma M$~\cite{Masjuan:2012gc}. The solid magenta lines are fits from ref.~\cite{Anisovich:2000kxa}: $\sigma^2_\pi=\sigma ^2 _\eta = \sigma ^2 _{\eta ^\prime}=1.39$ GeV$^2$. The turquoise bands are fits from ref.~\cite{Masjuan:2012gc} which exclude the ground states of the pion and the $\eta$: $\sigma^2_\pi=1.27(27)$ GeV$^2$ and $\hat M_\pi=766$ MeV as in~\eqref{pionTrajectory}, $\sigma_\eta ^2=1.33(11)$ GeV$^2$ and $\hat M_\eta=591$ MeV as in~\eqref{etaTrajectory}, and $\sigma ^2_{\eta ^\prime}=1.36(14)\text{ GeV}^2$. 
The $\eta (1440)$, $X(1835)$, and $\eta ^\prime (2070)$ states (purple squares) correspond to a different assignment of $\eta^{(\prime)}$ excitations suggested in ref.~\cite[Table 27]{Klempt:2007cp}. The dot-dashed purple lines correspond to our fits of these alternative trajectories: $\sigma _\eta^2=1.38\GeV^2$ with $\hat{M}_\eta=0.652\text{ GeV}$, and $\sigma _{\eta ^\prime}^2=1.81\GeV^2$. }
\label{fig:TrajectoryPlotPion}
\end{figure}

Having fixed the masses of the towers of vector resonances, our model for the pion TFF still has an infinite number of parameters, namely the couplings $G_{\pi V_\rho V_\omega}$, $F_{V_\rho}$, and $F_{V_\omega}$. One could in principle reduce the number of free parameters to a finite one by imposing a certain algebraic dependence of these couplings on the excitation numbers $n_\rho$ and $n_\omega$, as has been done for the masses. In doing so one would have to be able to satisfy low- and high-energy constraints for the pion TFF, which we recollect here from~\eqref{AnomalySec2} and \eqref{BLOPESec2}:
\begin{align}
\text{chiral anomaly~\cite{Adler:1969gk,Bell:1969ts,Bardeen:1969md}:}\qquad& F_{\pi^0\gamma\gamma}=\frac{1}{4\pi^2F_\pi};\label{Constraint1}\\ 
\text{BL limit~\cite{Lepage:1980fj,Brodsky:1981rp}:}\qquad& \lim_{Q^2\rightarrow \infty} Q^2  F_{\pi^0 \ga\ga^*}(-Q^2)= 2F_\pi  ; \label{Constraint3}\\
\text{symmetric pQCD limit~\cite{Novikov:1983jt}:}\qquad& \lim_{Q^2\rightarrow \infty} Q^2  F_{\pi^0 \ga^*\ga^*}(-Q^2,-Q^2)=\frac{2F_\pi}{3} \label{Constraint2}.
\end{align}
One immediately notices that while the $Q^2$ dependence of each individual term in~\eqref{GeneralVMD} is compatible with the Brodsky--Lepage (BL) limit, the symmetric pQCD limit can only be satisfied after resumming the series of vector resonances. To this end, the coupling constants must be arranged in such a way that the $Q^{-4}$ behavior of the individual terms becomes a $Q^{-2}$ behavior after resummation. That this is possible was shown in refs.~\cite{RuizArriola:2006jge,Arriola:2010aq}.

In addition, the pion TFF has been measured quite well in the singly-virtual case~\cite{Behrend:1990sr,Gronberg:1997fj,Aubert:2009mc,Uehara:2012ag}, and our model for the TFF would have to describe the data. For the doubly-virtual case a recent dispersive analysis has shown that data for related processes and theoretical arguments constrain the behavior of the TFF in that kinematical region~\cite{Hoferichter:2014vra,Hoferichter:2018dmo,Hoferichter:2018kwz}---a constraint we will also take into account. 

Imposing all these constraints on the model~\eqref{GeneralVMD} by adjusting its free parameters is technically cumbersome, especially if we consider that we must still add a third sum over the tower of pseudoscalar mesons, $J^{PC}=0^{-+}$, cf.\ figure~\ref{fig:HLbLpi0}, with which we aim to change the large-$Q^2$ behavior of the whole HLbL tensor. In particular, we will implement the SDCs on $\hat \Pi_1$ introduced in~\eqref{pQCDloop} and~\eqref{MVconstraint}:\footnote{Note that while for the MV SDC, which is derived based on the $VVA$ triangle, the flavor decomposition into pion, $\eta$, and $\eta'$ is unambiguously given by $C_a^2$, see~\eqref{MVconstraint}, the decomposition presented here for the SDC in the asymptotic region~\eqref{pQCDloop} is not unique. We choose to adopt the same separation 
as for the MV constraint.}
\begin{align}
\text{SDC for the mixed region \cite{Melnikov:2003xd}:}&\qquad\lim_{Q_3^2 \rightarrow \infty}\lim_{ Q^2 \rightarrow \infty}\,\sum _{n=0}^\infty  \hat{\Pi} _1^{\pi(n)\text{-pole}}(-Q^2,-Q^2,-Q_3^2)\nn\\
&\hspace{-8pt}\quad=-\lim_{Q_3^2 \rightarrow \infty}\lim_{ Q^2 \rightarrow \infty}\,\sum _{n=0}^\infty\frac{ F_{\pi(n) \gamma^* \gamma^*}( -Q^2, -Q^2)\,F_{\pi(n) \gamma \gamma^*}(-Q_3^2)}{Q_3^2+M_{\pi(n)}^2}\nn\\
&\hspace{-8pt}\quad=-\frac{1}{6\pi^2}\frac{1}{ Q^2} \frac{1}{Q_3^2} ;\label{MVConstraintSimple}\\
\text{SDC for the asymptotic region:}&\qquad\lim_{ Q^2 \rightarrow \infty}\,\sum _{n=0}^\infty  \hat{\Pi} _1^{\pi(n)\text{-pole}}(-Q^2,-Q^2,-Q^2)\nn\\
&\hspace{-8pt}\quad=-\lim_{ Q^2 \rightarrow \infty}\,\sum _{n=0}^\infty\frac{ F_{\pi(n) \gamma^* \gamma^*}( -Q^2, -Q^2)\, F_{\pi(n) \gamma \gamma^*}( -Q^2)}{ Q^2+M_{\pi(n)}^2}\nn\\
&\hspace{-8pt}\quad=-\frac{4}{9\pi^2}\frac{C_3^2}{\sum_{a=0,3,8} C_a^2}\frac{1}{ Q^4}.\label{pQCDSimple}
\end{align}
Here, $F_{\pi(n) \gamma^* \gamma^*}$ is the TFF of the $n$-th radially-excited pion and a radial Regge model is assumed for the pion masses starting from the first excitation, see figure~\ref{fig:TrajectoryPlotPion}:
\beq
M_{\pi(n)}^2=\begin{cases}M_{\pi}^2& n=0,\\
\hat M_{\pi}^2+n\, \sigma_\pi^2 &n \geq 1,\end{cases} \label{pionTrajectory}
\eeq
where $M_{\pi}=134.9770(5)$ MeV is the $\pi^0$ mass~\cite{Tanabashi:2018oca}.\footnote{Note that the ground-state is treated separately because for the Goldstone bosons a strong non-linearity of the Regge trajectory is expected~\cite{Arriola:2011en}.}
Given the complexity of implementing all these constraints simultaneously in terms of the general couplings of the Regge model, we therefore adopted a different approach:
\begin{enumerate}
    \item we allow the ground-state pion to couple only to the ground-state $\rho$ and $\omega$ mesons, and the $n$-th pion excitation to couple only to the $n$-th $\rho$ and $\omega$ excitations;
    \item we subsume the effect of the vector-meson excitations that we have just eliminated into a $Q_i^2$ dependence of the numerator multiplying the resonance propagators;
    \item the latter $Q_i^2$ dependence will be parameterized in simple terms with as few free parameters as necessary to satisfy the constraints listed above.
\end{enumerate}
The first step is motivated by the fact that non-diagonal couplings are suppressed by the reduced overlap of radial wave functions with different numbers of nodes~\cite{RuizArriola:2006jge}. For the same reason we are only considering the leading $S$-wave vector-meson trajectories and neglecting the $D$-wave daughter trajectories.
In appendix~\ref{sec:PionAlternative} we will consider an alternative model that already for the pQCD limit of the TFF itself, cf.\ \eqref{Constraint2}, uses the Regge resummation from ref.~\cite{RuizArriola:2006jge}, 
but for the main text we restrict the presentation to the most economical form sufficient to fulfill all constraints simultaneously. This strategy leads us to
\begin{align}
 \label{TFFpi}
F_{\pi(n) \gamma^* \gamma^*}(-Q_1^2,-Q_2^2)
&=\frac{1}{8 \pi^2
    F_\pi}   \left\{ \left(\frac{M_\rho^2 M_\omega^2}{D_{\rho(n)}^1 D_{\omega(n)}^2} +
  \frac{M_\rho^2 M_\omega^2}{D_{\rho(n)}^2 D_{\omega(n)}^1}
\right)
\right. \\ 
 &\times\left[c_\mathrm{anom}+c_A \frac{M_{+,\,n}^2}{\Lambda^2}+c_B \frac{M_{-,\,n}^2}{\Lambda^2}
         + c_\mathrm{diag} 
\frac{Q_1^2 Q_2^2}{\Lambda^2(Q_+^2+M^2_\mathrm{diag})} \right]\nn\\
&+\frac{Q_-^2}{Q_+^2}
\left(\frac{M_\rho^2 M_\omega^2}{D_{\rho(n)}^1 D_{\omega(n)}^2} -
  \frac{M_\rho^2 M_\omega^2}{D_{\rho(n)}^2 D_{\omega(n)}^1}
\right)
\left.
\left[c_\mathrm{BL}+c_A \frac{M_{-,\,n}^2}{\Lambda^2}+c_B \frac{M_{+,\,n}^2}{\Lambda^2}
\right]
 \right\}, \nonumber
\end{align}
where
\begin{equation}
M_{\pm,\,n}^2=\frac{1}{2} \left(M_{\omega(n)}^2\pm M_{\rho(n)}^2 \right)\,, \quad
Q_\pm^2=Q_1^2\pm Q_2^2 \,, \label{def1}
\end{equation}
and $\Lambda = \mathcal{O}(1\, \text{GeV})$ is a typical QCD scale introduced to
make all model parameters ($c_\mathrm{anom}$, $c_A$, $c_B$, $c_\mathrm{diag}$, $c_\mathrm{BL}$) dimensionless. The second mass
scale $M_\mathrm{diag}$ is determined by fitting the experimental data, it parameterizes the doubly-virtual behavior of the TFF.

With this parameterization, the three conditions for the TFF of the ground-state pion from~\eqref{Constraint1}--\eqref{Constraint2} can be expressed as follows:
\begin{align}
\mbox{anomaly:}\qquad &1= c_\mathrm{anom}+\frac{1}{\Lambda^2}\left(
        c_A M_{+,\,0}^2+c_B M_{-,\,0}^2\right);\label{ConstraintEquations1} \\
\mbox{BL limit:}\qquad & 1= \frac{1}{8 \pi^2 F_\pi^2}
\left(c_\mathrm{anom}M_{+,\,0}^2-c_\mathrm{BL} M_{-,\,0}^2 + 
  c_A \frac{M_\omega^2 M_\rho^2}{\Lambda^2} \right); \label{ConstraintEquationsBL} \\
\mbox{symmetric pQCD limit:} \qquad &1= \frac{3 M_\omega^2 M_\rho^2}{16 \pi^2 F_\pi^2
  \Lambda^2}\, c_\mathrm{diag}.\label{ConstraintEquations}
\end{align}
Since the mass scales $M_\rho$, $M_\omega$, $M_{+,\,0}$, and $\Lambda$ as well as
$\pi^2 F_\pi$ are of about the same order, all coupling constants that
appear in the constraint equations~\eqref{ConstraintEquations1}--\eqref{ConstraintEquations} multiplied by ratios of these mass scales are
expected to be of $\mathcal{O}(1)$. $M_{-,\,0}$, on the other hand, is much
smaller, so that the coupling constants multiplied by it ($c_\mathrm{BL}$
and $c_B$) are expected to be of $\mathcal{O}(\Lambda^2/M_{-,\,0}^2)\sim 100$,
otherwise their role in the equations would become irrelevant.

Of course these three conditions are not sufficient to determine all five
model parameters. Two more constraints follow from resumming the contributions of all
excited pseudoscalars to the HLbL amplitude, see~\eqref{MVConstraintSimple} and \eqref{pQCDSimple}. Details on the evaluation of infinite sums over rational functions can be found in appendix \ref{sec:PionAlternativeProof}. The MV SDC for the mixed region translates into:
\begin{equation}
\label{MV}
1=\frac{M_\omega^2 M_\rho^2}{2 \Lambda^2} \left[ (c_A+c_B)
  \frac{L_{\rho\pi}}{\Delta_{\rho\pi}}
  +(c_A-c_B)\frac{L_{\omega\pi}}{\Delta_{\omega\pi}}  \right],  
\end{equation}
where $c_\mathrm{diag}$ from~\eqref{ConstraintEquations} has been used and
\begin{equation}
L_{ij}=\log \frac{\sigma_i^2}{\sigma_j^2},\quad
\Delta_{ij}:=\sigma_i^2-\sigma_j^2.\label{def2}
\end{equation}
The second SDC concerns the limit $Q_i^2=Q^2 \to \infty$, for all
$i=1,2,3$. It also involves $c_{A}$ and $c_{B}$, but now in a different combination 
together with $c_\mathrm{diag}$:
\begin{align}
\label{pQCD3}
1&=\frac{9}{64 \pi^2 F_\pi^2\Lambda^4}\frac{M_\rho^4
 M_\omega^4}{\Omega_{\rho\omega\pi}^2}
\Bigg\{c_A^2\,\Delta_{\rho\omega} \Sigma_{\rho\omega}\Big[\sigma_\pi^2 \left( 
      \Delta_{\omega\pi}^2 L_{\rho\pi} -\Delta_{\rho \pi}^2
      L_{\omega\pi}\right) + \Omega_{\rho\omega\pi} 
     \Big]  \nonumber\\
&+c_B^2\, \Delta_{\rho\omega}\Big[\left(\Sigma_{\rho\omega}\sigma_\pi^2
    -2 \sigma_\rho^4\right) \Delta_{\omega\pi}^2 L_{\rho\pi} -
     \left(\Sigma_{\rho\omega} \sigma_\pi^2
    -2 \sigma_\omega^4\right)\Delta_{\rho\pi}^2 L_{\omega\pi}
 +(\Delta_{\rho\pi}+\Delta_{\omega\pi} )\,\Omega_{\rho\omega\pi} \Big] \nonumber \\ 
 &-2c_A c_B \Big[ \sigma_\pi^2
     \left(\sigma_\rho^4+\sigma_\omega^4\right)\left(\Delta_{\omega\pi}^2L_{\rho\pi}-\Delta_{\rho\pi}^2L_{\omega\pi} \right)+\Sigma_{\rho\omega}\left(\sigma_\omega^4\Delta_{\rho\pi}^2L_{\omega\pi}-\sigma_\rho^4\Delta_{\omega\pi}^2L_{\rho\pi}\right)\nn\\
 &-\Omega_{\rho\omega\pi}\left(\sigma_\pi^2
   \Sigma_{\rho\omega}-\sigma_\rho^4-\sigma_\omega^4\right) \Big]-c_\mathrm{diag}c_B
\Big[ \sigma_\rho^2\left\{\sigma_\rho^4+\sigma_\omega^2(\sigma_\rho^2 
-2\sigma_\pi^2)\right\} \Delta_{\omega\pi}^2 L_{\rho\pi} \nonumber \\
&-
  \sigma_\omega^2\left\{\sigma_\omega^4+\sigma_\rho^2(\sigma_\omega^2- 
  2 \sigma_\pi^2) \right\}\Delta_{\rho\pi}^2 L_{\omega\pi} +
\left(\sigma_\pi^2 \Sigma_{\rho \omega} -2 \sigma_\rho^2
  \sigma_\omega^2\right) \Omega_{\rho\omega\pi}  \Big]\nn\\
&+ c_\mathrm{diag}c_A\, \Delta_{\rho\omega}\Big[ \sigma _\rho ^2(\sigma ^2_\rho -2\sigma_\pi
  ^2)\Delta _{\omega \pi}^2L_{\rho \pi}-\sigma _\omega ^2 (\sigma _\omega
  ^2-2\sigma _\pi ^2)\Delta _{\rho  \pi}^2 L_{\omega \pi }-\sigma _\pi ^2\,
  \Omega_{\rho\omega\pi} \Big] \Bigg\}, 
\end{align}
with
\begin{equation}
\Omega_{ijk}\coloneqq(\sigma_i^2-\sigma_j^2)(\sigma_k^2-\sigma_i^2)(\sigma_k^2-\sigma_j^2),
\quad \Sigma_{ij}\coloneqq\sigma_i^2+\sigma_j^2 .\label{def3}
\end{equation}

In appendix~\ref{sec:PionModel1CouplingParameters},
this system of equations is solved analytically. Here, we discuss numerical
values for all parameters, also summarized in table~\ref{TableModel1}, based on the following choice of
Regge slopes~\cite{Masjuan:2012gc}:\footnote{From figure~5 of ref.~\cite{Masjuan:2012gc} we extracted $\hat M_\pi=766\, \mathrm{MeV}$.}
\begin{equation}
\sigma_\pi^2 = 1.27(27) \, \mathrm{GeV}^2,\quad \sigma_\rho^2= 1.43(13) \, 
\mathrm{GeV}^2, \quad \sigma_\omega^2= 1.50(12) \, \mathrm{GeV}^2. \label{ReggeSlopes1}
\end{equation}
Furthermore, we use $F_\pi=92.28$ MeV, $\Lambda=1$ GeV, and other input from the PDG~\cite{Tanabashi:2018oca}. 
\begin{table}[t]
\begin{center}
\begin{tabular}{crrr}
\toprule
&$\pi$& $\eta$&$\eta'$ \\ 
 \midrule
$c_\mathrm{anom}$&$-1.670$&---&---\\
$c_A$&$6.794$&$2.542$&$2.635$\\
$c_B$&$-252.346$&$-23.535$&$-23.706$\\
$c_\mathrm{diag}$&$1.218$&$0.401$&$0.502$\\
$c_\mathrm{BL}$&$141.688$&$18.721$&$18.877$\\
$M_\mathrm{diag}$&$1.519$&$0.898$&$0.898$\\
 \bottomrule 
\end{tabular}
\end{center}
\caption{``Natural'' model parameters of the large-$N_c$ Regge models for the pion, $\eta$, and $\eta'$ TFFs. Note that here we rescaled the $\eta^{(\prime)}$ parameters $c_A$, $c_B$, and $c_\mathrm{BL}$ with a factor of $C_{\phi \omega}^{\eta^{(\prime)}}/\mathcal{N}$. \label{TableModel1}}
\end{table}

The constant $c_\mathrm{diag}$ is independent of all the others and is
directly determined by~\eqref{ConstraintEquations}:
\begin{equation}
c_\mathrm{diag}=1.218.
\end{equation}
Once this is fixed, equations~\eqref{MV} and \eqref{pQCD3} determine $c_A$ and
$c_B$. Since the second equation is quadratic it has two solutions, but one can
be readily discarded because the two-photon couplings of the excited pions become unreasonably large, and so do the values of the constants $c_A$ and $c_B$. 
The physical solution gives:
\begin{equation}
c_A=6.794, \qquad  c_B=-252.346 .
\end{equation}
Having determined $c_A$ and $c_B$, \eqref{ConstraintEquations1} determines
$c_\mathrm{anom}$ to the value:
\begin{equation}
c_\mathrm{anom}=-1.670,
\end{equation}
and finally~\eqref{ConstraintEquationsBL} fixes the remaining parameter:
\begin{equation}
c_\mathrm{BL}=141.688 .
\end{equation}
As expected, all constants are of $\mathcal{O}(1)$, with the exception of $c_\mathrm{BL},
\; c_B \sim \mathcal{O}(100)$. 

\begin{figure}[t]
\centering
\includegraphics[width=0.9\linewidth]{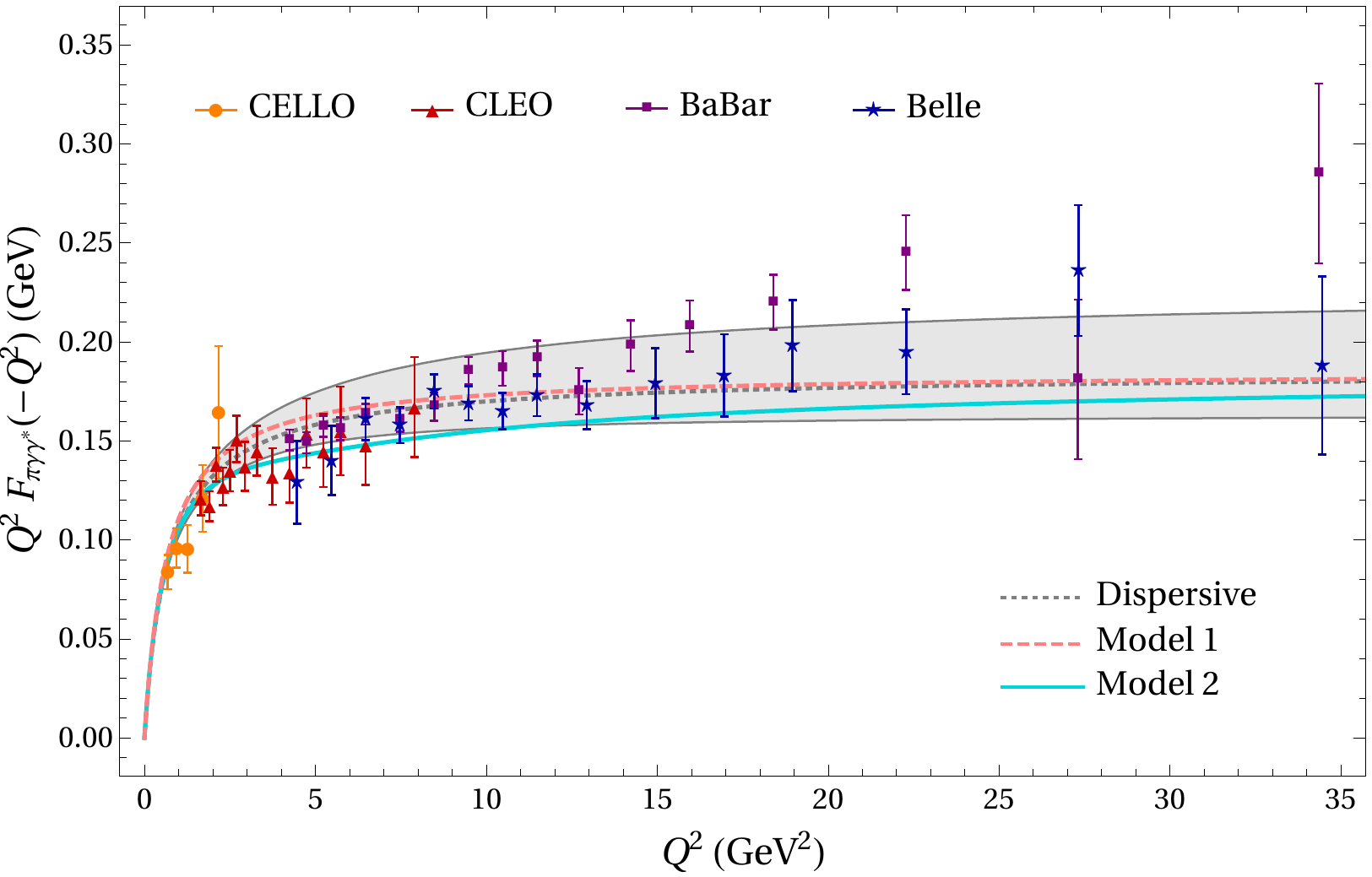}
\caption{Singly-virtual $\pi^0$ TFF. The large-$N_c$ Regge model, ``Model 1''~\eqref{TFFpi}, is indicated by the dashed pink curve. Our alternative TFF model, ``Model 2''~\eqref{MartinModel}, is indicated by the solid cyan curve. The gray band with the dotted curve is the dispersive result from refs.~\cite{Hoferichter:2018dmo,Hoferichter:2018kwz}. The data are from CELLO~\cite{Behrend:1990sr}, CLEO~\cite{Gronberg:1997fj}, BaBar~\cite{Aubert:2009mc}, and Belle~\cite{Uehara:2012ag}.}
\label{fig:SinglyVirtual}
\end{figure}

Since there is no direct empirical information on the doubly-virtual $\pi^0$ TFF available, we fit our model parameter $M_\text{diag}$ to 
the dispersive description of the $\pi^0$ TFF from refs.~\cite{Hoferichter:2018dmo,Hoferichter:2018kwz}. To find the best fit, we minimize the estimated variance: 
\beq
\chi^2=\frac{1}{j_\mathrm{max}-p}\sum_{j=1}^{j_\mathrm{max}}\left(\frac{f(-Q_{1,\,j}^2,-Q_{2,\,j}^2)-f^\text{data}(-Q_{1,\,j}^2,-Q_{2,\,j}^2)}{\Delta f^\text{data}(-Q_{1,\,j}^2,-Q_{2,\,j}^2)}\right)^2,
\eeq
where $j_\mathrm{max}$ is the length of the data set and $p$ is the number of fit parameters. Here,  $f$ is our model and $f^\mathrm{data}$ is the 
dispersive TFF.\footnote{Since the error band of the dispersive TFF is asymmetric, for each kinematic point its smallest value was extracted to obtain $\Delta f^\mathrm{data}(-Q_{1,\,j}^2,-Q_{2,\,j}^2)$.} The sum is over $j_\mathrm{max}= \mathcal{O}(2\!\times\!10^4)$ selected points in the region of $0<Q_1\leq Q_2$, where $Q_2^2 \in [0,40]\,\text{GeV}^2$. As a result, we obtain
$M_\mathrm{diag}=1.519\GeV$ with $\chi^2 \sim 0.37$.

The singly-virtual TFF of the ground-state pion is shown in figure~\ref{fig:SinglyVirtual}. The large-$N_c$ Regge model presented above is labeled as ``Model 1.'' In appendix~\ref{sec:PionAlternative}, 
an alternative TFF model, to which we refer as ``Model 2,'' is introduced, based on a Regge resummation for the TFF itself. 
Both models give a reasonable description of the experimental data, while Model 1 shows better agreement with the dispersive TFF in the intermediate-$Q$ region. In appendix~\ref{sec:plotsPion}, both models are shown also in the doubly-virtual region and further compared to the dispersive TFF~\cite{Hoferichter:2014vra,Hoferichter:2018dmo,Hoferichter:2018kwz}, a prediction from lattice QCD~\cite{Gerardin:2019vio}, and a result from DSE~\cite{Eichmann:2019tjk}. We stress that neither model should be evaluated for other than purely space-like virtualities, both are constructed in such a way as to provide an efficient implementation of all constraints relevant for 
the space-like region, but do not properly incorporate the analytic structure required to continue to time-like virtualities. In addition to our fits to the dispersive $\pi^0$ TFF, we also 
checked that the $\pi^0$ contribution to $(g-2)_\mu$ is reproduced correctly
\begin{align}
 a_\mu^{\pi^0\text{-pole}}\big|_\text{Model 1}&=64.3\times 10^{-11},\qquad a_\mu^{\pi^0\text{-pole}}\big|_\text{Model 2}= 64.5\times 10^{-11},\notag\\
 a_\mu^{\pi^0\text{-pole}}\big|_\text{\cite{Hoferichter:2018dmo,Hoferichter:2018kwz}}&=62.6^{\,+3.0}_{\,-2.5}\times 10^{-11}.\label{piong-2}
\end{align}
Finally, as detailed in ref.~\cite{Hoferichter:2018kwz}, effective-field-theory constraints on the pseudoscalar-pole contributions~\cite{Knecht:2001qg,RamseyMusolf:2002cy} are automatically encoded in the TFF phenomenology, for the leading constraint in its normalization, for the subleading one in the momentum dependence.

\subsection[Large-$N_c$ Regge model for the $\eta$ and $\eta'$ transition form factors]{Large-$\boldsymbol{N_c}$ Regge model for the $\boldsymbol{\eta}$ and $\boldsymbol{\eta'}$ transition form factors} 
\label{sec:ModelEta}

\begin{figure}[t]
\centering
\includegraphics[height=3cm]{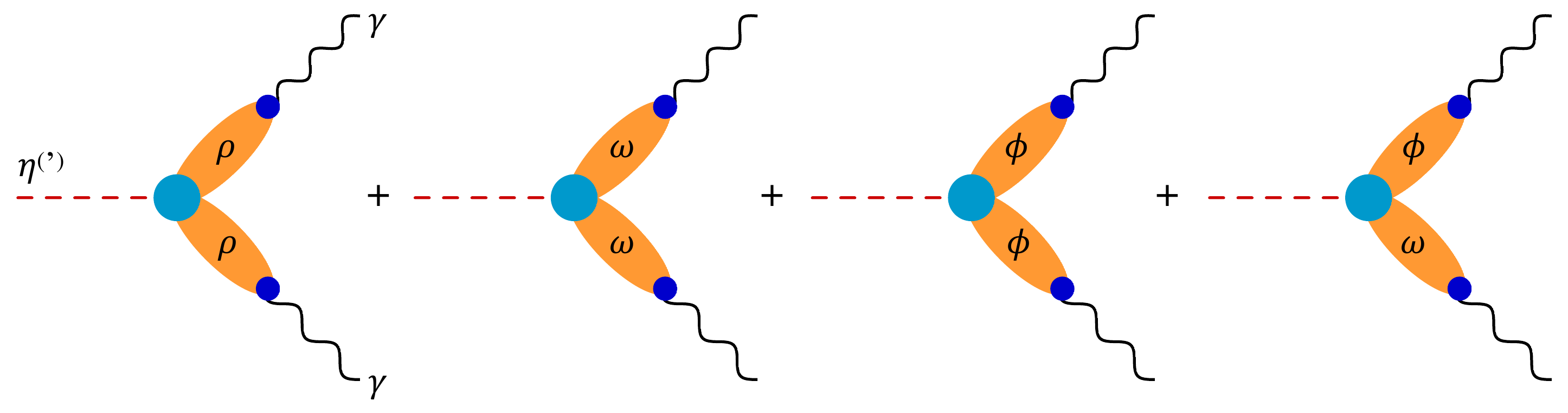}
\caption{$\eta$ and $\eta'$ TFFs in the large-$N_c$ limit. }
\label{fig:TFFlargeNceta}
\end{figure}

Analogously to the pion case, our large-$N_c$ Regge model for the $\eta$ and $\eta'$ TFFs shall satisfy the following five low- and high-energy constraints, cf.~\eqref{anomDef}, \eqref{BLOPESec2}, \eqref{pQCDloop}, and \eqref{MVconstraint}:
\begin{align}
\text{normalization:}&\qquad F_{\eta\ga\ga}^\mathrm{exp}=0.2739(48)\,\mathrm{GeV}^{-1}\;\text{\cite{Tanabashi:2018oca}},\nn\\
&\qquad F_{\eta'\ga\ga}^\mathrm{exp}=0.3413(76)\,\mathrm{GeV}^{-1}\;\text{\cite{Tanabashi:2018oca}};\label{Constraint1Eta}\\ 
\text{BL limit~\cite{Lepage:1980fj,Brodsky:1981rp}:}&\qquad \lim_{Q^2\rightarrow \infty} Q^2  F_{\eta \ga\ga^*}(-Q^2)= 12 \,C_8 F_\eta  ,\nn\\
&\qquad \lim_{Q^2\rightarrow \infty} Q^2  F_{\eta' \ga\ga^*}(-Q^2)= 12 \,C_0 F_{\eta'} ;\label{Constraint3Eta}\\
\text{symmetric pQCD limit~\cite{Novikov:1983jt}:}&\qquad \lim_{Q^2\rightarrow \infty} Q^2  F_{\eta \ga^*\ga^*}(-Q^2,-Q^2)=4\,C_8 F_\eta,\nn\\
&\qquad \lim_{Q^2\rightarrow \infty} Q^2  F_{\eta' \ga^*\ga^*}(-Q^2,-Q^2)=4\,C_0 F_{\eta'}\label{Constraint2Eta};\\
\text{SDC for the mixed region \cite{Melnikov:2003xd}:}&\qquad\lim_{Q_3^2 \rightarrow \infty}\lim_{ Q^2 \rightarrow \infty}\,\sum _{n=0}^\infty  \hat{\Pi} _1^{\eta^{(\prime)}(n)\text{-pole}}(-Q^2,-Q^2,-Q_3^2)\nn\\
&\hspace{-12pt}=-\lim_{Q_3^2 \rightarrow \infty}\lim_{ Q^2 \rightarrow \infty}\,\sum _{n=0}^\infty\frac{ F_{\eta^{(\prime)}(n) \gamma^* \gamma^*}( -Q^2, -Q^2)\,F_{\eta^{(\prime)}(n) \gamma \gamma^*}(-Q_3^2)}{Q_3^2+M_{\eta^{(\prime)}(n)}^2}\nn\\
&\hspace{-12pt}=-\frac{6\,C_{\eta^{(\prime)}}^2}{\pi^2}\frac{1}{ Q^2} \frac{1}{Q_3^2} ;\label{MVConstraintSimpleETA}\\
\text{SDC for the asymptotic region:}&\qquad\lim_{ Q^2 \rightarrow \infty}\,\sum _{n=0}^\infty  \hat{\Pi} _1^{\eta^{(\prime)}(n)\text{-pole}}(-Q^2,-Q^2,-Q^2)\nn\\
&\quad=-\lim_{ Q^2 \rightarrow \infty}\,\sum _{n=0}^\infty\frac{ F_{\eta^{(\prime)}(n) \gamma^* \gamma^*}( -Q^2, -Q^2)\, F_{\eta^{(\prime)}(n) \gamma \gamma^*}( -Q^2)}{ Q^2+M_{\eta^{(\prime)}(n)}^2}\nn\\
&\quad=-\frac{4}{9\pi^2}\frac{C_{\eta^{(\prime)}}^2}{\sum_{a=0,3,8} C_a^2}\frac{1}{ Q^4}.\label{pQCDSimpleETA}
\end{align}
Here, as compared to \eqref{BLOPESec2}, we now use the notation
\begin{align}
F_\eta&= \frac{1}{C_8}\sum_a C_a F^a_\eta, \qquad F_{\eta'}= \frac{1}{C_0}\sum_a C_a F^a_{\eta'}.
\end{align}
Furthermore, we introduced:
\begin{align}
C_\eta^2&=\frac{\left(F^8 \cos \theta_8-2\sqrt{2}F^0 \sin \theta_0\right)\left(F^0 \cos \theta_0-2\sqrt{2}F^8 \sin \theta_8\right)}{108F^0 F^8 \cos\left(\theta_0-\theta_8\right)},\nn\\
C_{\eta'}^2&=\frac{\left(2\sqrt{2}F^8 \cos \theta_8+F^0 \sin \theta_0\right)\left(2\sqrt{2}F^0 \cos \theta_0+F^8 \sin \theta_8\right)}{108F^0 F^8 \cos\left(\theta_0-\theta_8\right)}, 
\label{Ceta2}
\end{align}
as follow by separating the $\eta$ and $\eta'$ contributions to~\eqref{F_MV} according to~\eqref{eq:Pmixing}. These coefficients fulfill
$C_{\eta}^2+C_{\eta'}^2=C_0^2+C_8^2=1/12$.

Switching to the $\eta$ and $\eta'$ we face the problem that, since
these are $I=0$ mesons, they couple to isovector--isovector and
isoscalar--isoscalar vector mesons, so  to same-mass vector
mesons only (ignoring $\phi$--$\omega$ mixing), see figure~\ref{fig:TFFlargeNceta}. Taking the limit $M_{\omega(n)}=M_{\rho(n)}=M_{V(n)}$ in our parameterization of the pion TFF~\eqref{TFFpi}, we obtain a significant simplification: 
\begin{align}
F_{\pi(n)\ga^*\ga^*(-Q_1^2,-Q_2^2)}\propto    \frac{M_V^4 }{D_{V(n)}^1 D_{V(n)}^2}\left[c_\mathrm{anom}+c_A\frac{M_{V(n)}^2}{\Lambda^2}
         + c_\mathrm{diag} 
\frac{Q_1^2 Q_2^2}{\Lambda^2(Q_+^2+M^2_\mathrm{diag})} \right].
\end{align}
Since two free parameters dropped out, this parameterization cannot satisfy all relevant
low- and high-energy constraints.  

Fortunately, via vector-meson mixing in the isoscalar sector, there is a possible
contribution of a mixed $\phi$--$\omega$ term to the TFFs of the
$\eta^{(\prime)}$, which would be absent in the case of ideal mixing. The $\phi$--$\omega$
coupling to $\eta^{(\prime)}$ will certainly be small when compared to the same-mass vector-meson
couplings, see table~\ref{TableCouplings}, but since it contributes where the others
cannot, it is important to retain. 

\begin{table}[t]
\begin{center}
\begin{tabular}{lrrr}
\toprule
&$\pi$& $\eta$& $\eta'$ \\ 
 \midrule
$\rho \omega$&$1.154$&---&---\\
$\rho \phi$&$0.032$&---& --- \\
$\rho \rho$&---&$1.248$&$1.022$ \\
$\omega \omega$&---&$0.139$&$0.114$ \\
$\phi \phi$&---&$ -0.256$&$0.314$ \\
$\phi\omega $&---&$0.015$&$-0.002$ \\
 \bottomrule 
\end{tabular}
\end{center}
\caption{Pseudoscalar--vector--vector couplings derived in
appendix \ref{CouplingsApp}: $C^P_{V_1V_2}$ with $P=\pi,\eta,\eta'$ and $V_i=\rho,\omega,\phi$ as defined in~\eqref{CouplingsAnalytic} and \eqref{CouplingsAnalyticPion}. 
\label{TableCouplings}}
\end{table}

In summary, our large-$N_c$ Regge model for the $\eta^{(\prime)}$
TFFs reads:
\begin{equation}\label{eq:TFFetaandetap}
\hspace{-0.4cm}F_{\eta^{(\prime)}(n) \gamma^*
  \gamma^*}(-Q_1^2,-Q_2^2)=\frac{F_{\eta^{(\prime)}\gamma\gamma}}{\mathcal{N}}\left[F^{(a)}_{\eta^{(\prime)}(n) \gamma^*
  \gamma^*}(-Q_1^2,-Q_2^2)+F^{(b)}_{\eta^{(\prime)}(n) \gamma^*
  \gamma^*}(-Q_1^2,-Q_2^2)\right], 
\end{equation}
where the two parts parameterize the same-mass and mixed  vector-meson contributions, respectively:
\begin{align}
\label{TFFeta}
F^{(a)}_{\eta^{(\prime)} (n) \gamma^* \gamma^*}(-Q_1^2,-Q_2^2)&=
 \sum_{V=\rho,\omega,\phi}
 C^{\eta^{(\prime)}}_{VV} 
\left[1 + c_\mathrm{diag} \frac{\Lambda^2}{M_V^4}\frac{Q_1^2 
        Q_2^2}{(Q_+^2+M^2_\mathrm{diag})}
    \right]\frac{M_V^4}{D^1_{V(n)} D^2_{V(n)}},\\
    F^{(b)}_{\eta^{(\prime)}(n) \gamma^* \gamma^*}(-Q_1^2,-Q_2^2)&= 
 C^{\eta^{(\prime)}}_{\phi \omega}\left\{ \left[
        1+c_A \frac{M_{+,\,n}^2-M_{+,\,0}^2}{\Lambda^2}+c_B 
\frac{M_{-,\,n}^2-M_{-,\,0}^2}{ \Lambda^2}
         \right] \right.\notag\\
&\times\left(\frac{M_\phi^2 M_\omega^2}{D_{\omega(n)}^1 D_{\phi(n)}^2} +
  \frac{M_\phi^2 M_\omega^2}{D_{\omega(n)}^2 D_{\phi(n)}^1}
\right)+ \left[c_\mathrm{BL}+c_A \frac{M_{-,\,n}^2}{\Lambda^2}+c_B \frac{M_{+,\,n}^2}{\Lambda^2}
   \right] \nonumber \\
   & \times \left. \frac{Q_-^2}{Q_+^2}\left(\frac{M_\phi^2 M_\omega^2}{D_{\omega(n)}^1 D_{\phi(n)}^2} -
  \frac{M_\phi^2 M_\omega^2}{D_{\omega(n)}^2 D_{\phi(n)}^1}
\right) 
\right\}. \label{TFFeta-ophi}
\end{align}
Here we again use the short-hand notations from \eqref{def1}, with the modification that
\begin{equation}
M_{\pm,\,n}^2=\frac{1}{2} \left(M_{\phi(n)}^2\pm M_{\omega(n)}^2 \right).
\end{equation}
All meson spectra are assumed to follow a radial Regge model. For the $\phi$ meson, we use the analog of~\eqref{nonIdenticalSpectra}. For the $\eta$ and $\eta'$ mesons, we distinguish:
\beq
M_{\eta(n)}^2=\begin{cases}M_{\eta}^2& n=0,\\
\hat M_{\eta}^2+n\, \sigma_\eta^2 &n \geq 1, \end{cases},\label{etaTrajectory}
\eeq
and 
\beq
M_{\eta'(n)}^2=
M_{\eta'}^2+n\, \sigma_{\eta'}^2\label{etaPTrajectory}
\eeq
with the ground-state masses $M_{\eta}=547.862(17)$ MeV and $M_{\eta'}=957.78(6)$ MeV~\cite{Tanabashi:2018oca}.

The normalization coefficient is defined as:
\begin{equation}
\mathcal{N}=C^\eta_{\rho
  \rho}+C^\eta_{\omega \omega}+ C^\eta_{\phi \phi}+2 C^\eta_{\phi\omega
  },\label{NormalizationEta}
\end{equation}
where $C^P_{V_1 V_2}$ are the pseudoscalar--vector--vector couplings derived in
appendix \ref{CouplingsApp}, and $C^\eta_{\phi \omega}$ is the parameter that measures the deviation
from ideal mixing. By construction, each vector-meson pair contributes (up to normalization) exactly $C^P_{V_1 V_2}$ to $F_{\eta^{(\prime)} \gamma
  \gamma}$.

To simplify the parameterization, equation~\eqref{TFFeta-ophi} only contains terms which are unique to the $\phi$--$\omega$ contribution, and the $n$-dependence has been removed from the numerator of \eqref{TFFeta}. In this way, \eqref{TFFeta-ophi} is used to satisfy the BL limit for the ground-state $\eta^{(\prime)}$ TFF as well as the 
two SDCs on the HLbL tensor. 

\begin{figure}[t]
\centering
\includegraphics[width=0.7\linewidth]{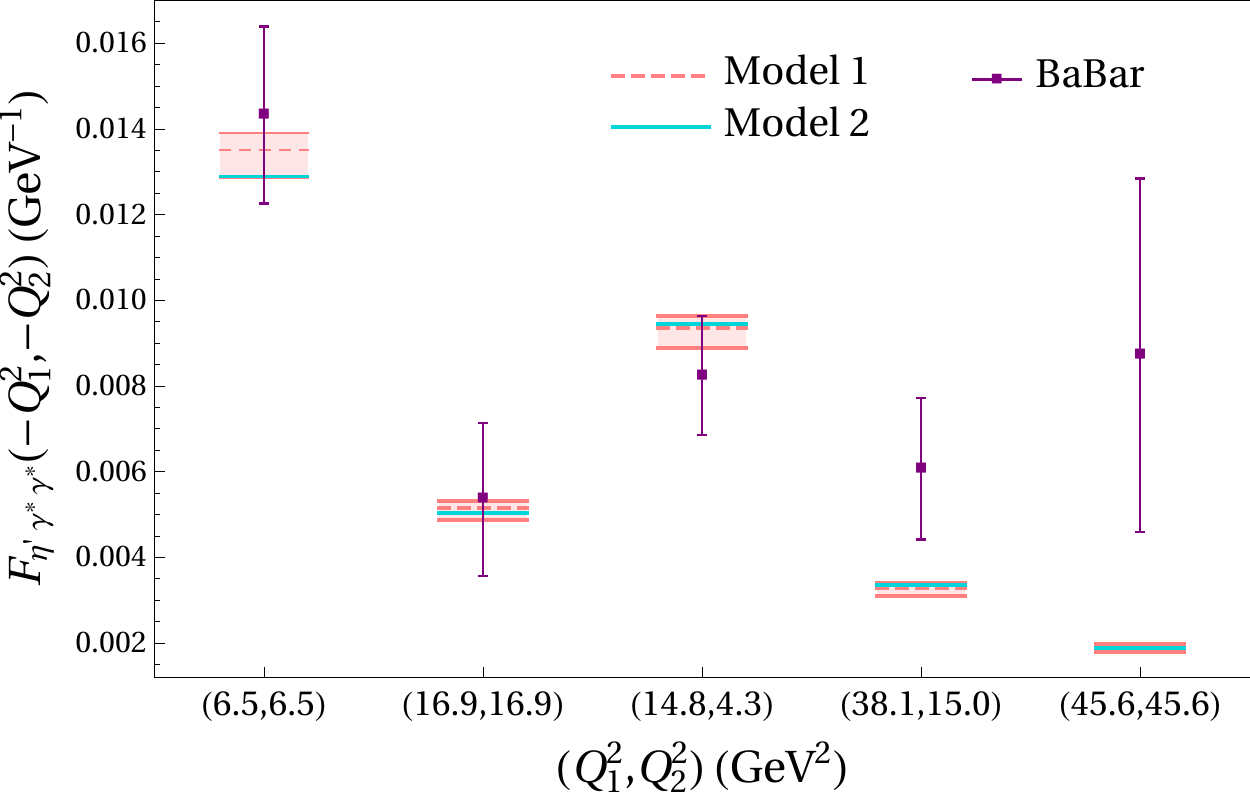}
\caption{Comparison to the doubly-virtual $\eta ^\prime$ TFF data from BaBar \cite{BaBar:2018zpn}. The large-$N_c$ Regge model, ``Model 1''~\eqref{eq:TFFetaandetap}, is indicated by the  pink bands with the dashed lines. Our alternative TFF model, ``Model 2''~\eqref{MartinModel}, is indicated by the solid cyan lines.}
\figlab{EtaPDataDV}
\end{figure}

The constraint equations following from~\eqref{Constraint3Eta} and \eqref{Constraint2Eta} read:
\begin{align}
\mbox{BL limit:}\qquad  1&=\frac{1}{\mathcal{N}}\frac{F_{\eta\ga\ga}}{12C_8F_\eta} \Bigg[C^\eta_{\rho
  \rho}M_\rho^2+C^\eta_{\omega \omega}M_\omega^2+ C^\eta_{\phi \phi}M_\phi^2 \label{ConstraintEquationsBLEta}\\
  &+2C^\eta_{ \phi\omega}\bigg(M^2_{+,\,0}
  -c_\mathrm{BL}M^2_{-,\,0}-c_A \frac{M_{-,\,0}^4}{\Lambda^2}-c_B \frac{M_{+,\,0}^2M_{-,\,0}^2}{\Lambda^2}\bigg)\Bigg] ; \notag \\
\mbox{symmetric pQCD limit:} \qquad 1&= \frac{C^\eta_{\rho
  \rho}+C^\eta_{\omega \omega}+ C^\eta_{\phi \phi}}{ \mathcal{N}}\frac{\Lambda^2F_{\eta\ga \ga}}{8C_8 F_\eta }\, c_\mathrm{diag};\label{ConstraintEquationsEta}
\end{align}
where the same equations hold for the $\eta'$ with the obvious replacements (including $C_8\rightarrow C_0$). The MV SDC for the HLbL tensor in the mixed region translates to:
\begin{align}
     1= \frac{C^\eta_{\phi
  \omega}}{\mathcal{N}} \frac{2\pi^2C_8 F_\eta F_{\eta\ga\ga} M_\phi^2 M_\omega^2}{3 C_\eta^2 \Lambda^2}\left[(c_A+c_B)\frac{L_{\omega\eta}}{\Delta_{\omega\eta}} + (c_A-c_B)\frac{L_{\phi\eta}}{\Delta_{\phi\eta}}\right],\label{MVEta}
 \end{align}
 where $c_\mathrm{diag}$ has already been inserted.
The SDC for the HLbL tensor in the asymptotic region becomes more complicated due to the presence of additional mass 
 scales:
\begin{align}
     1&=\frac{C^\eta_{\phi
  \omega}}{\mathcal{N}^2}\frac{\pi^2F_{\eta\ga\ga}^2 M_\phi^2 M_\omega^2 }{16C_\eta^2 \Lambda^4}\Bigg[\frac{4 C^\eta_{\phi
  \omega} M_\phi^2 M_\omega^2 }{\Omega_{\phi\omega \eta}^2} \Bigg\{ c_A^2 \Delta_{\phi\omega}\Sigma_{\phi\omega} \Big[\Omega_{\phi\omega \eta}+\sigma_\eta^2 \left(\Delta_{\omega\eta}^2L_{\phi\eta}-\Delta_{\phi\eta}^2L_{\omega\eta}\right)\Big]\nn\\
  &+c_B^2 \Delta_{\phi\omega}\Big[\left(\Sigma_{\phi\omega}\sigma_\eta^2
    -2 \sigma_\phi^4\right) \Delta_{\omega\eta}^2 L_{\phi\eta} -
     \left(\Sigma_{\phi\omega} \sigma_\eta^2
    -2 \sigma_\omega^4\right)\Delta_{\phi\eta}^2 L_{\omega\eta}
 +(\Delta_{\phi\eta}+\Delta_{\omega\eta} )\,\Omega_{\phi\omega\eta} \Big] \nn\\  
  &+2c_A c_B  \Big[ \sigma_\eta^2
     \left(\sigma_\phi^4+\sigma_\omega^4\right)\left(\Delta_{\omega\eta}^2L_{\phi\eta}-\Delta_{\phi\eta}^2L_{\omega\eta} \right)+\Sigma_{\phi\omega}\left(\sigma_\omega^4\Delta_{\phi\eta}^2L_{\omega\eta}-\sigma_\phi^4\Delta_{\omega\eta}^2L_{\phi\eta}\right)\nn\\
 &-\Omega_{\phi\omega\eta}\left(\sigma_\eta^2
   \Sigma_{\phi\omega}-\sigma_\phi^4-\sigma_\omega^4\right) \Big]\Bigg\}+c_\mathrm{diag}c_A\,\Lambda^4\Bigg(\frac{2C^\eta_{\rho\rho}}{\Delta_{\rho\phi}\Delta_{\rho\omega}}\Bigg\{\frac{\sigma_\rho^2 \left(\Delta_{\rho\phi}+\Delta_{\rho\omega}\right)}{\Delta_{\rho\eta}}\nn\\
   &+\frac{1}{\Omega_{\rho\phi\eta}\Omega_{\rho\omega\eta}\Omega_{\phi\omega\eta}} \Big[  \Omega_{\phi\omega\eta}\sigma_\eta^4\Delta_{\rho\phi}^2\Delta_{\rho\omega}^2\left(\Delta_{\phi\eta}+\Delta_{\omega\eta}\right)L_{\rho\eta}-\Delta_{\phi\omega}\left(\Omega_{\rho\omega\eta}^2\sigma_\phi^4  \Delta_{\phi\eta}L_{\rho\phi}\right.\nn\\
   &\left.+ \Omega_{\rho\phi\eta}^2\sigma_\omega^4 \Delta_{\omega\eta}L_{\rho\omega}\right)\Big]\Bigg\}+\frac{C^\eta_{\phi\phi}\Delta_{\omega\eta}}{\Omega^2_{\phi\omega\eta}\Delta_{\phi\eta}}\Big[2 \sigma_\eta^4 \Delta_{\phi\omega}^2 \left(\Delta_{\phi\eta}+\Delta_{\omega\eta}\right)L_{\phi\eta}-2 \sigma_\omega^4 \Delta_{\phi\eta}^3L_{\phi\omega}\nn\\
   &+\Omega_{\phi\omega\eta}\left(3 \sigma_\phi^4-\sigma_\phi^2\sigma_\omega^2-5 \sigma_\phi^2 \sigma_\eta^2 +3 \sigma_\omega^2 \sigma_\eta^2\right)\Big]
   +\frac{C^\eta_{\omega\omega}\Delta_{\phi\eta}}{\Omega^2_{\phi\omega\eta}\Delta_{\omega\eta}}\Big[2 \sigma_\phi^4 \Delta_{\omega\eta}^3L_{\phi\omega}\nn\\
   &+2 \sigma_\eta^4 \Delta_{\phi\omega}^2 \left(\Delta_{\phi\eta}+\Delta_{\omega\eta}\right)L_{\omega\eta}+\Omega_{\phi\omega\eta}\left(\sigma_\phi^2 \sigma_\omega^2-3 \sigma_\omega^4-3 \sigma_\phi^2 \sigma_\eta^2+5 \sigma_\omega^2 \sigma_\eta^2\right)\Big]\Bigg)\nn\\
   &-c_\mathrm{diag}c_B\,\Lambda^4\Bigg(\frac{2C^\eta_{\rho\rho}}{\Delta_{\rho\phi}\Delta_{\rho\omega}}\Bigg\{\frac{\sigma_\rho^2 \Delta_{\phi\omega}}{\Delta_{\rho\eta}}-\frac{1}{\Omega_{\rho\phi\eta}\Omega_{\rho\omega\eta}\Omega_{\phi\omega\eta}} \Big[  \Omega^2_{\rho\omega\eta}\sigma_\phi^4\Delta_{\phi\omega}\Delta_{\phi\eta}L_{\rho\phi}\nn\\
   &-
   \Omega^2_{\rho\phi\eta}\sigma_\omega^4\Delta_{\phi\omega}\Delta_{\omega\eta}L_{\rho\omega}+
   \Omega^2_{\rho\phi\omega}\sigma_\eta^4\Delta_{\phi\eta}\Delta_{\omega\eta}L_{\rho\eta}
   \Big]\Bigg\}-\frac{C^\eta_{\omega\omega}\Delta_{\phi\eta}}{\Omega^2_{\phi\omega\eta}\Delta_{\omega\eta}}\Big[2\left(\sigma_\phi^4 \Delta_{\omega\eta}^3+\sigma_\eta^4 \Delta_{\phi\omega}^3\right) L_{\omega\eta}\nn\\
   &-2 \sigma_\phi^4 \Delta_{\omega\eta}^3L_{\phi\eta}+\Omega_{\phi\omega\eta}\left(\sigma_\phi^2\sigma_\omega^2+\sigma_\omega^4-3 \sigma_\phi^2\sigma_\eta^2+\sigma_\omega^2 \sigma_\eta^2\right)\Big]-\frac{C^\eta_{\phi\phi}\Delta_{\omega\eta}}{\Omega^2_{\phi\omega\eta}\Delta_{\phi\eta}}\Big[2 \sigma_\omega^4 \Delta_{\phi\eta}^3L_{\omega\eta}\nn\\
   &-2\left(\sigma_\omega^4 \Delta_{\phi\eta}^3-\sigma_\eta^4 \Delta_{\phi\omega}^3\right) L_{\phi\eta}+\Omega_{\phi\omega\eta}\left(\sigma_\phi^4+\sigma_\phi^2\sigma_\omega^2+\sigma_\phi^2 \sigma_\eta^2-3 \sigma_\omega^2\sigma_\eta^2\right)\Big]
  \Bigg)\Bigg].\label{pQCD3Eta}
 \end{align}

In appendix~\ref{sec:EtaModel1CouplingParameters},
the above system of equations is solved analytically. In the following, we discuss numerical
values for all input parameters. The couplings $C^{\eta^{(\prime)}}_{V_1 V_2}$ are collected in table~\ref{TableCouplings}.
They are calculated based on~\eqref{CouplingsAnalytic} with the phenomenological $\eta$--$\eta'$ mixing parameters~\cite{Feldmann:1998vh,Feldmann:1999uf}:
\beq
F^8 = 1.26(4) F_\pi, \quad
F^0 = 1.17(3) F_\pi,\quad
\theta_8 = -21.2(1.6)^\circ,\quad
\theta_0 = -9.2(1.7)^\circ,\label{FKS98input}
\eeq
and the $\phi$--$\omega$ mixing angle $\theta_V= 36.4^\circ$ \cite{Tanabashi:2018oca}. The parameters $C_{\eta^{(\prime)}}^2$, which describe our choice for the splitting of the SDCs on the HLbL tensor into $\eta$ and $\eta'$ contributions, evaluate to:
\beq
C_\eta^2\sim 0.027\qquad C_{\eta'}^2\sim 0.057,
\eeq
as follows from~\eqref{Ceta2} with the $\eta$--$\eta'$ mixing parameters in~\eqref{FKS98input}. The decay constants $F_{\eta^{(\prime)}}$, on the other hand, are not deduced from the $\eta$--$\eta'$ mixing parameters, but fit to experimental data for the singly-virtual $\eta^{(\prime)}$ TFFs:
\begin{align}
F_{\eta}&=139^{\,+27}_{\,-2}\,\text{MeV} ,\nn\\
F_{\eta'}&=79^{\,+3}_{\,-5}\,\text{MeV}, \label{Fetafit}
\end{align}
with an estimated variance of $\chi^2\sim1.1$ and $\chi^2\sim0.9$, respectively. The errors are increased in order to cover the Pad\'{e} approximant predictions from refs.~\cite{Masjuan:2017tvw} and \cite{Escribano:2015yup} for $\eta$ and $\eta'$, respectively.  The large error on $F_{\eta}$ may be partly due to the fact that it is not clear 
when the asymptotic BL limit sets in, accordingly, we will keep the full range in the error analysis.
$M_\mathrm{diag}$ is fit to the recent BaBar data for the doubly-virtual $\eta'$ TFF \cite{BaBar:2018zpn}, see figure \ref{fig:EtaPDataDV}. The resulting value $M_\mathrm{diag}=898$ MeV (with $\chi^2\sim1.6$) is used for both the $\eta$ and $\eta'$ large-$N_c$ Regge model. Furthermore, we use the
Regge slopes collected in \eqref{ReggeSlopes1} as well as~\cite{Masjuan:2012gc}:\footnote{From figure~3 of ref.~\cite{Masjuan:2012gc} we extracted $\hat M_\eta=591\, \mathrm{MeV}$.}
\begin{equation}
\sigma_\eta^2 = 1.33(11) \, \mathrm{GeV}^2,\quad \sigma_{\eta'}^2= 1.36(14) \, 
\mathrm{GeV}^2, \quad\sigma_\phi^2= 1.84(6) \, \mathrm{GeV}^2.\label{ReggeSlopes2}
\end{equation}
The final model parameters ($c_A$, $c_B$, $c_\mathrm{diag}$, $c_\mathrm{BL}$) are summarized in table~\ref{TableModel1}, where we rescaled the numerical values 
with $C_{\phi \omega}^{\eta}/\mathcal{N}\sim0.0129$ and $C_{\phi \omega}^{\eta'}/\mathcal{N}\sim -0.0017$, respectively, to show that all parameters are of ``natural'' size.

\begin{figure}[t]
\centering
\includegraphics[width=0.9\linewidth]{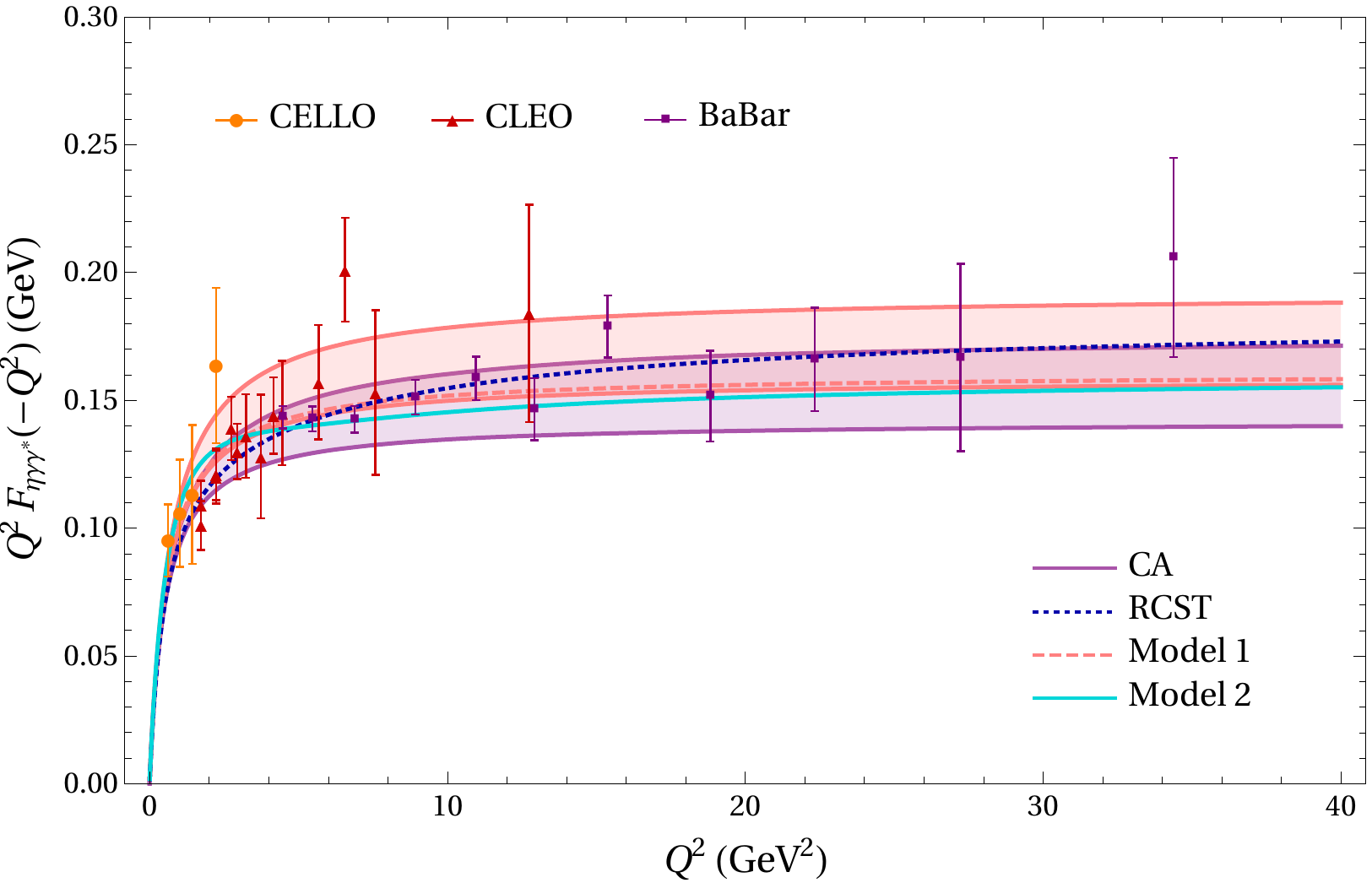}
\caption{Singly-virtual $\eta$ TFF. The large-$N_c$ Regge model, ``Model 1''~\eqref{eq:TFFetaandetap}, is indicated by the pink band with the dashed curve. Our alternative TFF model, ``Model 2''~\eqref{MartinModel}, is indicated by the solid cyan curve. The purple band is the CA result from ref.~\cite{Masjuan:2017tvw}. The dark blue dotted curve is the RCST result from ref.~\cite{Czyz:2017veo}. The data are from CELLO~\cite{Behrend:1990sr}, CLEO~\cite{Gronberg:1997fj}, and BaBar~\cite{BABAR:2011ad}.}
\label{fig:SinglyVirtualEta}
\end{figure}

\begin{figure}[t]
\centering
\includegraphics[width=0.9\linewidth]{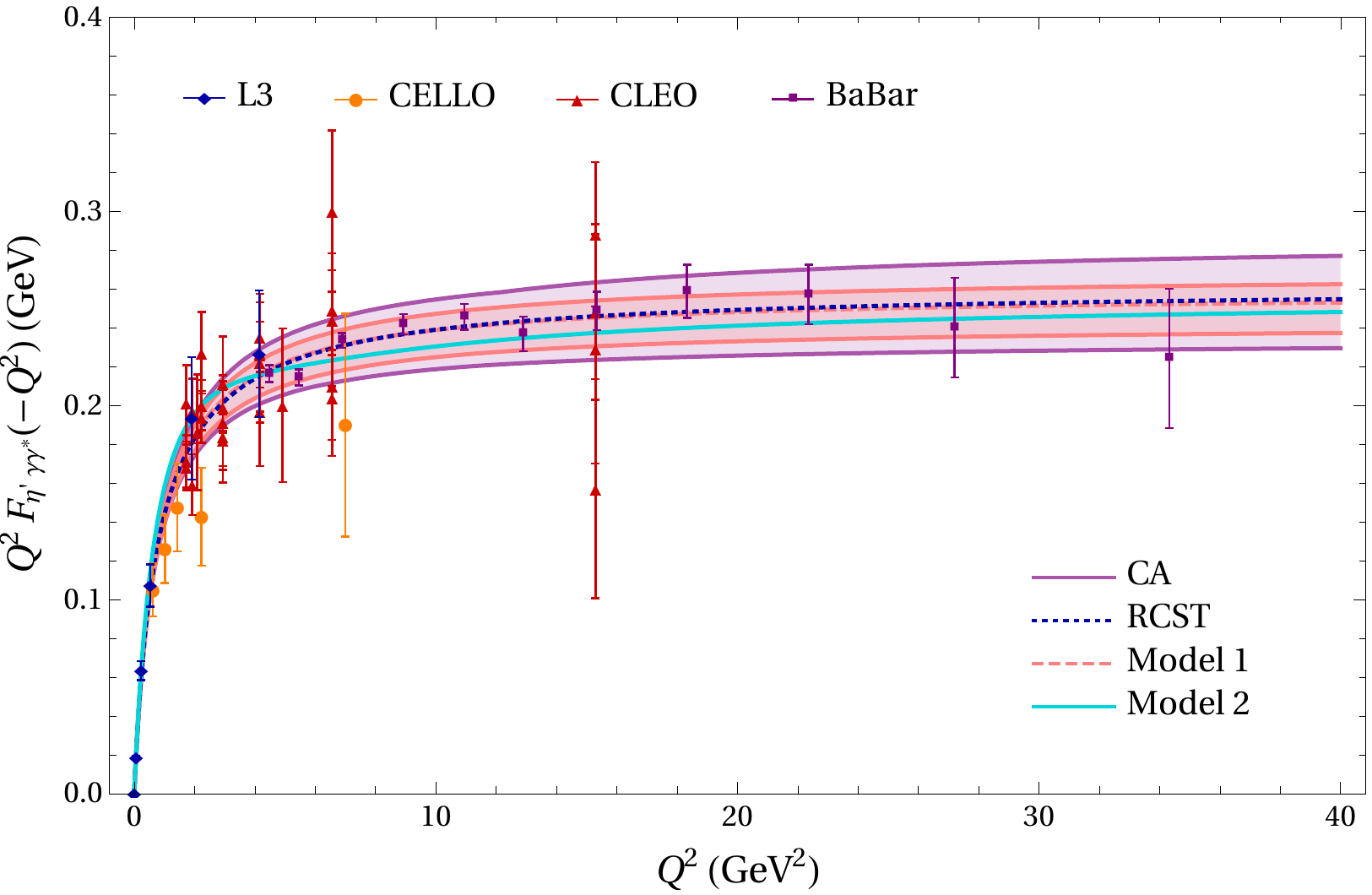}
\caption{Singly-virtual $\eta^\prime$ TFF. The large-$N_c$ Regge model, ``Model 1''~\eqref{eq:TFFetaandetap}, is indicated by the pink band with the dashed curve. Our alternative TFF model, ``Model 2''~\eqref{MartinModel}, is indicated by the solid cyan curve. The purple band is the CA result from ref.~\cite{Masjuan:2017tvw}. The dark blue dotted curve is the RCST result from ref.~\cite{Czyz:2017veo}. The data are from L3~\cite{Acciarri:1997yx}, CELLO~\cite{Behrend:1990sr}, CLEO~\cite{Gronberg:1997fj}, and BaBar~\cite{BABAR:2011ad}.}
\label{fig:SinglyVirtualEtaPrime}
\end{figure}

The singly-virtual TFFs of the ground-state $\eta$ and $\eta'$ are shown in figures \ref{fig:SinglyVirtualEta} and \ref{fig:SinglyVirtualEtaPrime}. The large-$N_c$ Regge model presented above is labeled as ``Model 1.'' The alternative TFF model, introduced in appendix~\ref{sec:PionAlternative}, is referred to as ``Model 2.'' The model error of the large-$N_c$ Regge TFFs is propagated from the errors of the input parameters $\sigma_P$, $\sigma_V$, $F_{P\ga\ga}$, $F_P$, as well as the $\eta$--$\eta'$ mixing parameters, see~\eqref{ReggeSlopes1}, \eqref{Constraint1Eta}, \eqref{ReggeSlopes2}, \eqref{FKS98input}, and \eqref{Fetafit}. While Model 2, for which we do not provide an error estimate, runs outside the error band of the Model 1 $\eta'$ ($\eta$) TFF for some low (intermediate) values of $Q^2$, both models give a good description of the experimental data and, thus, come out close to the results from CAs~\cite{Masjuan:2017tvw} and fits within resonance chiral symmetric theory (RCST) \cite{Czyz:2017veo}.\footnote{The RCST result is reproduced from fit 2 in ref.~\cite{Czyz:2017veo} and PDG input for the masses of $\rho(770)$, $\rho(1450)$, $\rho(1700)$, $\omega(782)$, $\omega(1420)$, $\omega(1650)$, $\phi(1020)$, $\phi(1680)$, and $\phi(2170)$~\cite{Tanabashi:2018oca}.} In appendices~\ref{sec:EtaPlots} and \ref{sec:EtaPrimePlots}, both models are further compared to CA, RCST, and DSE~\cite{Eichmann:2019tjk}, and the decomposition of Model 1 into $2\rho$, $2\phi$, $2\omega$, and $\phi\omega$ contributions is illustrated. We stress again that neither model should be evaluated for other than purely space-like virtualities.

Both the ground-state $\eta$ contribution to $(g-2)_\mu$,
\begin{align}
 a_\mu^{\eta\text{-pole}}\big|_\text{Model 1}&=16.4^{\,+1.3}_{\,-0.5}\times 10^{-11},\qquad a_\mu^{\eta\text{-pole}}\big|_\text{Model 2}=17.8 \times 10^{-11},\notag\\
 a_\mu^{\eta\text{-pole}}\big|_\text{\cite{Masjuan:2017tvw,PabloSanchezPrivateCom}}&=16.3(1.4)\times 10^{-11},\label{aGroundStateEta}
\end{align}
and the ground-state $\eta'$ contribution to $(g-2)_\mu$
\begin{align}
 a_\mu^{\eta'\text{-pole}}\big|_\text{Model 1}&=14.8^{\,+0.6}_{\,-0.7}\times 10^{-11},\qquad a_\mu^{\eta'\text{-pole}}\big|_\text{Model 2}=16.1 \times 10^{-11},\notag\\
 a_\mu^{\eta'\text{-pole}}\big|_\text{\cite{Masjuan:2017tvw,PabloSanchezPrivateCom}}&=14.5(1.9)\times 10^{-11},\label{aGroundStateEtaP}
\end{align}
are reproduced correctly with our $\eta^{(\prime)}$ TFF models.

\subsection{Comparison of two-photon couplings}
\label{sec:two_photon_couplings}

\begin{table}[t]
\begin{center}
\begin{tabular}{ ccccccccc}
\toprule
$n$&\multicolumn{4}{c}{$\eta$}&\multicolumn{4}{c}{$\eta'$}\\
 & \multicolumn{2}{c}{Assignment 1}&\multicolumn{2}{c}{Assignment 2}& \multicolumn{2}{c}{Assignment 1}&\multicolumn{2}{c}{Assignment 2}\\ 
 \midrule
$1$& $\eta(1295)$&$0.0354$&$\eta(1440)$&$0.0351$&$\eta'(1475)$&$0.0594$&$X(1835)$&$0.0561$\\
$2$& $\eta(1760)$&$0.0171$&$\eta(1760)$&$0.0169$&$\eta'(2010)$&$0.0305$&$\eta'(2070)$&$0.0281$\\
$3$& $\eta(2100)$&$0.0111$&&$0.0110$&$\eta'(2225)$&$0.0203$&&$0.0185$\\
$4$& $\eta(2320)$&$0.0082$&&$0.0081$&&$0.0151$&&$0.0137$\\
 \bottomrule
\end{tabular}
\caption{Two-photon couplings, $F_{P\gamma\gamma}$, of excited $\eta^{(\prime)}$ states from the large-$N_c$ Regge model \eqref{eq:TFFetaandetap}, in units of GeV$^{-1}$. ``Assignment 1'' and ``Assignment 2'' refer to the assignments of $\eta^{(\prime)}$ excitations shown in figure \ref{fig:TrajectoryPlotPion} and suggested in refs.~\cite{Masjuan:2012gc,Tanabashi:2018oca} and \cite[Table 27]{Klempt:2007cp}, respectively.
\label{TabTrajectories}}
\end{center}
\end{table}

Apart from the Regge slopes for the trajectories of pion, $\eta$, $\eta'$, as well the vector mesons, phenomenological input for the excited-pseudoscalar contributions 
could in principle be provided by their TFFs. Even though the normalizations are poorly known, it is still important to verify that the two-photon couplings implied by our Regge models compare reasonably 
to the available phenomenological constraints. For the first excited state in the pion trajectory, there is a limit 
\begin{equation}
 F_{\pi(1300)\ga\ga}<0.0544(71)\GeV^{-1},
\end{equation}
see appendix~\ref{app:photon_couplings}, which is indeed satisfied by $F_{\pi(1300)\ga\ga}=0.050\GeV^{-1}$ from our Regge model.
Nothing is known about the two-photon coupling of the $\pi(1800)$ and even heavier excited pions. 

The situation is more involved in the $\eta^{(\prime)}$ sector. As alluded to in the caption of figure~\ref{fig:TrajectoryPlotPion}, the spectroscopy of excited $\eta^{(\prime)}$ states is contentious, especially regarding 
the role of the states below $1500\MeV$. Table~\ref{TabTrajectories} collects two possible assignments of states to Regge trajectories. 
The first interpretation, favored by ref.~\cite{Tanabashi:2018oca}, considers the $\eta(1295)$ the lowest $\eta$ excitation 
and differentiates between $\eta(1405)$ and $\eta(1475)$ states. The latter is considered as the first $\eta'$ excitation, while the $\eta(1405)$ is described as a glueball candidate. 
In contrast, ref.~\cite{Klempt:2007cp} argues that there is only a single state below $1500\MeV$, the $\eta(1440)$, which should be interpreted as the first $\eta$ excitation. 
The $X(1835)$ is identified as suitable candidate for the first $\eta'$ excitation, although its quantum numbers are not yet established. 
In both cases, the $\eta(1760)$ emerges as the second $\eta$ excitation.

\begin{table}[t]
\begin{center}
\begin{tabular}{cccc}
\toprule 
& \multicolumn{3}{c}{$F_{P\gamma\gamma} [\GeV^{-1}]$} \\
$P$ & direct & indirect & assuming dominance of\\
 \midrule
$\eta(1295)$ & --- & $<0.030$ & $\eta\pi\pi$, $K\bar K\pi$\\
$\eta(1405)$ & $<0.122$ & $<0.033$ & $\eta\pi\pi$, $K\bar K\pi$\\
$\eta(1475)$ & $<0.195$, $>0.041(6)$ & $=0.041(6)$ & $K\bar K\pi$\\
$\eta(1760)$ & $>0.014(2)$ & $=0.014(2)$ & $\eta'\pi\pi$\\
$X(1835)$ & $<0.235$ & $<0.022$ & $\eta'\pi\pi$\\
\bottomrule
\end{tabular}
\caption{Constraints on the two-photon couplings of excited $\eta^{(\prime)}$ states, as collected in appendix~\ref{app:photon_couplings}. The column labeled ``direct'' includes 
constraints that follow directly from branching fractions listed in the PDG, while the column labeled ``indirect'' lists results obtained when assuming that the channels from the last column 
are dominant.
\label{TabPhotonCouplings}}
\end{center}
\end{table}

The available constraints on the two-photon couplings of $\eta^{(\prime)}$ are collected in table~\ref{TabPhotonCouplings}, see appendix~\ref{app:photon_couplings} for details. 
The results from the second column are valid under the assumption that the branching fractions listed in the PDG are accurate, while the third column assumes, in addition, 
dominance by some decay channels (given in the last column). For the $\eta(1295)$ only an indirect limit is available, in the first assignment the two-photon coupling of the $\eta(1295)$ comes out slightly larger. Note, however, that the very existence of the $\eta(1295)$ is called into question in ref.~\cite{Klempt:2007cp}, with the fact that in contrast to the $\eta(1475)$ 
this resonance has not been seen in the $\gamma\gamma$ reaction as one of the arguments.   
The $\eta(1405)$ is discarded in either assignment of Regge trajectories.
However, in the second assignment the $\eta(1440)$ would be interpreted as a single state instead of $\eta(1405)$ and $\eta(1475)$, see also ref.~\cite{Wu:2011yx}, 
in such a way that for the comparison the two-photon couplings of both states should be considered. 
Remarkably, the measured value for the $\eta(1475)$, $F_{\eta(1475)\gamma\gamma}=0.041(6)\GeV^{-1}$, agrees perfectly with $F_{\eta(1440)\gamma\gamma}=0.035\GeV^{-1}$ from Assignment 2. 
In Assignment 1, where the $\eta(1475)$ is considered the first $\eta'$ excitation, there is still reasonable agreement.
Next, the experimental result for the $\eta(1760)$ nicely confirms the two-photon coupling implied by both assignments, since a tiny correction beyond the 
dominant $\eta'\pi\pi$ channel would suffice to bring the numbers into complete agreement. 
Finally, the two-photon coupling of the $X(1835)$ in Assignment 2 fulfills the direct limit but not the one assuming dominance of $\eta'\pi\pi$, which may indicate that 
in case this assignment is correct, other channels besides $\eta'\pi\pi$ may play a role (as indeed suggested by other decay channels listed in the PDG). Moreover, the significance of the two-resonance fit from ref.~\cite{Zhang:2012tj} used to obtain the much stricter limit is only quoted at $2.8\,\sigma$. Taken together with the fact that not even the quantum numbers of the $X(1835)$ are firmly established, it thus seems difficult to draw meaningful conclusions on its two-photon coupling at this point.

Altogether, we conclude that the two-photon couplings implied by our large-$N_c$ Regge models are well compatible with the phenomenological constraints. In particular, for the cases 
where measurements and not just limits exist, the $\eta(1475)$ and the $\eta(1760)$, the resulting couplings are close to the ones that our large-$N_c$ Regge models would imply. The same is true for our alternative TFF model \eqref{MartinModel}, see figure \ref{fig:Coupling}, whose couplings are similar to the ones of the large-$N_c$ Regge models. We stress that the detailed comparison depends on the assignment of observed states to Regge trajectories, but in both variants considered there is reasonable agreement with the two-photon phenomenology of excited $\eta^{(\prime)}$ states.

\subsection[Excited-pseudoscalar contributions to $(g-2)_\mu$]{Excited-pseudoscalar contributions to $\boldsymbol{(g-2)_\mu}$}
\label{sec:excited_states_g_2}

\begin{figure}[t]
\centering
\includegraphics[width=0.8\linewidth]{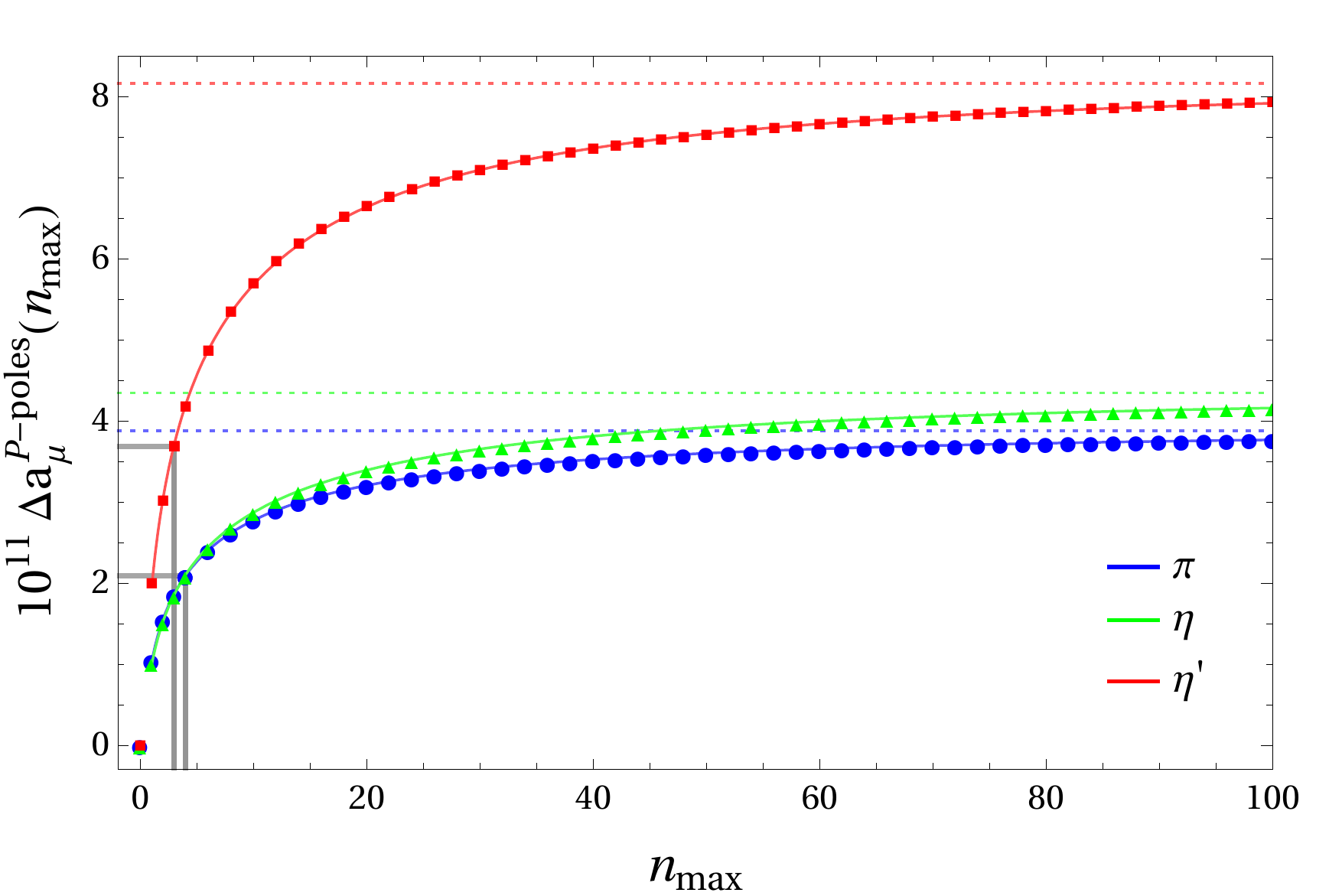}
\caption{Sum over radially-excited pion, $\eta$, and $\eta'$ contributions to $(g-2)_\mu$, as defined in~\eqref{DeltaADef}. The dotted curves are the extrapolated values $\Delta a_\mu^{P\text{-poles}}$ corresponding to the limit $n_\mathrm{max}\rightarrow \infty$. The gray lines indicate the contributions from the lowest (observed) pseudoscalar excitations shown in figure~\ref{fig:TrajectoryPlotPion}. All results are calculated with the Model 2 TFFs~\eqref{MartinModel}. }
\figlab{SaturationCurve}
\end{figure}

The ground-state pseudoscalar-pole contributions to $(g-2)_\mu$, calculated based on our TFF models, are given in~\eqref{piong-2}, \eqref{aGroundStateEta}, and \eqref{aGroundStateEtaP}. 
The uncertainty on the predictions from Model 1 is the propagated error from the input parameters $\sigma_P$, $\sigma_V$, $F_{P\ga\ga}$, $F_P$, and the $\eta$--$\eta'$ mixing parameters.
In all cases we observe good agreement with the literature~\cite{Hoferichter:2018dmo,Hoferichter:2018kwz,Masjuan:2017tvw,PabloSanchezPrivateCom},
which demonstrates that in addition to fulfilling the various SDCs, our Regge models capture the properties of the TFFs most relevant for the $g-2$ integral.

In the following, we derive the contribution to $(g-2)_\mu$ originating from radially-excited
pseudoscalar mesons. The large-$N_c$ Regge models introduced in the preceding sections and the alternative model discussed in appendix~\ref{sec:PionAlternative} 
are constructed in such a way as to describe not only the ground-state pseudoscalar TFFs, but also the TFFs of excited pseudoscalar mesons. 
Phenomenological input on these excited states enters mainly in terms of their masses as contained in the Regge parameters, 
while the infinite sum restores the correct asymptotic properties of the HLbL tensor,
which cannot be achieved with a finite number of pseudoscalar-pole contributions. 
Moreover, for some of the excited states limits on their two-photon couplings are available,  
see appendix~\ref{app:photon_couplings} as well as the discussion in the previous subsection, which shows that 
the couplings implied by our Regge model are consistent with the available constraints from phenomenology.

With the large-$N_c$ Regge model, we can calculate the pseudoscalar-meson tower exactly, i.e., we can perform the infinite sum over pseudoscalar-pole diagrams with excited pseudoscalars. For Model 2, we sum over the lowest $n=100$ radially-excited pseudoscalars ($P=\pi,\eta, \eta'$) and then fit a saturation curve,
\beq
\Delta a_\mu^{P\text{-poles}}(n_\mathrm{max})=\Delta a^{P\text{-poles}}-(\Delta a^{P\text{-poles}}-a_0)\, e^{-b \,(n_\mathrm{max})^c},
\eeq
in order to extrapolate to infinity. Here, we defined: 
\beq
\Delta a_\mu^{P\text{-poles}} (n_\mathrm{max})= \sum_{n=1}^{n_\mathrm{max}} a_\mu^{P(n)\text{-pole}},\label{DeltaADef}
\eeq
with the infinite-summation result denoted as: 
\beq
\Delta a_\mu^{P\text{-poles}}\coloneqq \Delta a_\mu^{P\text{-poles}} (\infty).
\eeq
The saturation curve procedure is illustrated in figure~\ref{fig:SaturationCurve}, where for reasons of clarity only  every other data point is plotted above $n=4$. The fits start from $n_\mathrm{max}=1$ and describe the data perfectly. 
The dotted lines indicate the extrapolated values for $\Delta a_\mu^{P\text{-poles}}$ and illustrate the good convergence of the summation already at $n_\mathrm{max}=100$. This procedure has been verified with the large-$N_c$ Regge model, for which the sum is already saturated at $n_\mathrm{max}=100$.

For the full pseudoscalar-pole contributions to $(g-2)_\mu$, we obtain:
\begin{align}
 &\sum_{n=0}^\infty a_\mu^{\pi(n)\text{-pole}}\big|_\text{Model 1}=67.1(0.4)\times 10^{-11},\;\qquad \sum_{n=0}^{\infty}a_\mu^{\pi(n)\text{-pole}}\big|_\text{Model 2}= 68.4\times 10^{-11},\notag\nn\\
&  \sum_{n=0}^\infty a_\mu^{\eta(n)\text{-pole}}\big|_\text{Model 1}=19.9^{\,+1.1}_{\,-0.9}\times 10^{-11},\qquad \sum_{n=0}^{\infty}a_\mu^{\eta(n)\text{-pole}}\big|_\text{Model 2}= 22.1\times 10^{-11},\notag\nn\\
 &\sum_{n=0}^\infty a_\mu^{\eta'(n)\text{-pole}}\big|_\text{Model 1}=21.3(1.2)\times 10^{-11},\qquad \sum_{n=0}^{\infty}a_\mu^{\eta'(n)\text{-pole}}\big|_\text{Model 2}=24.2 \times 10^{-11},\notag\nn\\
 &\sum_{n=0}^\infty a_\mu^{\pi(n)\text{-pole}}+a_\mu^{\eta(n)\text{-pole}}+a_\mu^{\eta'(n)\text{-pole}}\big|_\text{Model 1}=108.3^{\,+1.8}_{\,-1.7}\times 10^{-11},\label{FinalA}
\end{align}
where the uncertainty of the Model 1 prediction is solely estimated based on the error propagated from the input parameters on $\sum_{n=0}^{100}a_\mu^{P(n)\text{-pole}}$. Isolating the contribution from excited pseudoscalars, one finds:
\begin{align}
 \Delta a_\mu^{\pi\text{-poles}}\big|_\text{Model 1}&=2.7(0.4)\times 10^{-11},\qquad  &\Delta a_\mu^{\pi\text{-poles}}\big|_\text{Model 2}&=3.9 \times 10^{-11},\notag\nn\\
  \Delta a_\mu^{\eta\text{-poles}}\big|_\text{Model 1}&=3.4^{\,+0.9}_{\,-0.7}\times 10^{-11},\qquad  &\Delta a_\mu^{\eta \text{-poles}}\big|_\text{Model 2}&=4.3\times 10^{-11},\notag
\\
 \Delta a_\mu^{\eta'\text{-poles}}\big|_\text{Model 1}&=6.5(1.1)\times 10^{-11},\qquad  &\Delta a_\mu^{\eta'\text{-poles}}\big|_\text{Model 2}&=8.2 \times 10^{-11}.\label{DeltaA12}
\end{align}
 The difference between the $\Delta a_\mu^{P\text{-poles}}$ results from Model 1 and Model 2 can be used to quantify the systematic uncertainty of our prediction:
\begin{align}
 \Delta a_\mu^{\pi\text{-poles}}&=2.7\,(0.4)_\text{Model 1}\,(1.2)_\text{syst}\times 10^{-11}=2.7\,(1.3)\times 10^{-11},\nn \\
  \Delta a_\mu^{\eta\text{-poles}}&=3.4^{\,+0.9}_{\,-0.7}\big\vert_\text{Model 1}\,(0.9)_\text{syst}\times 10^{-11}=3.4^{\,+1.3}_{\,-1.1}\times 10^{-11},\nn \\
   \Delta a_\mu^{\eta'\text{-poles}}&=6.5\,(1.1)_\text{Model 1}\,(1.7)_\text{syst}\times 10^{-11}=6.5\,(2.0)\times 10^{-11}.\label{DeltaACombinedPionEtaP}
\end{align}
With the alternative assignment of $\eta^{(\prime)}$ excitations in the radial Regge trajectories~\cite{Klempt:2007cp}, see dot-dashed purple lines in figure~\ref{fig:TrajectoryPlotPion}, we obtain:
\begin{align}
  \Delta a_\mu^{\eta\text{-poles}}=3.4\times 10^{-11},\qquad
   \Delta a_\mu^{\eta'\text{-poles}}=6.4\times 10^{-11},
\end{align}
indicating that the net effect is remarkably insensitive to the assignment of the $\eta$, $\eta'$ Regge trajectories.  
Expressing $C_{\eta^{(\prime)}}^2$ through the experimental $F_{\eta^{(\prime)}\ga\ga}$ and $F_{\eta^{(\prime)}}$, see left-hand side of~\eqref{F_MV}, instead of the $\eta$--$\eta'$ mixing parameters, 
leads to a decrease of $a_\mu^{\eta^{(\prime)}\text{-poles}}$ that is well within the uncertainty quoted in~\eqref{DeltaACombinedPionEtaP}. Our final result for the sum of pion, $\eta$, and $\eta'$ states is: 
\begin{align}
 \Delta a_\mu^{\text{PS-poles}}&=\Delta a_\mu^{\pi\text{-poles}}+\Delta a_\mu^{\eta\text{-poles}}+\Delta a_\mu^{\eta'\text{-poles}}\notag\\
 &=12.6^{\,+1.6}_{\,-1.5}\big\vert_\text{Model 1}\,(3.8)_\text{syst}\times 10^{-11}\nn\\
 &=12.6(4.1)\times 10^{-11}.
 \label{final}
\end{align}
For Model 2 (Model 1) roughly $50\,\%$ ($80\,\%$) of  $\Delta a_\mu^{\text{PS-poles}}$ is generated by the lowest (observed) pseudoscalar excitations listed in figure \ref{fig:TrajectoryPlotPion}. This can be seen in figure \ref{fig:SaturationCurve} where $\Delta a_\mu^{P\text{-poles}}(n_\mathrm{max})$ is shown for Model 2.

\section{Matching quark loop and Regge model}
\label{sec:matching}

\subsection[Matching at the level of $(g-2)_\mu$]{Matching at the level of $\boldsymbol{(g-2)_\mu}$}

\begin{figure}[t]
 \centering
 \includegraphics[width=0.7\linewidth]{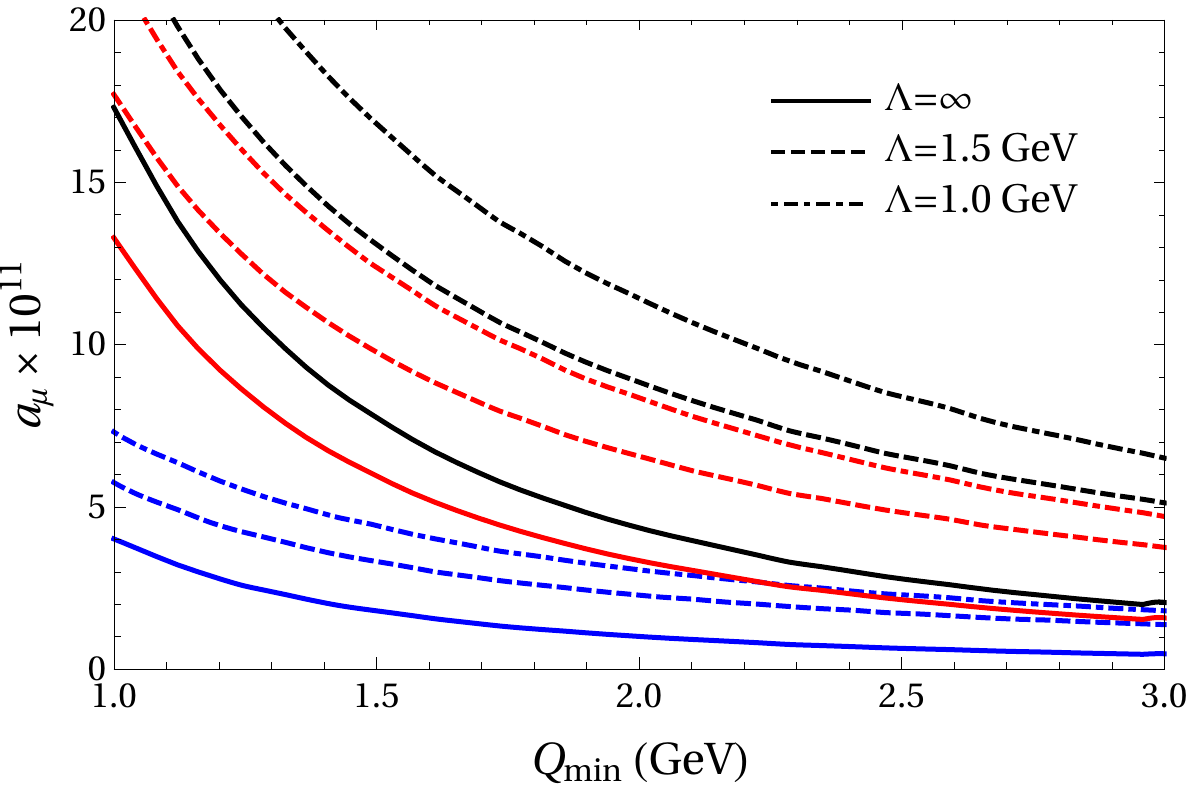}
	\caption{Contribution of the pQCD quark loop (with vanishing quark mass) to $a_\mu$  from the region $Q_{1,2}\geq Q_\text{min}$ with the contribution from $Q_3$ below $Q_\text{min}$ damped by $Q_3^2/(Q_3^2+\Lambda^2)$ (plus crossed). The total contribution from $\bar\Pi_{1\text{--}12}$ is shown in black, together with the partial ones from $\bar\Pi_{1\text{--}2}$ (red) and $\bar\Pi_{3\text{--}12}$ (blue).
	The pQCD contribution with common lower cutoff in all $Q_i$ is reproduced in the limit $\Lambda\to\infty$.}
	\label{quark_loop_MV}
\end{figure}

The simplest and most instructive matching to the massless pQCD quark loop proceeds at the level of the $(g-2)_\mu$ integral. 
The asymptotic pQCD region where all $Q_i$ are large can be captured by imposing the condition that all $Q_i$ be larger than $Q_\text{min}$. 
To be able to add the mixed regions, where one virtuality is smaller than $Q_\text{min}$, in the quark-loop integration, one needs to
dampen the contribution in the additional integration region, since it is already partly covered by the ground-state pseudoscalar poles.
To this end, we introduce the suppression factor $Q^2/(Q^2+\Lambda^2)$ for the virtuality $Q<Q_\text{min}$, while retaining the cut  
that at least two $Q_i\geq Q_\text{min}$. In this way, the limit $\Lambda\to\infty$ reproduces a common lower cutoff on all $Q_i$. 
The results are shown in figure~\ref{quark_loop_MV} for the total ($\bar\Pi_{1\text{--}12}$) as well as for longitudinal ($\bar\Pi_{1\text{--}2}$)
and transversal ($\bar\Pi_{3\text{--}12}$) contributions separately.

The Regge models in the preceding section predict a ratio $\Delta a_\mu^{\eta,\eta'}/\Delta a_\mu^{\pi^0}$
near the expectation $(C_0^2+C_8^2)/C_3^2=3$. 
Similarly, we obtain $\Delta a_\mu^{\eta'}/\Delta a_\mu^{\eta}$ close to $2$, as suggested by the scaling with $C_{\eta'}^2/C_\eta^2$~\eqref{Ceta2}.
To first approximation, the implementation of the various asymptotic constraints on the HLbL tensor thus reproduces the simple scaling that originates 
from the weight factors~\eqref{weights} appearing in the $VVA$ triangle.  
For the mixed regions this behavior is exact due to~\eqref{MVConstraintSimple} and \eqref{MVConstraintSimpleETA},  as long as the low-energy properties of the HLbL tensor are not disturbed, 
while for the asymptotic region it is a consequence of the flavor decomposition chosen in~\eqref{pQCDSimple} and~\eqref{pQCDSimpleETA}. 
The fact that the results from the summation of excited pseudoscalars confirm these expectations 
indicates that the pQCD quark loop dictates, if not the overall size of the effect, at least its decomposition in the various isospin channels. This is an encouraging sign that the model dependence which is intrinsic in the approach we are following here is mitigated by the QCD constraints. 
To understand even better the extent to which this mitigation occurs we analyze here in detail the matching between the Regge models and the quark loop integral, after introducing appropriate cutoffs.

For $\Delta a_\mu^{\text{PS-poles}}\sim 13\times 10^{-11}$ figure~\ref{quark_loop_MV} suggests scales $\Lambda$ and $Q_\text{min}$ around $1.4\GeV$. 
In addition, the pQCD quark loop would predict an additional increase from the transversal 
amplitudes around $4\times 10^{-11}$, but for these scales the interplay with axial-vector resonance contributions needs to be studied in more detail.
In the following, we will instead focus on the comparison of our Regge model and the pQCD quark loop
in the longitudinal amplitudes.

\subsection{Matching of short-distance contributions}
\label{sec:match-SDC}
\begin{figure}[t]
 \centering
 \includegraphics[width=0.7\linewidth]{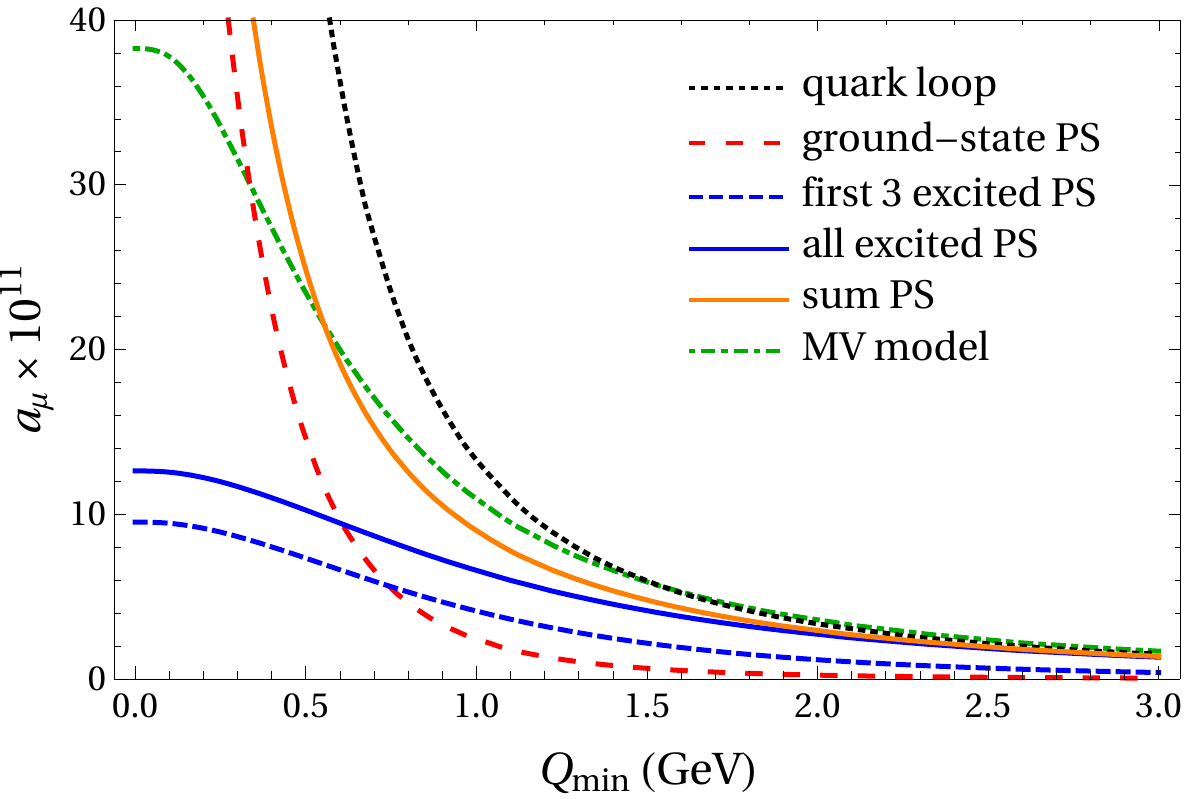}
	\caption{Contribution to $a_\mu$ for $Q_{i}\geq Q_\text{min}$: the longitudinal part of the massless pQCD quark loop (dotted black), the ground-state pseudoscalars (long-dashed red), their excitations from the large-$N_c$ Regge model (blue), the sum of both (orange),
	and the short-distance implementation from the MV model (dot-dashed green). The blue dot-dashed curve refers the sum of the first three excited pseudoscalars in each trajectory.}
	\label{SD_comparison}
\end{figure}

Beyond the matching at the level of $(g-2)_\mu$, it could be instructive to also compare the specific contributions to the BTT functions in the various kinematic domains. 
However, once the respective scaling with the virtualities is factored out, we find that the coefficient converges relatively slowly to its asymptotic value.
We conclude that it is rather the convolution with the kernel functions $T_i$
that becomes important to assess the relevant scales of the SDCs for the HLbL contribution.

\begin{figure}[t]
 \centering
 \includegraphics[width=0.7\linewidth]{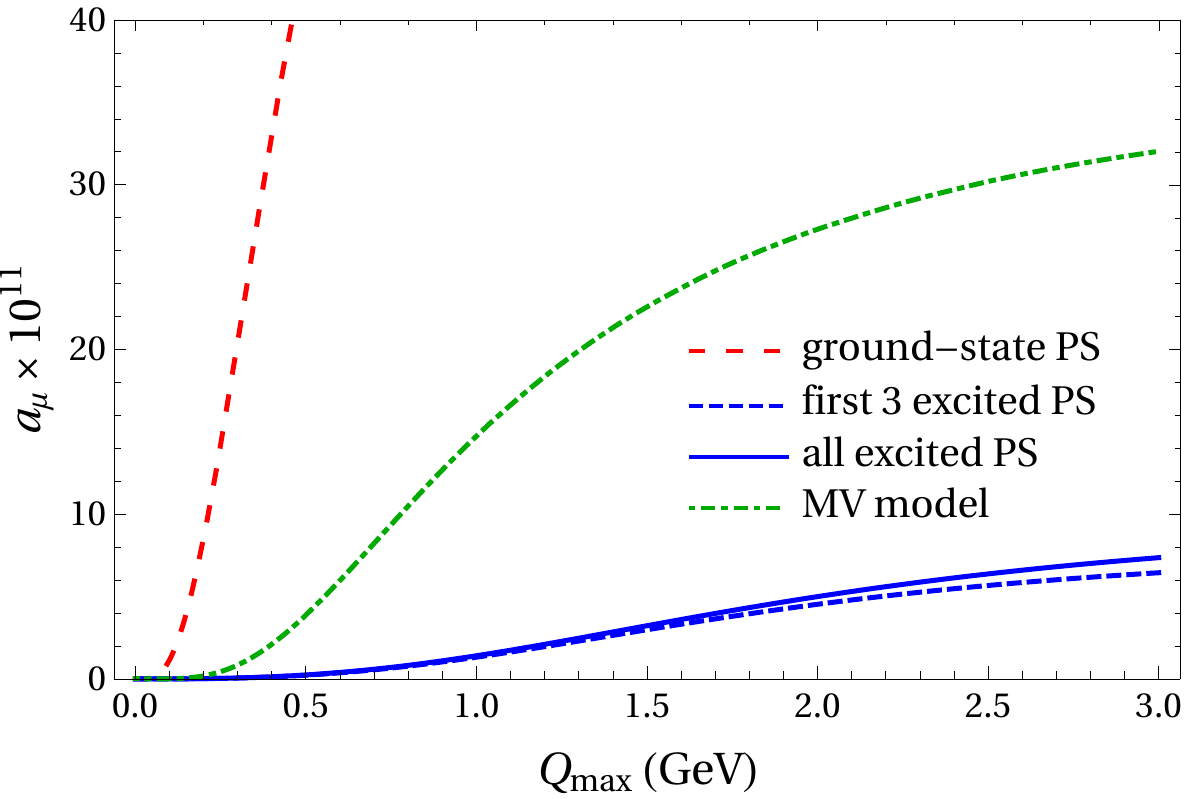}
	\caption{Same as figure~\ref{SD_comparison}, but for $Q_{i}\leq Q_\text{max}$.}
	\label{SD_comparison_max}
\end{figure}

This is illustrated in figure~\ref{SD_comparison}, which shows various contributions to $a_\mu$ as a function of a lower cutoff on all three virtualities $Q_i$, 
as well as in figure~\ref{SD_comparison_max}, which shows the opposite case of an upper cutoff on all three virtualities $Q_i$.
The ground-state pseudoscalars are saturated by $90\%$ for $Q_\text{max}=1.5\GeV$, while for the excited pseudoscalars only about $25\%$ of the total contribution
comes from this energy region. 
By construction, their contribution asymptotically matches onto the one from the pQCD quark loop, and figure~\ref{SD_comparison} shows how fast 
that asymptotic limit is reached after convolution with the $(g-2)_\mu$ integral kernels: at $1.5\GeV$ it is saturated by $70\%$, or about $80\%$ if the tail of the ground-state pseudoscalars is included.

\begin{figure}[t]
 \centering
 \includegraphics[width=0.7\linewidth]{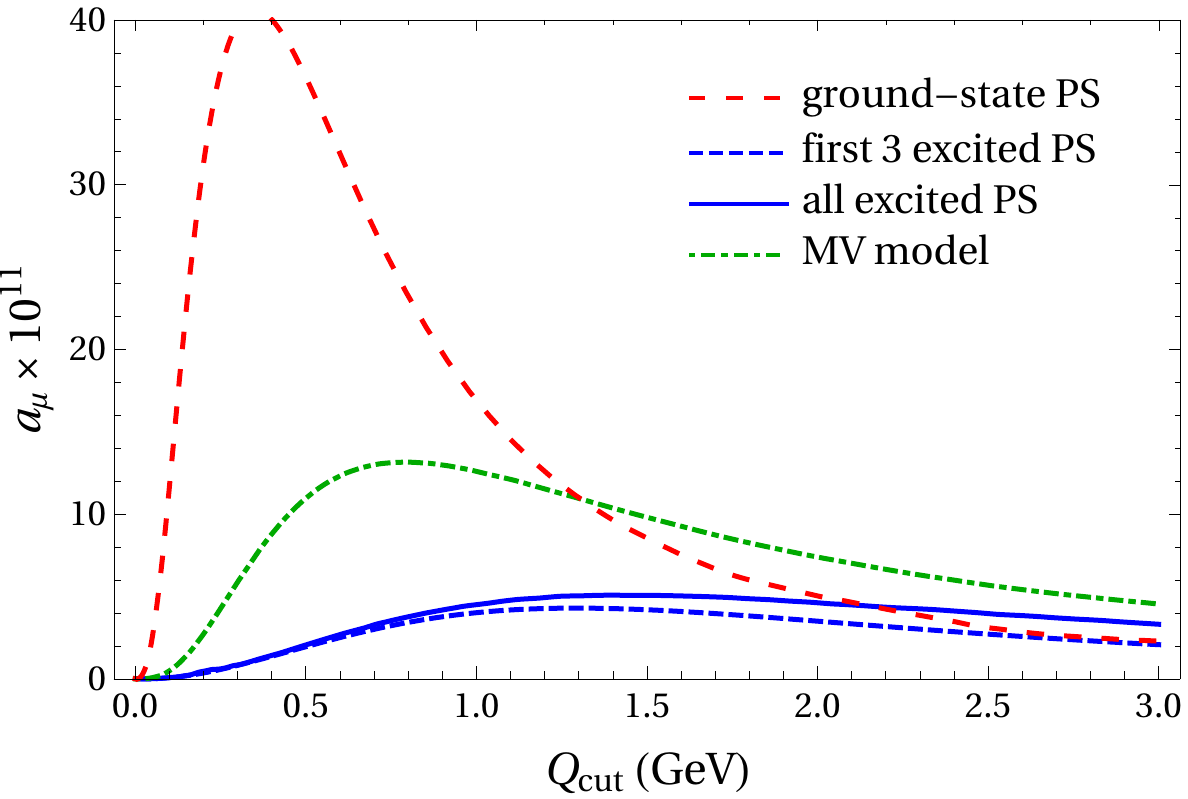}
	\caption{Mixed-region contribution to $a_\mu$, defined as the full integral minus the contributions from the low-energy (all $Q_{i}\leq Q_\text{cut}$) and high-energy (all $Q_{i}\geq Q_\text{cut}$)
	regions.}
	\label{SD_comparison_mixed}
\end{figure}

The mixed region is more difficult to illustrate, especially for the corresponding OPE constraint, 
because in addition to the hierarchy $Q_3^2\ll Q_1^2\sim Q_2^2$ the small virtuality still needs to be large compared to $\Lambda_\text{QCD}$, 
otherwise chiral corrections will become important. 
For that reason, the low-energy part of the integration region was suppressed by the second cutoff $\Lambda$ in figure~\ref{quark_loop_MV}. 
To obtain some measure of the size of the mixed-region contribution, figure~\ref{SD_comparison_mixed} shows the remainder if for a given cutoff $Q_\text{cut}$ both the regions 
where all $Q_{i}\leq Q_\text{cut}$ and all $Q_{i}\geq Q_\text{cut}$ are subtracted. For the ground-state pseudoscalars at $Q_\text{cut}=1.5\GeV$, 
this produces the remaining $10\%$ beyond the low-energy region, while the asymptotic region $Q_i\geq Q_\text{cut}$ is already largely negligible. 
For the sum of excited pseudoscalars, it is instructive to further scrutinize the decomposition at this scale into low-energy ($25\%$), mixed ($40\%$), and asymptotic ($35\%$) regions.
As concerns the contribution from the lowest excitations, in Model 1, the low-energy region is entirely saturated by the sum of the first three excitations, 
the mixed region by $80\%$, but for the asymptotic part of the integral the higher excitations make up about $50\%$. 
This pattern suggests the interpretation that indeed the lowest excitations are most important for the low-energy and mixed regions, while the infinite
tower of resonances restores the correct asymptotic behavior. In fact, we find that the numerical impact of the integration regions where the OPE constraint strictly applies, i.e., where both $Q_3\ll Q_{1,2}$ and $Q_i\gg\Lambda_\text{QCD}$, is already very small, so that in practice its main effect lies in constraining the TFF Regge models.

Altogether, this discussion indicates that at some point around $1.5\GeV$ the description of the HLbL tensor in terms of hadronic intermediate states should be matched onto the one from pQCD. 
In particular, the implementation of the SDCs in terms of excited pseudoscalars gives an indication how big an impact the intermediate regime may have (in the longitudinal amplitudes): 
while from pQCD alone one may have guessed a contribution around $5\times 10^{-11}$ from the asymptotic region, for a value of $Q_\text{min}$ chosen at $1.5\GeV$,
the excited pseudoscalars with masses in the same region will add a contribution of similar size, covering also the mixed regions of the $(g-2)_\mu$ integral.

To quantify the matching between the quark loop and the description in terms of hadronic states, one would need to define a concrete criterion for the matching scale. One way to define an optimal scale could be to consider the difference between Regge model and quark loop as a function of $Q_\text{min}$ in combination with the uncertainties of each description for a particular cutoff. For the Regge model, we can estimate this uncertainty as before, but for the quark loop one would need to know the $\alpha_s$ corrections and/or higher orders in the OPE, which when compared to the leading-order quark loop would already entail information about the scale where pQCD becomes an efficient description of the HLbL tensor. 
Absent such calculations, we may obtain a first estimate by comparison to similar pQCD uncertainties in inclusive $\tau$ decays~\cite{Davier:2008sk,Beneke:2008ad,Maltman:2008nf,Narison:2009vy,Caprini:2009vf,Pich:2013lsa}, given that we expect a matching scale 
not too far off the $\tau$ mass, which would suggest an uncertainty around $20\%$.  
Based on the combined uncertainties of the Regge model and the pQCD quark loop, we then find a preference for a matching scale around $Q_\text{match}=1.7\GeV$, leading to the decomposition
\begin{align}
    \Delta a_\mu^\text{PS-poles} - a_\mu^\text{PS-poles}\big|_{Q_\text{min}=Q_\text{match}}&=8.7 (3.6)\times 10^{-11},\notag\\
    a_\mu^{q\text{-loop}}\big|_{Q_\text{min}=Q_\text{match}}&=4.6(9)\times 10^{-11},
    \label{final_matching}
\end{align}
Note that, in the first line, we also subtracted the very small contribution from the ground-state pseudoscalars from the integration region $Q_i\geq Q_\text{match}$ to avoid any double counting. 
As expected, the comparison to~\eqref{final} confirms that for the asymptotic part of the integral it does not matter whether a description based on hadronic intermediate states or pQCD is employed: this means that about one third (i.e., the second line) of~\eqref{final_matching} is a model-independent part of the effect we have calculated. But how model dependent is the rest and can we adequately cover this model dependence with our uncertainty estimate? There are different uncertainties which need to be considered and we summarize all of them here:
\begin{itemize}
\item $3.6$ units coming from the uncertainties in the parameters of Model 1, as given in~\eqref{final_matching}, obtained by stretching 
the uncertainties in the Regge slopes by a factor three;\footnote{We thereby aim to cover scenarios in which other hadronic states could be used to implement the SDCs, in which case the Regge slopes would differ; e.g., according to ref.~\cite{Masjuan:2012gc}, the Regge slopes of the axial-vector $a_1$ and $f_1$ trajectories are $\sigma_{a_1}^2=1.36(49)\GeV^2$ and $\sigma_{f_1}^2=1.27(64)\GeV^2$.}
\item $1.7$ units are obtained by varying the matching point by $0.5\GeV$ (the main effect comes from the lowest $Q_\text{match}$, which we  vary to as low as $1.2\GeV$);
\item $3.8$ estimated from the difference between Model 1 and 2, cf.~\eqref{final}.
\end{itemize}
All these uncertainties concern essentially the contribution below the matching point of $Q_\text{match}=1.7\GeV$,
as estimated in the previous section.
The outcome of our analysis for this part therefore reads:
\begin{align}
     \Delta a_\mu^\text{PS-poles} - a_\mu^\text{PS-poles}\big|_{Q_\text{min}=Q_\text{match}}&=8.7 (3.6)_\text{excited PS}\,{}^{\,+1.7}_{\,-0.4}\big\vert_{Q_\text{match}}(3.8)_\mathrm{syst}\times 10^{-11}\nn\\
    &=8.7 (5.5)\times 10^{-11},
     \label{result_PS}
\end{align}
with a 65\% uncertainty, which we consider as sufficiently conservative and, in addition, covers the systematic effects related to the asymptotic behavior of the excited-state TFFs as discussed in appendix~\ref{sec:systematics}. Another observation corroborating this conclusion is that the contribution to the central value due to the first three pseudoscalar excitations (whose masses and, in part, two-photon couplings are constrained by phenomenology) amounts to $7.8$ units out of $8.7$. 
On the basis of these considerations we give as our final estimate 
\begin{equation}
 \Delta a_\mu^\text{LSDC}=\left[8.7(5.5)_{\text{PS-poles}}+4.6(9)_{q\text{-loop}}\right]\times 10^{-11} = 13(6)\times 10^{-11},
\label{final_result}
\end{equation}
and stress that the contribution of the higher excitations ($n>3$), which has been calculated with our Regge model and is the most uncertain and model-dependent part of our calculation, amounts to only less than $10\%$ of the total. We conclude that our final result~\eqref{final_result} has a generously estimated uncertainty that we expect to cover the remaining model dependence.

\subsection{Chiral limit and role of axial-vector mesons}
\label{sec:chiral_limit}

One may ask whether the implementation of the longitudinal SDCs adopted
here would work in the chiral limit: in this limit, excited pseudoscalars
have a vanishing coupling to the axial current and therefore would not be
able to contribute to the fulfillment of the OPE constraint.\footnote{We
  thank Arkady Vainshtein for calling our attention to this point.}
However, there are known cases in which the chiral and the large-$N_c$
limits do not commute, most notably in the context of baryon chiral
perturbation theory.  For instance, if one first takes $N_c\to\infty$ and
then $m_q\to 0$, the entire tower of excited baryons contributes to the
first non-analytic term in the quark-mass expansion of the nucleon mass,
while in the opposite order only nucleon intermediate states
appear~\cite{Gasser:1980sb,Dashen:1993jt}, and similar subtleties arise
elsewhere due to mass splittings of order $1/N_c$~\cite{Cohen:1996zz}.
Further subtleties in the order of the chiral limit have been pointed out
before even for the $VVA$ anomaly itself: the discontinuity of the fermion
triangle loop function in the axial-vector virtuality vanishes with the
fermion mass, but in a dispersion relation the mass dependence is canceled
and produces the anomaly that survives in the chiral
limit~\cite{Dolgov:1971ri,Zakharov:1990ef}. While these examples show that
care is required when exchanging the limits, at least the implementation of
the large-$N_c$ Regge models described here does not 
allow any such
subtleties to occur and is meant to be used only away from the chiral
limit.

If the excited pseudoscalar poles were to decouple in the chiral limit, an
alternative solution could be provided by axial-vector intermediate states,
which do contribute in the chiral and large-$N_c$ limits. For these mesons,
however, only model-dependent calculations are available in the literature
so far,\footnote{They are either based on the relation to transversal
  SDCs~\cite{Melnikov:2003xd,Jegerlehner:2017gek} or proceed in terms of
  Lagrangian models~\cite{Pauk:2014rta,Roig:2019reh}.} 
  whereas a calculation based on a dispersive framework is still lacking. Such a
framework would allow one to express the contribution to HLbL scattering in the most
general way in terms of all TFFs of the axial vectors (which admit three), as is
the case for the pseudoscalars (which admit only one). Another significant
difference is that while for pseudoscalars the sum rules that guarantee the
absence of ambiguities in the evaluation of the HLbL contribution~\cite{Colangelo:2017fiz} are
automatically satisfied, this is not the case for axial-vector mesons. We
believe that at present it is fair to say that even the ground-state
contributions of the latter are poorly understood.

Besides these theoretical reasons, there are also phenomenological ones
that favor a discussion in terms of pseudoscalars: while for the most
relevant excited pseudoscalar resonances, those in the energy range between
$1$--$2\GeV$, there is at least some information on the phenomenology 
relevant for HLbL, the situation is even worse for the known
axial-vector resonances in the same mass range. This is related to the fact
that for axial vectors a decay into two real photons is forbidden by the
Landau--Yang theorem~\cite{Landau:1948kw,Yang:1950rg}: 
hadronic channels such as three pions are dominant with
respect to suppressed decays to virtual photons, which have been observed
only for two ground-state axial-vector resonances~\cite{Achard:2001uu,Achard:2007hm}.

If a viable implementation of the longitudinal SDCs in terms of axial-vector resonances were possible, it
would have to look quite different from ours in terms of pseudoscalars excitations.
Besides the fact that different TFFs contribute, we observe that the axial-vector contribution to $\hat\Pi_{1\text{--}3}$ does not resemble 
the pseudoscalar-pole contribution~\eqref{Pi_PS}, in fact, both in a Lagrangian-based approach 
and in dispersion theory the pole in $q_3^2-m_A^2$ cancels in the longitudinal BTT amplitudes.
Based on what is known about the axial-vector TFFs, we cannot preclude the possibility 
that a finite number of axial vectors could be used to construct such a solution. 
If that were possible while being consistent with phenomenology and 
the SDCs on the axial-vector TFFs, this would be an appealing solution,
but the necessary theoretical framework for carrying out such an
analysis is not yet available. 
For the moment we took the pragmatic point of
view that we implement the longitudinal SDCs in terms of the hadronic
intermediate states that we can control best, both theoretically as well as
phenomenologically. Having adopted this strategy, we need to address the
question of whether our estimate of the systematic uncertainty is large
enough to cover the possibility of implementing the SDCs in terms of 
other hadronic intermediate states.

We believe that we can answer positively this question under the reasonable
assumption that even in an alternative scenario the matching to pQCD will occur in
the range we have considered. In this case the contribution from the quark
loop will remain unchanged and all we need to discuss is the excited-pseudoscalar-pole
contribution estimated as $8.7(5.5)\times 10^{-11}$. About
one unit out of nine comes from excited states with $n>3$: if these were
not needed to satisfy the SDCs, this contribution would have to be dropped,
a possibility amply covered by our uncertainties. 
The bulk of the
contribution comes from excited states with $n \leq 3$, and as we have
discussed in section~\ref{sec:two_photon_couplings} and in appendix~\ref{app:photon_couplings}, the estimate of
their two-photon couplings we have obtained by requiring that the
longitudinal SDCs be satisfied is compatible with what is know from
phenomenology. Our uncertainty estimate covers the present phenomenological
uncertainties on these couplings and could be reduced if the
phenomenological information on them were improved. In the end, 
even if the SDCs were to be implemented using axial-vector states, 
the first few pseudoscalar excitations would need to be included regardless,
it is just that the pseudoscalar TFFs would not be constrained by 
the HLbL SDCs. 

In conclusion, we believe that the uncertainty related to the nature of the hadronic states used in the implementation of the SDCs 
should be covered by the error assigned in~\eqref{result_PS}, an expectation that has been supported more recently by models in holographic QCD~\cite{Leutgeb:2019gbz,Cappiello:2019hwh}, in which the SDCs are implemented by summation of an infinite tower of axial-vector resonances instead.
Beyond the model context, assessing the role of axial-vector resonances in fulfilling the SDCs, especially the transversal ones, will first of all require 
an improved understanding of their ground-state contributions.

\section{Comparison to the Melnikov--Vainshtein model}
\label{sec:MV}

\begin{figure}[t]
 \centering
 \includegraphics[width=0.3\linewidth]{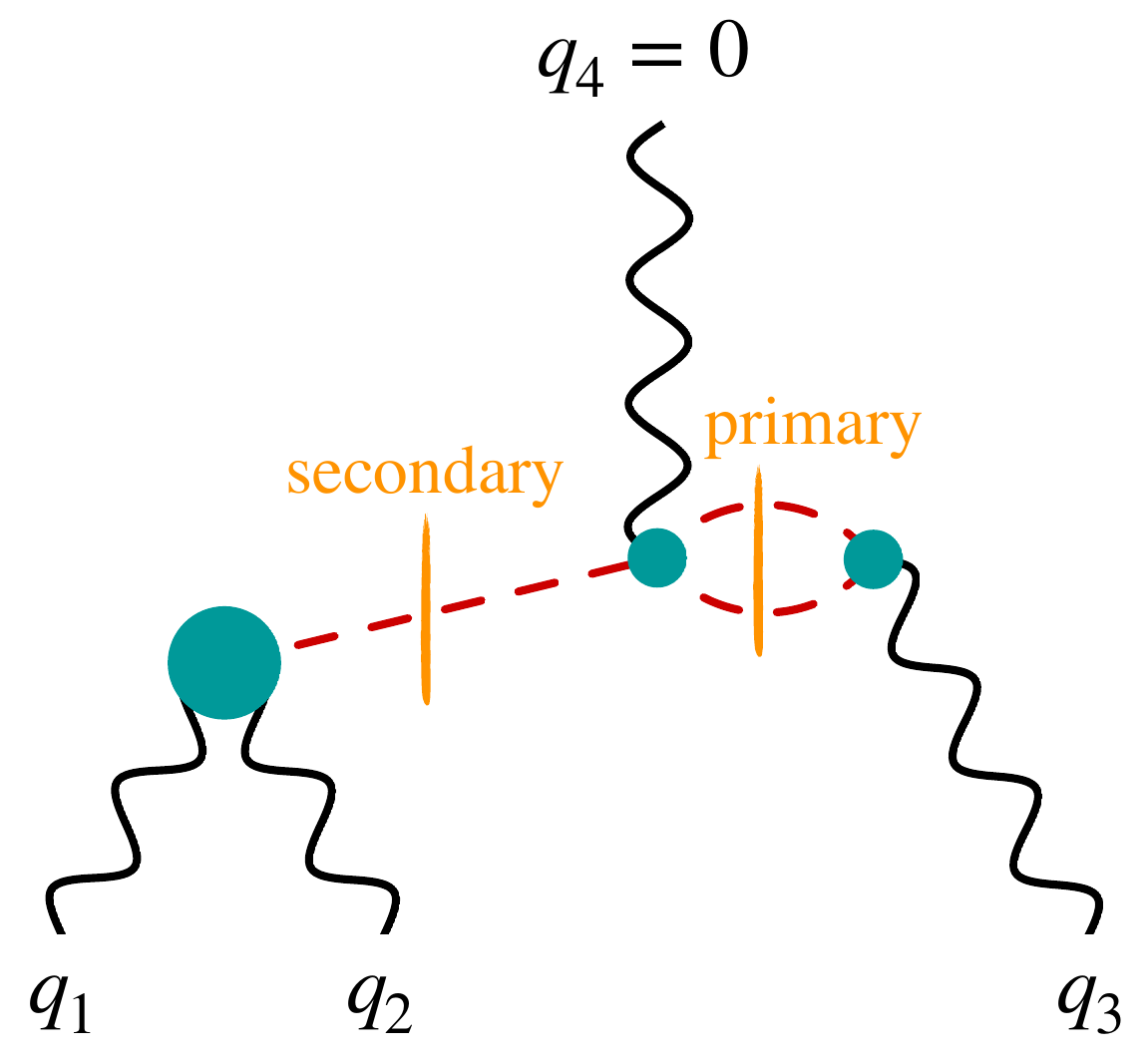}
	\caption{Dispersion relation for the HLbL tensor in the $(g-2)_\mu$
kinematic limit with singularities from primary and secondary channels ($2\pi$ state and  pseudoscalar pole).}
	\label{CutsDR}
\end{figure}

In this section, we compare our implementation of the longitudinal SDCs to the one from ref.~\cite{Melnikov:2003xd}, which is based on the observation that the modification
\beq
\label{MVprescription}
\hat\Pi_1^{P\text{-pole}}=\frac{F_{P\gamma^*\gamma^*}(q_1^2,q_2^2)F_{P\gamma\gamma^*}(q_3^2)}{q_3^2-M_P^2}
\to 
\hat\Pi_1\big|_\text{MV}=\frac{F_{P\gamma^*\gamma^*}(q_1^2,q_2^2)F_{P\gamma\gamma}}{q_3^2-M_P^2}
\eeq
of~\eqref{Pi_PS} ensures that both the normalization and the mixed-region OPE constraint~\eqref{OPE_MV} are fulfilled.\footnote{A reply to the preprint~\cite{Melnikov:2019xkq}, which appeared in response to our paper, is provided in appendix~\ref{app:MV_comment}.} 
Since the form~\eqref{Pi_PS} of the pseudoscalar poles is a direct consequence of the dispersion relation for the HLbL tensor, which we suggested in refs.~\cite{Colangelo:2014dfa,Colangelo:2014pva,Colangelo:2015ama,Colangelo:2017qdm,Colangelo:2017fiz} for the case of general four-point kinematics, this modification is not compatible with the description of other intermediate states in the same framework. However, the replacement~\eqref{MVprescription} could be justified by a dispersion relation for the HLbL amplitudes directly in the kinematic limit relevant for $(g-2)_\mu$, i.e., for $q_4=0$. In this limit it is not possible to work with a dispersion relation in the Mandelstam variables $s$, $t$, and $u$ at fixed $q_i^2$, because they cease to be independent: $s=q_3^2$, $t=q_2^2$, and $u=q_1^2$. This means that when writing dispersion relations in the $q_i^2$ for $g-2$ kinematics, there is no clear separation between the singularities of the HLbL amplitude and those generated by hadronic intermediate states directly coupling to individual electromagnetic currents, e.g., two-pion states as in figure \ref{CutsDR}:
in this framework the pseudoscalar TFFs can no longer be treated as external input quantities. 
The same holds for higher intermediate states, so that in general the factorization of form factors and 
scattering amplitudes of the intermediate state in question would cease to apply.

Nevertheless, we observe that, in principle, both forms of dispersion
relations are perfectly legitimate---the transition from the dispersion
relation for the four-point
function~\cite{Colangelo:2014dfa,Colangelo:2014pva,Colangelo:2015ama,Colangelo:2017qdm,Colangelo:2017fiz}
to a dispersion relation in the photon virtualities in the $(g-2)_\mu$
kinematic limit amounts to a relabeling of contributions from different
principal cuts. This is illustrated by writing the pseudoscalar
pole~\eqref{Pi_PS} in our framework as
\begin{align}
	\label{eq:ReshuffledPole}
	\hat\Pi_1^{P\text{-pole}} &= \frac{F_{P\gamma^*\gamma^*}(q_1^2,q_2^2)F_{P\gamma\gamma^*}(M_P^2)}{q_3^2-M_P^2} +  \frac{F_{P\gamma^*\gamma^*}(q_1^2,q_2^2)\big( F_{P\gamma\gamma^*}(q_3^2) - F_{P\gamma\gamma^*}(M_P^2) \big)}{q_3^2-M_P^2} ,
\end{align}
where the first term reproduces the pole in the alternative dispersive
framework for $(g-2)_\mu$ kinematics, while the second term does not
contain a pole at $q_3^2 = M_P^2$. More precisely, the second term is the
contribution due to intermediate states $X$ in a cut through the TFF, with
the discontinuity determined by the sub-processes $\gamma^*(q_3) \to X$ and
a pseudoscalar-pole contribution to $\gamma^*(q_1)\gamma^*(q_2) \to \gamma
X$, as illustrated in figure \ref{CutsDR}. This piece is present even in the alternative dispersive framework,
which demonstrates that changing the dispersive framework simply amounts to
a reshuffling of contributions between different principal cuts. Due to the
mixed-region SDC, for large $q_1^2 \sim q_2^2$ the second piece
in~\eqref{eq:ReshuffledPole} has to cancel against the contribution from
the infinite tower of higher intermediate states up to chiral corrections.
Since this is a key point in ref.~\cite{Melnikov:2003xd}, and the basis for
the construction of the MV model, it is worthwhile discussing in detail how
this cancellation has to work. For simplicity we concentrate on the
pion contribution only (isospin-triplet component) and include all other
contributions other than the pion pole to $\hat \Pi_1^{3}$ into a single
function $G$: 
\begin{equation}
\hat\Pi_1^{3}=\hat\Pi_1^{\pi\text{-pole}}(q_1^2,q_2^2,q_3^2)+G(q_1^2,q_2^2,q_3^2).
\end{equation}
From the requirement that~\eqref{MVconstraint} provide the leading term
for asymptotic $q_1^2 \sim q_2^2$, but that it be exact (in the chiral
limit) as far as the $q_3^2$ dependence is concerned, it follows that the
function $G$ must have the following asymptotic behavior:
\begin{align}
\lim_{\hat{q}^2 \to \infty} \hat{q}^2
G(\hat{q}^2,\hat{q}^2,q_3^2)\Big|_{m_q=0} =\frac{2 F_\pi}{3}\frac{F_{\pi
    \gamma \gamma^*}(q_3^2)-F_{\pi \gamma \gamma}}{q_3^2}\bigg|_{m_q=0}.
\label{eq:chiral}
\end{align}
We stress that~\eqref{eq:chiral} is exact in the chiral limit, a
property inherited from~\eqref{MVconstraint}, which is a remarkable and
interesting result. The MV model~\eqref{MVprescription} consists of taking
the pion pole as the only contribution to $\hat{\Pi}_1^3$. This effectively
amounts to promoting~\eqref{eq:chiral} to an equation valid for any
value of $q_1^2$ and $q_2^2$: 
\begin{align}
G(q_1^2,q_2^2,q_3^2)\Big|_{m_q=0}
=-F_{\pi \gamma^* \gamma^*}(q_1^2,q_2^2)\frac{F_{\pi \gamma \gamma^*}(q_3^2)-F_{\pi \gamma
    \gamma}}{q_3^2}\bigg|_{m_q=0}. 
\label{eq:MV1}
\end{align}
While this provides a simple way to exactly satisfy~\eqref{eq:chiral},
there is no physical justification in support of such a very strong
assumption. It is therefore not surprising that this leads to uncontrolled
numerical effects, which we have been able to quantify here. In 
addition, we note that away from the chiral limit the residue
in~\eqref{eq:ReshuffledPole} contains $F_{P\gamma\gamma^*}(M_P^2)$ instead
of $F_{P\gamma\gamma}$, which at least for $\eta^{(\prime)}$ entails
significant chiral corrections.

Numerically, ref.~\cite{Melnikov:2003xd} concluded an increase of $13.5\times 10^{-11}$ for the pion and $5\times 10^{-11}$ each for $\eta^{(\prime)}$, 
based on the modification in~\eqref{MVprescription} and the TFFs from ref.~\cite{Knecht:2001qf} (LMD+V for the pion and VMD for $\eta^{(\prime)}$):
\begin{align}
 \Delta a_\mu^{\text{PS-poles}}\big|_\text{MV}=23.5\times 10^{-11}.
\end{align}
However, we note that with modern input for the TFFs this number would increase substantially: for the pion, our Model 1 implies an increase of $16.2\times 10^{-11}$, which would increase to $17.3\times 10^{-11}$ if one used the dispersive TFF instead. Here, the change to the original MV number mainly reflects the differences 
between the LMD+V model from ref.~\cite{Knecht:2001qf} and the dispersive result for the $\pi^0$ TFF~\cite{Hoferichter:2018dmo,Hoferichter:2018kwz}. 
For $\eta^{(\prime)}$, the differences are more severe because the incorrect asymptotic behavior of the VMD form factors in the pQCD limit 
suppresses the impact of taking the singly-virtual form factor to a constant. We find $10.0 \times 10^{-11}$ and $12.1 \times 10^{-11}$ for $\eta$ and $\eta'$, respectively, 
which in total produces an increase of $38\times 10^{-11}$ beyond the pseudoscalar ground-state contributions, nearly three times the result
given in~\eqref{final}. 

Apart from the overall size, another key difference in our implementation concerns the 
hierarchy $\Delta a_\mu^{\pi\text{-poles}}<\Delta a_\mu^{\eta\text{-poles}}<\Delta a_\mu^{\eta'\text{-poles}}$ found with the excited pseudoscalars, see~\eqref{DeltaACombinedPionEtaP}, 
while in ref.~\cite{Melnikov:2003xd} the largest effect was found for the pion. 
The fact that  $\Delta a_\mu^{\eta\text{-poles}}$ comes out much smaller than $\Delta a_\mu^{\eta'\text{-poles}}$ can be partly explained by the two-photon couplings, $F_{\eta(n)\ga\ga}<F_{\eta'(n)\ga\ga}$, and also through the scaling of the excited state TFFs in the BL limit, see figure \ref{fig:Coupling}.

This observation also has consequences for the matching to the quark loop. While in our case the scaling of the flavor components 
follows essentially the expectation from the weights $C_a^2$, this is not the case for the MV model, and therefore it is less clear how the matching to pQCD 
should proceed. In fact, as shown in figure~\ref{SD_comparison}, 
despite not being implemented explicitly, the MV model also comes close to the pQCD asymptotics: 
the main difference to our Regge model occurs in the low-energy region below $1\GeV$.
This matching onto pQCD asymptotics is coincidental, however: by construction, the model saturates the MV constraint 
also in the limit in which all virtualities are large and therefore exceeds the proper pQCD limit by a factor of $3/2$. 
Since the asymptotic value is approached rather slowly, the resulting curve happens to be close to the pQCD quark loop for the range of 
$Q_\text{min}$ displayed in figure~\ref{SD_comparison}.

Figures~\ref{SD_comparison}--\ref{SD_comparison_mixed} also illustrate the origin of the difference between our implementations. 
For the reference scale of $1.5\GeV$, the low-, mixed-, and high-energy regions contribute $75\%$, $20\%$, and $5\%$, respectively, which demonstrates 
that indeed the approximations of the MV model manifest themselves primarily in the low-energy region, where the dispersive framework provides the best constraints and the contribution of higher states only leads to a moderate uncertainty.

\section{Summary and outlook}
\label{sec:summary}

In this work, we studied short-distance constraints (SDCs) for the hadronic light-by-light (HLbL) contribution to $(g-2)_\mu$. We concentrated on the longitudinal constraints that are intimately related to pseudoscalar-pole contributions.
Since the HLbL tensor can only be constrained from data in the low-energy region, but not in the mixed- and high-energy regions, 
SDCs are important for a model-independent approach towards HLbL scattering. 
In sections~\ref{sec:HLbL_tensor} and \ref{sec:OPE}, the Lorentz decomposition of the HLbL tensor from refs.~\cite{Colangelo:2015ama,Colangelo:2017fiz} 
was used to formulate the known expressions for the perturbative QCD (pQCD) quark loop and the operator-product-expansion (OPE) constraints on the HLbL tensor, respectively. 
The OPE constraint in the symmetric region with $Q_1^2=Q_2^2=Q_3^2\equiv Q^2$ is given in~\eqref{pQCDloop}, the Melnikov--Vainshtein constraint~\cite{Melnikov:2003xd} for the mixed region 
with $Q_3^2\ll Q_1^2\sim Q_2^2$ in~\eqref{OPE}. Both are implemented including the singlet component, for which in addition to chiral corrections also perturbative corrections arise.

Subsequently, we focused on the  longitudinal SDCs, related to the pseudoscalar-pole diagrams~\eqref{Pi_PS} by means of~\eqref{OPE_MV}. While a finite number of poles cannot saturate the SDCs, 
an infinite tower of them can \cite{Melnikov:2003xd}. To that end, we have constructed two models for the transition form factors (TFFs) of ground-state and radially-excited pseudoscalar mesons: our large-$N_c$ Regge model for pion, $\eta$, and $\eta'$ is described in sections~\ref{sec:Model}
and \ref{sec:ModelEta}, and an alternative model using the Regge resummation from ref.~\cite{RuizArriola:2006jge} is introduced in appendix \ref{sec:PionAlternative} to estimate the systematic uncertainty (see also appendix~\ref{sec:systematics}). 
While applicable only in the space-like region as relevant for $(g-2)_\mu$, both models satisfy all relevant low- and high-energy constraints for the TFFs---the chiral anomaly (normalization), the Brodsky--Lepage limit, and the symmetric pQCD limit, see~\eqref{ConstraintEquations1}--\eqref{ConstraintEquations} and \eqref{Constraint1Eta}--\eqref{Constraint2Eta}---give a good description of the experimental data, and reproduce the established results for the ground-state contributions to $(g-2)_\mu$.
In addition, with an infinite tower of excited pseudoscalars, they restore the correct asymptotic $Q^2$-behavior of the HLbL tensor in the mixed- and high-energy regions, 
see~\eqref{MVConstraintSimple} and~\eqref{pQCDSimple}, as well as~\eqref{MVConstraintSimpleETA} and~\eqref{pQCDSimpleETA}. 

Thus, it has been shown that the SDCs on the HLbL tensor, and in particular the MV constraint, can indeed be satisfied with an infinite sum over excited pseudoscalar-pole diagrams, 
while maintaining the correct low-energy behavior. Our result~\eqref{final}
\beq
 \Delta a_\mu^{\text{PS-poles}}=\Delta a_\mu^{\pi\text{-poles}}+\Delta a_\mu^{\eta\text{-poles}}+\Delta a_\mu^{\eta'\text{-poles}}=12.6(4.1)\times 10^{-11},
 \label{FinalConcl}
\eeq
derived from the large-$N_c$ Regge models alone, is significantly smaller than the original estimate $\Delta a_\mu\vert_\mathrm{MV} = 23.5\times 10^{-11}$ from ref.~\cite{Melnikov:2003xd}, 
which was obtained by removing the momentum dependence of the TFF at the external photon vertex. 
In fact, with modern input for the pseudoscalar TFFs this effect would increase further to $\Delta a_\mu\vert_\mathrm{MV}\sim 38\times 10^{-11}$, 
demonstrating the dangers of ad-hoc modifications of the low-energy properties of the HLbL tensor. Indeed, we observe that by far the main part of the difference to our implementation
originates from the low-energy part of the $g-2$ integral. 

Furthermore, in contrast to ref.~\cite{Melnikov:2003xd}, we find $\Delta a_\mu^{\pi\text{-poles}}<\Delta a_\mu^{\eta\text{-poles}}<\Delta a_\mu^{\eta'\text{-poles}}$. 
Accordingly, the flavor decomposition into excited $\pi^0$, $\eta$, $\eta'$ states follows roughly the expectation from the coefficients determining the SDCs, motivating a matching 
of our hadronic implementation onto a description in terms of the pQCD quark loop.
This matching, illustrated in figures~\ref{quark_loop_MV}--\ref{SD_comparison_mixed}, shows that, 
as expected, the ground-state pseudoscalars are relevant only at low energies, 
but about half the excited-state contribution comes from the integration region 
of $Q_i\geq 1\GeV$, while the other half could be interpreted as an estimate of the mixed regions.
Since, by construction, the excited-state contribution asymptotically matches onto the one from pQCD, 
we then replaced the hadronic formulation in favor of the quark loop in the asymptotic part of the integral, 
at a matching scale of $Q_\text{match}=1.7\GeV$ obtained from our best estimates of the uncertainties in the Regge models and pQCD corrections. 
Due to the assumed pQCD uncertainties and variation of the matching scale, as well as the inflated errors for the Regge slopes in Model 1, see discussion between  \eqref{final_matching} and \eqref{result_PS}, the uncertainty of our final result~\eqref{final_result}
\begin{equation}
\Delta a_\mu^\text{LSDC}=\left[8.7(5.5)_{\text{PS-poles}}+4.6(9)_{q\text{-loop}}\right]\times 10^{-11} = 13(6)\times 10^{-11},
\end{equation}
slightly increases with respect to~\eqref{FinalConcl}, the advantage being that the asymptotic part of the result is 
manifestly independent of the nature of the hadronic states in terms of which the correct asymptotic behavior was restored.
In this way, our final result mainly relies on the Regge models for an estimate of potential contributions 
for which the asymptotic constraints do not yet apply, and, while data is scarce, this is the energy region
where at least some phenomenological guidance for the excited pseudoscalar states is available. In particular, this strategy ensures that since the excited pseudoscalars decouple in the chiral limit, see section~\ref{sec:chiral_limit}, our implementation of the asymptotic 
part of the integral remains valid for vanishing quark masses, while for the low-energy phenomenology chiral corrections are essential.

In the future, the matching to pQCD could be improved if explicit calculations of pQCD corrections became available, a first step in this direction was already taken in ref.~\cite{Bijnens:2019ghy}.
Moreover, the phenomenological analysis would profit from further experimental information on the two-photon physics of hadronic resonances in the $1$--$2\GeV$ region, which holds true 
not only for the longitudinal amplitudes but in general. In fact, to address the transversal amplitudes, the effects of axial-vector resonances need to be understood in the context of 
dispersion relations, especially given that their masses are much closer to the typical matching scale found for the longitudinal SDCs in this paper. Work along these lines is in progress.

\section*{Acknowledgements}
\addcontentsline{toc}{section}{Acknowledgements}

We thank P.~Bickert, J.~Bijnens, S.~Eidelman, S.~Holz, B.~Kubis, S.~Leupold, J.~L{\"u}dtke, A.~Manohar, M.~Procura, E.~Ruiz Arriola, P.~Sanchez-Puertas, and A.~Vainshtein  for useful discussions; A.~G\'{e}rardin, C.~Hanhart, S.~Holz, P.~Sanchez-Puertas, and E.~Weil for providing us with their results; R.~Arnaldi, A.~Uras, G.~Usai, and V.~Metag for helping us with the experimental data sets. Financial support by  
the DOE (Grant Nos.\ DE-FG02-00ER41132 and DE-SC0009919)
and the Swiss National Science Foundation
is gratefully acknowledged. M.H.\ is supported by an SNSF Eccellenza Professorial Fellowship (Project No.\ PCEFP2\_181117).

\appendix

\section{Anomalous Pseudoscalar--Vector--Vector Coupling}
\label{CouplingsApp}

For the pion TFF, we only considered the coupling of the pion to a $\rho\omega$ pair, see figure~\ref{fig:TFFlargeNc}, and neglected the contribution given by a $\rho\phi$ pair. In the following, we motivate why the $\rho\phi$ pair can be neglected for the pion, and derive the relative strength of $2\rho$, $2\omega$, $2\phi$, and $\phi\omega$ contributions to the TFFs of the $\eta$ and $\eta'$, see figure~\ref{fig:TFFlargeNceta}.

In ref.~\cite{Meissner:1987ge}, we find the Lagrangians for the anomalous pseudoscalar--vector--vector coupling,
\beq
\mathcal{L}_\mathrm{VV\Phi}=-g_\mathrm{VV\Phi}\,\epsilon^{\mu \nu \al \beta} \,\Tr \left(\partial_\mu V_\nu \partial_\al V_\beta \Phi\right),
\eeq
with $g_\mathrm{VV\Phi}=3g^2/32\pi^2 F_\pi^2$,
and the electromagnetic photon--vector interaction, 
\beq
\mathcal{L}_\mathrm{em}=\frac{\sqrt{2}\,e}{g}A^\mu \left(m_\rho^2 \rho_\mu+\frac{1}{3}m_\omega^2 \omega_\mu-\frac{\sqrt{2}}{3}m_\phi^2 \phi_\mu\right)=A^\mu\! \left(g_{\rho\gamma }\, \rho_\mu+g_{\omega\gamma }\, \omega_\mu+g_{\phi\gamma }\, \phi_\mu\right),\label{Lag-em}
\eeq
with $g_{V\gamma}$ the individual coupling strengths. $\Phi$ stands for the neutral ground-state pseudoscalar mesons, denoted by $\pi^0$, $\eta^{(8)}$ and $\eta^{(0)}$:
\beq
\Phi=\frac{1}{\sqrt{2}}\left(\begin{array}{ccc}
   \pi^0 + \frac{1}{\sqrt{3}}\,\eta^{(8)} +\sqrt{\frac{2}{3}} \,\eta^{(0)}  & 0&0 \\
   0  &  -\pi^0 + \frac{1}{\sqrt{3}}\,\eta^{(8)} +\sqrt{\frac{2}{3}}\, \eta^{(0)}  &0\\
   0&0& -\frac{2}{\sqrt{3}}\,\eta^{(8)}+\sqrt{\frac{2}{3}} \,\eta^{(0)}
\end{array}\right),
\eeq
and $V_\mu$ stands for the neutral ground-state vector mesons, denoted by $\rho_\mu$, $\phi_\mu^{(8)}$, and $\phi_\mu^{(0)}$:
\beq
\hspace{-0.3cm}V_\mu=\frac{1}{\sqrt{2}}\left(\begin{array}{ccc}
   \rho_\mu + \frac{1}{\sqrt{3}}\,\phi_\mu^{(8)} +\sqrt{\frac{2}{3}} \,\phi_\mu^{(0)}  & 0&0 \\
   0  &  -\rho_\mu + \frac{1}{\sqrt{3}}\,\phi_\mu^{(8)} +\sqrt{\frac{2}{3}}\, \phi_\mu^{(0)}  &0\\
   0&0& -\frac{2}{\sqrt{3}}\,\phi_\mu^{(8)}+\sqrt{\frac{2}{3}} \,\phi_\mu^{(0)}
\end{array}\right).
\eeq
Note that the latter Lagrangian \eqref{Lag-em} for the neutral vector mesons is given in the ideal mixing situation: $\rho \sim 1/\sqrt{2}\left(u\bar{u}-d\bar{d}\right)$, $\omega \sim 1/\sqrt{2}\left(u\bar{u}+d\bar{d}\right)$ and $\phi \sim -s \bar{s}$. In general, we use a $\phi$--$\omega$ mixing:
\beq
\begin{pmatrix}\phi_8\\\phi_0\end{pmatrix}=\begin{pmatrix}\cos \theta_V&\sin \theta_V\\ -\sin \theta_V& \cos \theta_V \end{pmatrix}\begin{pmatrix}\phi\\\omega\end{pmatrix},
\eeq
with $\theta_V= 36.4^\circ$ \cite{Tanabashi:2018oca} which, however, almost corresponds to the ideal case ($ \theta_V^\mathrm{ideal}= \arctan 1/\sqrt{2}\sim 35.3^\circ$).
For the $\eta$--$\eta'$ mixing, we 
 use the short-hand notation:
\beq
\begin{pmatrix}\eta_8\\\eta_1\end{pmatrix}=T\begin{pmatrix}\eta\\\eta'\end{pmatrix}=\begin{pmatrix}T_{11}&T_{12}\\T_{21}&T_{22}\end{pmatrix}\begin{pmatrix}\eta\\\eta'\end{pmatrix},
\eeq
where the mixing
matrix in the standard two-angle mixing scheme is given by:
\begin{equation}
T\coloneqq F_\pi \left(\begin{array}{lr} F^8 \cos \theta_8 & -F^0 \sin
    \theta_0 \\ F^8 \sin \theta_8 & F^0 \cos \theta_0   \end{array} \right)^{-1} ,\label{TMatrix}
\end{equation}
with the mixing parameters introduced in \eqref{eq:Pmixing}.

The coupling strengths of the pseudoscalar meson to two-photon interactions in the VMD picture, see figures~\ref{fig:TFFlargeNc} and \ref{fig:TFFlargeNceta}, can be reconstructed from the above Lagrangians, taking into account the $\phi$--$\omega$ and $\eta$--$\eta'$ mixings:
\begin{align}
\Gamma_\mathrm{\Phi\ga\ga}&\propto\frac{3\al}{4\pi F_\pi^2}\,\epsilon^{\mu \nu \al \beta}\Bigg\{\frac{\sqrt{2}}{3\sqrt{3}}\,\pi^0\,\bigg[\partial_\mu\rho_\nu \,\partial_\al\omega_\be \,\left(\sqrt{2}\cos \theta_V+\sin \theta_V\right)\notag\\
&-\sqrt{2}\partial_\mu\rho_\nu \,\partial_\al\phi_\be \,\left(\cos \theta_V-\sqrt{2}\sin \theta_V\right)\bigg]\nn\\
&+\frac{1}{2\sqrt{6}}\,\eta \bigg[2\,\partial_\mu\rho_\nu \,\partial_\al\rho_\be\left(T_{11}+\sqrt{2}\,T_{21}\right)\notag\\
&+\frac{1}{9}\left(2\sqrt{2}\,T_{21}-T_{11}\right)\left(\partial_\mu\omega_\nu \,\partial_\al\omega_\be+2\partial_\mu\phi_\nu \,\partial_\al\phi_\be\right)\nn\\
&+\frac{1}{9}\,T_{11} \left(\cos 2\theta_V+2\sqrt{2} \sin 2 \theta_V\right)\left( \partial_\mu\omega_\nu \,\partial_\al\omega_\be-2\partial_\mu\phi_\nu \,\partial_\al\phi_\be\right)\nn\\
&- \frac{2\sqrt{2}}{9}\, T_{11} \left(2\sqrt{2}\cos 2\theta_V- \sin 2 \theta_V\right)\partial_\mu\omega_\nu \,\partial_\al\phi_\be\bigg]\nn\\
&+\frac{1}{2\sqrt{6}}\,\eta' \bigg[2\,\partial_\mu\rho_\nu \,\partial_\al\rho_\be\left(T_{12}+\sqrt{2}\,T_{22}\right)\notag\\
&+\frac{1}{9}\left(2\sqrt{2}\,T_{22}-T_{12}\right)\left(\partial_\mu\omega_\nu \,\partial_\al\omega_\be+2\partial_\mu\phi_\nu \,\partial_\al\phi_\be\right)\nn\\
&+ \frac{1}{9}\,T_{12} \left(\cos 2\theta_V+2\sqrt{2} \sin 2 \theta_V\right)\left(\partial_\mu\omega_\nu \,\partial_\al\omega_\be-2\partial_\mu\phi_\nu \,\partial_\al\phi_\be\right)\nn\\
&-\frac{2\sqrt{2}}{9}\, T_{12} \left(2\sqrt{2}\cos 2\theta_V- \sin 2 \theta_V\right)\partial_\mu\omega_\nu \,\partial_\al\phi_\be\bigg]
\Bigg\}.
\end{align}
A similar approach is chosen in ref.~\cite{Landsberg:1986fd}, where the contributions to the singly-virtual TFFs are analyzed through the combination of pseudoscalar--photon--vector and 
photon--vector interactions. 
The dependence of the electromagnetic photon--vector interactions~\eqref{Lag-em} on the vector-meson masses are canceled out by the vector-meson propagators. Our final couplings read, for $P=\eta, \eta'$:
\begin{align}
C^P_{\rho \rho}&=T_{1I_P}+\sqrt{2} T_{2I_P}, \nn\\
C^P_{\omega \omega}&=\frac{1}{18}\left[(\cos 2 \theta_V + 2 \sqrt{2} \sin 2
  \theta_V-1) T_{1I_P}+2 \sqrt{2} T_{2I_P} \right], \nn \\
C^P_{\phi \phi}&=-\frac{1}{9}\left[(\cos 2 \theta_V + 2 \sqrt{2} \sin 2
  \theta_V+1) T_{1I_P}-2 \sqrt{2} T_{2I_P} \right],\nn \\
C^P_{\omega \phi}&=\frac{\sqrt{2}}{9}(\sin 2 \theta_V-2 \sqrt{2} \cos 2
\theta_V ) T_{1I_P}, \label{CouplingsAnalytic}
\end{align}
with $I_\eta=1$, $I_{\eta'}=2$, and $T_{IJ}$ given in \eqref{TMatrix}. Similarly, we define for the pion:
\begin{align}
C^\pi_{\rho \omega}&=\frac{2}{3} \left(\sqrt{2}\cos \theta_V+\sin \theta_V\right), \nn\\
C^\pi_{\rho \phi}&=-\frac{2 \sqrt{2}}{3} \left(\cos \theta_V-\sqrt{2}\sin \theta_V\right).\label{CouplingsAnalyticPion}
\end{align}
Note that in~\eqref{CouplingsAnalytic} and \eqref{CouplingsAnalyticPion} we divided all couplings by a common factor: $\sqrt{3}\al/(4\sqrt{2}\pi F_\pi^2)$. This is allowed because the relative strength of the different couplings does not change. The large-$N_c$ Regge model for the $\eta^{(\prime)}$ TFFs is then constructed such that each vector-meson pair contributes exactly $C^P_{V_1 V_2}/\mathcal{N}$ to $F_{\eta^{(\prime)} \gamma
  \gamma}$, where $\mathcal{N}$ is the normalization~\eqref{NormalizationEta}.

Numerical values for $C_{V_1V_2}^P$ can be found in table \ref{TableCouplings}. One can clearly see that $C_{\rho\omega}^\pi \gg C_{\rho\phi}^\pi$, which is why we neglected the $\rho\phi$ contribution to the pion TFF. Furthermore, one can see that the ground-state $\eta^{(\prime)}$ TFFs are dominated by the $2\rho$, while the contribution from $\phi$--$\omega$ mixing is small. This is also illustrated in figures \ref{fig:ContributionsEta} and \ref{fig:ContributionsEtaP}, where we show the $2\rho$, $2\omega$, $2\phi$, and $\phi \omega$ contributions to the singly-virtual and doubly-virtual $\eta^{(\prime)}$ TFFs.

\section{Alternative model for pion, $\boldsymbol{\eta}$, and $\boldsymbol{\eta'}$ transition form factors}
\seclab{PionAlternative}

In this appendix, we present an alternative model for the pseudoscalar TFFs, which will help us to study the systematic uncertainty of our $g-2$ result. This alternative model uses the Regge resummation from ref.~\cite{RuizArriola:2006jge} to satisfy the pQCD limit of the TFF, cf.\ \eqref{BLOPESec2}. It reads:
\begin{align}
F_{P(n) \ga^*\ga^*}(-Q_1^2,-Q_2^2)\nn
&=\sum_{i=0}^\infty \,\int_0^1 d x\;\left\{ \frac{c_1 \,e^{-(Q_1^2+Q_2^2)/\Lambda^2}}{\left[M_{V_1(n+i)}^2+Q_1^2\, x+Q_2^2\, (1-x)\right]^2}\right.\\
&+\left.\frac{c_2\,\left[1-e^{-(Q_1^2+Q_2^2)/\Lambda^2}\right]x(1-x)}{\left[M_{V_2(n+i)}^2+Q_1^2\, x+Q_2^2\, (1-x)\right]^2}\right\},\label{MartinModel}
\end{align}
where $P=\pi, \eta, \eta'$ and the introduced mass spectra again follow a radial Regge ansatz:
\begin{align}
M_{V_1(i)}^2&=M_{V_1}^2+i\, \sigma_{V_1}^2,\nn\\
M_{V_2(i)}^2&=M_{V_2}^2+i\, \sigma_{V_2}^2.\label{MV2Regge}
\end{align}

\begin{table}[t]
\begin{center}
\begin{tabular}{ cccc}
\toprule
&$\pi$& $\eta$&$\eta'$ \\ 
 \midrule
$M_{V_1}$ [MeV]&$779$&$774$&$859$\\
$M_{V_2}$ [MeV]&$585$&$404$&$452$\\
$\si_{V_1}^2$ [GeV$^2$]&$1.252$&$1.593$ &$1.577$\\
$\si_{V_2}^2$ [GeV$^2$]&$0.076$&$0.034$&$0.060$\\
$\Lambda$ [GeV]&$1.318$&$1.318$&$1.318$\\
 \bottomrule
\end{tabular}
\caption{Parameters of the alternative model for pion, $\eta$, and $\eta'$ TFFs.\label{ParametersModel2}}
\end{center}
\end{table}

The first term in~\eqref{MartinModel}, proportional to $c_1$, corresponds to a variant of a large-$N_c$ Regge TFF model with equal mass spectra for all vector mesons. The second term in~\eqref{MartinModel}, proportional to $c_2$, has an additional factor of $x(1-x)$ in the numerator, originating from the asymptotic wave function, cf.~\eqref{FeynmanParameterNotation}. This term is crucial for the model to satisfy the BL limit~\eqref{BLOPESec2}: 
\beq
\label{BL2}
\lim_{Q^2\rightarrow \infty}Q^2F_{P \ga\ga^*}(-Q^2)=\frac{c_2}{2 \sigma_{V_2}^2},
\eeq
with 
\beq
c_2=\left\{\begin{array}{cc}
   4 F_\pi \,\sigma_{V_2}^2  & \qquad \text{for } \pi, \\
    24 \,C_8 F_\eta \,\sigma_{V_2}^2 & \qquad \text{for }\eta,\\
   24 \,C_0 F_{\eta'}\, \sigma_{V_2}^2   & \qquad \text{for }\eta',\\
\end{array}\right. 
\eeq
and the fit $F_\eta$, $F_{\eta'}$ from \eqref{Fetafit}.
The exponential functions in the numerator, $e^{-(Q_1^2+Q_2^2)/\Lambda^2}$, shall support the VMD in the region of small momentum transfers, and suppress the $x(1-x)$ correction which is only needed in the asymptotic region. Therefore, the real-photon limit~\eqref{anomDef} is proportional to $c_1$:
\beq
\label{Anomaly2}
F_{P\gamma\gamma}=\frac{c_1}{\sigma_{V_1}^4}\,\psi^{(1)}\left(\frac{M_{V_1}^2}{\sigma_{V_1}^2}\right),
\eeq
with $\psi^{(1)}$ the trigamma function and
\beq
c_1 =\left\{\begin{array}{cc}
   \si_{V_1}^4F_{\pi\ga\ga}\left[\psi^{(1)}\left(\frac{M_{V_1}^2}{\sigma_{V_1}^2}\right)\right]^{-1}  & \qquad \text{for }\pi, \\
    \si_{V_1}^4F_{\eta\ga\ga}\left[\psi^{(1)}\left(\frac{M_{V_1}^2}{\sigma_{V_1}^2}\right)\right]^{-1} & \qquad \text{for }\eta,\\
   \si_{V_1}^4F_{\eta'\ga\ga}\left[\psi^{(1)}\left(\frac{M_{V_1}^2}{\sigma_{V_1}^2}\right)\right]^{-1}   & \qquad \text{for }\eta',\\
\end{array}\right.
\eeq
whereas the symmetric pQCD limit \eqref{BLOPESec2}, to leading order in $Q^2$, is proportional to $c_2$:
\beq
\label{OPE2}
\lim_{Q^2\rightarrow \infty}Q^2F_{P \ga^*\ga^*}(-Q^2,-Q^2)=\frac{c_2}{6 \sigma_{V_2}^2}.
\eeq
In this way, $c_1$ and $c_2$ are fixed and the TFF model reproduces the chiral anomaly, the BL limit, and the symmetric pQCD limit exactly.

Evaluating the SDCs of the HLbL tensor, cf.~\eqref{pQCDloop} and \eqref{MVconstraint}, we obtain for the mixed region:
\begin{align}
&-\lim_{Q_3^2 \rightarrow \infty}\,\lim_{Q^2 \rightarrow \infty}\,\sum _{n=0}^\infty \frac{F_{P(n) \gamma^* \gamma^*}(-Q^2,- Q^2)F_{P(n) \gamma \gamma^*}(-Q_3^2)}{Q_3^2+M_{P(n)}^2}\nn\\
&=\frac{1}{Q^2Q_3^2}\frac{c_2^2}{24 \sigma _P ^6 \sigma _{V_2}^4} \Bigg\{2\si_P^2 \left(2\si_{V_2}^2 -\si_P^2\right) L_{P V_2}-\si_P^2 \left(4\si_{V_2}^2-3\si_P^2\right) \nn\\
 &-4  \sigma_{V_2}^2\Delta_{P V_2} \left[\frac{\pi^2}{6}-\text{Li}_2 \left(1-\frac{\si_P^2}{\si_{V_2}^2}\right)\right]\Bigg\},\label{aMV}
\end{align}
with $\text{Li}_2$ the dilogarithm, and for the asymptotic region:
\begin{align}
&-\lim_{Q^2 \rightarrow \infty}\,\sum _{n=0}^\infty \frac{F_{P(n) \gamma^* \gamma^*}(-Q^2,- Q^2)F_{P(n) \gamma \gamma^*}(-Q^2)}{Q^2+M_{P(n)}^2}\nn\\
&=\frac{1}{Q^4}\frac{c_2^2}{12 \sigma _P ^4 \sigma _{V_2}^4\Delta_{P V_2}} \Bigg\{\si_P^2 \left(2\si_{V_2}^2 -\si_P^2\right) L_{P V_2}+2\si_P^2 \Delta_{P V_2} \nn\\
 &-2  \sigma_{V_2}^2\Delta_{P V_2} \left[\frac{\pi^2}{6}-\text{Li}_2 \left(1-\frac{\si_P^2}{\si_{V_2}^2}\right)\right]\Bigg\}.\label{apQCD}
\end{align}
Thus, both the mixed region and the asymptotic region acquire the correct $Q^2$ behavior, as is discussed in detail in appendix \ref{sec:PionAlternativeProof}.

The model parameters $M_{V_i}$, $\sigma_{V_i}$, and $\Lambda$ are determined as follows, see table \ref{ParametersModel2}:
\begin{itemize}
    \item For the pion TFF we use $M_{V_1}=\frac{1}{2}\left[M_{\rho(770)}+M_{\omega(782)}\right]=0.779$ GeV; for the $\eta$ and $\eta'$ TFFs we use for $M_{V_1}$ the pole-mass parameters of a VMD ansatz fit to the CLEO data \cite{Gronberg:1997fj}. 
    \item For reasons of comparison, $\sigma_{V_1}$ is chosen to reproduce the values of the two-photon couplings to the first excited pseudoscalars obtain with our large-$N_c$ Regge model:
    \begin{equation}
      F_{\pi(1)\ga\ga}\sim 0.0500\GeV^{-1},\quad 
      F_{\eta(1)\ga\ga}\sim 0.0354\GeV^{-1},\quad 
    F_{\eta'(1)\ga\ga}\sim 0.0594\GeV^{-1}.\notag
     \end{equation}
      Alternatively, one could use the phenomenological constraints on the two-photon couplings listed in table \ref{TabPhotonCouplings};
    \item $\sigma_{V_2}$ is chosen to satisfy the MV SDC;
    \item For the pion TFF $\Lambda$ and $M_{V_2}$ are adjusted to bring the model in line with the dispersive description of the $\pi^0$ TFF \cite{Hoferichter:2014vra,Hoferichter:2018dmo,Hoferichter:2018kwz};\footnote{We find $\Lambda=1.318$ GeV and $M_{V_2}=585$ MeV with estimated variance $\chi^2\sim0.33$ for a fit of $\mathcal{O}(2\!\times\!10^4)$ selected points in the region of $Q_1\leq Q_2$ and $Q_2^2 \in [0,40]\,\text{GeV}^2$. } for the $\eta$ and $\eta'$ TFFs the same $\Lambda$ as in the pion case is used, while $M_{V_2}$ is fit to the available experimental data.
\end{itemize}

With the parameters in table \ref{ParametersModel2}, the MV SDC is satisfied to about $\sim 2 \times 10^{-3}$ relative accuracy or better, and the two-photon couplings of our large-$N_c$ Regge model are reproduced to about $\sim 3 \times 10^{-4}$ relative accuracy or better. The SDC for the asymptotic region, cf.\ \eqref{pQCDSimple} and \eqref{pQCDSimpleETA}, is not implemented in our alternative TFF model, however, even without further adjustments it is reproduced to $117\,\%$ for the pion, $124\,\%$ for the $\eta$, and $120\,\%$ for the $\eta'$. Of course, one could also choose the model parameters differently and implement the SDC for the asymptotic region precisely and the MV limit approximately.
Note that the parameters  $M_{V_1}$ and $\sigma_{V_1}$ are close to the physical values for the masses of the lightest vector mesons and the slopes of their radial Regge trajectories, cf.\ figure~\ref{fig:TrajectoriesPlot}.
These physical values assure that the first term in~\eqref{MartinModel} indeed resembles a large-$N_c$ Regge model. 

\begin{figure}[t]
\includegraphics[width=0.48\linewidth]{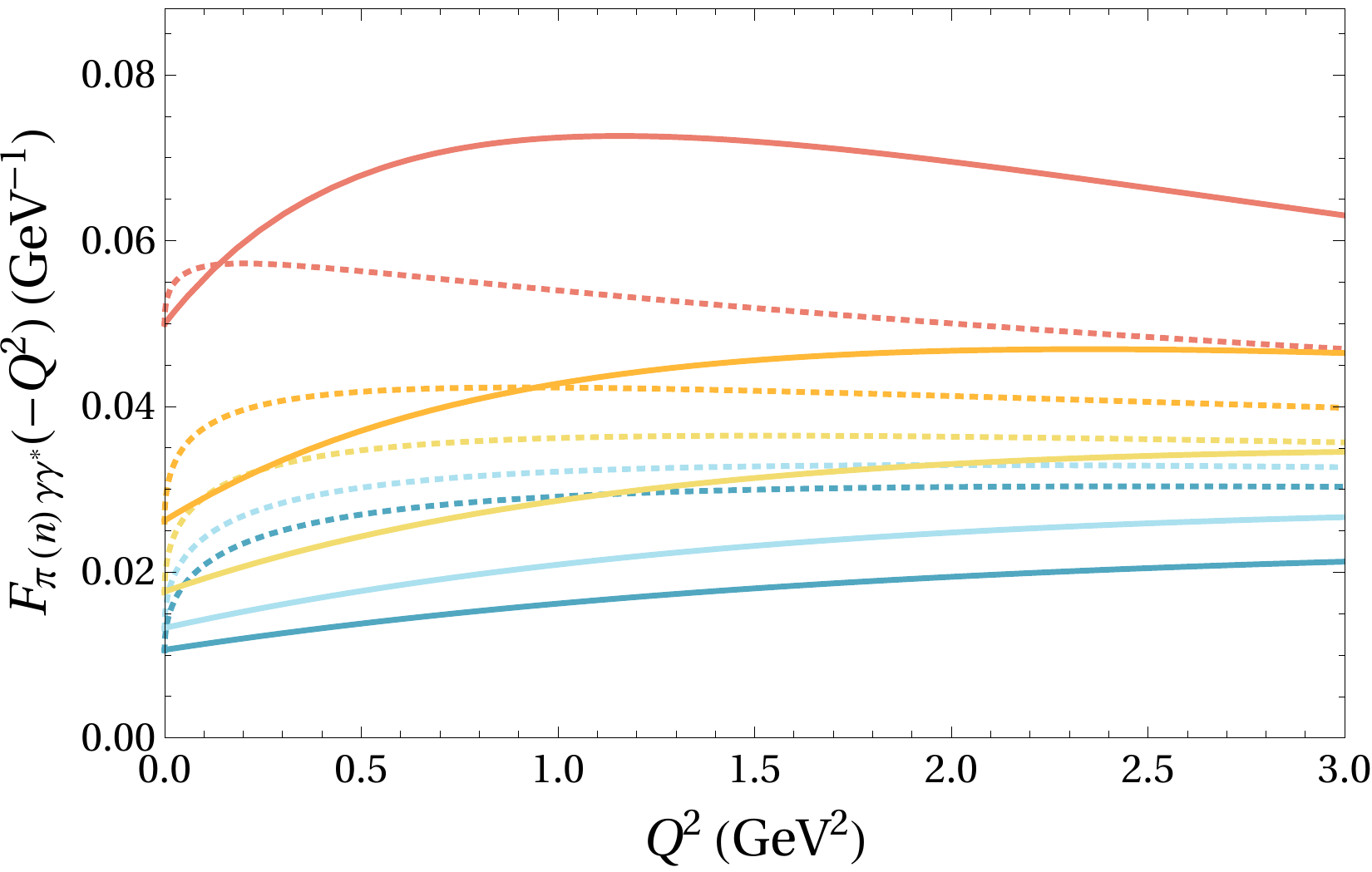}\hfill
\includegraphics[width=0.48\linewidth]{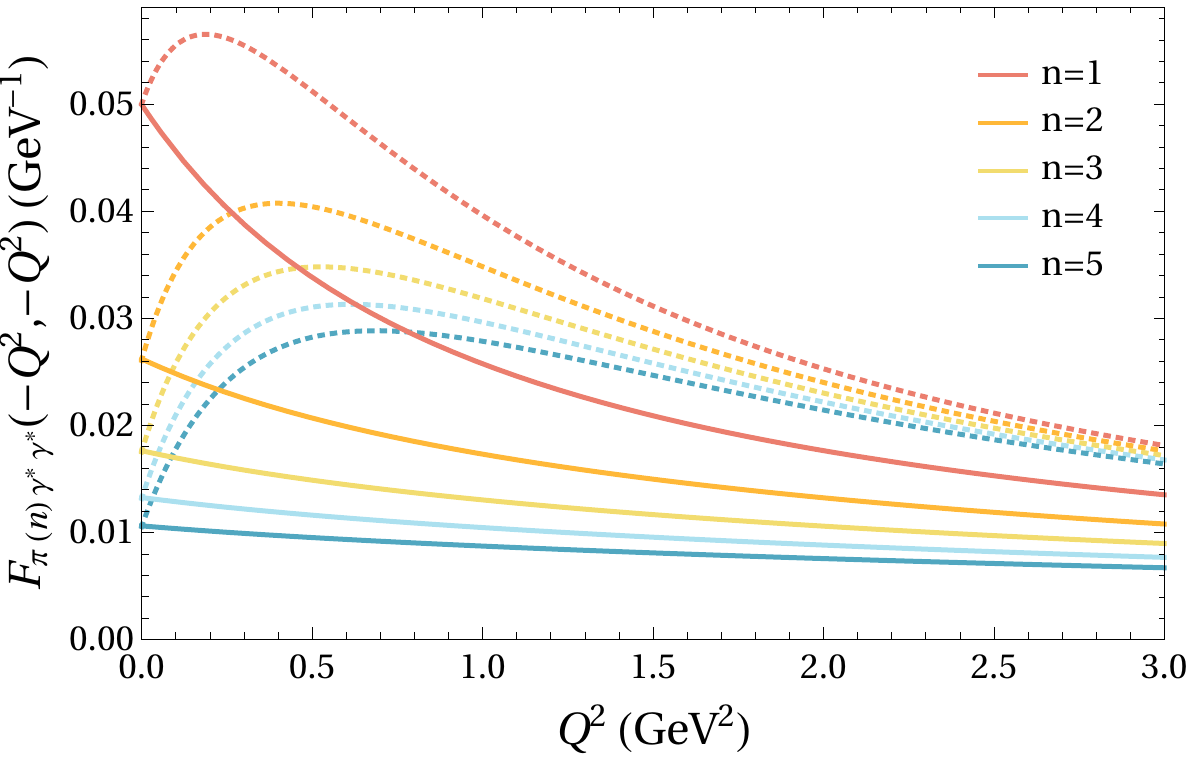}\\[0.2cm]
\includegraphics[width=0.48\linewidth]{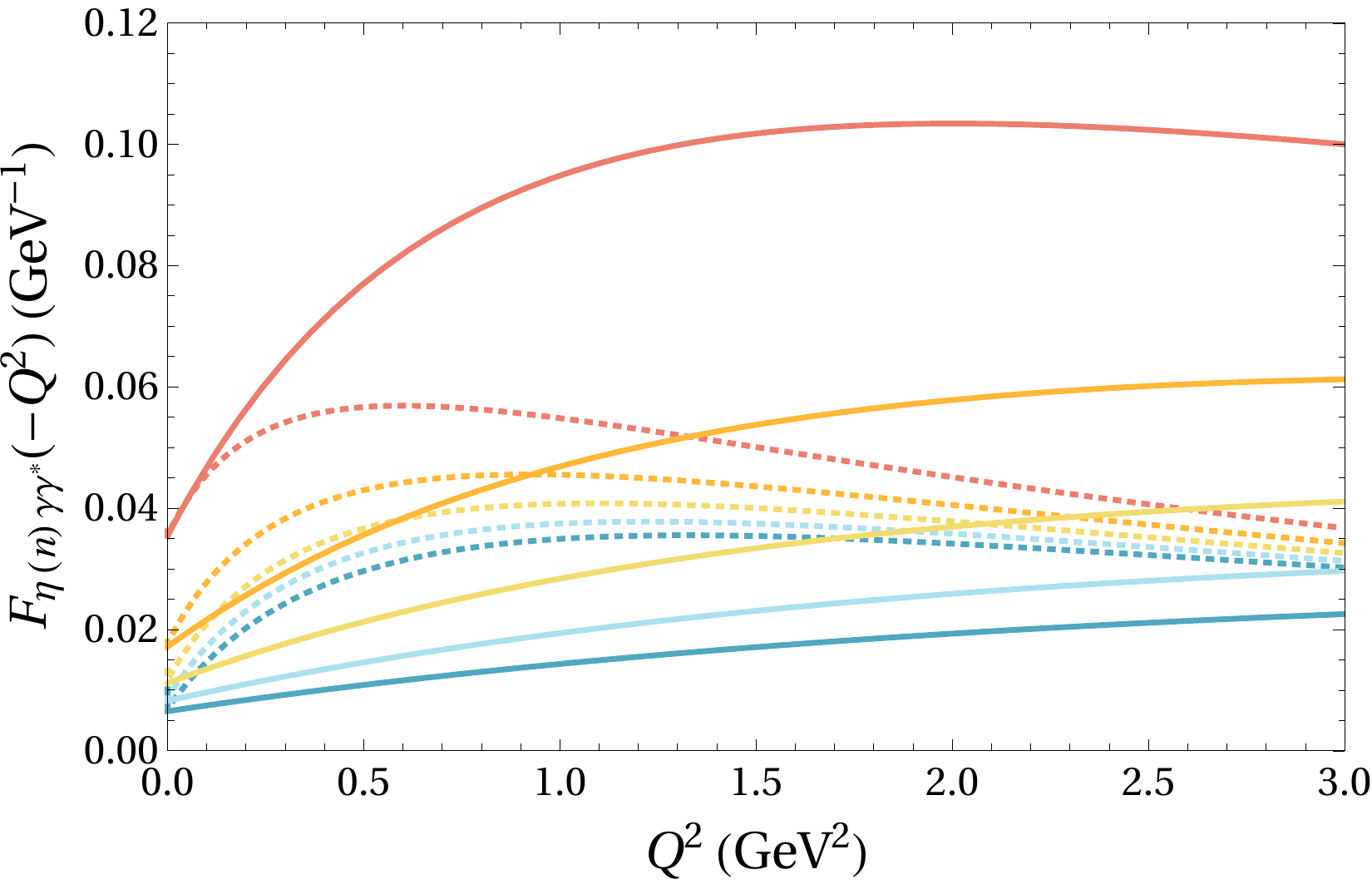}\hfill
\includegraphics[width=0.48\linewidth]{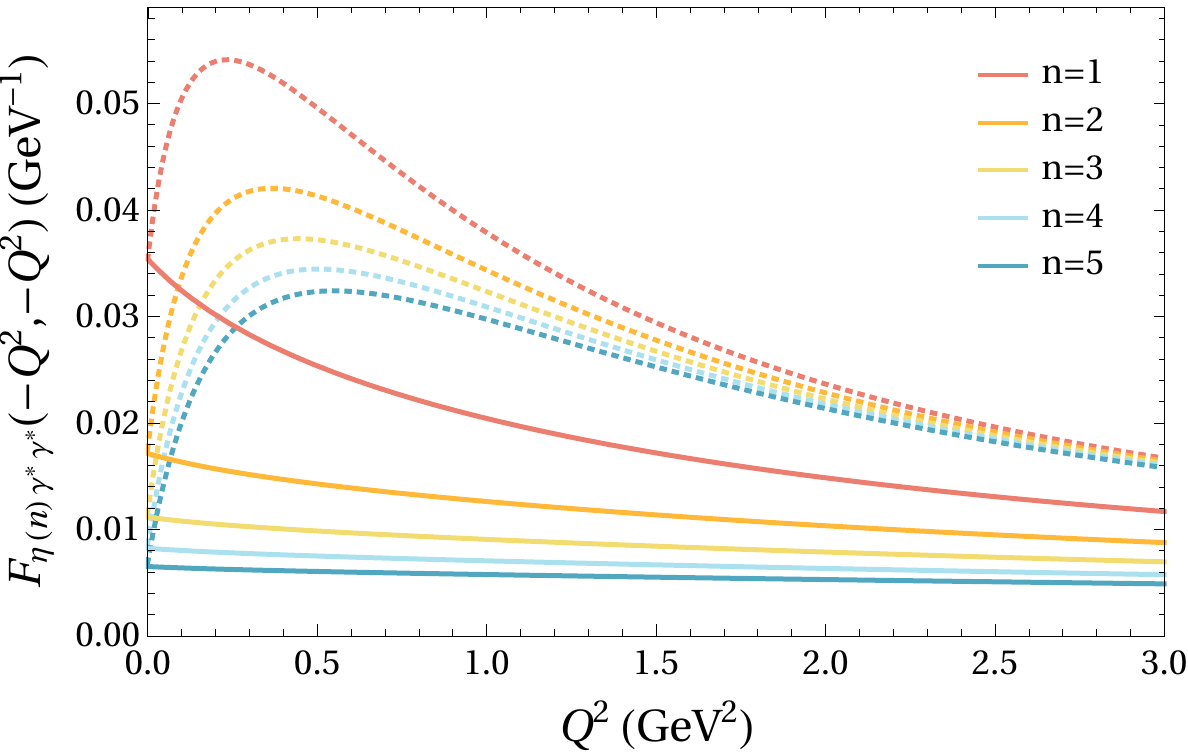}\\[0.2cm]
\includegraphics[width=0.48\linewidth]{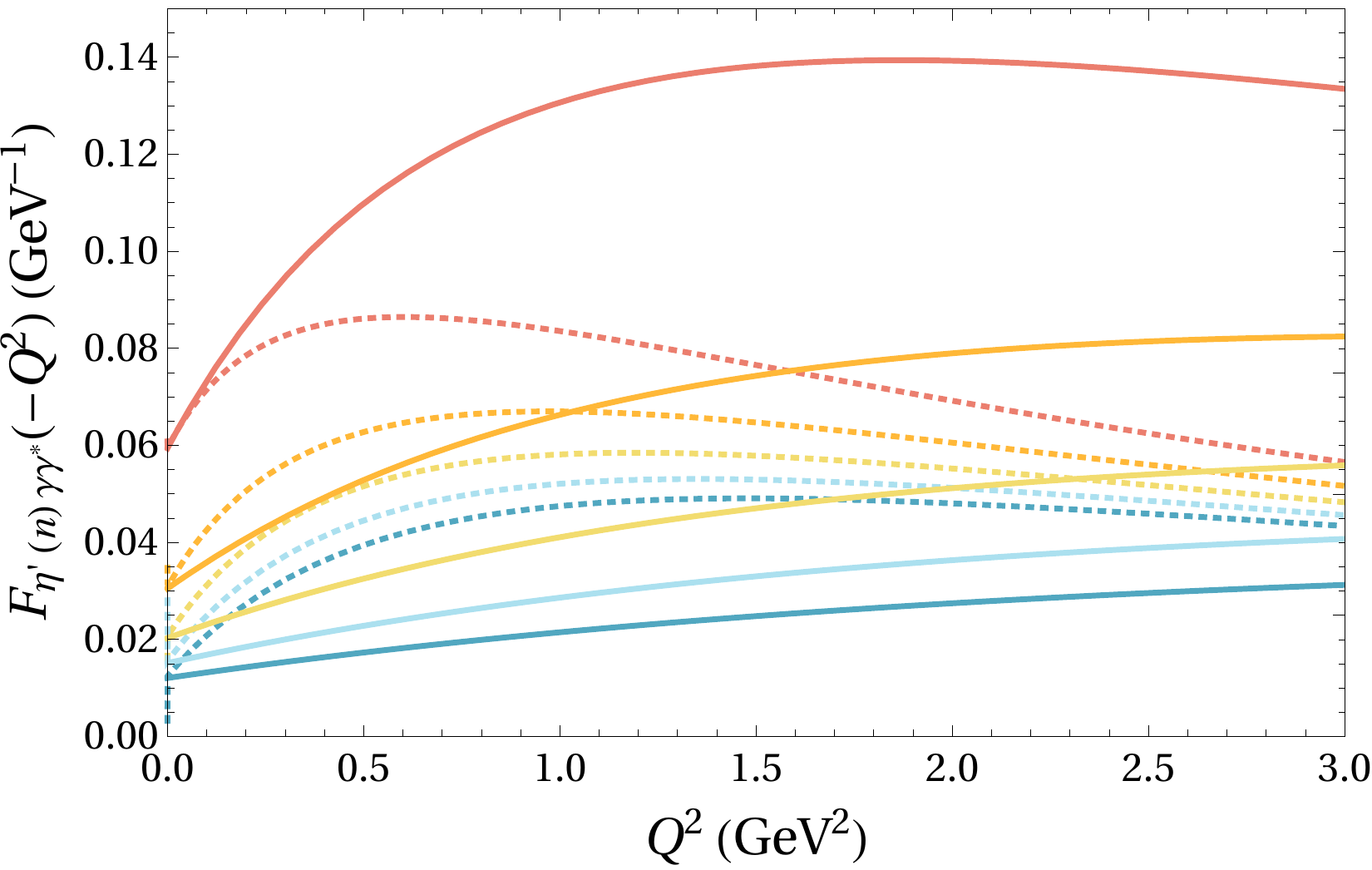}\hfill
\includegraphics[width=0.48\linewidth]{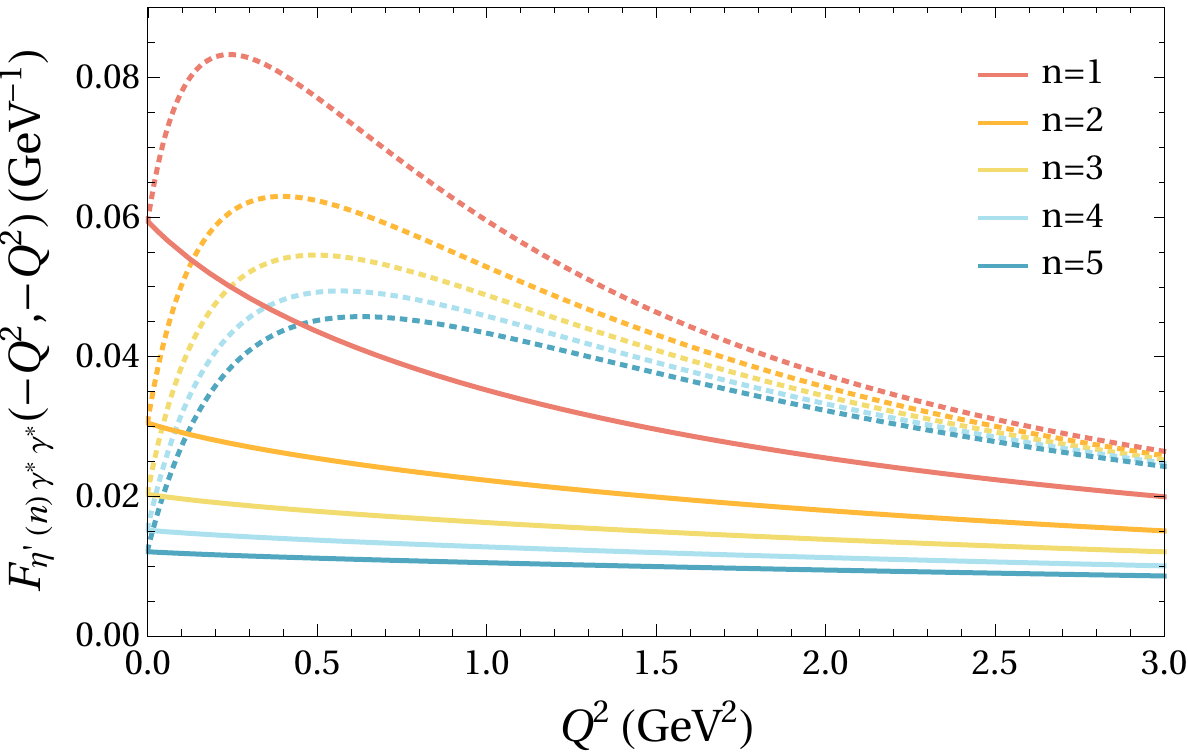}
\caption{TFFs of the first $n=1,\dots 5$ radially excited pion, $\eta$, and $\eta'$ states. Comparison of the large-$N_c$ Regge models from section \ref{sec:Regge}, indicated by the solid curves, and our alternative TFF model \eqref{MartinModel}, indicated by the dotted curves. The left panel shows the TFFs in the singly-virtual limit, the right panel shows the doubly-virtual region with $Q_1^2=Q_2^2=Q^2$.}
\figlab{ComparisonModels}
\end{figure}

\begin{figure}[t]
\centering 
 \includegraphics[height=4.6cm]{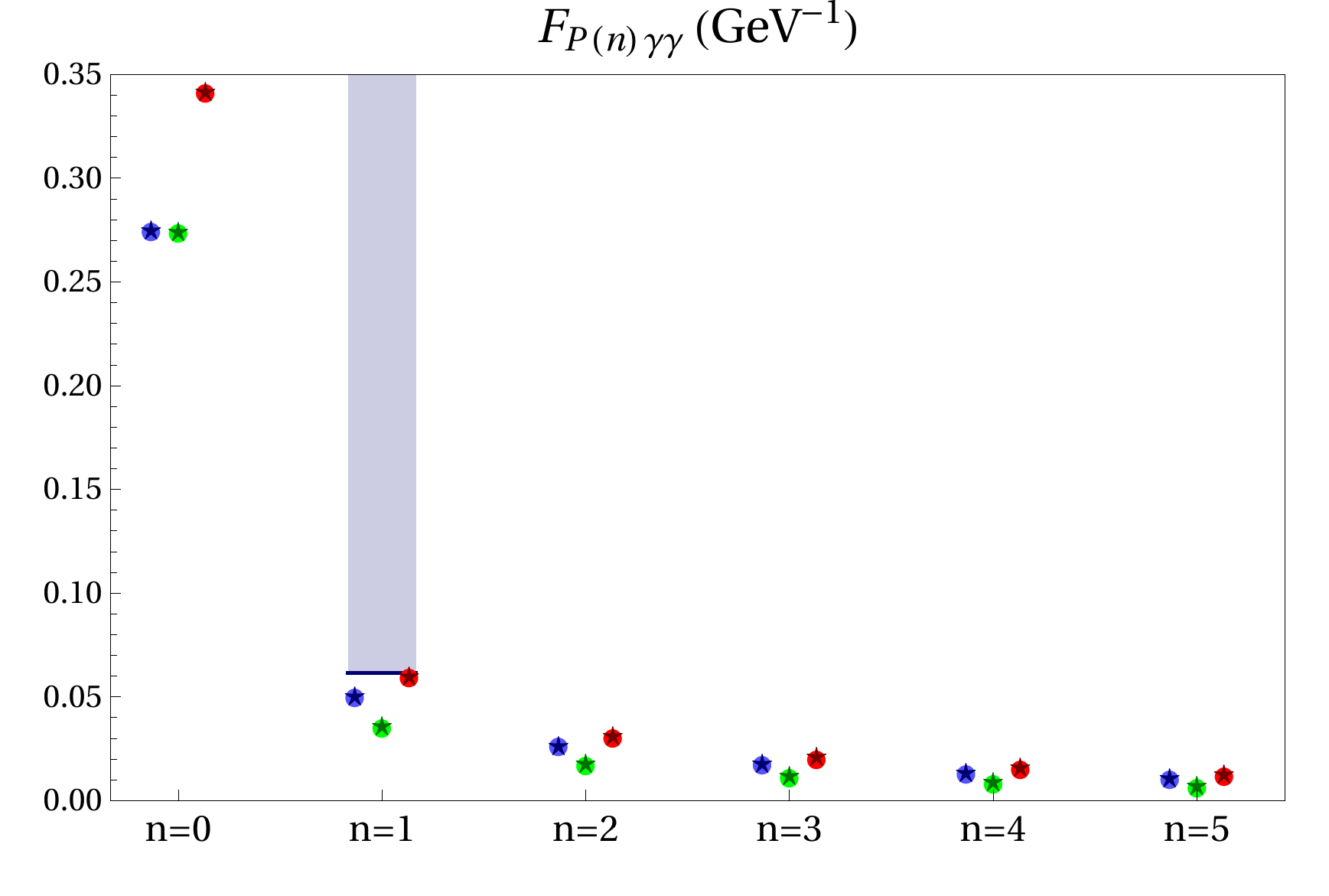}\hfill\includegraphics[height=4.8cm]{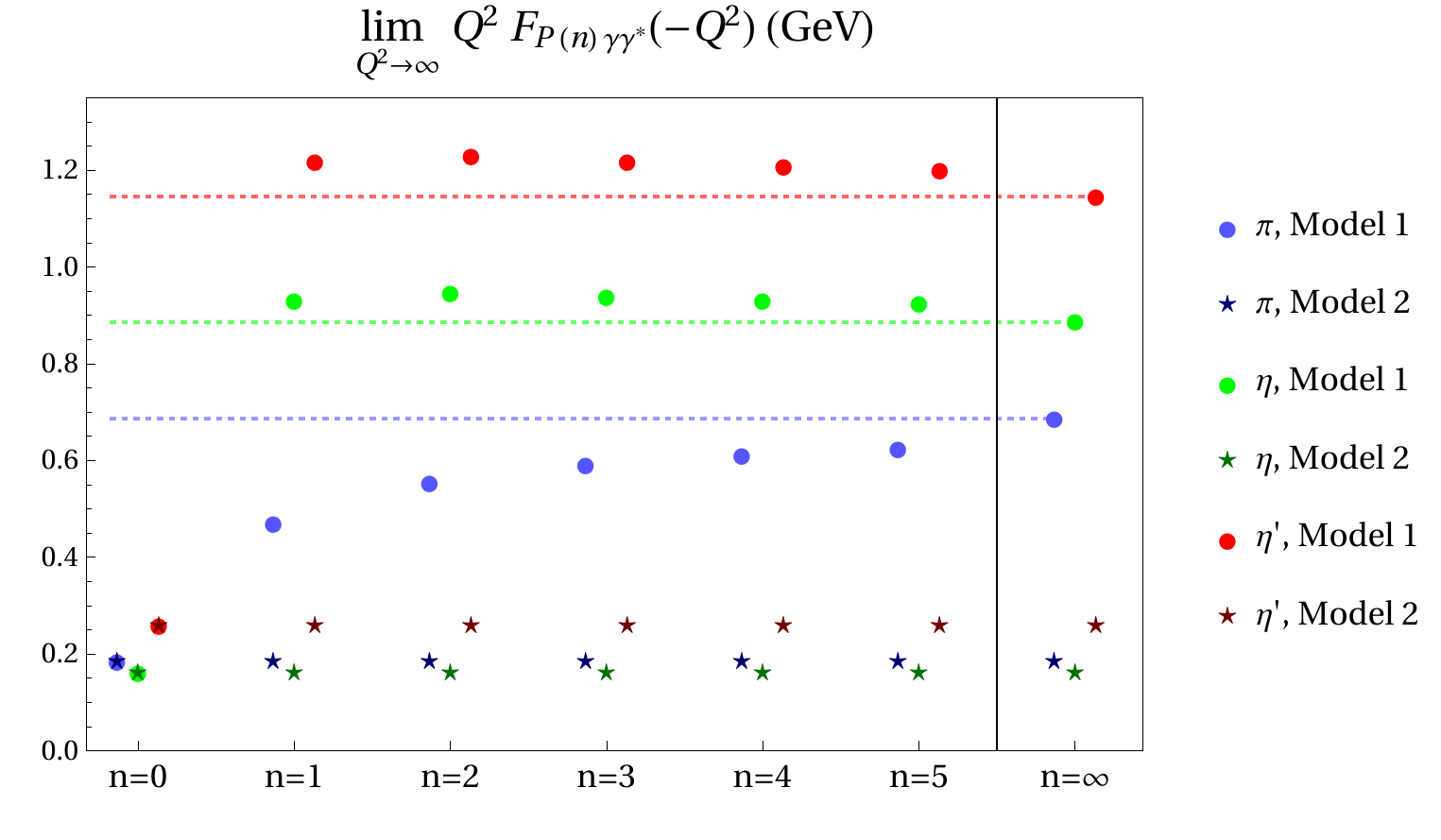}
\caption{Two-photon couplings and BL limits of the excited-state pion, $\eta$, and $\eta ^\prime$ TFFs. The blue, green, and red points (stars) are from the large-$N_c$ Regge model (alternative model) for pion, $\eta$, and $\eta ^\prime$, respectively.  The blue bar indicates values excluded for $F_{\pi(1)\ga\ga}$ by the empirical estimate~\eqref{couplingconstraint}.}
\label{fig:Coupling}
\end{figure}

In figure~\ref{fig:ComparisonModels}, the TFF presented in this appendix (Model 2) is compared to the large-$N_c$ Regge model (Model 1) from section~\ref{sec:Regge} for the lowest radial excitations of pion, $\eta$, and $\eta'$. A comparison to experimental data for the ground-state pseudoscalars is postponed to appendices~\ref{sec:plotsPion}, \ref{sec:EtaPlots}, and \ref{sec:EtaPrimePlots}.  The two-photon couplings $F_{P(n)\ga\ga}$ of the excited states come out in close agreement between both models, see also figure~\ref{fig:Coupling}. For Model 2, we observe an enhancement of the excited-state TFFs in the low-$Q$ region, especially for the doubly-virtual kinematics. This enhancement becomes weaker with increasing $\Lambda$, since it is an artefact of the interplay between the two terms in~\eqref{MartinModel}. Fitting both $M_{V_2}$ and $\Lambda$ to data for the ground-state $\eta$ and $\eta'$ TFFs would lead to $\Lambda<1$ GeV, and thus, exacerbate the enhancement of the excited-state TFFs at low $Q$. Therefore, we decided to use $\Lambda=1.318\GeV$, as obtained for the pion, also for $\eta$ and $\eta'$.

Note that for Model 1 the derivatives of the TFFs in the limit of zero momentum transfer are not unique but depend on the direction, 
a consequence of the construction in terms of $Q_-^2/Q_+^2$ in~\eqref{TFFpi} as a minimal way to 
implement the different asymptotic limits. This can be seen when comparing the slopes of the singly-virtual and symmetric doubly-virtual TFFs in the 
left and right panels of figure~\ref{fig:ComparisonModels}. The modifications of Model 1 described in \eqref{cBLpionnew} and \eqref{cBLetanew} reduce the direction-dependence of the derivative at the origin.
However, the derivative of the TFFs is not needed for the evaluation of $(g-2)_\mu$ and the alternative implementation in Model 2
does not exhibit this issue: in figure~\ref{fig:ComparisonModels}, the slopes for Model 2 are always positive, but for Model 1 they change sign
between the left and right panels.  Accordingly, this will be another systematic effect estimated by the comparison of the two models. 
We stress again that neither model has the required good analytic properties to remain valid outside the space-like region relevant for 
$(g-2)_\mu$, of which the zero-momentum-transfer limit of the derivatives in Model 1 is one particular manifestation.

In the right panel of figure \ref{fig:Coupling}, the BL limits of the excited-state TFFs are shown. For Model 1 
this limit increases with the excitation number $n$ until it reaches an asymptotic value, but for Model 2 it remains constant. 
Since the true asymptotic behavior  for radially-excited pseudoscalar TFFs in the BL limit is unknown, the two models with different asymptotics will allow us to understand the systematic uncertainty of our prediction for the excited-state contributions to $(g-2)_\mu$. The symmetric pQCD limit of the TFFs, on the other hand, is independent of the excitation number $n$ for both models. The two-photon couplings, which enter dominantly into the $(g-2)_\mu$ integral, agree by default for $n=0$ and $n=1$, and also match perfectly for $n>2$.

\section{Verifying short-distance constraints for the HLbL tensor}
\seclab{PionAlternativeProof}

In this appendix, the mathematical formalism used to derive the behavior of the HLbL tensor in the mixed-energy region, cf.\ \eqref{MV}, \eqref{MVEta}, and \eqref{aMV}, and the high-energy region, cf.\ \eqref{pQCD3}, \eqref{pQCD3Eta}, and \eqref{apQCD},  is presented.

\subsection{Polygamma functions and infinite sums over rational functions}

The gamma function is defined on $\mathbb{R}^*_+$ as~\cite{HandbookMath}:
\begin{equation}
\Gamma (z)=\int _0^\infty dt \,t^{z-1} e^{-t}. 
\end{equation}
It can be analytically continued to a meromorphic function in the complex plane, with poles at non-positive integers. In order to deal with the infinite sums over pseudoscalar and vector-meson poles, we  use the polygamma functions, which are defined on $\mathbb{C}$ as derivatives of the logarithm of the gamma function:
\begin{equation}\label{eq:defPolygamma}
    \psi^{(m)} (z):=\frac{d ^{m+1}}{d z^{m+1}}\log \Gamma (z).
\end{equation}
They are meromorphic in the complex plane and admit the following series representation~\cite{HandbookMath}:
\begin{equation}\label{eq:seriespolygamma}
    \psi ^{(n)} (z)=\begin{cases}
    -\gamma +\sum_{k=0}^\infty\, \left(\frac{1}{k+1}-\frac{1}{k+z}\right)&n=0,\\
    (-1)^{n+1} \,n! \sum _{k=0}^\infty \frac{1}{(z+k)^{n+1}}&n>0,
    \end{cases}
\end{equation}
which is converging for any $z\in \mathbb{C}$ except negative integers. With this, we can express an infinite sum over rational functions. Let $\{f_n\}_{n\in  \mathbb{N}}$ be a sequence of the form $f_n=\frac{p(n)}{q(n)}$ where $p(n)$ and $q(n)$ are polynomials in $n$ with $\deg (p(n))<\deg (q(n))$. Let $\alpha _k$ be the roots of the denominator $q(n)$. If all the roots are simple, the fraction $f_n$ can be written as (partial fraction decomposition):
\begin{equation}
    f_n=\sum _{k=1}^m \frac{A_k}{n-\alpha _k},
\end{equation}
where $m=\deg(q(n))$. In general, if one or more roots $\alpha _k$ have multiplicity $m_k\geq 2$, the formula becomes:
\begin{equation}
f_n=\sum _{k=1}^{\tilde{m}} \sum_{r_k=1}^{m_k} \frac{A_{k,r_k}}{(n-\alpha _k)^{r_k}},
\end{equation}
where $\tilde{m}\leq \deg(q(n))$ is the number of distinct roots. It follows 
that:
\begin{equation}
\sum_{n=0}^\infty f_n=\sum _{n=0}^\infty \sum _{k=1}^{\tilde{m}} \sum_{r_k=1}^{m_k} \frac{A_{k,r_k}}{(n-\alpha _k)^{r_k}}=\sum_{k=1}^{\tilde{m}}  \sum_{r_k=1}^{m_k} \frac{(-1)^{r_k}}{(r_k-1)!} A_{k,r_k}\,  \psi ^{(r_k-1)}(-\alpha _k),
\end{equation}
provided that the series based on the sequence $\{f_n\}_{n\in  \mathbb{N}}$ is converging. This can be used to compute the infinite sum over pseudoscalar poles, $\sum _{n=0}^\infty  \hat{\Pi} _1^{P(n)\text{-pole}}(-Q_1^2,-Q_2^2,-Q_3^2)$, within our large-$N_c$ Regge model for the TFFs.

\subsection{Euler--Maclaurin summation formula}

A key ingredient in the discussion of the SDCs for the HLbL tensor  is the Euler--Maclaurin summation formula, which describes the difference between an integral and a related sum, see for instance ref.~\cite[chapter 8]{MathBookOlver}. Notably, it can be used to derive asymptotic expansions. 

Let $a<b$ and $m>0$ be integers, and $f$ be a function whose derivatives $f^{(2m)}(x)$ are absolutely integrable over the interval $(a,b)$. Then the \textbf{Euler--Maclaurin formula} reads:
\begin{equation}\label{eq:Euler--Maclaurin}
    \hspace{-0.2cm}\sum _{k=a}^b f(k)=\int_a^bd x\, f(x) +\frac{1}{2}\left[f(a)+f(b)\right]+\sum _{s=1}^{m-1}\frac{B_{2s}}{(2s)!}\left[ f^{(2s-1)}(b)-f^{(2s-1)}(a) \right]+R_m(b),
\end{equation}
where the remainder is given by:
\begin{equation}
    R_m(n)=\int _a ^b d x\, \frac{B_{2m}-B_{2m}(x-\floor*{x})}{(2m)!}\,f^{(2m)}(x),
\end{equation}
where $\floor{x}$ is the greatest integer smaller or equal to $x$, $B_s$  are the Bernoulli numbers, and $B_s(x)$ are the Bernoulli polynomials.\footnote{The Bernoulli numbers can be generated through:
\begin{equation}
B_s=\begin{cases}
    1\,&s=0,\\
    -\frac{1}{s+1}\sum _{j=0}^{s-1}{s+1 \choose j}B_j\,&\,s \geq 1,
\end{cases}
\end{equation}
and the Bernoulli polynomials can be constructed according to:
\begin{equation}
    B_s(x)=\sum _{j=0}^\infty {s \choose j}B_{s-j}x^j.
\end{equation}
For $0\leq x \leq 1$, they satisfy \cite{MathBookOlver}:
\begin{equation}\label{ineqB}
    |B_{2s}(x)-B_{2s}|\leq (2-2^{1-2s})|B_{2s}|.
\end{equation}}
Using \eqref{ineqB}, we can find a bound for the remainder:
\begin{equation}
\label{eq:RemainderBound}
 |R_m (b)|\leq \left(2-2^{1-2m}\right)\frac{|B_{2m}|}{(2m)!}\int _a^bd x  \,|f^{(2m)}(x)|.
\end{equation}
In particular, if $f^{(2m)}(x)$ does not change sign in the considered interval, the remainder is bounded by $\left(2-2^{1-2m}\right)$ times the first neglected term in~\eqref{eq:Euler--Maclaurin}.

The Euler--Maclaurin formula can be used to derive the asymptotic expansion of the polygamma functions~\eqref{eq:defPolygamma} at large $z\in \mathbb{R}$.
To illustrate, consider the trigamma function 
$\psi ^{(1)}(z)=\sum _{k=0}^\infty \frac{1}{(z+k)^2}$.
Inserting $f(x)=\frac{1}{(z+x)^2}$, $a=0$, $b=\infty$, and $m=1$ into~\eqref{eq:Euler--Maclaurin} leads to the asymptotic expansion:
\begin{equation}\label{Psi1EMC}
        \psi ^{(1)}(z)=\int _0^\infty d x \,\frac{1}{(z+x)^2} + \frac{1}{2z^2}+R_1(\infty,z)= \frac{1}{z}+ \frac{1}{2z^2}+R_1(\infty,z),
\end{equation}
where the notation of the remainder has been slightly modified compared to~\eqref{eq:Euler--Maclaurin} in order to highlight the additional $z$-dependence. The derivatives $f^{(2m)}(x)=\frac{(2m+1)!}{(x+z)^{2m+2}}$ do not change sign and the first neglected term in~\eqref{Psi1EMC} is given by $\frac{B_2}{2!}(0+\frac{2}{z^3})=\frac{1}{6z^3}$. This implies that $|R_1(\infty,z)|\leq (2-2^{1-2})\frac{1}{6z^3}=\frac{1}{4z^3} $. In the next subsection, we will be interested in the remainder generated by truncating the asymptotic expansion in \eqref{Psi1EMC} after the first term:
\begin{equation}\label{eq:usefulPsi1}
     \psi ^{(1)}(z)=:\frac{1}{z}+R_0(\infty, z).
\end{equation}
It follows from~\eqref{Psi1EMC} that:
\begin{align}
  |R_0(\infty,z)|&=\left|\frac{1}{2z^2}+R_1(\infty,z)\right|\leq  \left|\frac{1}{2z^2}\right|+\left|R_1(\infty,z)\right|\nn\\
  &\leq \frac{1}{2z^2}+\frac{1}{4z^3}\leq  \frac{1}{2z^2}+\frac{1}{4z^2}=\frac{3}{4z^2},\label{eq:boundR0}
\end{align}
where the last inequality holds when $z\geq 1$. For a general $n\in \mathbb{N}$, a similar procedure leads to \cite{HandbookMath}:
\begin{equation}\label{eq:polygammaAsymptotic}
    \psi ^{(n)} (z) \sim \begin{cases}
   \log (z) -\frac{1}{2z}-\sum _{k=2}^\infty \frac{B_k}{k z^k}&n=0,\\
    (-1)^{n+1}  \left( \frac{(n-1)!}{z^n}+\frac{n!}{2z^{n+1}}+ \sum _{k=2}^\infty \frac{(k+n-1)!}{k!}\frac{B_k}{z^{k+n}}\right)&n>0.
    \end{cases}
\end{equation}

The asymptotic expansion~\eqref{eq:polygammaAsymptotic} in combination with~\eqref{eq:seriespolygamma} is sufficient to derive~\eqref{MV}, \eqref{pQCD3}, \eqref{MVEta}, and \eqref{pQCD3Eta}, and thereby fix the parameters of the large-$N_c$ Regge models to satisfy the required SDCs on the HLbL tensor. 

\subsection{Short-distance constraints for the alternative transition form factor model}

For the alternative TFF model, introduced in appendix~\ref{sec:PionAlternative}, the situation is more complicated, because the summation over the pseudoscalar-pole diagrams involves three infinite sums---one additional sum over vector-meson towers per TFF~\eqref{MartinModel}---and only two of them can be performed analytically. Therefore, one needs to use the Euler--Maclaurin formula to extract the asymptotic behavior.

We first consider the mixed-energy region $Q_1^2\approx Q_2^2 \gg Q_3^2$. The terms in our TFF model~\eqref{MartinModel} with exponential weights in the numerator are suppressed and do not contribute to the MV limit:
\begin{align} 
   & \sum _{n=0}^\infty \hat{\Pi}_1^{P(n)\text{-pole}}(-Q^2,-Q^2,-Q_3^2) \nn \\
     &\sim c_2^2\,\sum _{n=0}^\infty \sum _{i=0}^\infty \sum _{j=0}^\infty \int _0^1 d y \int_0^1 d x \frac{ y(1-y)\,x(1-x)}{\left[Q_3^2+M_{P(n)} ^2\right]\left[ M_{V_2(n+i)}^2+  Q^2 \right] ^2 \left[ M_{V_2(n+j)}^2 +Q_3^2 \,y\right] ^2 }\nn\\
        &=\frac{c_2^2}{6}\int _0^1 d y \,\sum _{n=0}^\infty \sum _{i=0}^\infty \sum _{j=0}^\infty \underbrace{ \frac{ y(1-y)\,x(1-x)}{\left[Q_3^2+M_{P(n)} ^2\right]\left[ M_{V_2(n+i)}^2+  Q^2\right] ^2 \left[ M_{V_2(n+j)}^2+Q_3^2 \,y\right] ^2 }}_{:=f_{nij}(y)}.  \label{eq:MVproof}
\end{align}
The integration over the Feynman parameter $x$ is trivial. Since the $f_{nij}(y)$ in \Eqref{MVproof} do not contain any singularities in the space-like region and the integration domains are bounded, the convergence is uniform and the commutation of integrations and summations is justified.\footnote{Formally, we use the
\textbf{dominated convergence theorem }(in the setting of Riemann integrals) \cite[p.~54]{MathBookOlver}: 
\\Let  $(a,b) \subset \mathbb{R}$ be an open, finite or infinite interval. Let $\{ f_n \} _{n \in \mathbb{N}}$ be a sequence of real or complex functions which are continuous on $(a,b)$ and satisfy:
\begin{enumerate}
\item The series $\sum _{n=1}^\infty f_n(x)$ converges uniformly in any compact interval in $(a,b)$
\item Either $\int _a^b d x \sum _{n=1}^\infty \left| f_n (x) \right|   < \infty$ or  $\sum _{n=1}^\infty \int _a^b d x\left| f_n (x) \right|  < \infty$
\end{enumerate}
Then $\int _a^b d x \sum _{n=1}^\infty  f_n (x)  =\sum _{n=1}^\infty \int _a^b d x \,f_n (x)\, $.} Using~\eqref{eq:seriespolygamma}, we can express the sums over $i$ and $j$ in~\eqref{eq:MVproof} 
with trigamma functions:
\begin{align} 
&\sum _{n=0}^\infty \hat{\Pi}_1^{P(n)\text{-pole}}(-Q^2,-Q^2,-Q_3^2) \nn \\
&\sim\frac{c_2^2}{6\sigma _{V_2}^8}\int_0^1 d y \sum _{n=0}^\infty\, \frac{y(1-y)}{Q_3^2+M_{P(n)} ^2}\;\psi ^{(1)}\!\left( \frac{M_{V_2(n)}^2+ Q^2}{\sigma _{V_2}^2} \right)\psi ^{(1)}\!\left( \frac{M_{V_2(n)}^2+Q_3^2\,y}{\sigma _{V_2}^2} \right),\label{eq:MV31}
\end{align}
where we use the notations from~\eqref{pionTrajectory}, \eqref{etaTrajectory}, \eqref{etaPTrajectory}, and \eqref{MV2Regge}, assuming for simplicity that $\hat M_{P}=M_P$. The remaining sum over $n$ cannot be performed analytically. We can, however, rewrite the trigamma function as the first term in its asymptotic expansion and a remainder, see~\eqref{eq:usefulPsi1}:
\begin{align}
&\sum _{n=0}^\infty \hat{\Pi}_1^{P(n)-\text{pole}}(-Q^2,-Q^2,-Q_3^2)\nn \\
&=\frac{c_2^2}{6\sigma _{V_2}^8}\int_0^1 d y \sum _{n=0}^\infty\, \frac{y(1-y)}{Q_3^2+M_{P(n)} ^2}\left\{ \frac{\sigma _{V_2}^2}{M_{V_2(n)}^2+ Q^2} +R_0\bigg(\infty,\frac{M_{V_2(n)}^2+ Q^2}{\sigma _{V_2}^2}\bigg)\right\} \nn \\
    & \times \left\{\frac{\sigma _{V_2}^2}{M_{V_2(n)}^2+Q_3^2\,y} +R_0\bigg(\infty,\frac{M_{V_2(n)}^2+Q_3^2\,y}{\sigma _{V_2}^2}\bigg) \right\}\nn \\
    &=:H^{P}_\text{MV}(Q^2,Q_3^2)+\delta H^P_\text{MV}(Q^2,Q_3^2).\label{eq:MVproofBis}
\end{align}
Here, we defined:
\begin{equation}\label{Fmv}
    H^{P}_\text{MV}(Q^2,Q_3^2)=\frac{c_2^2}{6\sigma _{V_2}^4}\int_0^1 d y \sum _{n=0}^\infty \frac{y(1-y)}{\left[M_{V_2(n)}^2+  Q^2\right]\left[Q_3^2+M_{P(n)} ^2\right]\left[M_{V_2(n)}^2+Q_3^2\,y\right]},
\end{equation}
and included the remaining terms of~\eqref{eq:MVproofBis} in $\delta H^P_{\text{MV}}$. The sum over $n$ in~\eqref{Fmv} can now be expressed in terms of polygamma functions, but the integral over the Feynman parameter $y$ is difficult to perform analytically. Therefore, we expand in $  Q^2$ and $Q_3^2$ before integrating over $y$. The assumption that the two operations commute will  be checked a posteriori. We find ($ {Q}^2\gg Q_3^2$):
\beq
\label{eq:FMV}
 H^P_\text{MV}(Q^2,Q_3^2)=\int _0^1 d y \left\{ \frac{f^P_\text{MV}(y)}{  Q^2Q_3^2}+\mathcal{O}\left( \frac{1}{  Q^2 Q_3^4} \right) \right\}=\frac{\,a^P_\text{MV}}{  Q^2Q_3^2}+\mathcal{O}\left( \frac{1}{  Q^2 Q_3^4} \right),
\eeq
with 
\begin{align}
   a^P_\text{MV}&= \frac{c_2^2}{24 \sigma _P ^6 \sigma _{V_2}^4} \Bigg\{2\si_P^2 \left(2\si_{V_2}^2 -\si_P^2\right) L_{P V_2}-\si_P^2 \left(4\si_{V_2}^2-3\si_P^2\right) \nn \\
 &-4  \sigma_{V_2}^2\Delta_{P V_2} \left[\frac{\pi^2}{6}-\text{Li}_2 \left(1-\frac{\si_P^2}{\si_{V_2}^2}\right)\right]\Bigg\}.
\end{align}
An appropriate choice of $\sigma _{V_2}$ in $a^{P}_\text{MV}$ therefore reproduces the MV limit. 

\begin{figure}[t]
\centering
\begin{subfigure}[b]{0.48\textwidth}
\includegraphics[width=\textwidth]{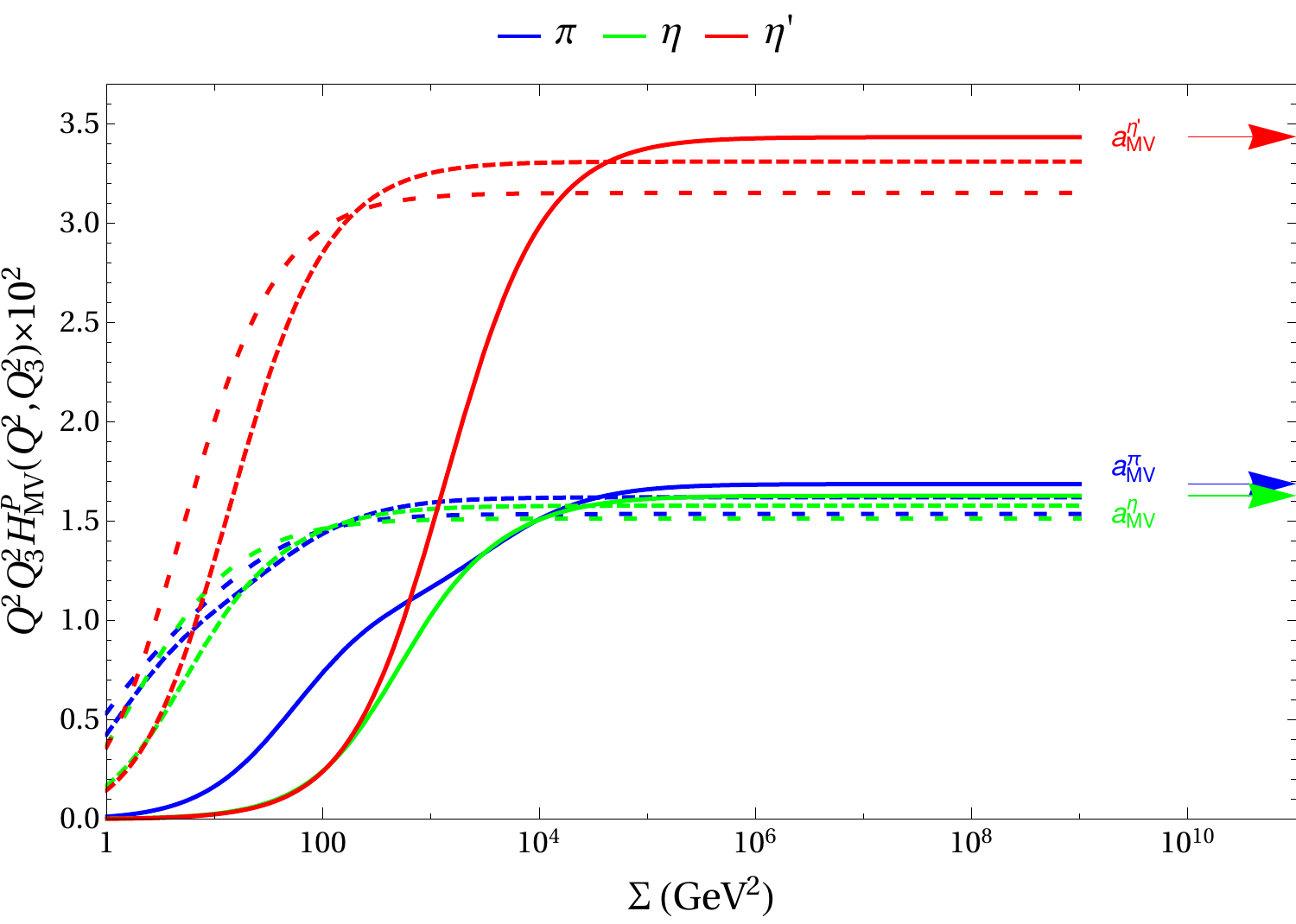}
\caption{Numerical check of~\eqref{eq:FMV}.}
\label{fig:checknumMV00}
\end{subfigure}
\hfill
\begin{subfigure}[b]{0.48\textwidth}
\includegraphics[width=\textwidth]{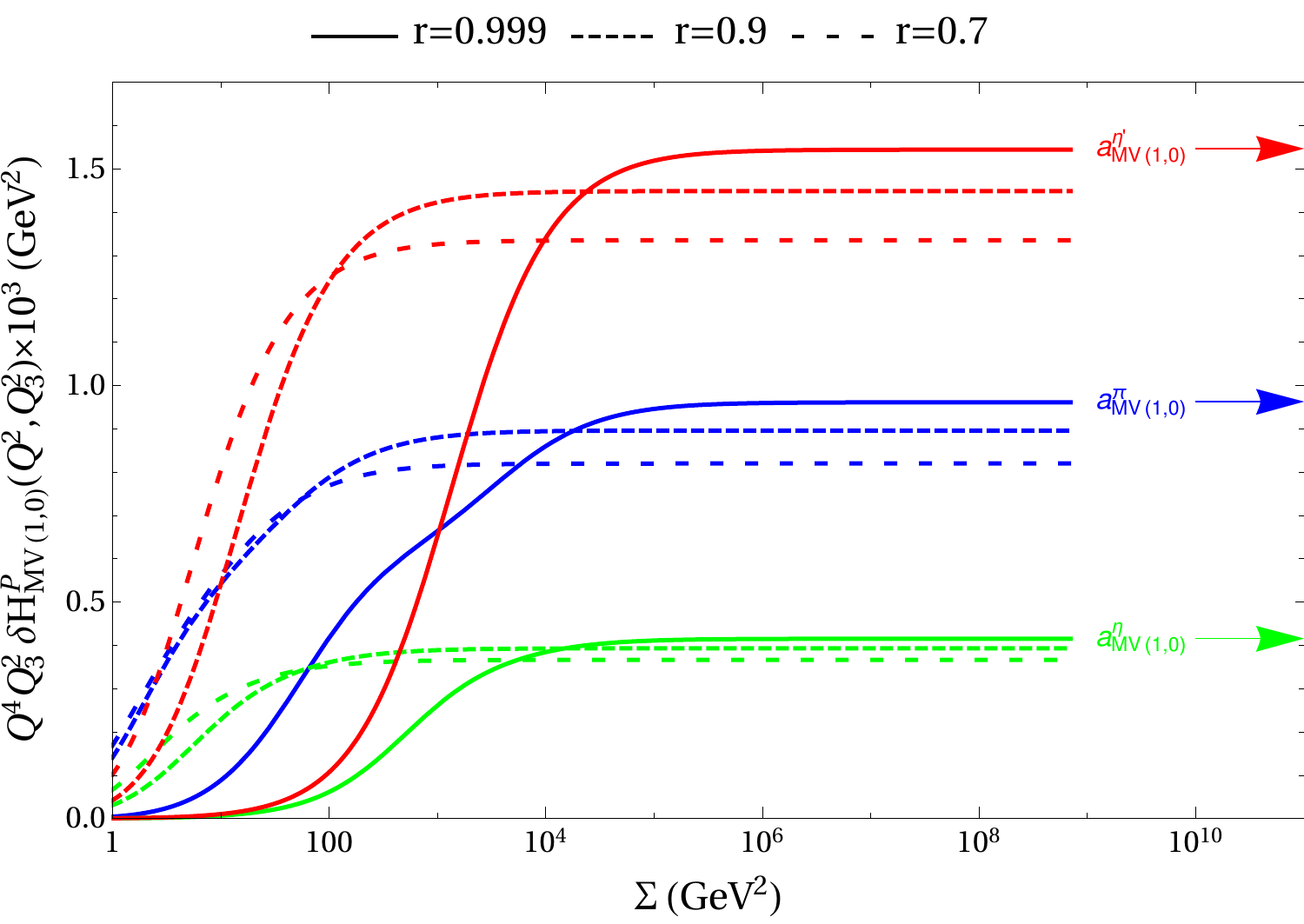}
\caption{Numerical check of~\eqref{eq:F01check}.}
\label{fig:checknumMV01}
\end{subfigure}\\[0.5cm]
\begin{subfigure}[b]{0.48\textwidth}
\includegraphics[width=\textwidth]{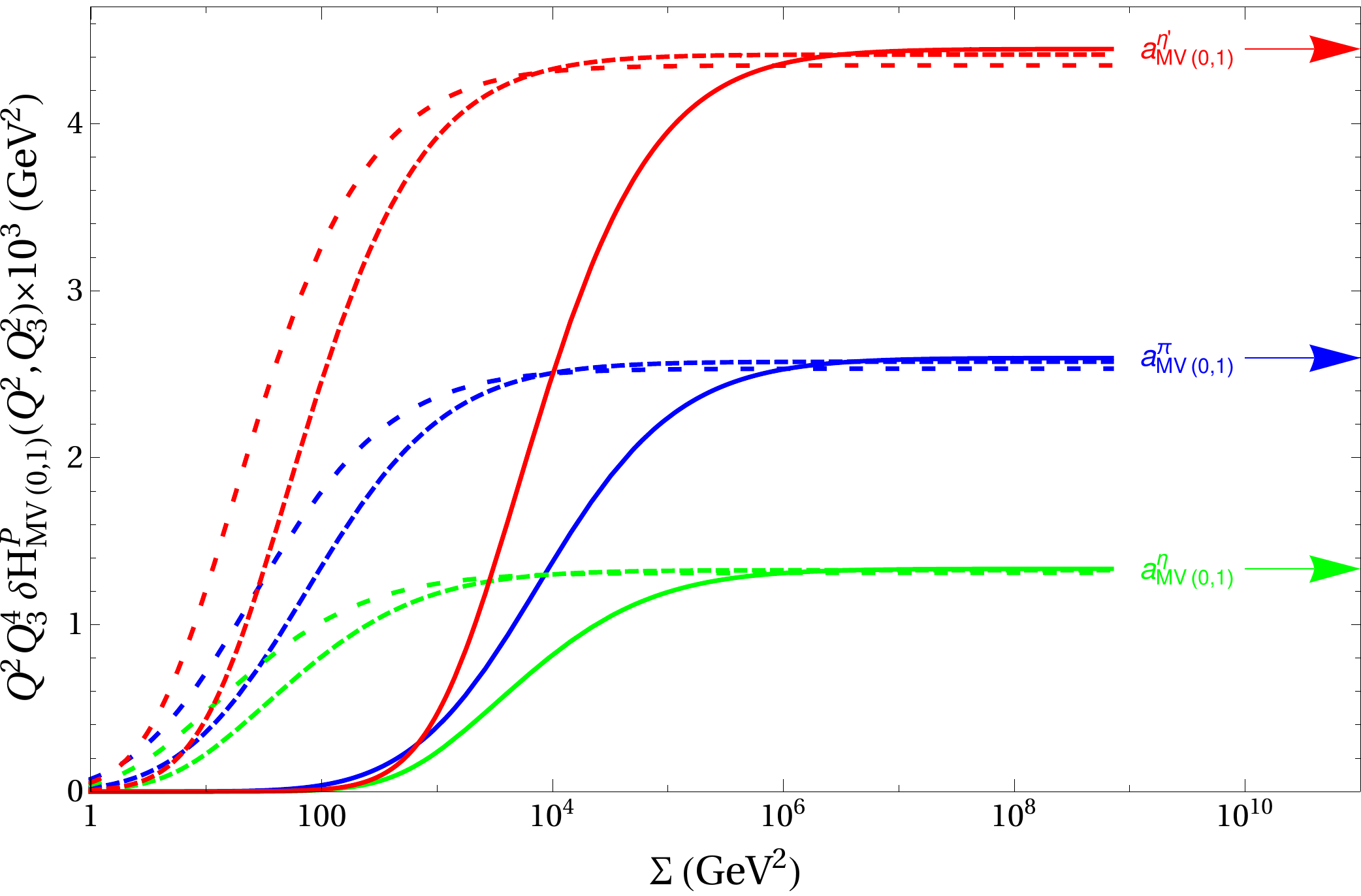}
\caption{Numerical check of~\eqref{eq:F10check}.}
\label{fig:checknumMV10}
\end{subfigure}
\hfill
\begin{subfigure}[b]{0.48\textwidth}
\includegraphics[width=\textwidth]{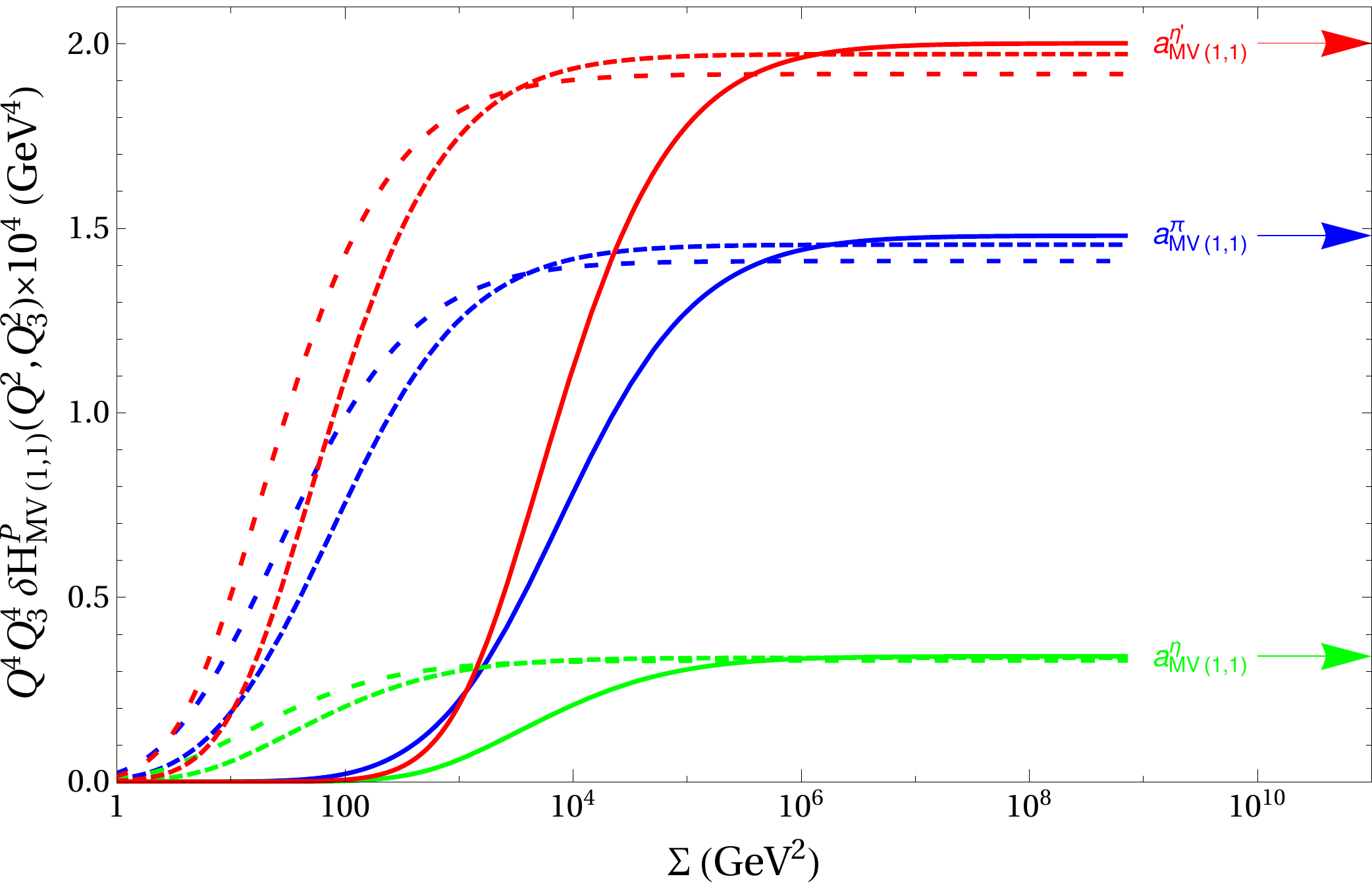}
\caption{Numerical check of~\eqref{eq:F11check}.}
\label{fig:checknumMV11}
\end{subfigure}
\caption{Numerical checks for the validity of commuting the expansion in $Q^2$ and $Q_3^2$, and the integration over $y$ in~\eqref{eq:MVproofBis}. In all cases, we can see that when $r\to 1$ (MV kinematics), the curves tend to the coefficients obtained by expanding in the virtualities first and then integrating over the Feynman parameter $y$. Note that the scales on the $y$-axis vary between the plots.}
\label{fig:checknumMV}
\end{figure}

Let us now verify that expanding before integrating in \Eqref{FMV} was justified and that $\delta H^P_\text{MV}(Q^2,Q_3^2)$ is subleading. The first issue can be addressed numerically. We use the coordinates defined in~\eqref{eq:g2coordinates}. The kinematics corresponding to the MV limit can be expressed in those coordinates by setting $\phi =\pi$,
\begin{equation}
    Q_1^2=Q_2^2=  Q^2=\frac{\Sigma}{3}\left( 1+\frac{r}{2} \right),\qquad
    Q_3^2=\frac{\Sigma}{3}\left( 1-r \right),
\end{equation}
then considering $r$ close to $1$, and finally taking the limit $\Sigma \to \infty$. In figure~\ref{fig:checknumMV00}, we study the function $Q^2 Q_3^2 H^{P}_\text{MV}(Q^2,Q_3^2)$ in \eqref{Fmv} for different values of $r$ using a numerical integration over $y$. One can see that with $r$ getting closer to $1$ the curves tend to $a^{P}_\text{MV}$ and this justifies the commutation of expansion and integration in this case. 

We are left to study the error made by considering only $H^P_{\text{MV}}$ and not the remainder $\delta H^{P}_\text{MV}$, which  can be decomposed into three terms:
\begin{equation}
   \delta H^{P}_\text{MV}(Q^2,Q_3^2)=\delta H^{P}_{\text{MV}(0,1)}( Q^2,Q_3^2)+\delta H^{{P}}_{\text{MV}(1,0)}(Q^2,Q_3^2)+\delta H^{{P}}_{\text{MV}(1,1)}(Q^2,Q_3^2).\label{HNotation}
\end{equation}
The notation can be understood as follows. From \eqref{eq:MV31} to \eqref{eq:MVproofBis}, two trigamma functions, which stem  from the doubly-virtual TFF $F_{P\ga^*\ga^*}(Q^2,Q^2)$ (first index) and the singly-virtual TFF $F_{P\ga\ga^*}(Q_3^2)$ (second index), were expanded in a leading piece ($0$) and a remainder ($1$). In other words, $\delta H^{P}_{\text{MV}(1,0)}$ combines the remainder of the trigamma function in $Q^2$ with the leading term in the expansion of the trigamma function in $Q_3^2$:
\begin{align} \label{eq:F01check}
&\delta H^{P}_{\text{MV}(1,0)}(Q^2,Q_3^2)\nn \\
&=\frac{c_2^2}{6\sigma _{V_2}^6}\int_0^1 d y \sum _{n=0}^\infty R_0\left( \infty, \frac{M_{V_2(n)}^2+  Q^2}{\sigma ^2 _{V_2}}\right)\, \frac{y(1-y)}{\left[Q_3^2+M_{P(n)} ^2\right]\left[M_{V_2(n)}^2+Q_3^2\,y\right]} \nn \\
  &\leq \frac{c_2^2}{8 \sigma _{V_2}^6}\int_0^1 d y \sum _{n=0}^\infty \frac{\sigma _{V_2}^4}{\left[M_{V_2(n)}^2+  Q^2\right]^2}\frac{y(1-y)}{\left[Q_3^2+M_{P(n)} ^2\right]\left[M_{V_2(n)}^2+Q_3^2\,y\right]}  \nn \\
  &=\int _0^1 d y \left\{ \frac{f^P_{\text{MV}(0,1)}(y)}{Q^4Q_3^2}+\mathcal{O}\left( \frac{1}{Q^4 Q_3^4} \right) \right\}=\frac{a^P_{MV(0,1)}}{  Q^4Q_3^2}+\mathcal{O}\left( \frac{1}{Q^4 Q_3^4} \right).
\end{align}
Here, we used \Eqref{RemainderBound} to show that the term is bounded from above, as can be done for each term in \eqref{HNotation}. 
As before, it can be checked numerically that $H^{P}_{\text{MV}(0,1)}$ indeed tends to the result obtained by expanding first and integrating second, see figure \ref{fig:checknumMV01}. We proceed analogously for the two remaining terms:
\begin{align} \label{eq:F10check}
 &\delta H^P_{\text{MV}(0,1)}(Q^2,Q_3^2)\nn \\
 &=\frac{c_2^2}{6\sigma _{V_2}^6}\int_0^1 d y \sum _{n=0}^\infty \frac{y(1-y)}{\left[M_{V_2(n)}^2+  Q^2\right]\left[Q_3^2+M_{P(n)} ^2\right]}\, R_0\left( \infty, \frac{M_{V_2(n)}^2+Q_3^2\,y}{\sigma _{V_2}^2} \right) \nn \\
  &\leq \frac{c_2^2}{8\sigma _{V_2}^6}\int_0^1 d y \sum _{n=0}^\infty \frac{y(1-y)}{\left[M_{V_2(n)}^2+ Q^2\right]\left[Q_3^2+M_{P(n)} ^2\right]} \frac{\sigma _{V_2}^4}{\left[M_{V_2(n)}^2+Q_3^2\,y\right]^2} \nn \\
  &=\int _0^1 d y \left\{ \frac{f^P_{\text{MV}(1,0)}(y)}{  Q^2Q_3^4}+\mathcal{O}\left( \frac{1}{  Q^2 Q_3^6} \right) \right\}=\frac{a^P_{\text{MV}(1,0)}}{  Q^2Q_3^4}+\mathcal{O}\left( \frac{1}{Q^2 Q_3^6} \right),
\end{align}
which is checked numerically in figure \ref{fig:checknumMV10} and:
\begin{align} \label{eq:F11check}
 &\delta H^P_{\text{MV}(1,1)}(Q^2,Q_3^2)\nn \\
 &=\frac{c_2^2}{6\sigma _{V_2}^8}\int_0^1 d y \sum _{n=0}^\infty R_0\left( \infty, \frac{M_{V_2(n)}^2+  Q^2}{\sigma _{V_2}^2}\right) \frac{y(1-y)}{Q_3^2+M_{P(n)} ^2} \,R_0\left( \infty, \frac{M_{V_2(n)}^2+Q_3^2\,y}{\sigma _{V_2}^2} \right)  \nn \\
  &\leq \frac{3c_2^2}{32\sigma _{V_2}^8}\int_0^1 d y \sum _{n=0}^\infty \frac{y(1-y)}{Q_3^2+M_{P(n)} ^2}   \frac{ \sigma _{V_2}^8}{\left[M_{V_2(n)}^2+  Q^2\right]^2\left[M_{V_2(n)}^2+Q_3^2\,y\right]^2}\nn \\
  &=\int _0^1 d y \left\{ \frac{f^P_{\text{MV}(1,1)}(y)}{  Q^4Q_3^4}+\mathcal{O}\left( \frac{1}{Q^4 Q_3^6} \right) \right\}=\frac{a^P_{\text{MV}(1,1)}}{Q^4Q_3^4}+\mathcal{O}\left( \frac{1}{Q^4 Q_3^6} \right),
\end{align}
see figure \ref{fig:checknumMV11}. The above considerations add up to:
\begin{equation}
    \delta H^P_\text{MV}(Q^2,Q_3^2)=\mathcal{O} \left( \frac{1}{Q^2Q_3^4}\right),
\end{equation}
i.e., the error we make by keeping only the leading term in the expansion of the $\psi ^{(1)}$ in~\eqref{eq:MVproofBis} is subdominant.

When considering the high-energy region $Q_1^2\approx Q_2^2 \approx Q_3^2=Q^2$, the same technique can be applied, but the situation simplifies slightly, since there is only one large scale. The pQCD constraint on the HLbL tensor reads:
\begin{align}\label{eq:pQCD31}
&\sum _{n=0}^\infty \hat{\Pi}_1^{P(n)\text{-pole}}(-Q^2,-Q^2,-Q^2)\nn \\
&\sim\frac{c_2^2}{6\sigma _{V_2}^8}\int_0^1 d y \sum _{n=0}^\infty\, \frac{y(1-y)}{Q^2+M_{P(n)} ^2}\;\psi ^{(1)}\!\left( \frac{M_{V_2(n)}^2+ Q^2}{\sigma _{V_2}^2} \right)\psi ^{(1)}\!\left( \frac{M_{V_2(n)}^2+Q^2\,y}{\sigma _{V_2}^2} \right).
\end{align}
Similarly to the previous case, since the sum over $n$ cannot be performed analytically, we rewrite the polygamma function as the first term in its asymptotic expansion and a remainder. This leads to:
\begin{align}
 H^P_{\text{pQCD}}(Q^2)&:= \frac{c_2^2}{6\sigma _{V_2}^4}\int_0^1 d y \sum _{n=0}^\infty \frac{y(1-y)}{\left[Q^2+M_{P(n)} ^2\right]\left[M_{V_2(n)}^2+  Q^2\right]\left[M_{V_2(n)}^2+Q^2\,y\right]}, \nn\\
\delta H^P_{\text{pQCD}(1,0)}(Q^2)&:= \frac{c_2^2}{6\sigma _{V_2}^6}\int_0^1 d y \sum _{n=0}^\infty R_0\bigg( \infty, \frac{M_{V_2(n)}^2+  Q^2}{\sigma ^2 _{V_2}}\bigg)\frac{y(1-y)}{\left[Q^2+M_{P(n)} ^2\right]\left[M_{V_2(n)}^2+Q^2\,y\right]} , \nn \\
\delta H^P_{\text{pQCD}(0,1)}(Q^2)&:= \frac{c_2^2}{6\sigma _{V_2}^6}\int_0^1 d y \sum _{n=0}^\infty \frac{y(1-y)}{\left[Q^2+M_{P(n)} ^2\right]\left[M_{V_2(n)}^2+  Q^2\right]}\, R_0\bigg( \infty, \frac{M_{V_2(n)}^2+Q^2\,y}{\sigma _{V_2}^2} \bigg), \nn \\
\delta H^P_{\text{pQCD}(1,1)}(Q^2)&:=\frac{c_2^2}{6\sigma _{V_2}^8}\int_0^1 d y \sum _{n=0}^\infty \frac{y(1-y)}{Q^2+M_{P(n)} ^2} R_0\bigg( \infty, \frac{M_{V_2(n)}^2+  Q^2}{\sigma _{V_2}^2}\bigg) \nn\\
&\times R_0\bigg( \infty, \frac{M_{V_2(n)}^2+Q^2\,y}{\sigma _{V_2}^2} \bigg) . \label{HPQCD} 
\end{align}
Analogously to $H^P_{\text{MV}}$, the term $H^P_{\text{pQCD}}$ is treated as follows: the sum over $n$ is performed, the expression is expanded in $Q^2$, and the integration over $y$ is carried out. The commutation of the expansion and integration is checked numerically in figure~\ref{fig:checknumpQCD}. For the other terms, $\delta H^P_{\text{pQCD}(i,j)}$, we use~\eqref{eq:boundR0} and then proceed analogously to the leading term, see figure~\ref{fig:checknumpQCD} for the numerical checks.

\begin{figure}[t]
\includegraphics[width=0.48\textwidth]{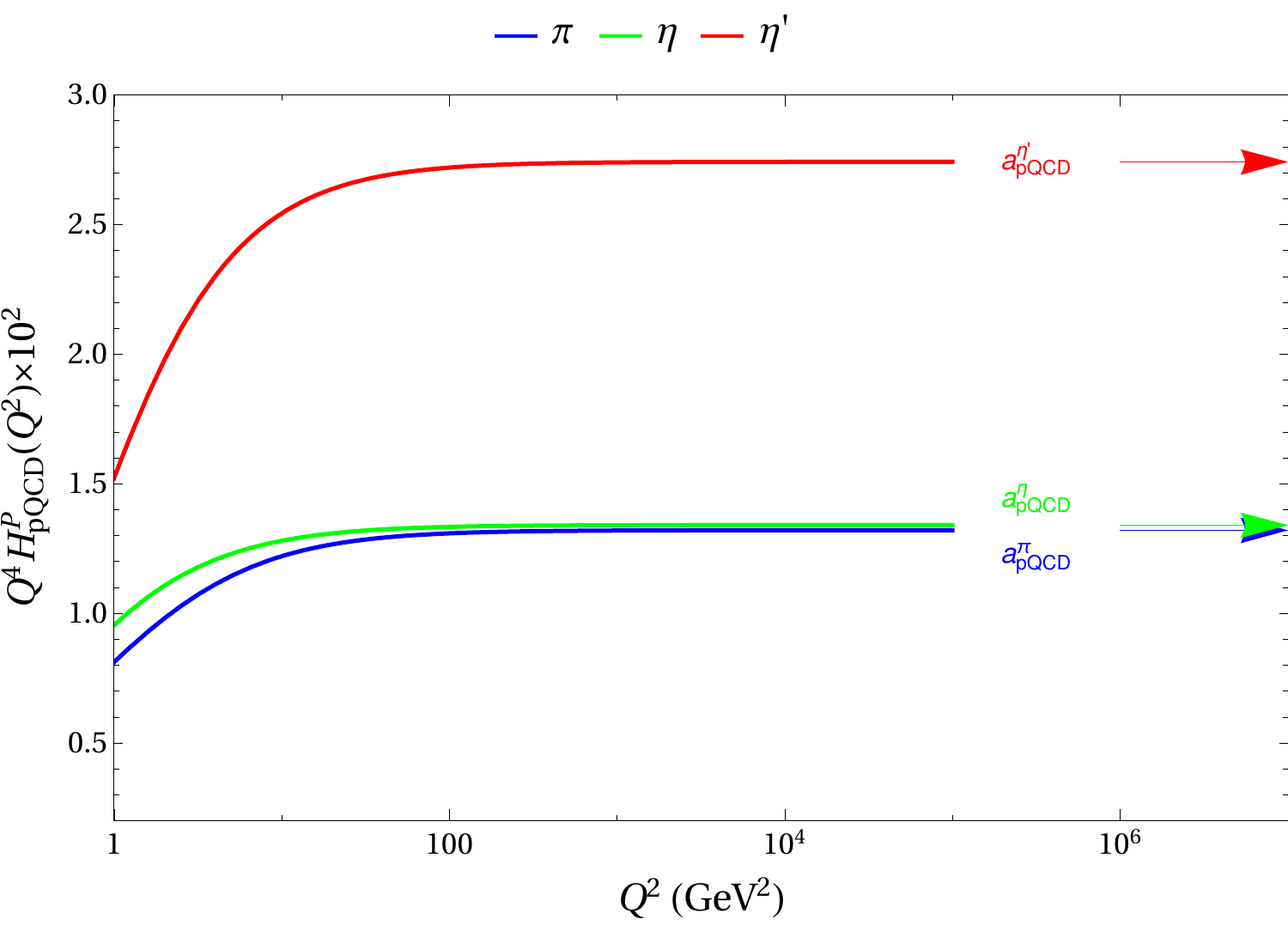}
\hfill
\includegraphics[width=0.48\textwidth]{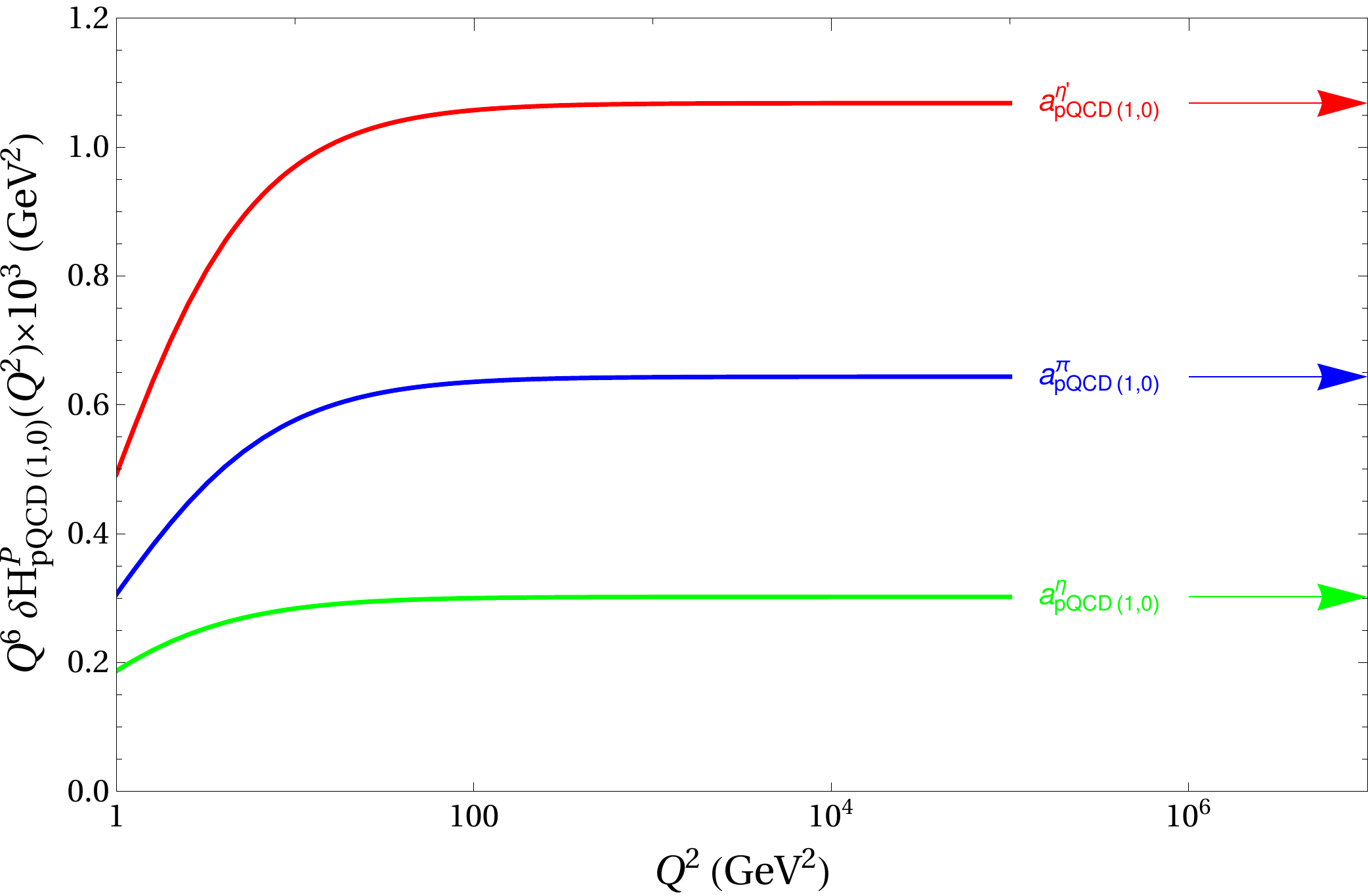}\\[0.2cm]
\includegraphics[width=0.48\textwidth]{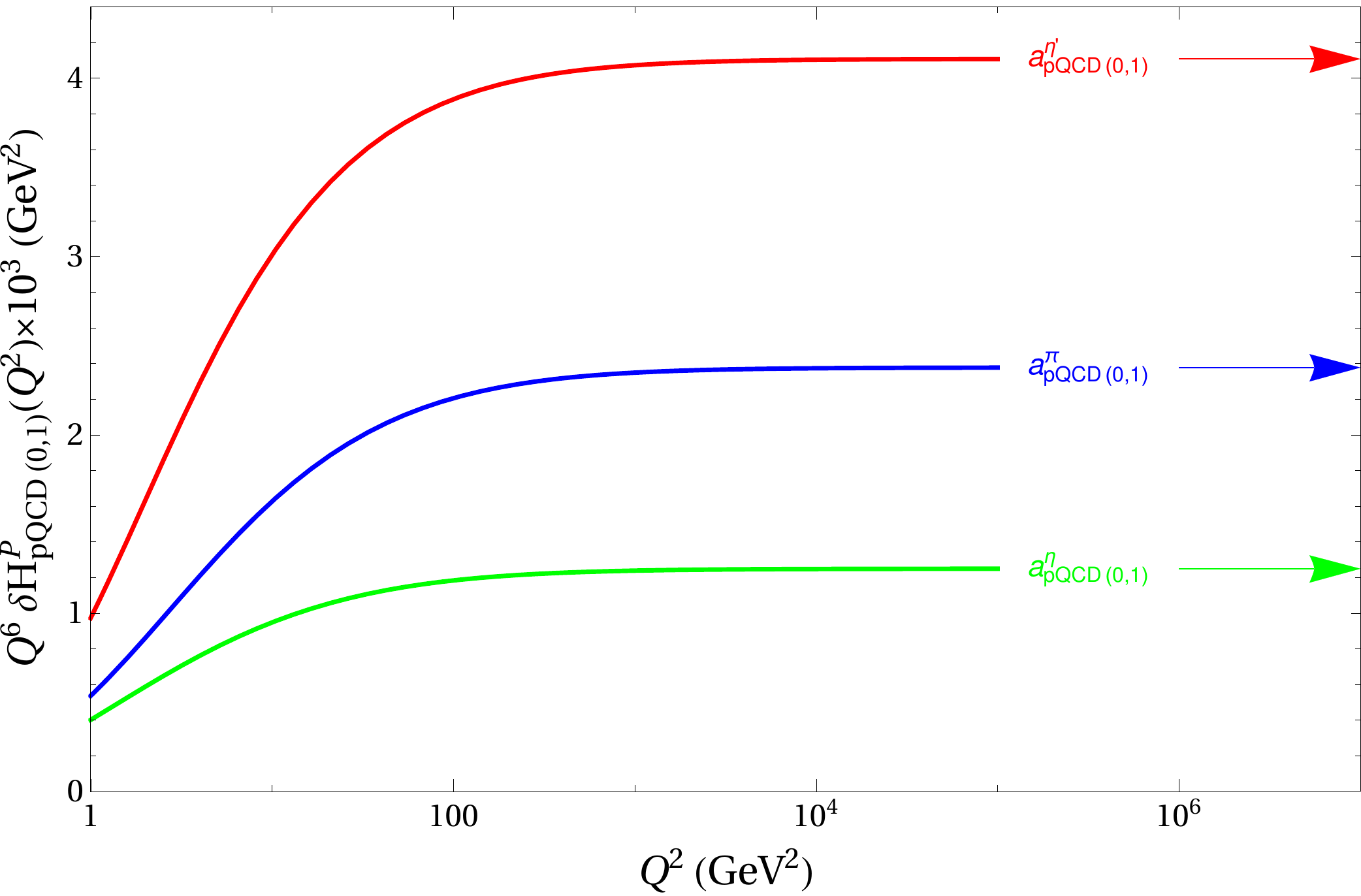}\hfill
\includegraphics[width=0.48\textwidth]{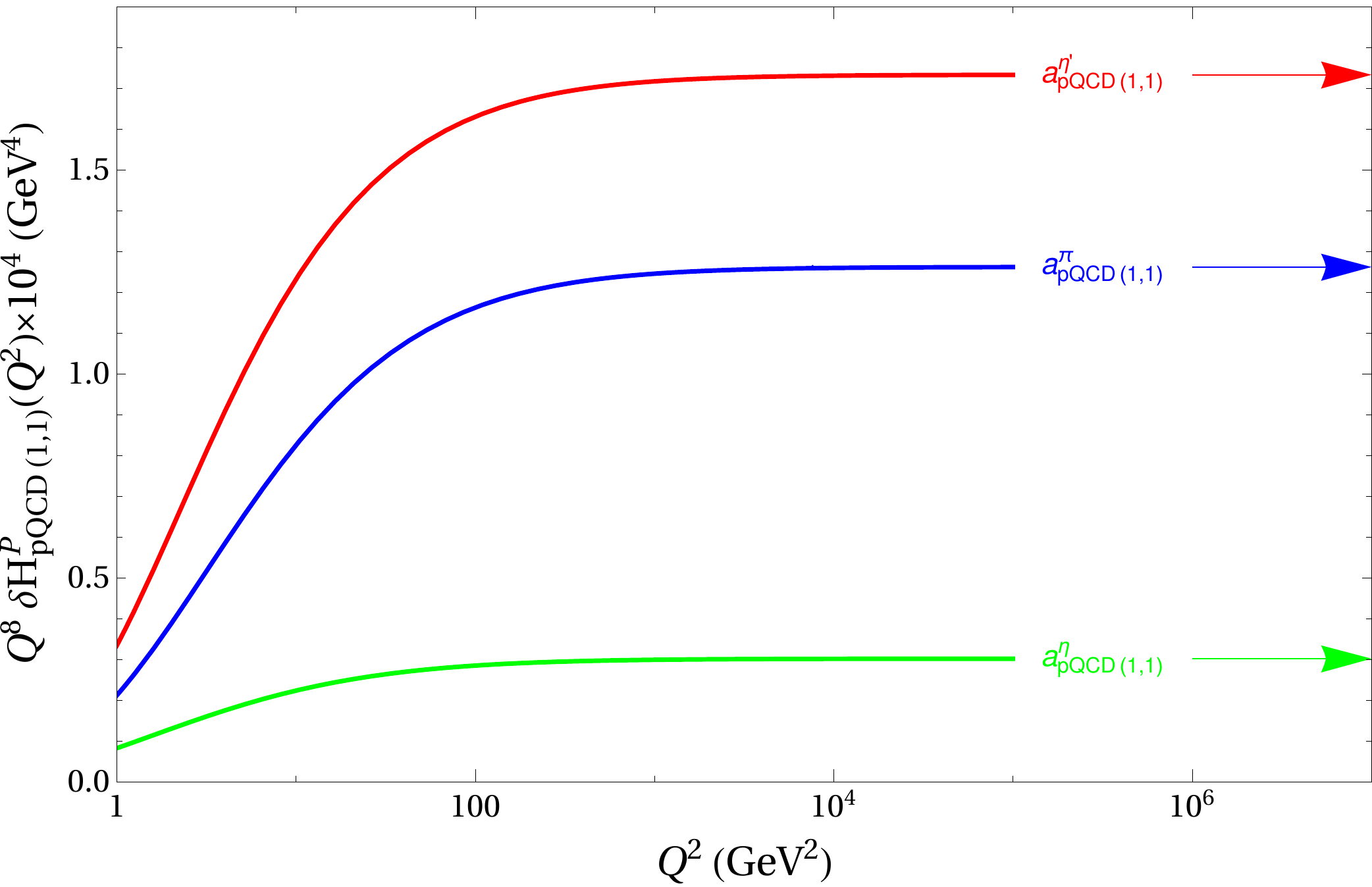}
\caption{Numerical checks for the validity of commuting the expansion in $Q^2$ and the integration over $y$ in \eqref{HPQCD}. In all cases, we can see that the curves tend to the coefficients obtained by expanding in the virtualities first and then integrating over the Feynman parameter $y$. Note that the scales on the $y$-axis vary between the plots.
}
\figlab{checknumpQCD}
\end{figure}

\section{Two-photon couplings of excited pseudoscalars}
\label{app:photon_couplings}

In this appendix we collect the phenomenological information that is available on the two-photon couplings of the excited pseudoscalars listed in the PDG~\cite{Tanabashi:2018oca}, in comparison to 
the two-photon couplings of the first radially-excited pion, $\eta$, and $\eta'$ states as shown in figure~\ref{fig:Coupling} for both the large-$N_c$ Regge and the alternative TFF model. 

Phenomenologically, the two-photon couplings of the excited pseudoscalars are unknown, but for many states some information on these couplings can be extracted either from direct limits on the two-photon channel or from measurements of particular branching fractions. For the excited pion states, the only available information concerns the $\pi(1300)$. 
The blue bar in the left panel of figure~\ref{fig:Coupling} indicates values excluded by the limit
\beq
\label{couplingconstraint}
 F_{\pi(1300)\ga\ga}<0.0544(71)\GeV^{-1}.
\eeq
Since at present there is no measurement of the $\pi(1300)$ width and the two-photon branching ratio, the above bound is an estimate based on the available empirical information. The $\pi(1300)$  decays predominantly into $3\pi$, e.g., into $\rho \pi$:\footnote{Note that the role of the $S$-wave component is not settled:  
while $\Gamma(\pi(\pi \pi)_{S\text{-wave}}/\Gamma(\rho \pi)=2.12$ from ref.~\cite{Aaron:1980zk} agrees with ref.~\cite{Salvini:2004gz}, ref.~\cite{Abele:2001js} found a negligible $S$-wave component $\Gamma(\pi(\pi \pi)_{S\text{-wave}}/\Gamma(\rho \pi)<0.15$. In the latter case the limit on the two-photon decay width $F_{\pi(1300)\ga\ga}$ would become stricter by a factor $\sqrt{3.2}\sim 1.8$, 
in which case there would be some mild tension with $F_{\pi(1300)\ga\ga}$ implied by our Regge models.
}
\begin{equation}
\frac{\Gamma(\ga \ga)\Gamma(\rho \pi)}{\Gamma_\mathrm{total}}<85\eV\, \text{\cite{Acciarri:1997rb}},\qquad 
\frac{\Gamma(\pi(\pi \pi)_{S\text{-wave}})}{\Gamma(\rho \pi)}=2.2(4)\,\text{\cite{Salvini:2004gz}}, \label{channelratio}
\end{equation}
whereas the $\pi f_0(1300)$ and $\gamma\gamma$ decays are suppressed~\cite{Barnes:1996ff}. Assuming 
\beq
\Gamma_\mathrm{total}\sim \Gamma(\rho \pi)+\Gamma(\pi(\pi \pi)_{S\text{-wave}}),\label{dominantChannels}
\eeq
this leads to:
\beq
\label{result22}
\Gamma(\pi(1300) \rightarrow \ga \ga)< 272(34)\eV,\\
\eeq
where the error is propagated from~\eqref{channelratio}. The bound in~\eqref{couplingconstraint} follows from~\eqref{anomDef} with~\eqref{result22} and  $M_{\pi(1300)}=1300(100)\MeV$. 
As one can see from figure~\ref{fig:Coupling}, the large-$N_c$ Regge model indeed satisfies the bound~\eqref{couplingconstraint}, it has $F_{\pi(1300)\ga\ga}=0.0500\GeV^{-1}$.
Thus, even though it is not possible to generate a suppression by inverse powers of the pseudoscalar mass in our models, as seen in~\eqref{anomDef}, the two-photon couplings of the excited pseudoscalars are sufficiently suppressed by inverse powers of the excited vector-meson masses.

Similar constraints exist for several excited $\eta$, $\eta'$ states. As discussed in the main text, the assignment of Regge trajectories is not settled, so here we simply reproduce 
the listing according to the PDG, see section~\ref{sec:two_photon_couplings} for a discussion of the phenomenological implications.
We stress that given that even the identification of states is contentious, the experimental limits should be treated with caution and mainly serve as guidance that 
our Regge models do not assume implausible values for the two-photon couplings. 
For the $\eta(1295)$, $M_{\eta(1295)}=1294(4)\MeV$, we have
\begin{equation}
\frac{\Gamma(\ga \ga)\Gamma(\eta\pi\pi)}{\Gamma_\mathrm{total}}<66\eV\, \text{\cite{Acciarri:2000ev}},\qquad
\frac{\Gamma(\ga \ga)\Gamma(K\bar K \pi)}{\Gamma_\mathrm{total}}<14\eV\, \text{\cite{Ahohe:2005ug}}. 
\end{equation}
Assuming that the branching fraction into other channels can be neglected,\footnote{The limit from ref.~\cite{Acciarri:2000ev} already includes the conversion factor 
$\Gamma(\eta\pi^+\pi^-)/(\Gamma(\eta\pi^+\pi^-)+\Gamma(\eta\pi^0\pi^0))=2/3$, which emerges from the 
combination of isospin and symmetry factors in
$\Gamma(\eta\pi^0\pi^0)/\Gamma(\eta\pi^+\pi^-)=1/2$.} we would conclude 
$\Gamma(\ga \ga) < 80\eV$ and thus  
\begin{equation}
 F_{\eta(1295)\ga\ga}<0.030\GeV^{-1}.
\end{equation}

For the $\eta(1405)$, $M_{\eta(1405)}=1408.8(2.0)\MeV$, we have 
\begin{align}
\frac{\Gamma(\ga \ga)\Gamma(K\bar K\pi)}{\Gamma_\mathrm{total}}&<35\eV\, \text{\cite{Ahohe:2005ug}},\qquad
\frac{\Gamma(\ga \ga)\Gamma(\eta\pi\pi)}{\Gamma_\mathrm{total}}<95\eV\, \text{\cite{Acciarri:2000ev}}, 
 \notag\\
 \frac{\Gamma(\ga \ga)}{\Gamma(K\bar K\pi)}&<1.78\times 10^{-3}\, \text{\cite{Ablikim:2018ajr}}.
\end{align}
Using the total width $\Gamma_{\eta(1405)}=48(4)\MeV$ as measured in the $K\bar K\pi$ channel, the two limits involving this channel imply $\Gamma(\ga \ga) < 1.73\keV$, 
while assuming that $K\bar K\pi$ and $\eta\pi\pi$ constitute the dominant decay channels would lead to a much stronger constraint 
$\Gamma(\ga \ga) < 130\eV$. The two limits on the two-photon coupling are
\begin{equation}
 F_{\eta(1405)\ga\ga}<0.122\GeV^{-1}, \qquad F_{\eta(1405)\ga\ga}<0.033\GeV^{-1},
\end{equation}
respectively. Similarly, for the $\eta(1475)$, $M_{\eta(1475)}=1475(4)\MeV$, the PDG lists 
\begin{equation}
\frac{\Gamma(\ga \ga)\Gamma(K\bar K\pi)}{\Gamma_\mathrm{total}}=230(71)\eV\, \text{\cite{Achard:2007hm}},\qquad
 \frac{\Gamma(\ga \ga)}{\Gamma(K\bar K\pi)}<1.27\times 10^{-3}\, \text{\cite{Ablikim:2018ajr}},
\end{equation}
which together with $\Gamma_{\eta(1475)}=90(9)\MeV$ implies the limit $\Gamma(\ga \ga) < 5.13\keV$ or, assuming  the $K\bar K\pi$ channel to be dominant, 
$\Gamma(\ga \ga) = 230(71)\eV$, leading to
\begin{equation}
 F_{\eta(1475)\ga\ga}<0.195\GeV^{-1}, \qquad F_{\eta(1475)\ga\ga}=0.041(6)\GeV^{-1}.
\end{equation}

Next, for the $\eta(1760)$, $M_{\eta(1760)}=1751(15)\MeV$, we have
\begin{equation}
\frac{\Gamma(\ga \ga)\Gamma(\eta'\pi^+\pi^-)}{\Gamma_\mathrm{total}}=28.2(8.7)\eV\, \text{\cite{Zhang:2012tj}},
\end{equation}
which, assuming dominance of $\eta'\pi\pi$ and including the neutral channel 
by means of the relation $\Gamma(\eta'\pi^0\pi^0)/\Gamma(\eta'\pi^+\pi^-)=1/2$, translates into
\begin{equation}
 F_{\eta(1760)\ga\ga}=0.014(2)\GeV^{-1}.
\end{equation}
In case other channels do contribute, this number would have to be considered a lower limit.

Finally, there is some information available on the two-photon couplings of the $X(1835)$, 
\begin{equation}
\frac{\Gamma(\ga \ga)\Gamma(\eta'\pi^+\pi^-)}{\Gamma_\mathrm{total}}<83\eV\, \text{\cite{Zhang:2012tj}},\qquad
 \frac{\Gamma(\ga \ga)}{\Gamma(\eta'\pi^+\pi^-)}<9.80\times 10^{-3}\, \text{\cite{Ablikim:2018ajr}},
\end{equation}
where the two-resonance fit from ref.~\cite{Zhang:2012tj} only quotes a significance of $2.8\,\sigma$.  
The combination of the two produces the limit $\Gamma(\ga \ga) < 14.0\keV$, again a lot weaker than the limit $\Gamma(\ga \ga) < 124.5\eV$ obtained when assuming dominance of the $\eta'\pi\pi$ channel. The resulting two-photon couplings are
\begin{equation}
 F_{X(1835)\ga\ga}<0.235\GeV^{-1}, \qquad F_{X(1835)\ga\ga}<0.022\GeV^{-1}.
\end{equation}

The above phenomenological constraints on the two-photon couplings are collected in table \ref{TabPhotonCouplings}, while the couplings from our large-$N_c$ Regge models with different Regge trajectory assignments are listed in table \ref{TabTrajectories}.

\section{Systematic uncertainties and decay constants of excited pseudoscalars}
\label{sec:systematics}

The systematic errors quoted for $\Delta a_\mu$ in \secref{match-SDC} are based on comparing results from two different models for the pseudoscalar TFFs  introduced in section \ref{sec:Regge} and appendix \ref{sec:PionAlternative}, respectively. This has then been added to a conservatively estimated uncertainty coming directly from the parameters of our models. By construction, our models link the TFFs of the different pion, $\eta$, or $\eta'$ states such that a resummation is at all possible, but it is clear that the details of the TFFs so obtained may turn out not to be realistic, at least for the lowest lying excited pseudoscalars. Here, we explore this specific question in particular for the first excited pseudoscalars, on which some information from the phenomenology is indeed available. As we will show, if we adapt the parameters of our models presented in sections \ref{sec:Model} and \ref{sec:ModelEta} to be in agreement with phenomenology, or theoretical expectations, we obtain shifts in our results which are well covered by the present error budget.

In \secref{two_photon_couplings} and
appendix \ref{app:photon_couplings}, we confirmed that our TFF models are well compatible with phenomenological constraints for the two-photon couplings of $\eta(1295)$, $\eta(1405)$, $\eta(1475)$, $\eta(1760)$, and $X(1835)$. In the following, we will be interested in the leptonic decay constants, $F_{P(n)}$, of the excited pseudoscalars, limiting our analysis to $n \leq 3$ states. Since all pseudoscalar mesons, except for the Goldstone mode, decouple from the axial-vector current in the chiral limit, the corresponding decay constants, defined in \eqref{decayconstantsDef}, are suppressed. Note that contrary to the decay constants, the two-photon couplings are non-vanishing in the chiral limit. At low $Q^2$ it is the latter that are most relevant.

There are several theoretical studies of the leptonic decay constants of excited pions, e.g., based on lattice QCD \cite{McNeile:2006qy, Mastropas:2014fsa}, QCD sum rules \cite{ Elias:1997ya}, quark models \cite{Volkov:1996br,Andrianov:1998kj}, or finite-energy sum rules \cite{Maltman:2001gc,Kataev:1982xu}. Experimentally, 
$F_{\pi(1300)}$ can be measured in $\tau$ decays. Presently, there is only an upper bound $\vert F_{\pi(1300)}\vert <8.4$ MeV \cite{Diehl:2001xe}, deduced from the branching fraction $\mathrm{Br}(\tau \rightarrow \pi(1300)\nu_\tau)< 10^{-4}$ \cite{Groom:2000in}. The predictions in refs.~\cite{ Elias:1997ya,Volkov:1996br,Andrianov:1998kj,Maltman:2001gc,Kataev:1982xu} are all in agreement with this bound. In the following, we will work with $F_{\pi(1300)}=2.20(46)\MeV$ \cite{Maltman:2001gc}, which implies $F_{\pi(1300)}/F_\pi \approx 2\%$. For the $\eta^{(\prime)}$, the suppression is expected to be weaker as it is given by the $SU(3)$ chiral limit, thus, we assume $F_{\eta^{(\prime)}(1)}/F_{\eta^{(\prime)}} \approx 20\%$. Furthermore, it is expected that the decay constants of the excited states, $F_{P(n)}$, are inversely proportional to the excited-state masses, $M_{P(n)}$, or the masses squared \cite{ Elias:1997ya,Kataev:1982xu}, which generates an additionally suppression for the decay constants of the higher excitations.

We start by considering the symmetric pQCD limit of the TFFs, which for the ground-state pseudoscalars is given in~\eqref{Constraint2} and \eqref{Constraint2Eta}. The same relations also hold for the excited pseudoscalars, replacing only the  ground-state decay constants, i.e., $F_P \rightarrow F_{P(n)}$. To change the asymptotic limit of $F_{P(n)\ga^*\ga^*}(-Q^2,-Q^2)$ for $n=1,2,3$, we modify $c_\mathrm{diag}$ by replacing \eqref{cdiagpion} with
\beq
\label{changecdiagpion}
c_\mathrm{diag}\rightarrow c_\mathrm{diag}+\frac{16\Lambda^2\pi^2 F_\pi}{3M_\rho^2M_\omega^2}\left[F_{\pi(1300)}-F_\pi\right],
\eeq
and \eqref{cdiageta}
with
\begin{align}
c_\mathrm{diag}&\rightarrow c_\mathrm{diag}+\frac{ \mathcal{N}}{C^\eta_{\rho
  \rho}+C^\eta_{\omega \omega}+ C^\eta_{\phi \phi}}\frac{8C_8 }{\Lambda^2F_{\eta\ga \ga}} \left[F_{\eta(1)}-F_\eta\right]\label{changecdiageta},
\end{align}
and analogously for $\eta'$, while keeping all other model parameters the same, cf.\ table~\ref{TableModel1}. Varying only the TFFs of the lowest excitations $n\leq 3$, it is ensured that the SDCs remain intact. Applying this modification for $n=1$ decreases $\Delta a_\mu$ by $(0.22+0.27+0.45)\times 10^{-11}=0.94\times 10^{-11}$, where the individual numbers are the pion, $\eta$, and $\eta'$ contributions, respectively. Note that we are removing the contribution from $Q_i \geq Q_\mathrm{match}$, as this region is described by the pQCD quark loop, cf.\ \secref{match-SDC}.

The BL limit of the pseudoscalar TFFs is not known for general $n$. Presently, this limit is therefore not fine-tuned in our large-$N_c$ Regge models, as can be seen in the right panel of figure \ref{fig:Coupling}. Here, we want to assume a BL limit constant in $n$, $\lim_{Q^2\rightarrow \infty} F_{P(n)\ga\ga^*}= \lim_{Q^2\rightarrow \infty} F_{P\ga\ga^*}$, which is also found for the alternative TFF model introduced in appendix \ref{sec:PionAlternative}. Replacing \eqref{cBLpion} with
\beq
c_\mathrm{BL}= \frac{1}{M_{-,\,n}^2}\left[c_\mathrm{anom}M_{+,\,n}^2+M_{\rho(n)}^2M_{\omega(n)}^2\left(\frac{c_A}{\Lambda^2}-\frac{8\pi^2F_\pi^2}{M_\rho^2 M_\omega^2}\right)\right],\label{cBLpionnew}
\eeq
and \eqref{cBLeta} with
\begin{align}
c_\mathrm{BL}&=\left(\frac{1}{M_{\phi(n)}^2}-\frac{1}{M_{\omega(n)}^2}\right)^{-1}\left\{\frac{1}{C_{\phi\omega}^\eta M_\phi^2 M_\omega^2}\left[\frac{12C_8F_{\eta}\,\mathcal{N}}{F_{\eta\ga\ga}}-\frac{C_{\rho\rho}^\eta M_\rho^4}{M_{\rho(n)}^2}-\frac{C_{\omega\omega}^\eta M_\omega^4}{M_{\omega(n)}^2}-\frac{C_{\phi\phi}^\eta M_\phi^4}{M_{\phi(n)}^2}\right]\right.\nn\\
&\left.- \frac{2c_A}{\Lambda^2}-\left(\frac{1}{M_{\phi(n)}^2}+\frac{1}{M_{\omega(n)}^2}\right)\left(1+\frac{c_B}{2\Lambda^2}(M_\omega^2-M_\phi^2)-\frac{c_A}{2\Lambda^2}(M_\phi^2+M_\omega^2)\right)\right\},\label{cBLetanew}
\end{align}
while keeping all other model parameters the same, this is achieved without changing the two-photon couplings or other SDCs. Looking at the $n=1$ case, $\Delta a_\mu$ decreases by $(0.49+1.18+1.92)\times 10^{-11}=3.59\times 10^{-11}$. This shift predominantly comes from a change of the low-$Q^2$ shape of the TFFs and not from the different asymptotic limits.

Combining the changes in \eqref{changecdiagpion} and \eqref{cBLpionnew}, as well as \eqref{changecdiageta} and \eqref{cBLetanew}, the total decrease of $\Delta a_\mu$ amounts to $(0.62+1.27+2.10)\times 10^{-11}=3.99\times 10^{-11}$ if only $n=1$ is modified, $5.26\times 10^{-11}$ if also $n=2$ is modified, and $5.82\times 10^{-11}$ if $n=1,2,3$ are modified.

As a last point, we study the effect of non-diagonal couplings. To be more precise, in our large-$N_c$ Regge models from section \ref{sec:Regge}, we allowed the $n$-th pion, $\eta$, or $\eta'$ excitation to couple only to the $n$-th $\rho$, $\omega$, and $\phi$ excitations, whereas now we allow the first excited pseudoscalars to couple to the ground-state vector mesons. For the $\pi(1300)$, we modify 
\begin{align}
F_{\pi(1)\ga^*\ga^*}(-Q_1^2,-Q_2^2)&\rightarrow F_{\pi(1)\ga^*\ga^*}(-Q_1^2,-Q_2^2)-\frac{F_{\pi(1)\ga\ga}}{2} \left[M_\rho^2 M_\omega^2\left(\frac{1}{D_\rho^1 D_\omega^2}+\frac{1}{ D_\omega^1 D_\rho^2}\right)\right.\nn\\
&\left.-M_{\rho(1)}^2 M_{\omega(1)}^2\left(\frac{1}{D_{\rho(1)}^1 D_{\omega(1)}^2}+\frac{1}{ D_{\omega(1)}^1 D_{\rho(1)}^2}\right)\right],
\end{align}
and generate the two-photon coupling through $\rho(770)$ and $\omega(782)$. For the first excited $\eta^{(\prime)}$, we only express the dominant isovector--isovector part of the two-photon coupling through $\rho(770)$:
\begin{align}
F_{\eta^{(\prime)}(1)\ga^*\ga^*}(-Q_1^2,-Q_2^2)&\rightarrow F_{\eta^{(\prime)}(1)\ga^*\ga^*}(-Q_1^2,-Q_2^2)\nn\\
&-\frac{F_{\eta^{(\prime)}(1)\ga\ga}C^{\eta^{(\prime)}}_{\rho
  \rho}}{\mathcal{N}} \left[\frac{M_\rho^4}{D_\rho^1 D_\rho^2}-\frac{M_{\rho(1)}^4}{D_{\rho(1)}^1 D_{\rho(1)}^2}\right].
\end{align}
These modifications lead to a 
decrease of $(0.53+0.54+0.72)\times 10^{-11}=1.79\times 10^{-11}$.
Such non-diagonal couplings are also present in the alternative TFF model introduced in appendix \ref{sec:PionAlternative}.

All these modifications affect in one way or another the low-$Q^2$ behavior of the excited-state TFFs, which could be constrained more rigorously if data were available. At present, we observe that the corresponding changes, which tend to lower $\Delta a_\mu$, are well covered by our final uncertainty estimate in~\eqref{result_PS}. Since, on the other hand, the consideration of Model 2 suggests systematic effects in the opposite direction, we leave the central value as derived from Model 1, with uncertainties as assigned in~\eqref{result_PS}.

\section{Pion transition form factors $\boldsymbol{F_{\pi(n)\ga^*\ga^*}}$}
\label{app:TFF_pion}

\subsection[Large-$N_c$ Regge model]{Large-$\boldsymbol{N_c}$ Regge model} 
\seclab{PionModel1CouplingParameters}

\begin{figure}[t]
\includegraphics[width=0.48\linewidth]{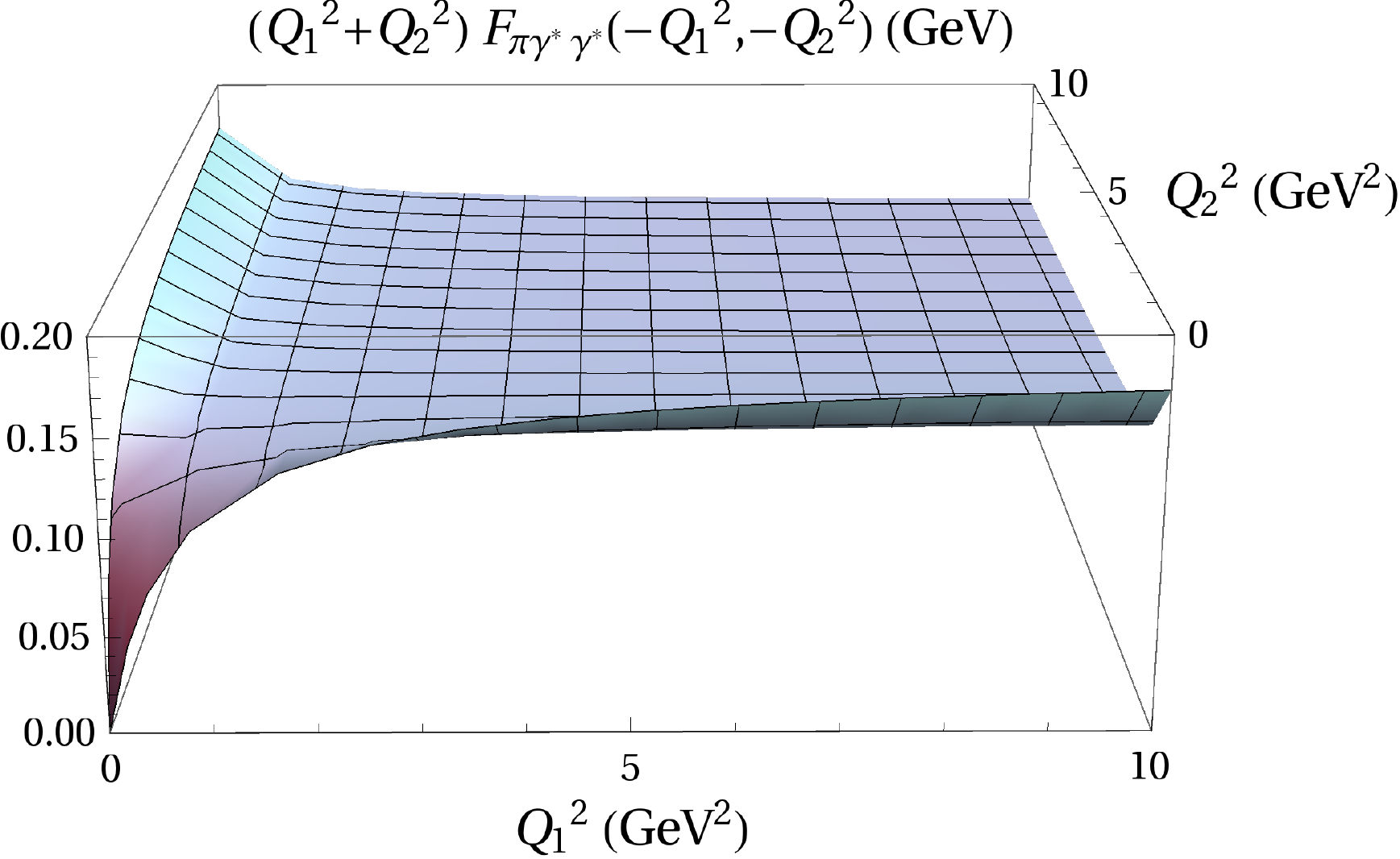}\hfill\includegraphics[width=0.48\linewidth]{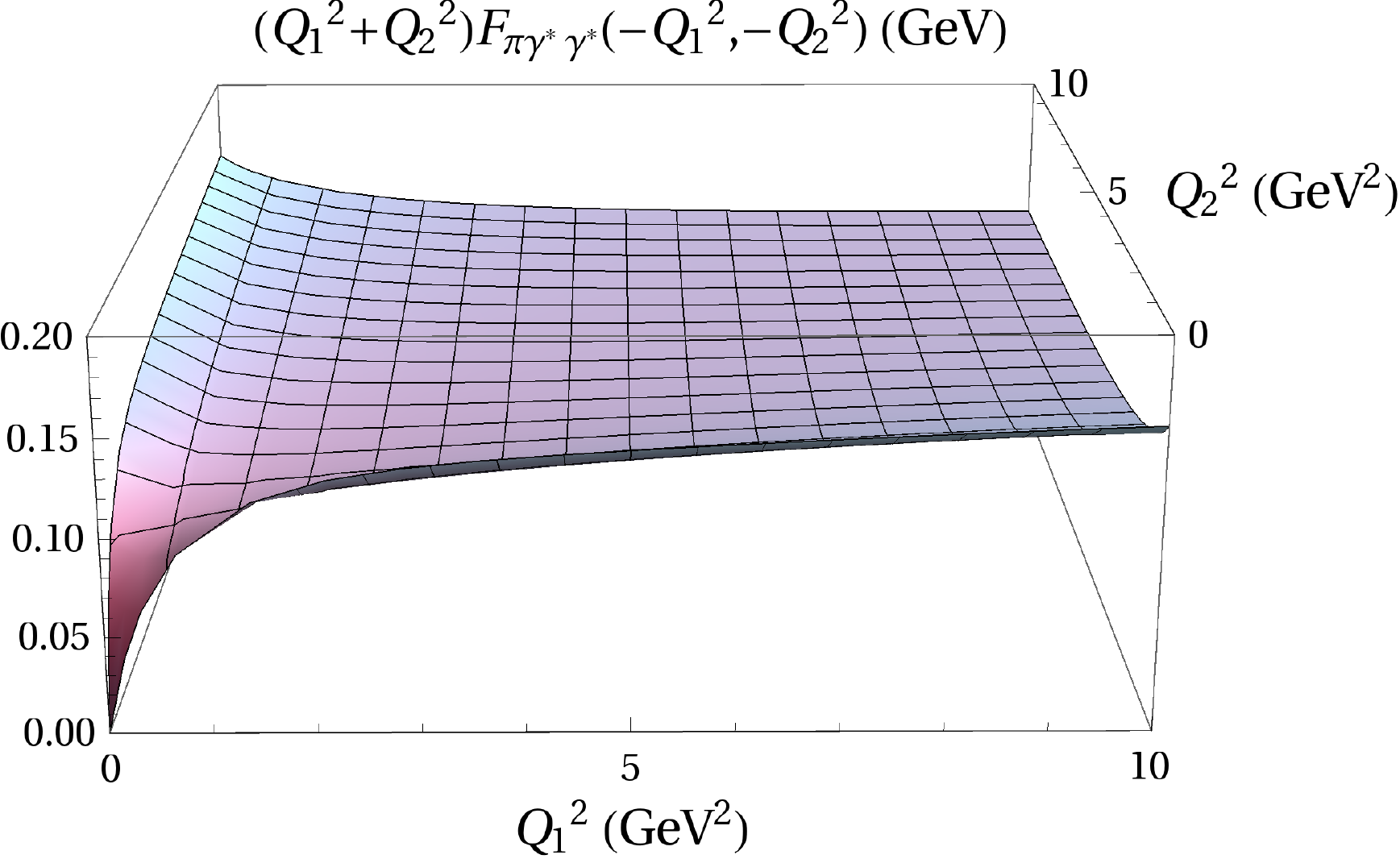}
\caption{$\pi^0$ TFF in the full space-like region for $Q_1^2,Q_2^2<10$ GeV$^2$. The large-$N_c$ Regge model, \eqref{TFFpi}, is shown in the left panel, and our alternative model, \eqref{MartinModel}, is shown in the right panel.}
\figlab{3DPlot}
\end{figure}

In this appendix, we describe the large-$N_c$ Regge model for the pion TFF, given in~\eqref{TFFpi}, in more details. A comparison to experimental data and other parameterizations available from the literature is postponed to \secref{plotsPion}. Based on the constraint equations in \eqref{ConstraintEquations}, \eqref{MV}, and \eqref{pQCD3}, the model parameters should be replaced by:
\begin{align}
c_\mathrm{diag}&=\frac{16 \pi^2 F_\pi^2
  \Lambda^2}{3 M_\omega^2 M_\rho^2},\label{cdiagpion}\\
c_\mathrm{anom}&=1-\frac{1}{\Lambda^2}\left(
        c_A M_{+,\,0}^2+c_B M_{-,\,0}^2\right),\\
c_\mathrm{BL}&=-\frac{8 \pi^2 F_\pi^2}{M_{-,\,0}^2}+\frac{M_{+,\,0}^2}{M_{-,\,0}^2}-\frac{1}{\Lambda^2}\left(
        c_A M_{-,\,0}^2+c_B M_{+,\,0}^2\right),\label{cBLpion}\qquad
\end{align}
where the parameters related to the asymptotic limits of the HLbL tensor simplify to:
\beq
c_{\bfrac{A}{B}}=\frac{\Lambda^2}{M_\omega^2 M_\rho^2}\Bigg[\frac{\Delta_{\pi\rho}}{L_{\pi\rho}}-4\pi^2 F_\pi\left(\frac{\Delta_{\pi\rho}L_{\pi\omega}}{\Delta_{\pi\omega}L_{\pi\rho}}\mp 1\right)
\bigg(\frac{b}{2a}+ \sqrt{\bigg(\frac{b}{2a}\bigg)^2-\frac{1}{a}\bigg(c-\frac{1}{9\pi^2}\bigg)}\Bigg)\Bigg],
\eeq
with
\begin{align}
a&=\frac{\Delta_{\pi\rho}}{\Omega_{\rho\omega\pi}^2}\left[f_2(\sigma_\pi,\sigma_\rho,\sigma_\omega)-\frac{L_{\pi\omega}}{L_{\pi \rho}}f_2(\sigma_\pi,\sigma_\omega,\sigma_\rho) \right]\left(\sigma_\rho^2-\sigma_\omega^2\,\frac{ \Delta_{\pi\rho}L_{\pi \omega}}{\Delta_{\pi\omega}L_{\pi \rho}}\right),\nn\\
b&=\frac{\Delta_{\pi \rho}}{\Omega_{\rho\omega\pi}^2}\bigg\{-\frac{2F_\pi }{3}\left[ f_1(\sigma_\pi,\sigma_\omega,\sigma_\rho)-\frac{ L_{\pi \omega}}{L_{\pi\rho}}f_1(\sigma_\pi,\sigma_\rho,\sigma_\omega)\right]\nn\\
&-\frac{1}{4\pi^2F_\pi}\frac{\Delta_{\pi\omega}}{L_{\pi\rho}}\left[\sigma_\rho^2\,f_2(\sigma_\pi,\sigma_\omega,\sigma_\rho)+ \frac{\Delta_{\pi \rho}}{\Delta_{\pi\omega}}\sigma_\omega^2\left(f_2(\sigma_\pi,\sigma_\rho,\sigma_\omega)-\frac{2L_{\pi\omega}}{L_{\pi\rho}}f_2(\sigma_\pi,\sigma_\omega,\sigma_\rho)\right)\right]\bigg\},\nn\\
c&=\frac{\Delta_{\pi \rho}}{\Omega_{\rho\omega\pi}^2}\frac{\Delta_{\pi \omega}}{L_{\pi \rho}}\left[\frac{1}{(4\pi^2 F_\pi)^2}\frac{\Delta_{\pi\rho}\sigma_\omega^2}{L_{\pi \rho}}f_2(\sigma_\pi,\sigma_\omega,\sigma_\rho)+\frac{f_1(\sigma_\pi,\sigma_\rho,\sigma_\omega)}{6\pi^2}\right],
\end{align}
and the
auxiliary functions
\begin{align}
\label{auxFunc}
g_1(\si_P,\si_{V_1},\si_{V_2})&\coloneqq\sigma_P^2\left(\si^4_{V_1}-\si^4_{V_2}\right)-\Delta_{V_1V_2}\left(\sigma_P^4+\si^2_{V_1}\si^2_{V_2}\right)+\si_{V_2}^4\si_{V_1}^2 L_{V_1V_2},\\
g_2(\si_P,\si_{V_1},\si_{V_2})&\coloneqq\sigma_P^2 \Delta_{V_1V_2}^2 L_{PV_1},\nn\\
f_1(\si_P,\si_{V_1},\si_{V_2})&\coloneqq\sigma_{V_1}^2\,g_1(\si_P,\si_{V_1},\si_{V_2})+\sigma_{P}^2\,g_2(\si_P,\si_{V_1},\si_{V_2})+\left(\sigma_{P}^2 -2 \sigma_{V_1}^2\right) \sigma_{V_2}^4 \sigma_{P}^2\, L_{V_1V_2}\nn,\\
f_2(\si_P,\si_{V_1},\si_{V_2})&\coloneqq g_1(\si_P,\si_{V_1},\si_{V_2})-g_2(\si_P,\si_{V_1},\si_{V_2})
+\left(\sigma_{P}^2\sigma_{V_1}^2-\sigma_{V_1}^4-\sigma_{V_2}^4\right)\sigma_{P}^2\, L_{V_1V_2}. \nn
\end{align}
Here, the definitions from~\eqref{def2} and \eqref{def3} are used.

\subsection[{Comparison of  data and literature: $F_{\pi(n)\ga^*\ga^*}$}]{Comparison of  data and literature: $\boldsymbol{F_{\pi(n)\ga^*\ga^*}}$}
\seclab{plotsPion}

\begin{figure}[t]
\centering
\includegraphics[width=0.85\linewidth]{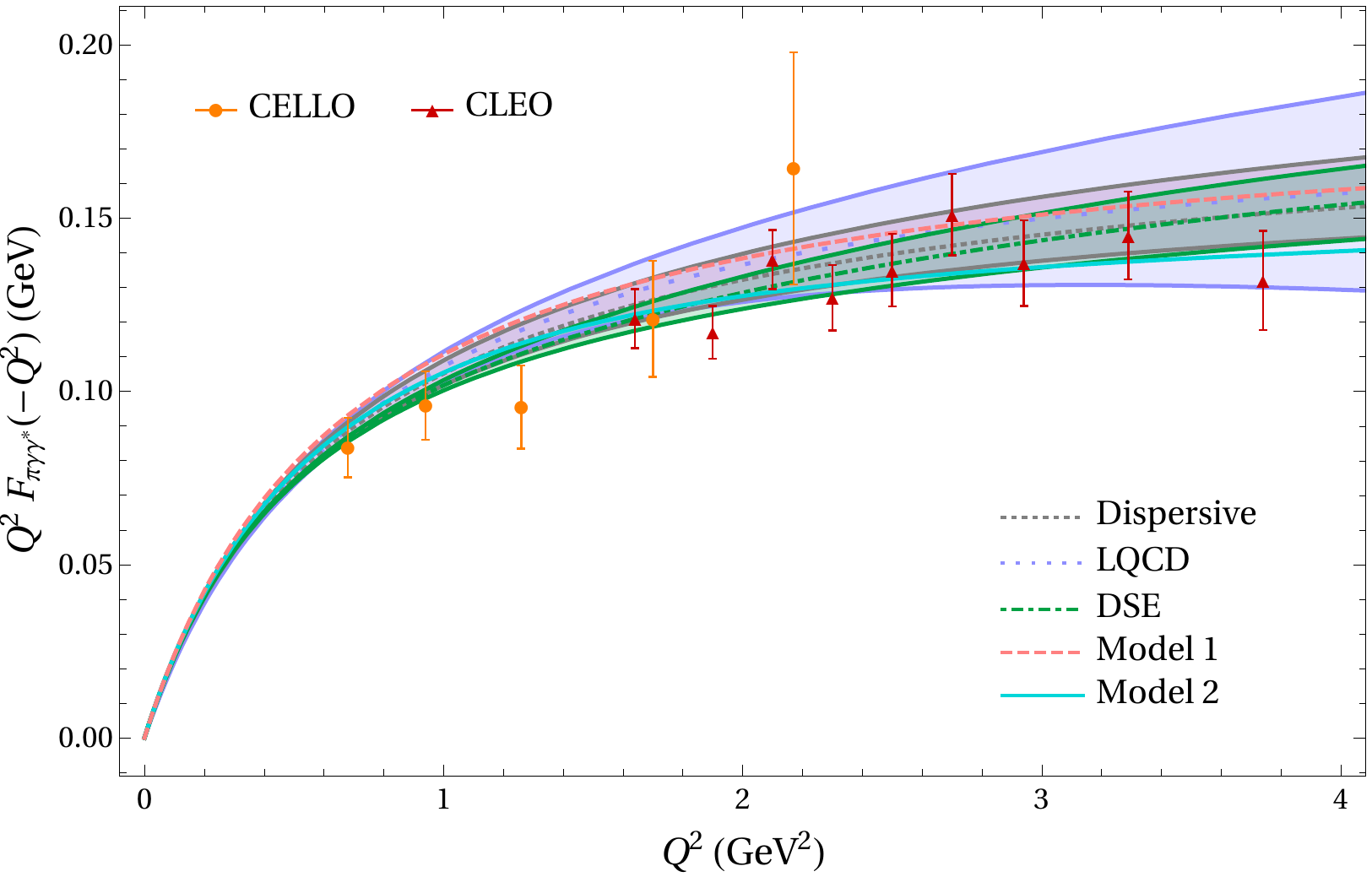}
\caption{Singly-virtual $\pi^0$ TFF in the low-$Q$ region. The large-$N_c$ Regge model, ``Model 1''~\eqref{TFFpi}, is indicated by the dashed pink curve. Our alternative TFF model, ``Model 2''~\eqref{MartinModel}, is indicated by the solid cyan curve. The gray band with the dotted curve is the dispersive result from refs.~\cite{Hoferichter:2018dmo,Hoferichter:2018kwz}. The blue band with the long-dotted curve is the lattice QCD result from ref.~\cite{Gerardin:2019vio}. The green band with the dot-dashed curve is the DSE result from ref.~\cite{Eichmann:2019tjk}. The data are from CELLO~\cite{Behrend:1990sr} and CLEO~\cite{Gronberg:1997fj}.}
\figlab{DataZoom}
\end{figure}

In this appendix, we compare
our large-$N_c$ Regge model, ``Model 1''~\eqref{TFFpi}, and our alternative model, ``Model 2''~\eqref{MartinModel}, for $F_{\pi(n)\ga^*\ga^*}$ to data and other parameterizations available from the literature.

In figure \ref{fig:SinglyVirtual}, the singly-virtual $\pi^0$ TFF is shown for $Q^2 \in [0,35]\,\text{GeV}^2$. In figure \ref{fig:DataZoom}, we focus on the low-$Q$ region and include a comparison to the recent lattice QCD \cite{Gerardin:2019vio} and DSE~\cite{Eichmann:2019tjk} results. Our $\pi^0$ TFF models, for which we do not display error estimates, are in good agreement with the dispersive and lattice QCD TFFs, while we observe some deviation of our Model 1 from the DSE prediction. However, the error quoted for the DSE result in ref.~\cite{Eichmann:2019tjk}, as pointed out therein, is only a rough estimate based on the variation of their one model parameter and does not account for the total truncation error. Therefore, we conclude that our $\pi^0$ TFF models also agree with the DSE prediction in the singly-virtual region. 

In figure \ref{fig:OPELattice}, we show the doubly-virtual $\pi^0$ TFF in the low-$Q$ region. Both lattice QCD and DSE are able to give much more accurate predictions of the (pseudoscalar) TFFs for doubly-virtual than for singly-virtual kinematics, as is obvious by comparing the error bands in figures \ref{fig:DataZoom} and \ref{fig:OPELattice}. In the symmetric region, $Q_1^2=Q_2^2=Q^2$, starting from $\sim 1$ GeV$^2$, the DSE predict a slightly larger $\pi^0$ TFF than lattice QCD, see left panel in figure \ref{fig:OPELattice}. Our models for the $\pi^0$ TFF run just between these DSE and lattice QCD predictions. Note, however, that the discrepancy in figure \ref{fig:OPELattice} is visually enhanced by showing $Q^2 F_{\pi^0\ga^*\ga^*}(-Q^2,-Q^2)$ instead of $F_{\pi^0\ga^*\ga^*}(-Q^2,-Q^2)$. For doubly-virtual kinematics away from the symmetric limit, see for instance $Q_1^2=Q^2$ and $Q_2^2=2Q^2$ in the right panel of figure \ref{fig:OPELattice}, our $\pi^0$ TFF models are in closer agreement with the lattice QCD prediction.

\begin{figure}[t]
\centering
\includegraphics[width=0.48\linewidth]{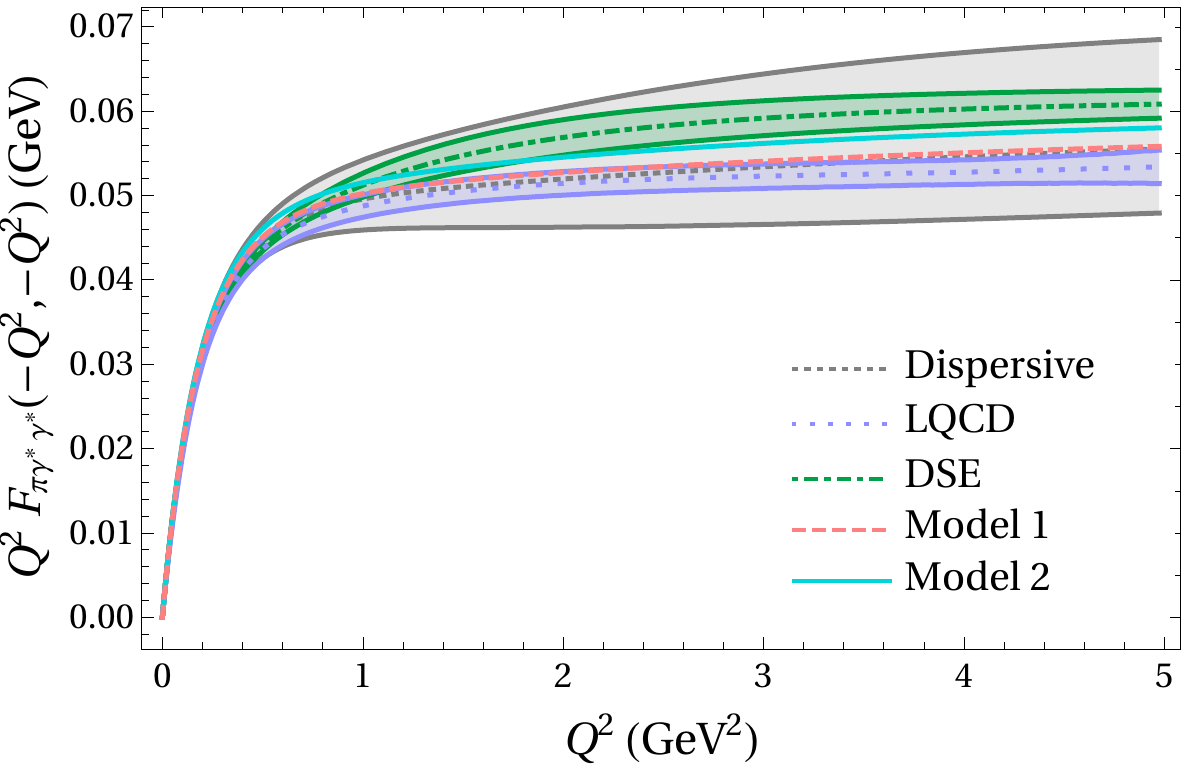}\hfill\includegraphics[width=0.48\linewidth]{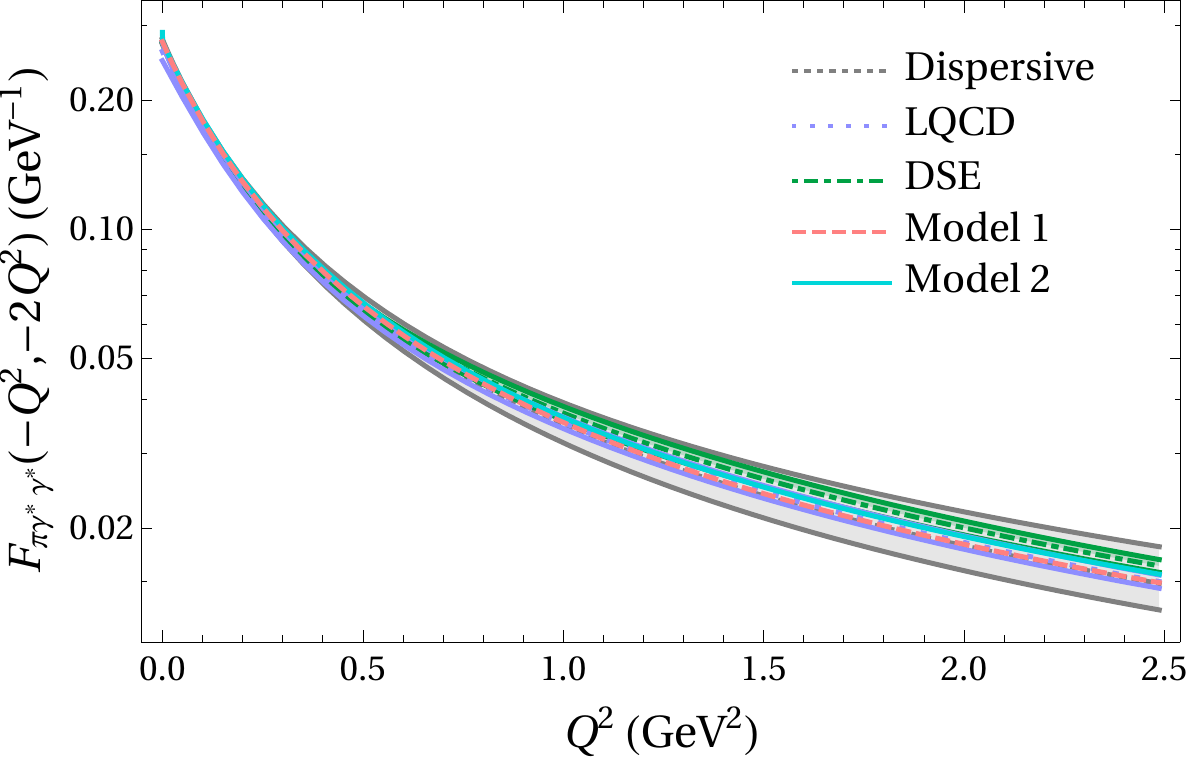}
\caption{Doubly-virtual $\pi^0$ TFF in the symmetric region $Q_1^2=Q_2^2=Q^2$ (left) and in the region where $Q_1^2=Q^2$ and $Q_2^2=2Q^2$ (right). Legend is the same as in figure \ref{fig:DataZoom}.}
\figlab{OPELattice}
\end{figure}

\begin{figure}[t]
\centering
\includegraphics[width=0.475\linewidth]{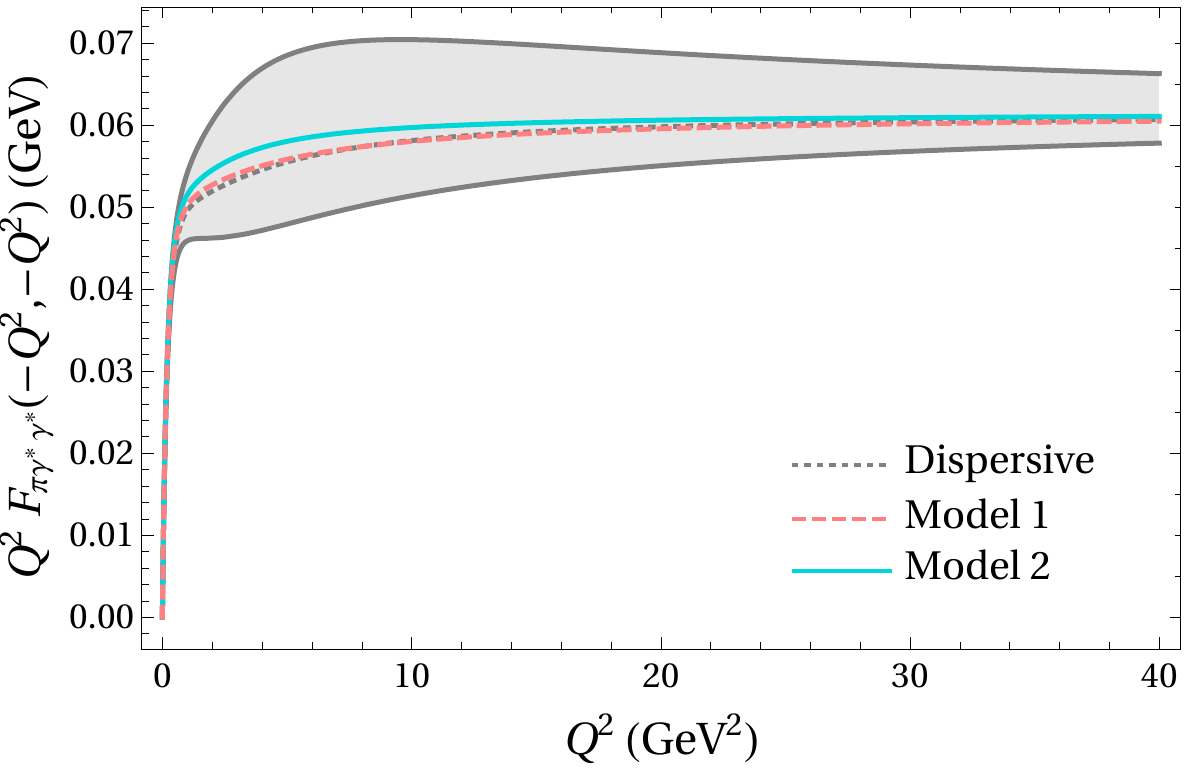}\hfill\includegraphics[width=0.485\linewidth]{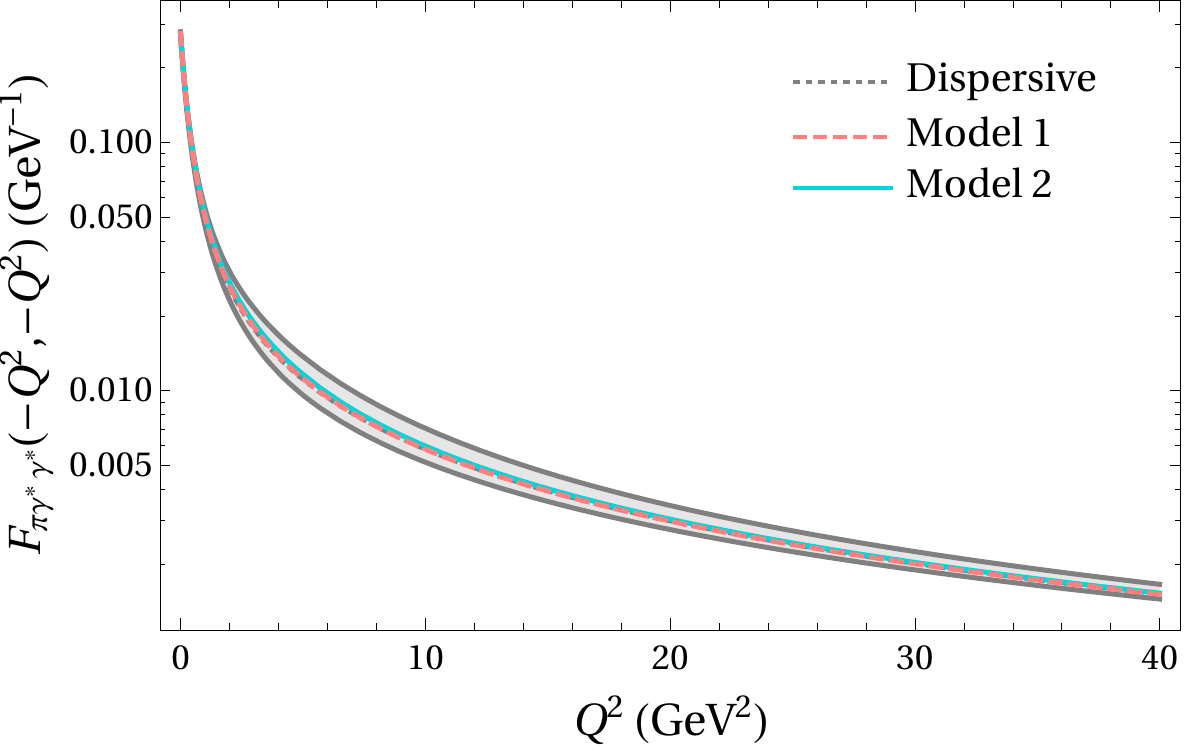}
\caption{Doubly-virtual $\pi^0$ TFF in the symmetric region $Q_1^2=Q_2^2=Q^2$ for $Q^2 \in [0,40]\,\text{GeV}^2$. Legend is the same as in figure \ref{fig:DataZoom}.}
\figlab{OPEFitHoferichter}
\end{figure}

In figure~\ref{fig:OPEFitHoferichter}, we show the doubly-virtual $\pi^0$ TFF in the symmetric region for $Q^2 \in [0,40]\,\text{GeV}^2$. Both models, but in particular Model 1, are in perfect agreement with the dispersive description~\cite{Hoferichter:2018dmo,Hoferichter:2018kwz}. In figure~\ref{fig:3DPlot}, Model 1 and 2 are shown in the full space-like region for $Q_1^2,Q_2^2<10$ GeV$^2$. One can see that their main difference is in the regions where at least one of the photon virtualities is small.

\section{$\boldsymbol{\eta}$ and $\boldsymbol{\eta'}$ transition form factors $\boldsymbol{F_{\eta^{(\prime)}(n)\ga^*\ga^*}}$}
\label{app:TFF_eta}

\subsection[Large-$N_c$ Regge model]{Large-$\boldsymbol{N_c}$ Regge model} \seclab{EtaModel1CouplingParameters}

In this appendix, we describe the large-$N_c$ Regge model for the $\eta$ and $\eta'$ TFFs, introduced in section~\ref{sec:ModelEta}, in more details. A comparison to experimental data and other parameterizations available from the literature is postponed to appendices~\ref{sec:EtaPlots} and~\ref{sec:EtaPrimePlots}.

\begin{figure}[t]
\centering
\includegraphics[width=0.48\linewidth]{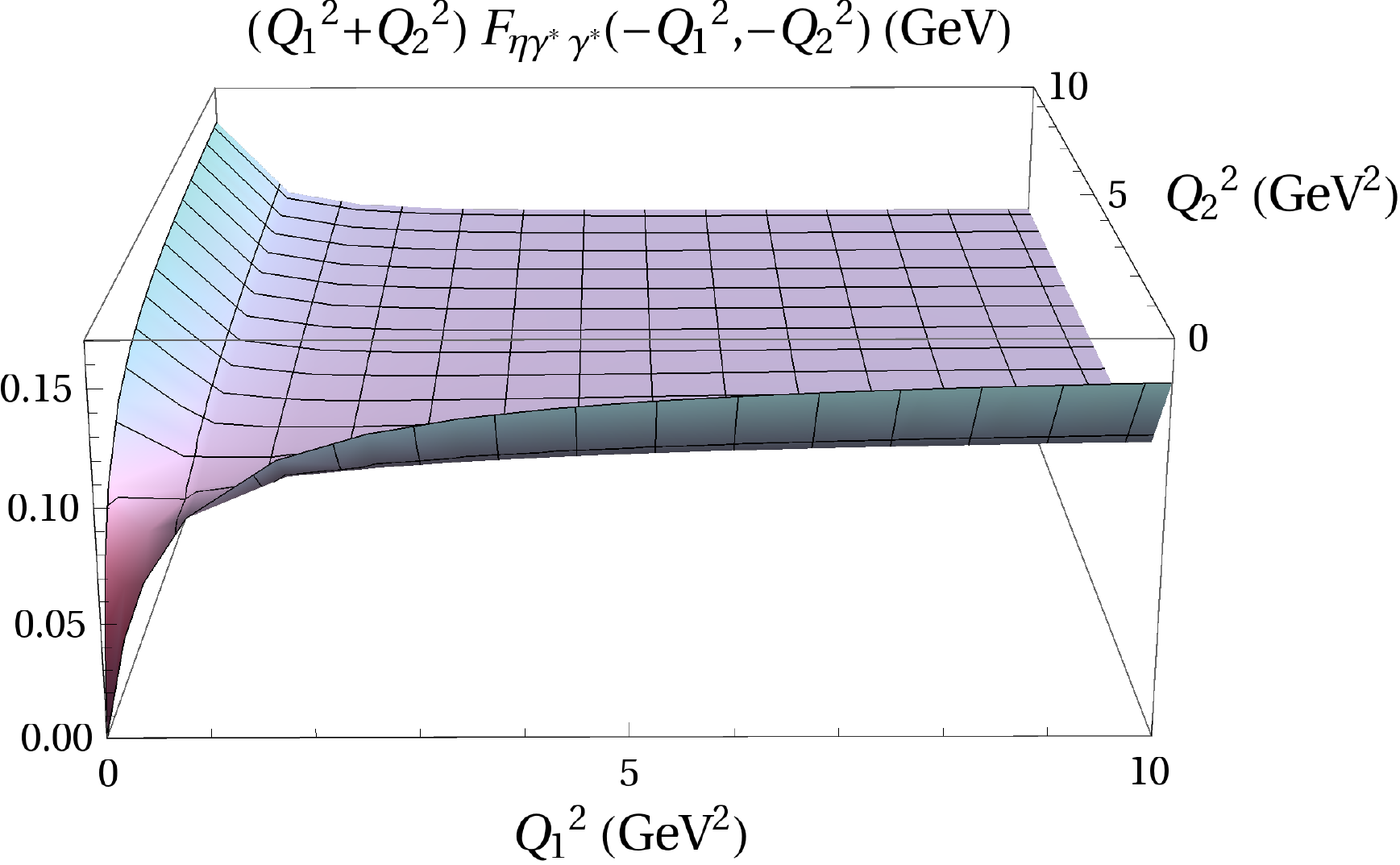}\hfill\includegraphics[width=0.48\linewidth]{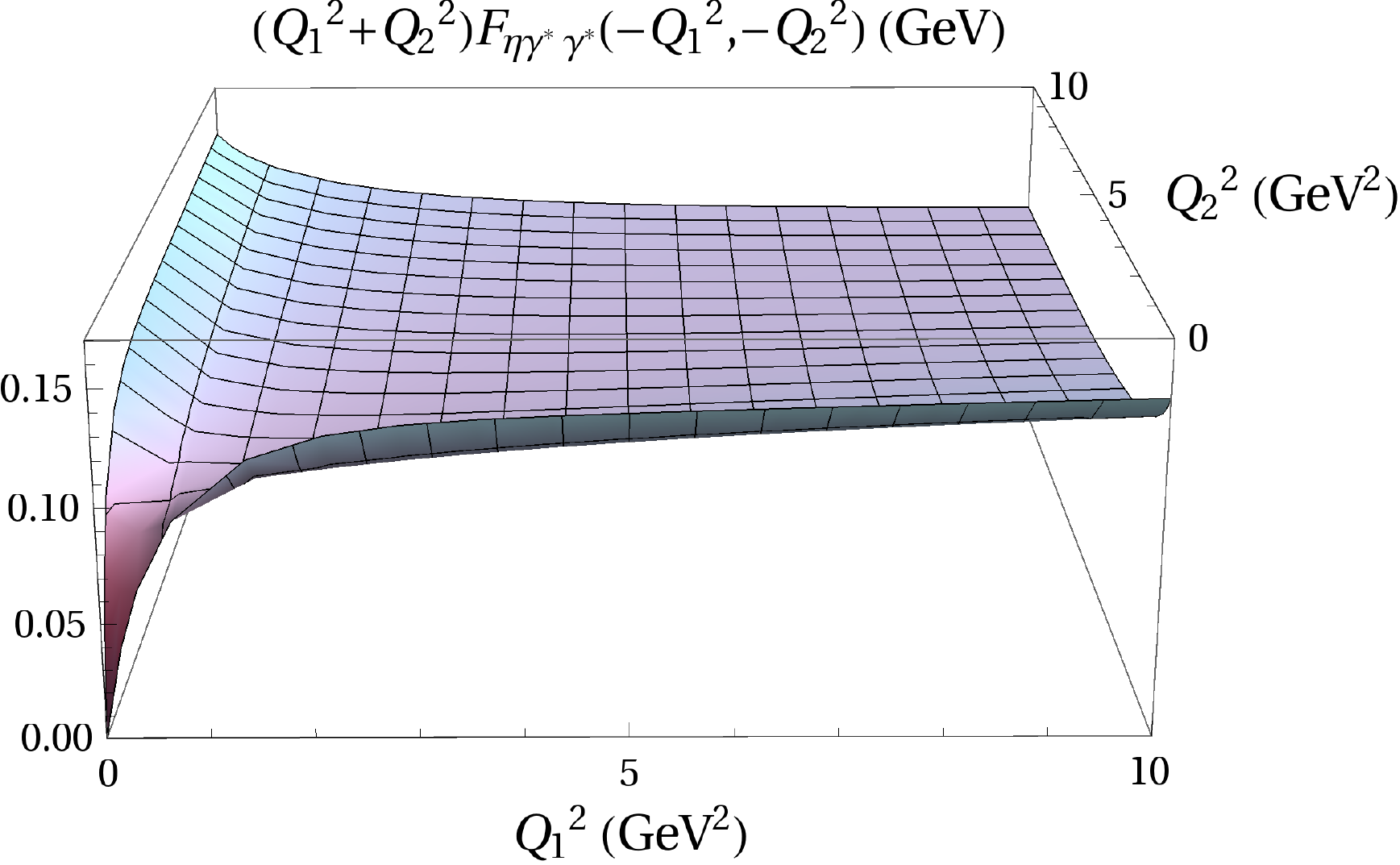}
\caption{$\eta$ TFF in the full space-like region for $Q_1^2,Q_2^2<10$ GeV$^2$. The large-$N_c$ Regge model, \eqref{eq:TFFetaandetap}, is shown in the left panel, and our alternative model, \eqref{MartinModel}, is shown in the right panel.}
\figlab{Eta3DPlot}
\end{figure}

\begin{figure}[t]
\includegraphics[width=0.48\linewidth]{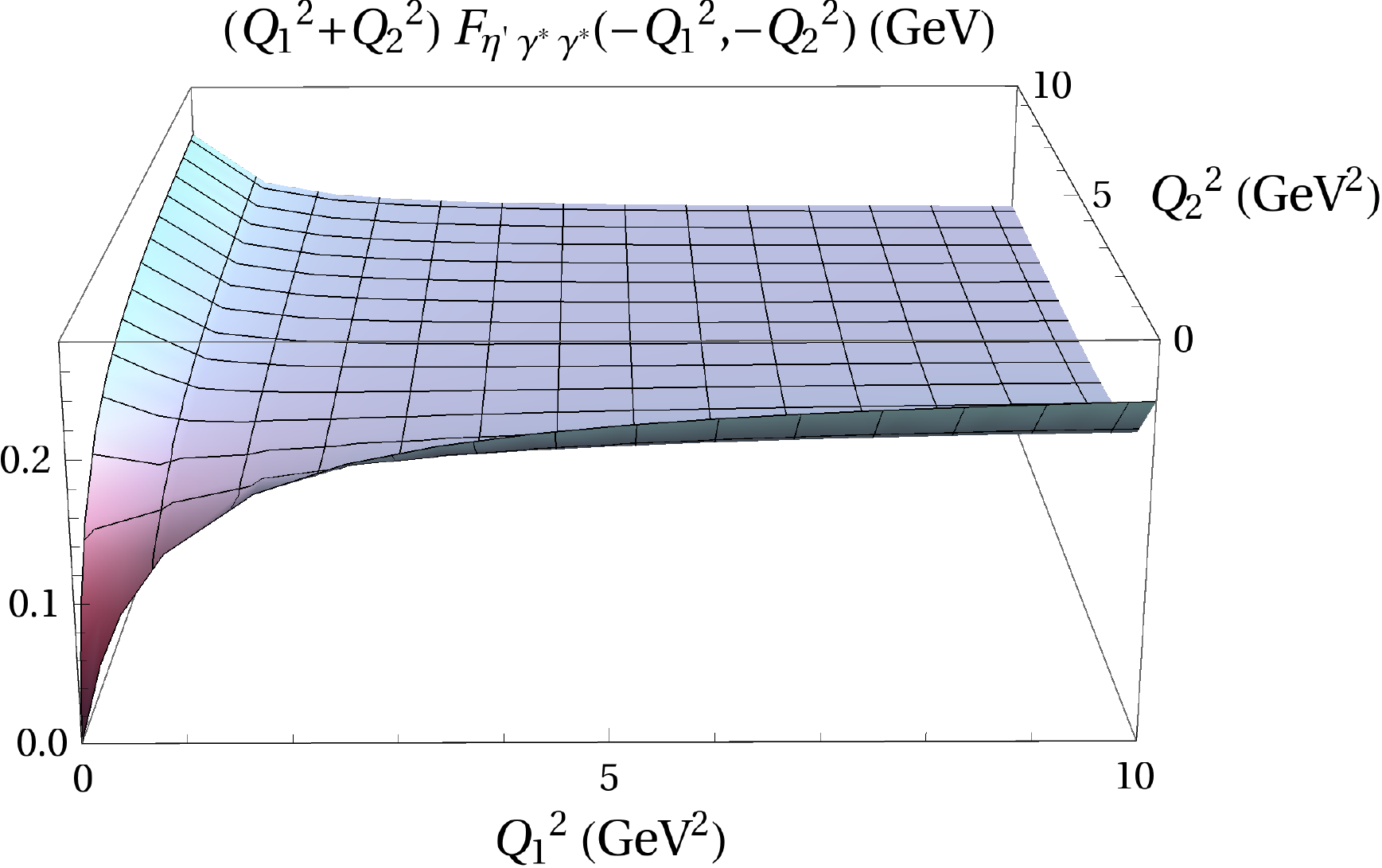}\hfill\includegraphics[width=0.48\linewidth]{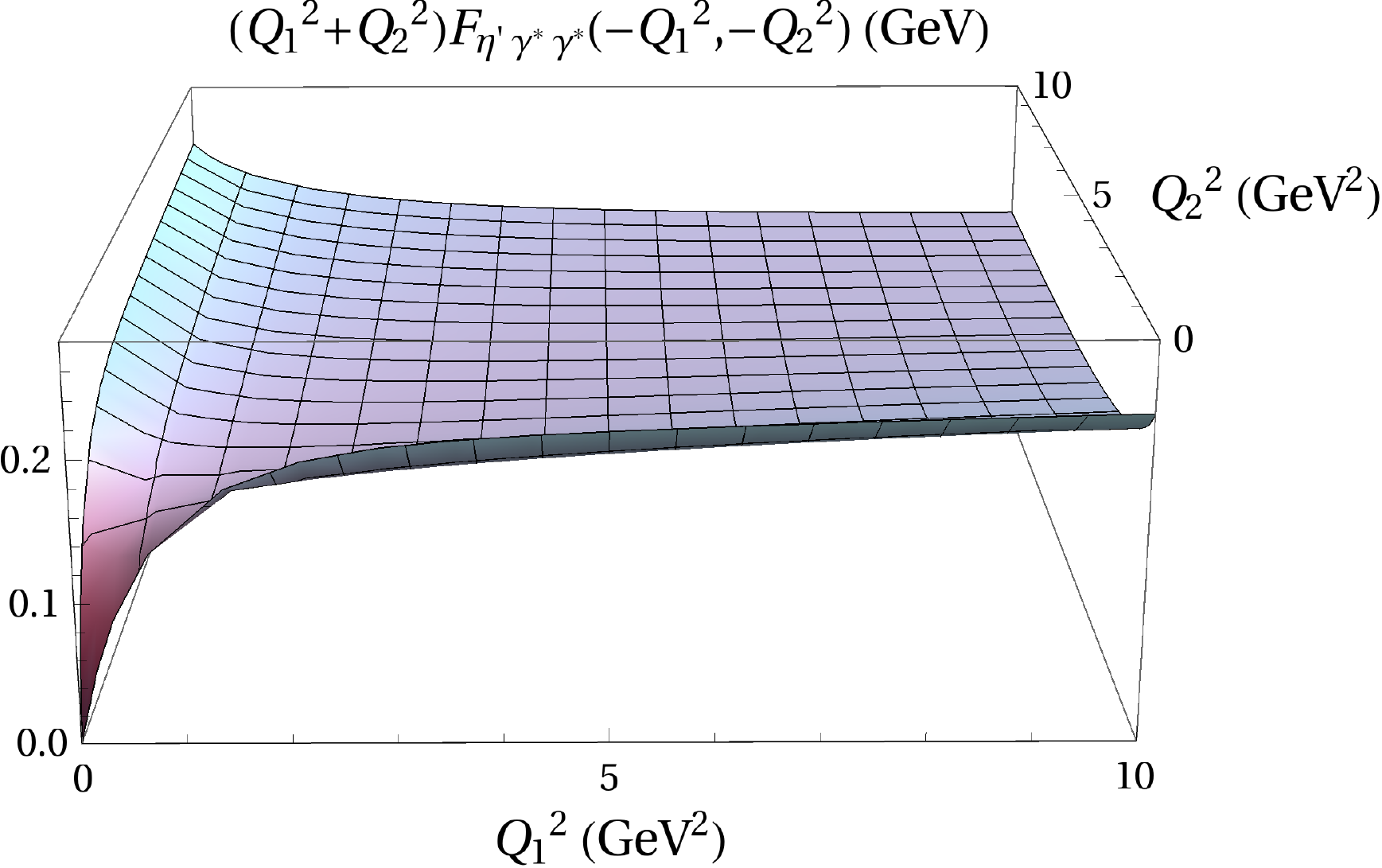}
\caption{$\eta^\prime$ TFF in the full space-like region for $Q_1^2,Q_2^2<10$ GeV$^2$. The large-$N_c$ Regge model, \eqref{eq:TFFetaandetap}, is shown in the left panel, and our alternative model, \eqref{MartinModel}, is shown in the right panel.}
\figlab{EtaP3DPlot}
\end{figure}

All expressions are given for the $\eta$, but hold as well for the $\eta'$ after obvious replacements (including $C_8 \rightarrow C_0$).
Based on the constraint equations in \eqref{ConstraintEquationsEta}, \eqref{MVEta}, and \eqref{pQCD3Eta} the model parameters should be replaced by:
\begin{align}
c_\mathrm{diag}&= \frac{ \mathcal{N}}{C^\eta_{\rho
  \rho}+C^\eta_{\omega \omega}+ C^\eta_{\phi \phi}}\frac{8C_8 F_\eta }{\Lambda^2F_{\eta\ga \ga}},\label{cdiageta}\\
c_\mathrm{BL}&=\frac{1}{M^2_{-,\,0}}\Bigg[\frac{C^\eta_{\rho
  \rho}M_\rho^2+C^\eta_{\omega \omega}M_\omega^2+ C^\eta_{\phi \phi}M_\phi^2}{2C^\eta_{ \phi\omega}}-\frac{\mathcal{N}}{C^\eta_{ \phi\omega}}\frac{6C_8F_\eta}{F_{\eta\ga\ga}}+M^2_{+,\,0}-c_A \frac{M_{-,\,0}^4}{\Lambda^2}\nn\\
  &-c_B \frac{M_{+,\,0}^2M_{-,\,0}^2}{\Lambda^2}\Bigg],\qquad\label{cBLeta}
\end{align}
where the parameters related to the asymptotic limits of the HLbL tensor simplify to:
\begin{align}
c_{\bfrac{A}{B}}&=\frac{\Lambda^2}{2}\Bigg[\frac{\Delta_{\omega\eta}}{L_{\omega\eta}}\frac{\mathcal{N}}{C^\eta_{\phi
  \omega}} \frac{3 C_\eta^2}{2\pi^2C_8 F_\eta F_{\eta\ga\ga} M_\phi^2 M_\omega^2}\notag\\
  &+\left(\frac{\Delta_{\omega\eta}L_{\phi\eta}}{\Delta_{\phi\eta}L_{\omega\eta}}\mp 1\right)
  \Bigg(\frac{b}{2a}- \sqrt{\bigg(\frac{b}{2a}\bigg)^2-\frac{1}{a}\bigg(c-\frac{4C_{\eta}^2}{\pi^2}\bigg)}\Bigg)\Bigg],
\end{align}
with
\begin{align}
a&=\left(\frac{C^\eta_{\phi
  \omega} F_{\eta\ga\ga} M_\phi^2 M_\omega^2}{\mathcal{N}}\right)^2\frac{\Delta_{\omega\eta}}{\Omega_{\phi\omega\eta}^2}\left[f_2(\sigma_\eta,\sigma_\omega,\sigma_\phi)-\frac{L_{\phi\eta}}{L_{\omega\eta}}f_2(\sigma_\eta,\sigma_\phi,\sigma_\omega) \right]\left(\sigma_\phi^2\, \frac{\Delta_{\omega\eta}L_{\phi\eta}}{\Delta_{\phi\eta}L_{\omega\eta}}-\sigma_\omega^2\right),\nn\\
  b&=-\frac{C^\eta_{\phi
  \omega} F_{\eta\ga\ga} M_\phi^2 M_\omega^2}{\mathcal{N}\Delta_{\phi\eta}L_{\omega\eta}}\Bigg\{\frac{-4C_8F_\eta}{C^\eta_{\rho\rho}+C^\eta_{\phi\phi}+C^\eta_{\omega\omega}}\Bigg[\frac{C^\eta_{\rho\rho}}{ \Delta_{\rho\eta}^2}\left(\frac{L_{\phi\eta}}{\Delta_{\rho\omega}^2}f_1(\sigma_\eta,\sigma_\rho,\sigma_\omega)-\frac{L_{\omega\eta}}{\Delta_{\rho\phi}^2}f_1(\sigma_\eta,\sigma_\rho,\sigma_\phi)\right)\nn\\
  &+\frac{C^\eta_{\phi\phi}}{ \Delta_{\phi\eta}^2}\left(\frac{L_{\phi\eta}}{\Delta_{\phi\omega}^2}f_1(\sigma_\eta,\sigma_\phi,\sigma_\omega)+\frac{L_{\omega\eta}}{2}f_3(\sigma_\eta,\sigma_\phi)\right)-\frac{C^\eta_{\omega\omega}}{ \Delta_{\omega\eta}^2}\left(\frac{L_{\omega\eta}}{\Delta_{\phi\omega}^2}f_1(\sigma_\eta,\sigma_\omega,\sigma_\phi)\right.\nn\\
  &+\left.\frac{L_{\phi\eta}}{2}f_3(\sigma_\eta,\sigma_\omega)\right)\Bigg]-\frac{3C_\eta^2}{2\pi^2C_8F_\eta\Omega_{\phi\omega\eta}\Delta_{\phi\omega}}\Bigg[\left(2 \sigma_\phi^2 \Delta_{\omega\eta}\frac{L_{\phi\eta}}{L_{\omega\eta}}-\sigma_\omega^2 \Delta_{\phi\eta}\right)f_2(\sigma_\eta,\sigma_\phi,\sigma_\omega)\nn\\
  &-\sigma_\phi^2\Delta_{\omega\eta}f_2(\sigma_\eta,\sigma_\omega,\sigma_\phi)\Bigg] \Bigg\},\nn\\
  c&=-\frac{3C_\eta^2}{2\pi^2C_8F_\eta\Delta_{\phi\eta}L_{\omega\eta}}\Bigg[\frac{3C_\eta^2}{2\pi^2C_8F_\eta}\frac{\sigma_\phi^2}{\Delta_{\phi\omega}^2L_{\omega\eta}}f_2(\sigma_\eta,\sigma_\phi,\sigma_\omega)-\frac{4C_8F_\eta \Delta_{\phi\eta}}{C^\eta_{\rho\rho}+C^\eta_{\phi\phi}+C^\eta_{\omega\omega}}\nn\\
  &\times\Bigg(\frac{C^\eta_{\omega\omega}}{2\Delta_{\omega\eta}^2}f_3(\sigma_\eta,\sigma_\omega)-\frac{C^\eta_{\rho\rho}}{\Delta_{\rho\omega}^2\Delta_{\rho\eta}^2}f_1(\sigma_\eta,\sigma_\rho,\sigma_\omega)-\frac{C^\eta_{\phi\phi}}{\Delta_{\phi\omega}^2\Delta_{\phi\eta}^2}f_1(\sigma_\eta,\sigma_\phi,\sigma_\omega)\Bigg)\Bigg].
\end{align}
Here, the definitions from~\eqref{def2} and \eqref{def3}, the auxiliary functions from \eqref{auxFunc}, as well as 
\beq
f_3(\si_P,\si_{V})\coloneqq 3 \sigma_P^4-4\sigma_P^2\sigma_V^2+\sigma_V^4-2\sigma_P^4\,L_{PV},
\eeq
are used. Note that for the $\eta'$ one has to instead choose: 
\begin{align}
c_{\bfrac{A}{B}}&=\frac{\Lambda^2}{2}\Bigg[\frac{\Delta_{\omega\eta'}}{L_{\omega\eta'}}\frac{\mathcal{N}}{C^{\eta'}_{\phi
  \omega}} \frac{3 C_{\eta'}^2}{2\pi^2C_0 F_{\eta'} F_{\eta'\ga\ga} M_\phi^2 M_\omega^2}
  \notag\\ 
 &+
\left(\frac{\Delta_{\omega\eta'}L_{\phi\eta'}}{\Delta_{\phi\eta'}L_{\omega\eta'}}\mp 1\right)
\Bigg(\frac{b}{2a}+ \sqrt{\bigg(\frac{b}{2a}\bigg)^2-\frac{1}{a}\bigg(c-\frac{4C_{\eta'}^2}{\pi^2}\bigg)}\Bigg)\Bigg],
\end{align}
as the physical solution for the quadratic equation~\eqref{pQCD3Eta}.

\subsection[{Comparison of   data and literature: $F_{\eta(n)\ga^*\ga^*}$}]{Comparison of   data and literature: $\boldsymbol{F_{\eta(n)\ga^*\ga^*}}$}
\seclab{EtaPlots}

\begin{figure}[t]
\centering
\includegraphics[width=0.85\linewidth]{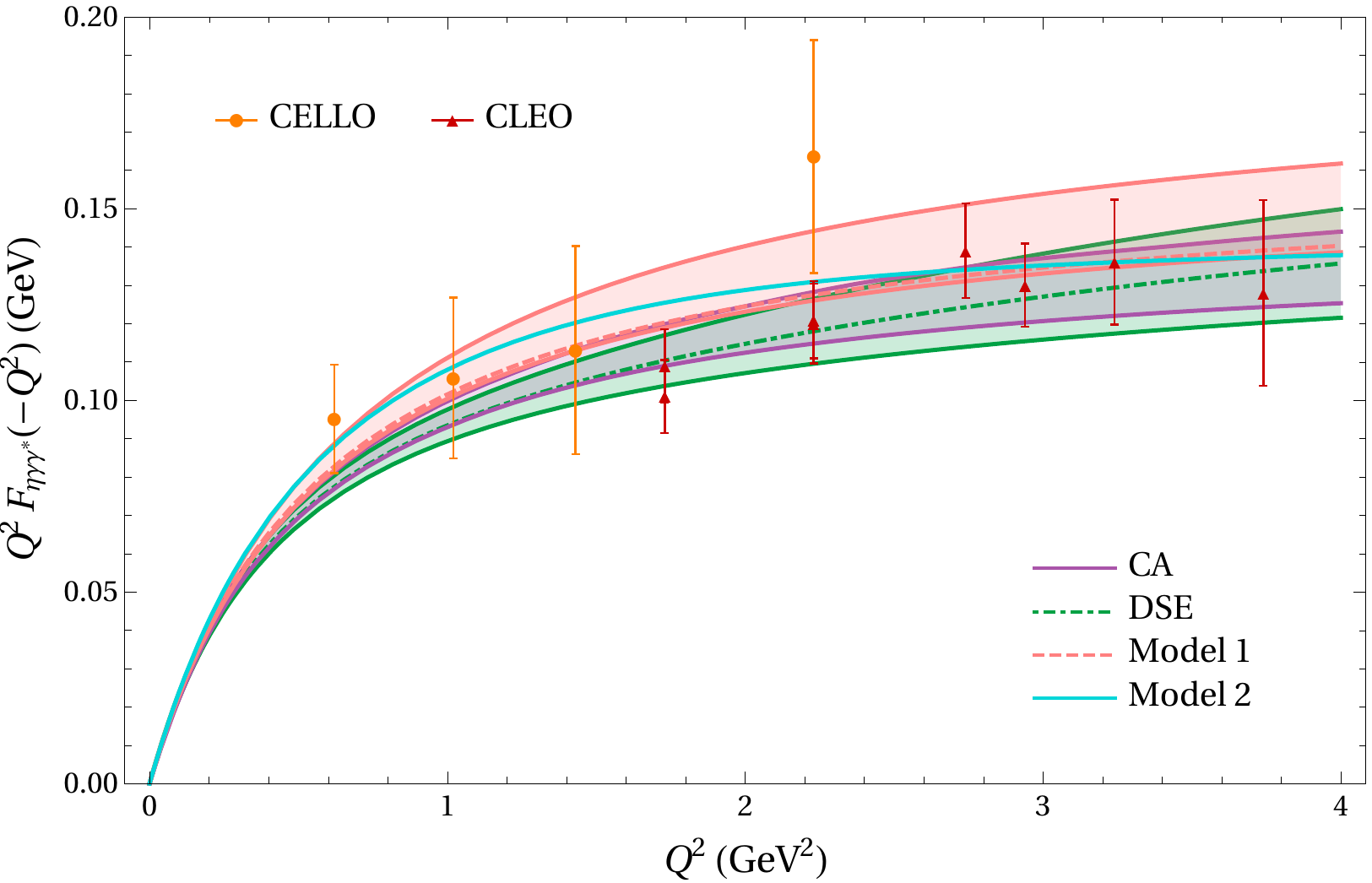}
\caption{Singly-virtual $\eta$ TFF in the low-$Q$ region. The large-$N_c$ Regge model, ``Model 1''~\eqref{eq:TFFetaandetap}, is indicated by the  pink band with the dashed curve. Our alternative TFF model, ``Model 2''~\eqref{MartinModel}, is indicated by the solid cyan curve. The purple band is the CA result from ref.~\cite{Masjuan:2017tvw}. The green band with the dot-dashed curve is the DSE result from ref.~\cite{Eichmann:2019tjk}. The data are from CELLO~\cite{Behrend:1990sr} and CLEO~\cite{Gronberg:1997fj}.}
\figlab{EtaDataZoom}
\end{figure}

\begin{figure}[t]
\centering
\includegraphics[width=0.48\linewidth]{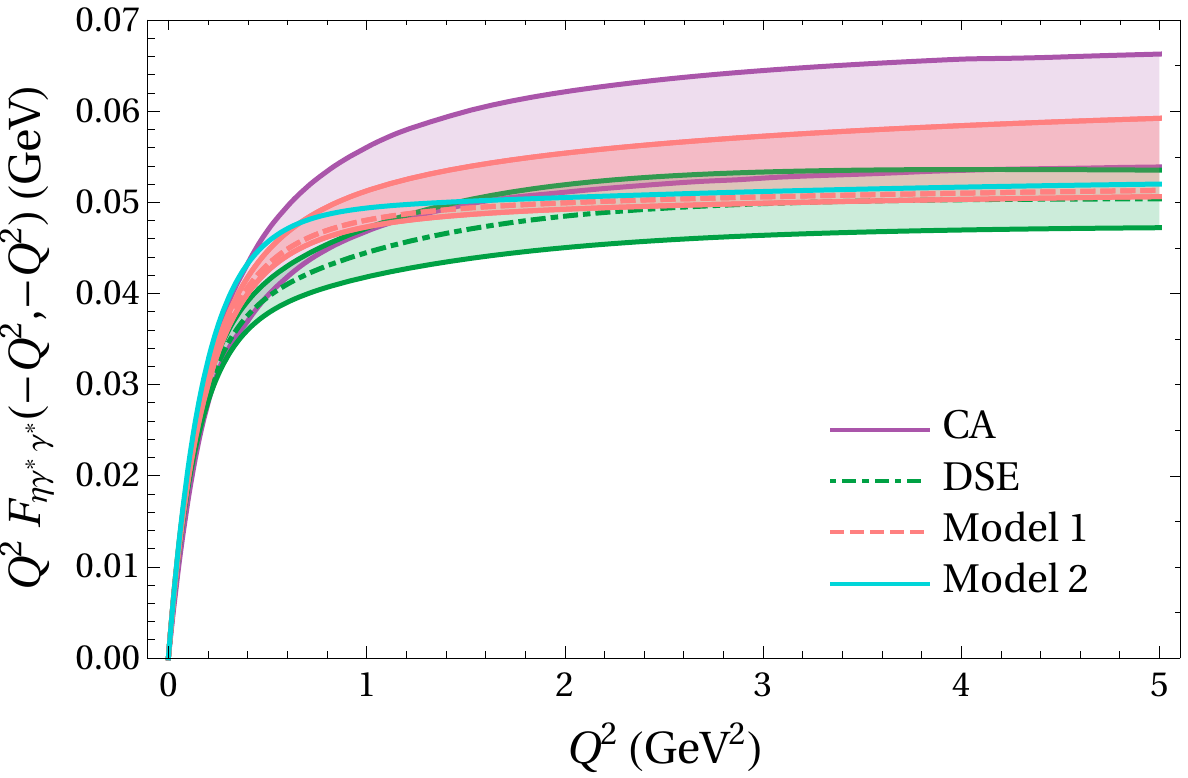}\hfill\includegraphics[width=0.48\linewidth]{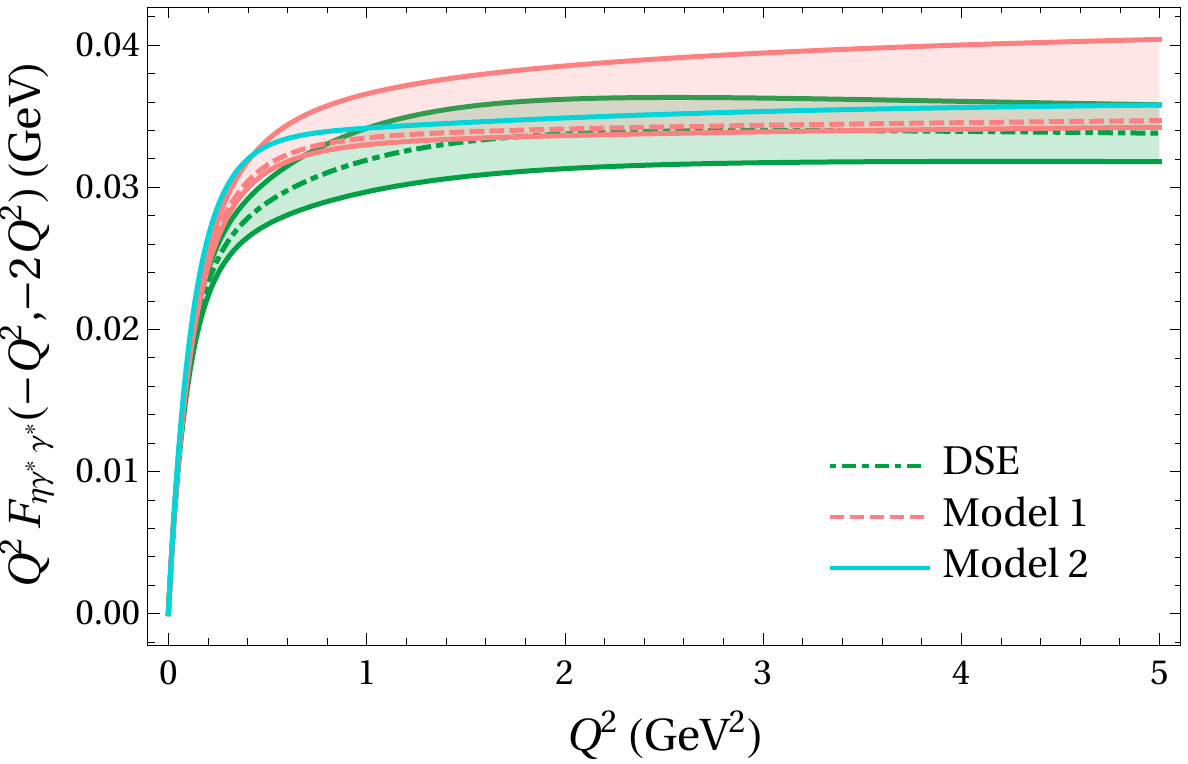}
\caption{Doubly-virtual $\eta$ TFF in the symmetric region $Q_1^2=Q_2^2=Q^2$ (left) and in the region where $Q_1^2=Q^2$ and $Q_2^2=2Q^2$ (right). Legend is the same as in figure \ref{fig:EtaDataZoom}.}
\figlab{EtaDV}
\end{figure}

In this appendix, we compare
our large-$N_c$ Regge model, ``Model 1''~\eqref{eq:TFFetaandetap}, and our alternative model, ``Model 2''~\eqref{MartinModel}, for $F_{\eta(n)\ga^*\ga^*}$ to data and other parameterizations available from the literature. The error band shown for Model 1 is generated by propagating the errors of the input parameters $\sigma_P$, $\sigma_V$, $F_{\eta\ga\ga}$, $F_\eta$, $F^8$, $F^0$, $\theta_8$, $\theta_0$.

In figure \ref{fig:SinglyVirtualEta}, the singly-virtual TFF of the ground-state $\eta$ is shown for $Q^2 \in [0,40]\,\text{GeV}^2$. In figure \ref{fig:EtaDataZoom}, we focus on the low-$Q$ region and include a comparison to the DSE result \cite{Eichmann:2019tjk}. One can see that our models agree with the experimental data from CELLO~\cite{Behrend:1990sr} and CLEO~\cite{Gronberg:1997fj}, but tend to a larger $\eta$ TFF than CA \cite{Masjuan:2017tvw} and DSE. In addition, Model 2 is larger than Model 1 for $Q^2<2.4$ GeV$^2$. This low-$Q$ enhancement explains why  $a_\mu^{\eta\text{-pole}}\vert_\text{Model 2}>a_\mu^{\eta\text{-pole}}\vert_\text{Model 1}$, see \eqref{aGroundStateEta}.

\begin{figure}[t]
\includegraphics[width=0.48\linewidth]{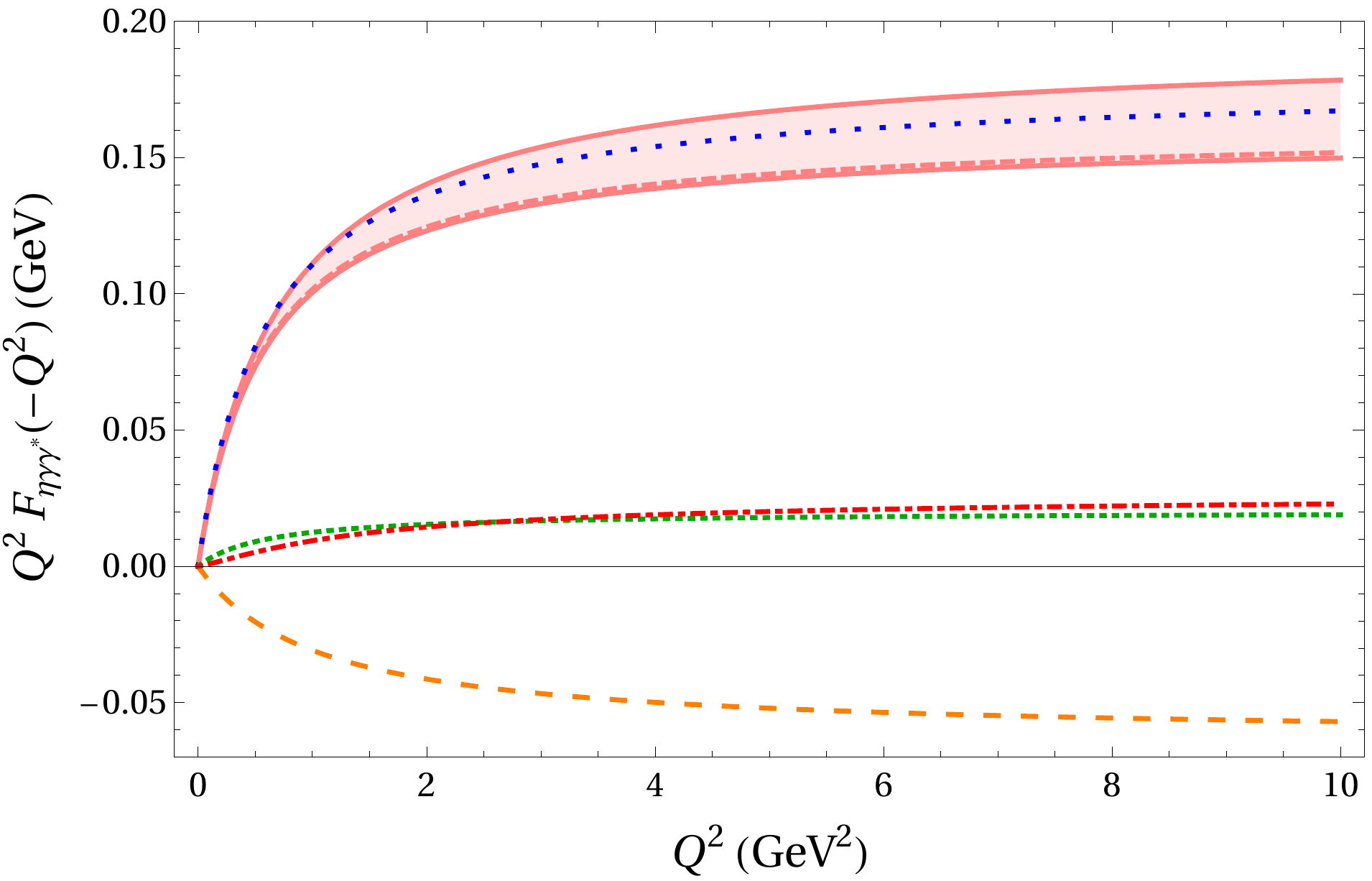}\hfill
\includegraphics[width=0.48\linewidth]{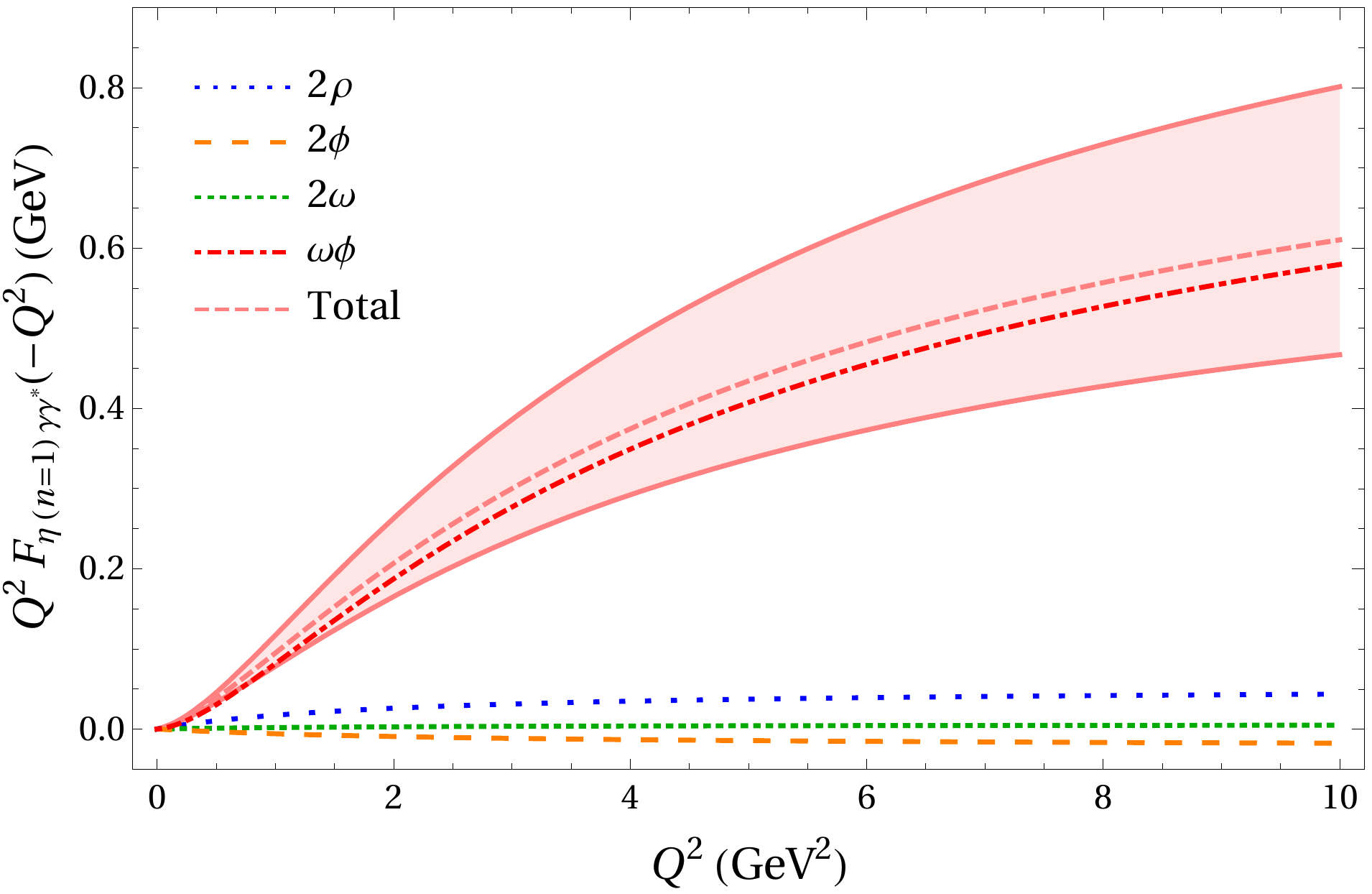}\\[0.2cm]
\includegraphics[width=0.48\linewidth]{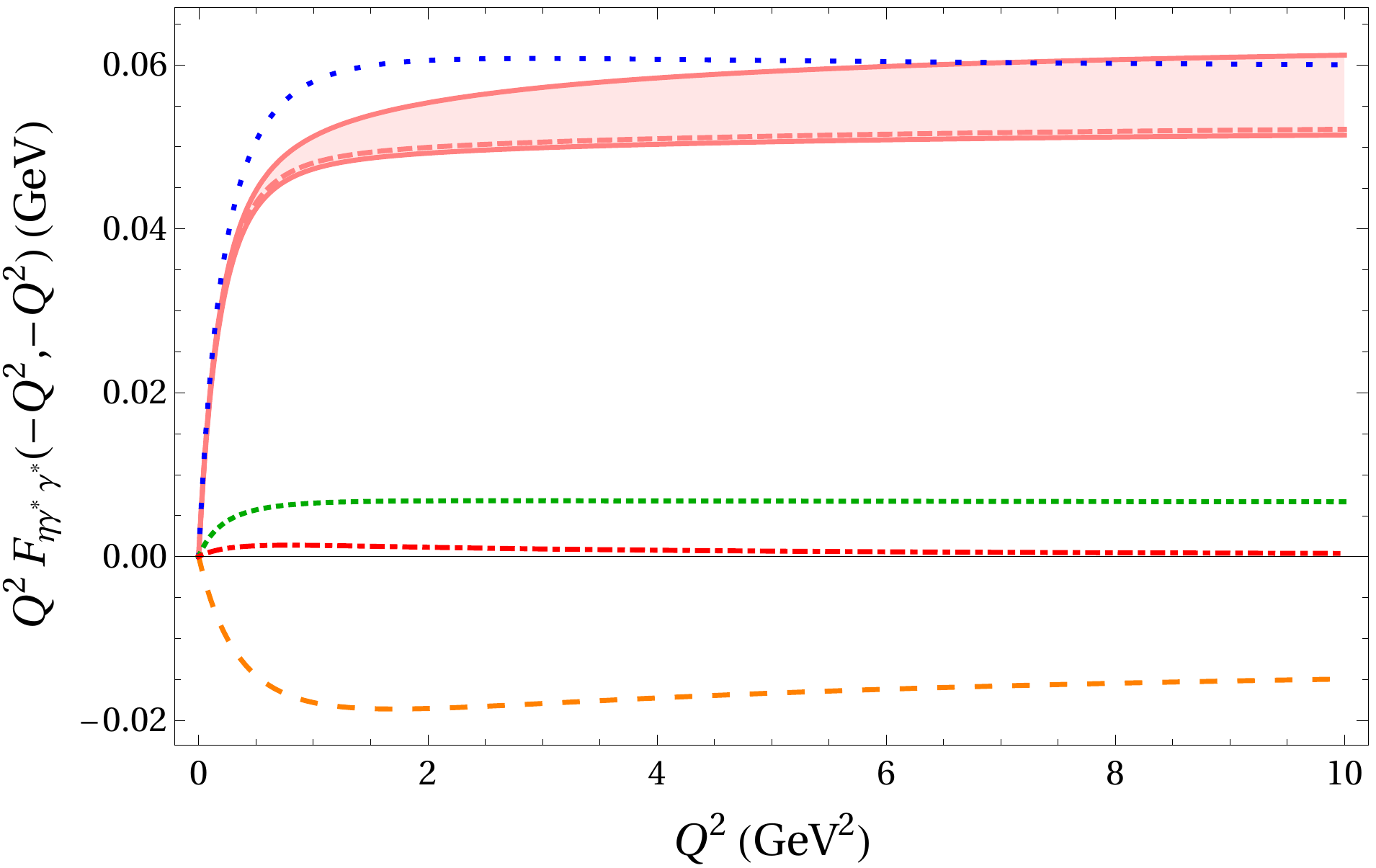}\hfill
\includegraphics[width=0.48\linewidth]{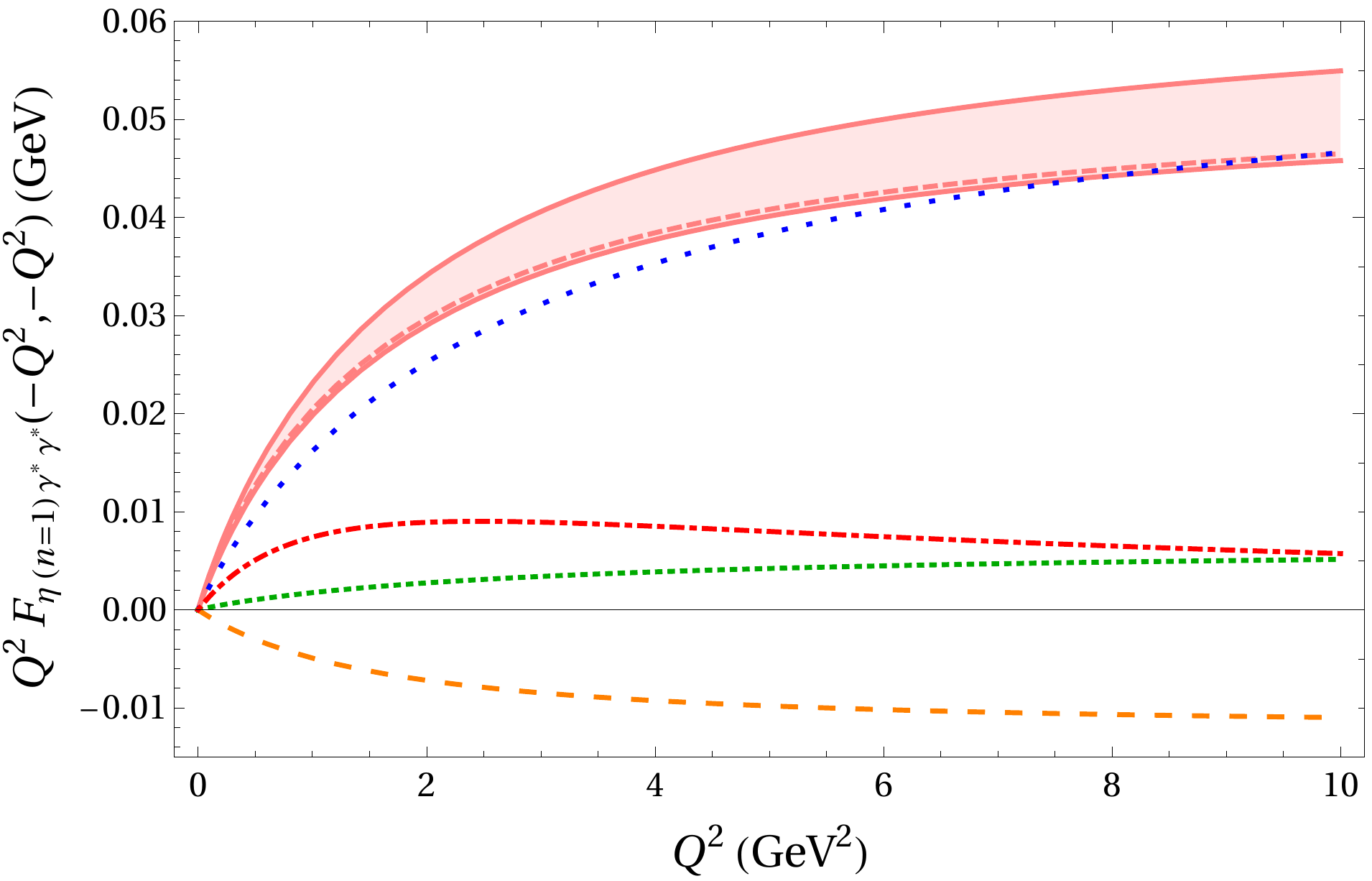}
\caption{$2\rho$, $2\omega$, $2\phi$, and $\phi \omega$ contributions to the singly-virtual (top) and doubly-virtual (bottom) $\eta$ ground state (left) and first excited state (bottom). }
\figlab{ContributionsEta}
\end{figure}

In figure~\ref{fig:EtaDV}, the doubly-virtual $\eta$ TFF is shown for two kinematic situations: symmetric momenta, and $Q_1^2=Q^2$ and $Q_2^2=2Q^2$. Considering Model 1, we observe a slight tension with the DSE prediction in the region of
$Q^2 \in [0.2,0.8]\,\text{GeV}^2$. This tension should, however, not be taken too serious, because both the DSE and our error band are only based on the variation of input parameters and do not take into account all possible error sources.

In figure~\ref{fig:Eta3DPlot}, Model 1 and 2 are shown in the full space-like region for $Q_1^2,Q_2^2<10$ GeV$^2$. One can see that their main difference lies, similarly as for the $\pi^0$ TFF, in the regions where at least one of the photon virtualities is small.

In the left panel of figure~\ref{fig:ContributionsEta}, the ground-state $\eta$ TFF is decomposed into the contributions from $2\rho$, $2\omega$, $2\phi$, and $\phi \omega$ vector mesons. As expected, the TFF is dominated by the isovector--isovector $2\rho$ contribution, followed by the isoscalar--isoscalar $2\phi$ contribution. The $\phi\omega$ contribution~\eqref{TFFeta-ophi}, which was needed to generate enough freedom in our large-$N_c$ Regge model to satisfy the BL limit of the TFF and the two SDCs on the HLbL tensor, is small. 

In the right panel of figure~\ref{fig:ContributionsEta}, we show the TFF of the first ($n=1$) radially-excited $\eta$ state. In the doubly-virtual region, the relative strength of vector-meson pairs is comparable to what one finds for the ground-state $\eta$. The $2\rho$ contribution is now slightly smaller than the total TFF, and the $\phi\omega$ contribution is now larger than the $2\omega$ contribution. In contrast, the singly-virtual TFFs of the radially-excited $\eta$ states will be dominated by the $\phi\omega$ contribution. This enhancement is generated by the $n$-dependence in the numerator of~\eqref{TFFeta-ophi} through terms proportional to $M_{+,\,n}$. The two-photon couplings and BL limits of the excited-state $\eta$ TFFs are shown in figure \ref{fig:Coupling}.

\begin{figure}[t]
\centering
\includegraphics[width=0.85\linewidth]{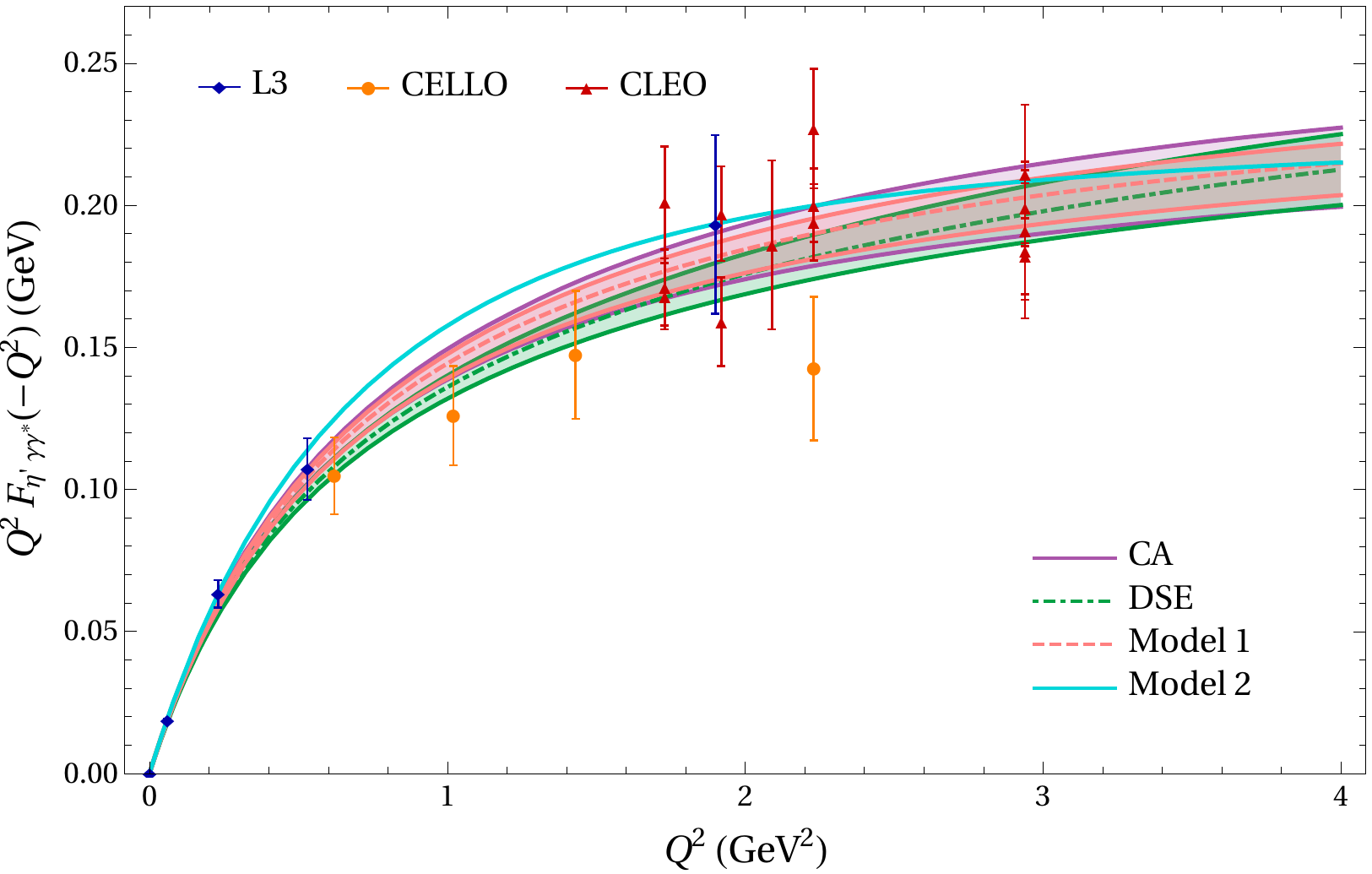}
\caption{Singly-virtual $\eta^\prime$ TFF in the low-$Q$ region. The large-$N_c$ Regge model, ``Model 1''~\eqref{eq:TFFetaandetap}, is indicated by the  pink band with the dashed curve. Our alternative TFF model, ``Model 2''~\eqref{MartinModel}, is indicated by the solid cyan curve. The purple band is the CA result from ref.~\cite{Masjuan:2017tvw}. The green band with the dot-dashed curve is the DSE result from ref.~\cite{Eichmann:2019tjk}. The data are from L3~\cite{Acciarri:1997yx}, CELLO~\cite{Behrend:1990sr}, and CLEO~\cite{Gronberg:1997fj}.}
\figlab{EtaPDataZoom}
\end{figure}

\subsection[{Comparison of data and literature: $F_{\eta^{\prime}(n)\ga^*\ga^*}$}]{Comparison of data and literature: $\boldsymbol{F_{\eta^{\prime}(n)\ga^*\ga^*}}$}
\seclab{EtaPrimePlots}

\begin{figure}[t]
\centering
\includegraphics[width=0.48\linewidth]{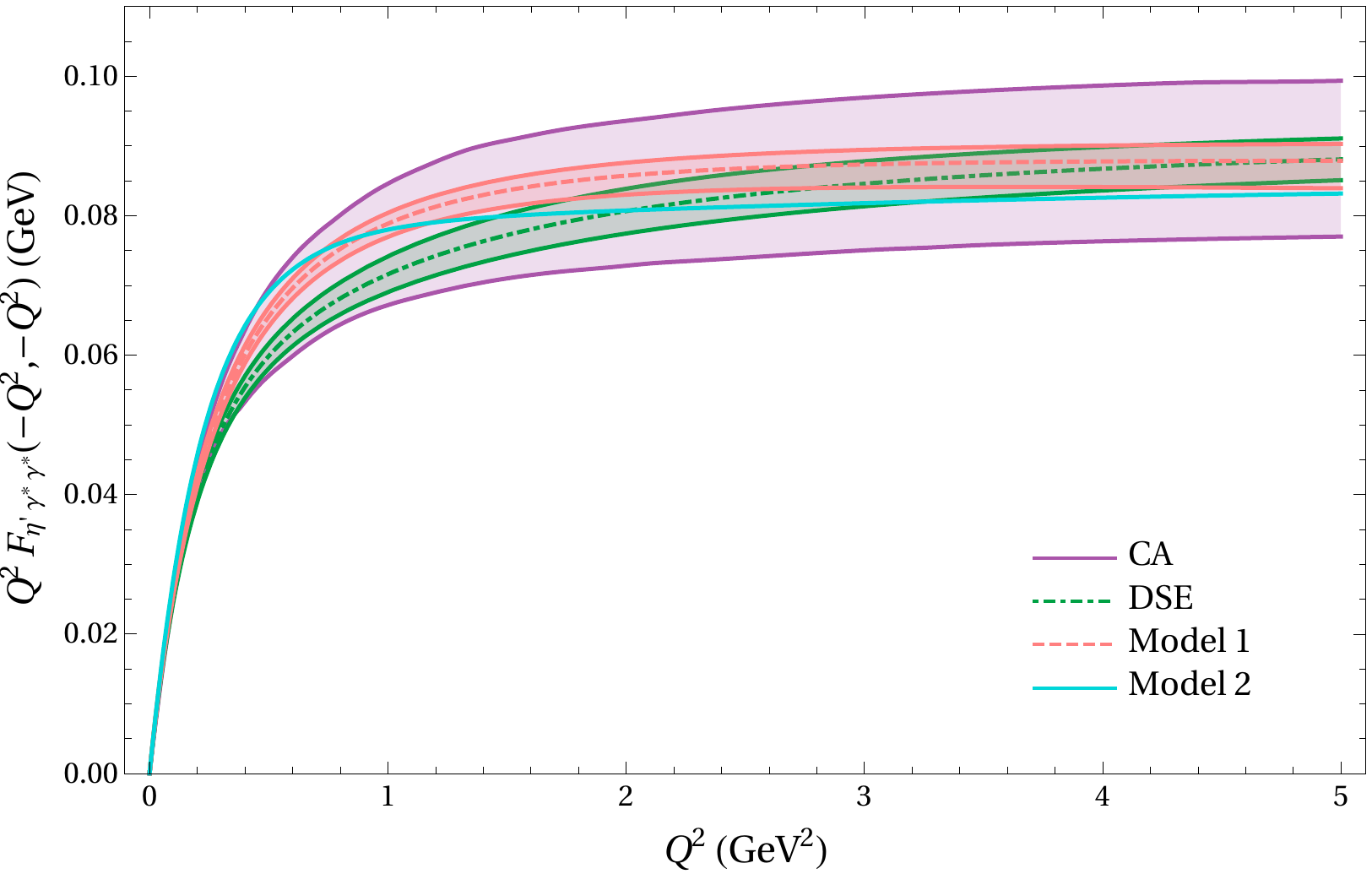}\hfill\includegraphics[width=0.48\linewidth]{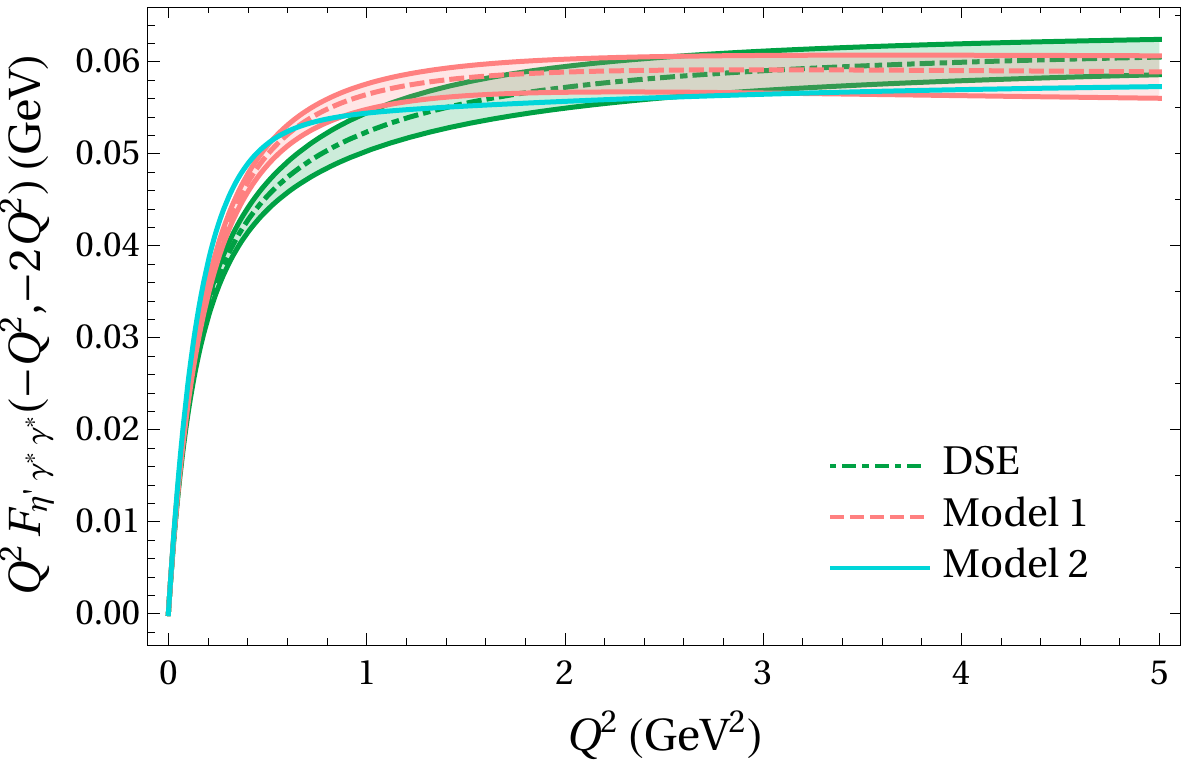}
\caption{Doubly-virtual $\eta'$ TFF in the symmetric region $Q_1^2=Q_2^2=Q^2$ (left) and in the region where $Q_1^2=Q^2$ and $Q_2^2=2Q^2$ (right). Legend is the same as in figure \ref{fig:EtaPDataZoom}.}
\figlab{EtaPDV}
\end{figure}

In this appendix, we compare
our large-$N_c$ Regge model, ``Model 1''~\eqref{eq:TFFetaandetap}, and our alternative model, ``Model 2''~\eqref{MartinModel}, for $F_{\eta(n)\ga^*\ga^*}$ to data and other parameterizations available from the literature. The error band shown for Model 1 is generated by propagating the errors of the input parameters $\sigma_P$, $\sigma_V$, $F_{\eta'\ga\ga}$, $F_{\eta'}$, $F^8$, $F^0$, $\theta_8$, $\theta_0$.

In figure \ref{fig:SinglyVirtualEtaPrime}, the singly-virtual TFF of the ground-state $\eta'$ is shown for $Q^2 \in [0,40]\,\text{GeV}^2$. In figure \ref{fig:EtaPDataZoom}, we focus on the low-$Q$ region and include a comparison to the DSE result \cite{Eichmann:2019tjk}. One can see that Model 1 agrees with the experimental data from L3~\cite{Acciarri:1997yx}, CELLO~\cite{Behrend:1990sr}, and CLEO~\cite{Gronberg:1997fj}, as well as the CA \cite{Masjuan:2017tvw} and DSE results. Model 2 tends to a larger $\eta$ TFF for $Q^2<2$ GeV$^2$. This low-$Q$ enhancement explains why  $a_\mu^{\eta'\text{-pole}}\vert_\text{Model 2}>a_\mu^{\eta'\text{-pole}}\vert_\text{Model 1}$, see~\eqref{aGroundStateEtaP}.

In figure \ref{fig:EtaPDV}, the doubly-virtual $\eta'$ TFF is shown for two kinematic situations: symmetric momenta, and $Q_1^2=Q^2$ and $Q_2^2=2Q^2$. Model 1 is in slight tension with the DSE prediction for
$Q^2 \in [0.2,1.6]\,\text{GeV}^2$. This tension should, however, not be taken too serious. A comparison of our models with the CA result shows perfect agreement for symmetric momenta. For large photon virtualities, both models agree with each other and give a reasonably good description of the recent doubly-virtual $\eta ^\prime$ TFF data from BaBar \cite{BaBar:2018zpn}, see figure \ref{fig:EtaPDataDV}.

In figure \ref{fig:EtaP3DPlot}, Model 1 and 2 are shown in the full space-like region for $Q_1^2,Q_2^2<10$ GeV$^2$. One can see that their main difference is, similar as for the $\eta$ TFF, in the regions where at least one of the photon virtualities is small.

\begin{figure}[t]
\includegraphics[width=0.48\linewidth]{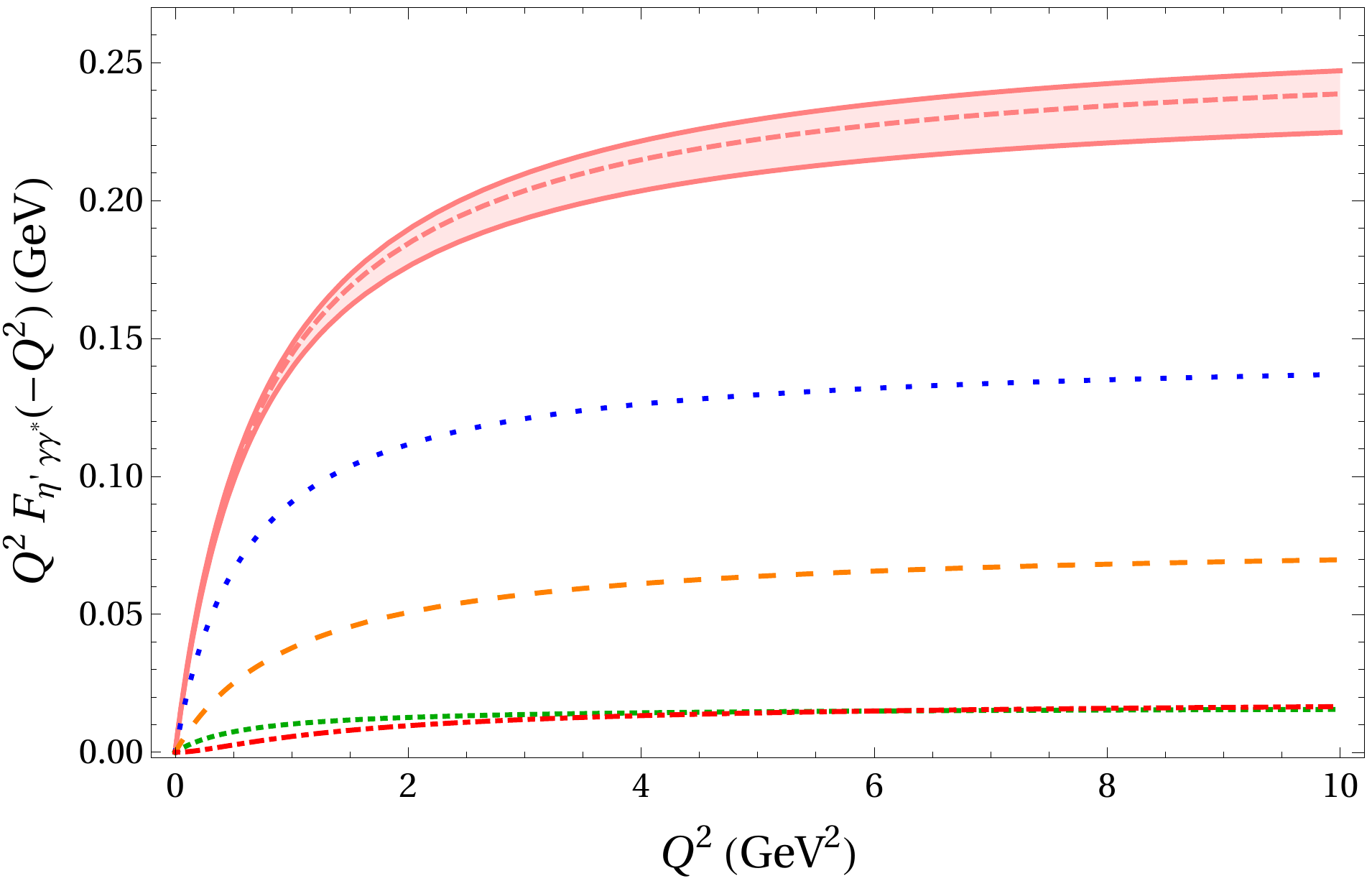}\hfill
\includegraphics[width=0.48\linewidth]{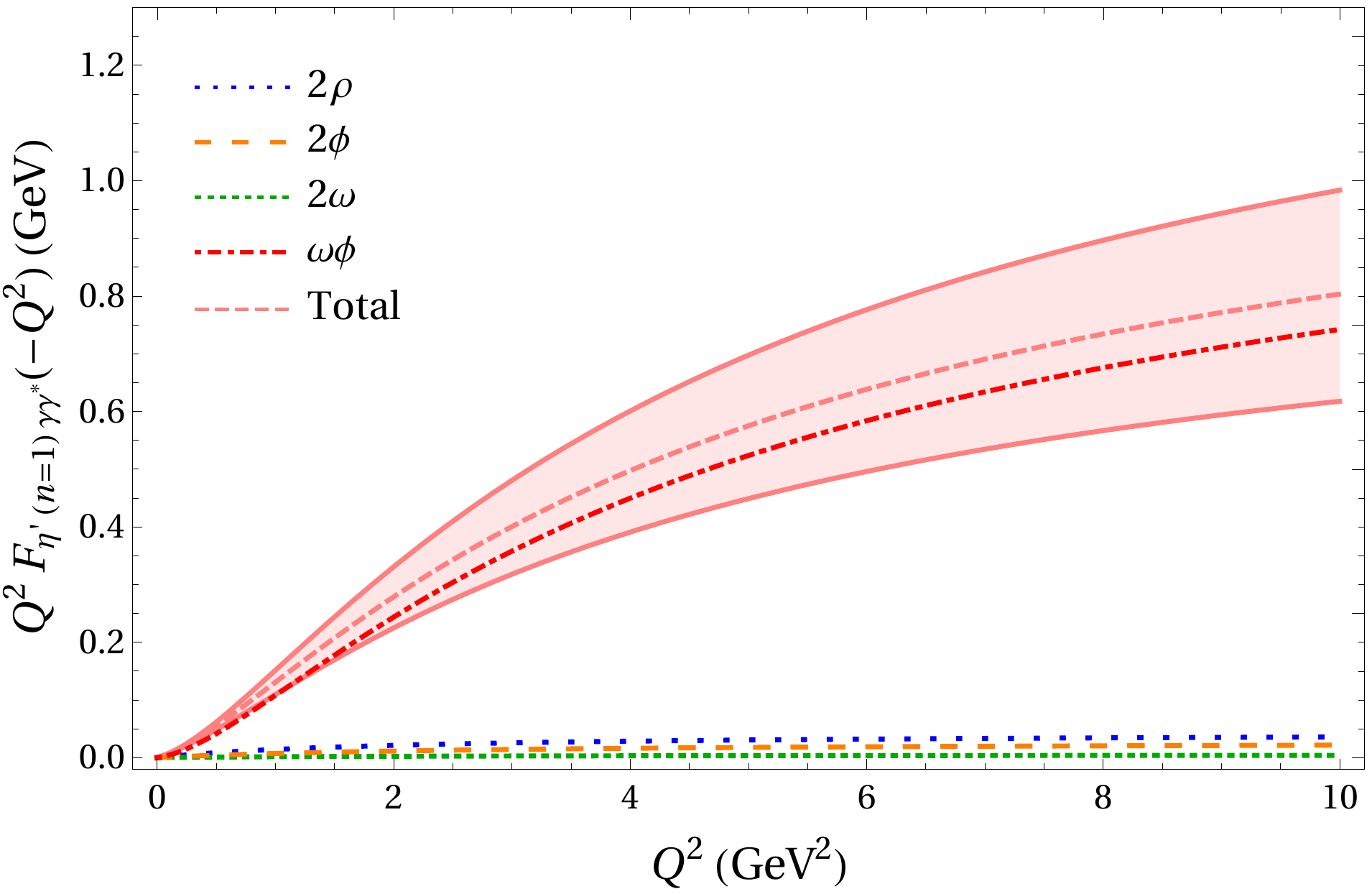}\\[0.2cm]
\includegraphics[width=0.48\linewidth]{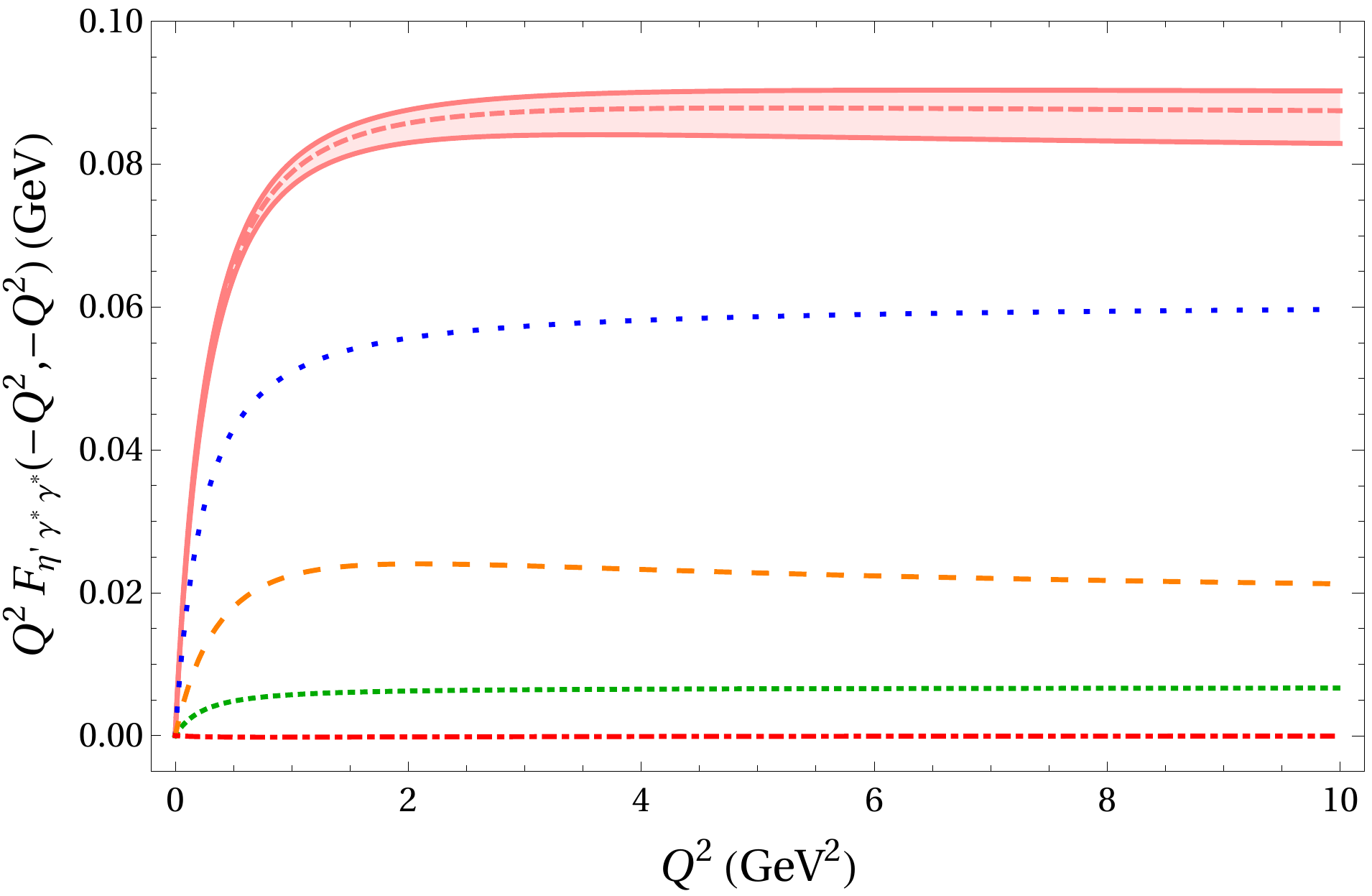}\hfill
\includegraphics[width=0.48\linewidth]{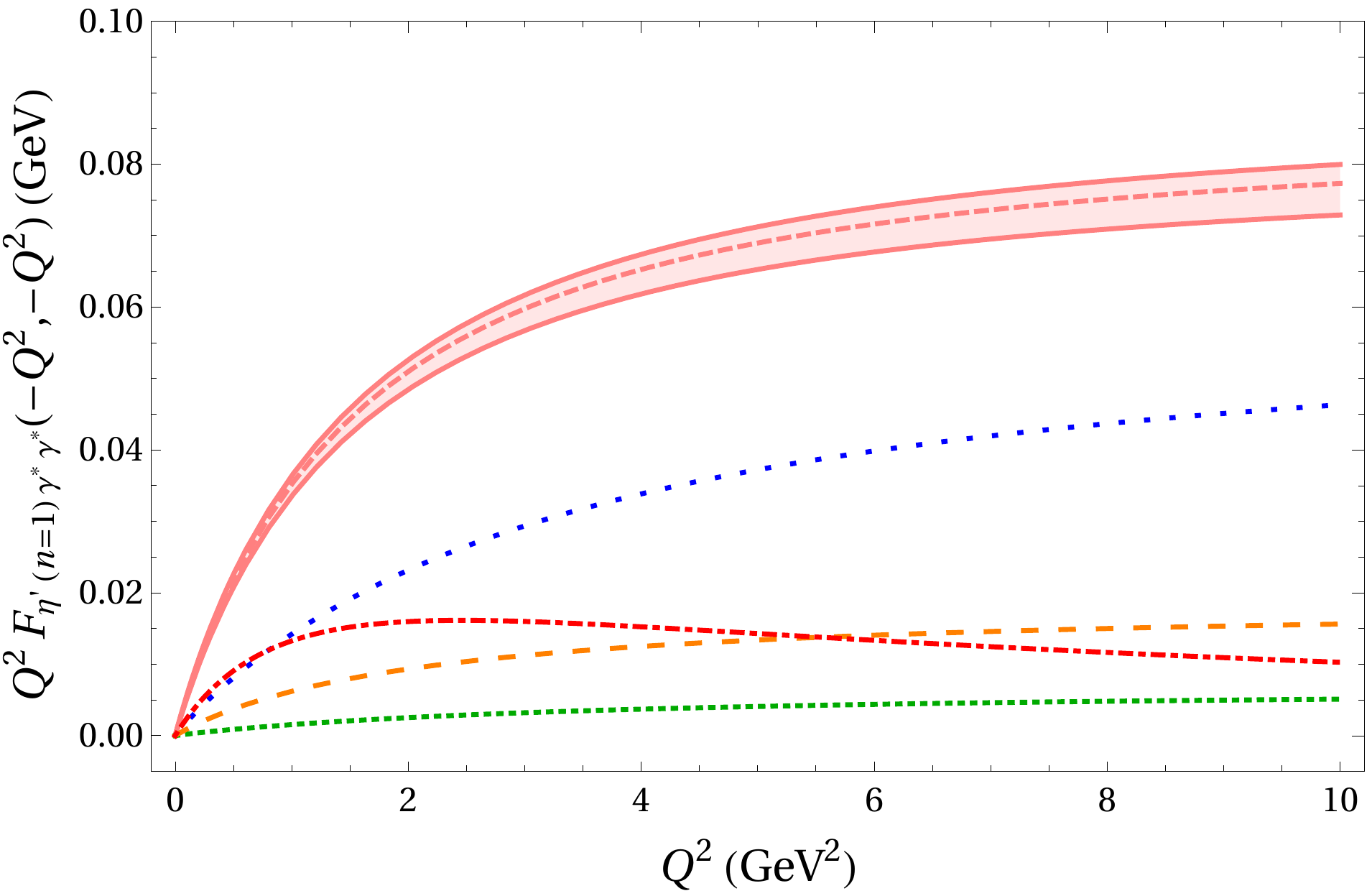}
\caption{$2\rho$, $2\omega$, $2\phi$, and $\phi \omega$ contributions to the singly-virtual (top) and doubly-virtual (bottom) $\eta^\prime$ ground state (left) and first excited state (bottom). }
\figlab{ContributionsEtaP}
\end{figure}

In the left panel of figure \ref{fig:ContributionsEtaP}, the ground-state $\eta'$ TFF is decomposed into the contributions from $2\rho$, $2\omega$, $2\phi$, and $\phi \omega$ vector mesons. As expected, the largest contribution to the TFF is coming from the isovector--isovector $2\rho$ mesons, followed by the isoscalar--isoscalar $2\phi$ mesons. Unlike in the case of the $\eta$ TFF, the $2\phi$ mesons gives a positive contribution to the $\eta'$ TFF, just like the $2\rho$, $2\omega$, and $\phi \omega$ mesons. Thus, since the $2\rho$ contribution does not need to cancel out a negative $2\phi$ contribution as it does in the $\eta$ TFF, it appears to be smaller than the total $\eta'$ TFF. The $\phi\omega$ contribution \eqref{TFFeta-ophi}, generated through $\phi$--$\omega$ mixing, is small. 

In the right panel of figure \ref{fig:ContributionsEtaP}, we show the TFF of the first ($n=1$) radially-excited $\eta'$ state. In the doubly-virtual region, the relative strength of vector-meson pairs is similar to what one finds for the ground-state $\eta'$. The $\phi\omega$ contribution is now larger than the $2\omega$ contribution, and the $2\phi$ contribution at low $Q$. In contrast, the singly-virtual TFFs of the radially-excited $\eta'$ states will be dominated by the $\phi\omega$ contribution, while all other contributions are of negligible size. This enhancement is generated by the $n$-dependence in the numerator of \eqref{TFFeta-ophi} through terms proportional to $M_{+,\,n}$.  The two-photon couplings and BL limits of the excited-state $\eta'$ TFFs are shown in figure \ref{fig:Coupling}.

\section{Reply to {\tt arXiv:1911.05874}}
\label{app:MV_comment}

The analysis presented here has been criticized in a preprint by MV~\cite{Melnikov:2019xkq}, which appeared during the review process of this manuscript. In this appendix we refute the main objections
raised therein: 
\begin{enumerate}
\item We disagree with the claim that in the
  pion-pole contribution to HLbL the second form factor has to
  be taken at $q_3^2=M_\pi^2$, see discussion in the paragraphs from~(3) to~(5). The claim is a consequence of the statement in the paragraph
  after~(5) in ref.~\cite{Melnikov:2019xkq}: ``with obvious
  constraints on $q_{1,\ldots,4}$ in the form factors
  $s=(q_1+q_2)^2=(q_3+q_4)^2=M_\pi^2$.'' This statement is incorrect. We recall that in the definition of the TFF~\eqref{eq:defpiTFF}, translation invariance has already been applied and the resulting overall delta function is not part of the TFF.
  In the unitarity relation for HLbL, the single-pion intermediate state generates an imaginary part proportional to $\delta(s-M_\pi^2)$, which however disappears when it is put into the dispersion integral to generate the pion pole, i.e., no constraint on $s$ is left. The independent variables in the HLbL process are $q_1^2$, $q_2^2$, $q_3^2$, $q_4^2$, $s$, and $t$ and the residue of the pion pole, i.e., the product of
  two pion TFFs, can only depend on the first four.
  If one takes the limit $q_4\to 0$ this implies $s=q_3^2$ and $t=q_2^2$,
  but by no means does it imply $q_3^2=M_\pi^2$. Of course one is free at
  that point to separate the pure pole in $q_3^2$ (with only its residue in
  the numerator) from non-pole terms. Between the two different dispersive representations, a simple reshuffling takes place, see~\eqref{eq:ReshuffledPole} and the whole discussion in section~\ref{sec:MV}. 
\item
As discussed in section~\ref{sec:MV}, the MV model is based on an
unjustified extrapolation to low $q_{1,2}^2$ of the constraint at high
$q_{1,2}^2$. We have called this a ``distortion'' of the low-energy behavior of
the HLbL tensor in ref.~\cite{Colangelo:2019}, a description considered unjustified
in ref.~\cite{Melnikov:2019xkq}. Figures~\ref{SD_comparison}
and~\ref{SD_comparison_max} very clearly illustrate this distortion. An
alternative solution to the SDCs based on a tower of 
axial-vector mesons in holographic QCD has been presented in two papers~\cite{Leutgeb:2019gbz,Cappiello:2019hwh}, which  appeared after ref.~\cite{Melnikov:2019xkq}. These alternative solutions to the SDCs have a very similar behavior as the curves corresponding to our model in figures~\ref{SD_comparison}
and~\ref{SD_comparison_max}, and confirm that the MV model~\cite{Melnikov:2003xd} leads to a
low-$q^2$ behavior of the HLbL amplitude that cannot be explained in terms
of any other physical states---in other words, a ``distortion.''
\item
After~(18) in ref.~\cite{Melnikov:2019xkq} it is stated that: the
model in the present manuscript ``violates the above equation and claims,
effectively, that $c_L^\rho \sim 1$ also in the chiral limit.'' This
statement is incorrect: in our model we are not able to take the chiral
limit simply because it is formulated in terms of effective parameters
that are fit to data or theoretical constraints. It is not the point of
the model to make any claim about the behavior in the chiral limit. The
underlying philosophy is to fulfill the SDCs only for large $q_3^2$, where the chiral limit becomes
irrelevant, and not to rely on it at low $q_3^2$, because it
would be a bad approximation and we can instead use known
phenomenological constraints. Such a strategy is best carried out with excited pseudoscalars because, unlike for axial-vector resonances, there are no ambiguities regarding their dispersive definition and because at least some information from phenomenology is available. Section~\ref{sec:chiral_limit} explicitly discusses the issue of the chiral limit in our model.
\item 
  The conclusion of ref.~\cite{Melnikov:2019xkq} contains the following three
  statements: ``there is no doubt that: (a) this region ($Q_{1,2}^2 \gg Q_3^2$) provides
  the largest contribution to $a_\mu^\mathrm{HLbL}$; (b) it allows for an
  exact non-perturbative analysis of the longitudinal structure function in
  the chiral limit and (c) it supplies strong evidence that corrections to
  the chiral limit are small.''  The first statement is plainly wrong, in particular for the
  model by MV, which actually receives most of its corrections from the
  region $Q_{1,2,3}^2< Q_\mathrm{match}^2$, as can be clearly seen from
  figure~\ref{SD_comparison}.  It is precisely this observation that leads to
  the conclusion that the modifications in the low-$q^2$ region are unphysical. Point (b) is correct, by
  construction, but the question is whether the chiral limit is a useful
  approximation at low $q^2$, which is the most important region for
  $a_\mu$. This is claim (c), which, unfortunately, is also not correct: the difference between
  the original MV model~\eqref{MVprescription} and the first term in~\eqref{eq:ReshuffledPole}  is such a quark-mass correction. In the case of the pion the two
  expressions give contributions to $a_\mu^\mathrm{HLbL}$ that indeed
  differ by a small amount (about $10\%$), but in the case of $\eta$ and
  $\eta'$ the difference is much larger, about $100\%$, as anticipated in section~\ref{sec:MV}.
\end{enumerate}

\bibliographystyle{h-physrev}
\bibliography{AMM}

\end{document}